**Kurdistan Regional Government**

**Ministry of Higher Education & Scientific Research**

**University of Sulaimani**

**College of Agricultural Engineering Sciences**

# EVALUATION OF DROUGHT TOLERANCE OF SOME ALMOND GENOTYPES BY MORPHOLOGICAL, PHYTOCHEMICAL AND MOLECULAR MARKERS IN SULAYMANIYAH GOVERNORATE

**A Dissertation**

**Submitted to the Council of the College of Agricultural Engineering Sciences at the University of Sulaimani in Partial Fulfillment of the Requirements for the Degree of Doctor of Philosophy**

**in**

**Horticulture - Dry Farming Fruit Production**

By

**Anwar Mohammed Raouf Mahmood**

B.Sc. Agricultural Sciences / Horticulture (1999)
M.Sc. Agricultural Sciences / Horticulture / Dry Farming Fruit Production (2010)

Supervisor

**Dr. Fakhraddin Mustafa Hamasalih**

Assistant Professor

**2719 K.** **2020 A. D.**

بِسۡمِ ٱللَّهِ ٱلرَّحۡمَٰنِ ٱلرَّحِيمِ

# وَجَعَلۡنَا مِنَ ٱلۡمَآءِ كُلَّ شَيۡءٍ حَيٍّ

سورة الانبياء الآية 30

صدق الله العظيم

## Supervisor Certification

I certify that this dissertation was prepared under my supervision at the University of Sulaimani, College of Agricultural Engineering Sciences, as partial fulfillment of the requirements for the degree of **Doctor of Philosophy** in [**Dry Farming Fruit Production**].

**Dr. Fakhraddin Mustafa Hamasalih**

Supervisor: Assistant Prof.

Date: / / 2020

In view of the available recommendation, I forward this dissertation for debate by the examining committee.

**Dr. Rupak Towfiq Abdulrazaq**

Assistant Prof.

Head of Horticulture Department

**College of Agricultural Engineering Sciences**

Date: / / 2020

## Examining Committee Certification

We Chairman and Members of the Examining Committee have read this dissertation and discussed the candidate (**Anwar Mohammed Raouf Mahmood**) in its contents on **(12-5-2020)**. Accordantly, we found this dissertation is accepted as a partial of the fulfillment of the requirements for the degree of Doctor of Philosophy in **[Horticulture, Dry Farming Fruit Production]**.

**Dr. Nawroz Abdulrazak Tahir**
**Professor**
University of Sulaimani
/ / 2020
**(Chairman)**

**Dr. Shler Mahmood Taha**
**Professor**
University of Salahaddin
/ / 2020
**(Member)**

**Dr. Mohammed Qader Khursheed**
**Assistant Professor**
University of Salahaddin
/ / 2020
**(Member)**

**Dr. Jassim Mohammad Khalaf**
**Assistant Professor**
University of Kirkuk
/ / 2020
**(Member)**

**Dr. Ibrahim Maaroof Noori**
**Assistant Professor**
University of Sulaimani
/ / 2020
**(Member)**

**Dr. Fakhraddin Mustafa Hamasalih**
**Assistant Professor**
University of Sulaimani
/ / 2020
**(Supervisor - Member)**

Approved by the Council of the College of Agricultural Engineering Sciences

**Dr. Karzan Tofiq Mahmood**
**Assistant Professor**
/ / 2020
**(The Dean)**

# Dedication

*I WOULD LIKE TO DEDICATE THIS THESIS TO:*

***MY DEARS FATHER WHO ENCOURAGED ME.***

***SOURCE OF KINDNESS MY MOTHER.***

***MY BELOVED WIFE KANAR.***

***MY DEAR SISTERS AND BROTHER.***

***MY HANDSOME SONS ZHER AND ZHYAR.***

***MY COLLEAGUES AND FRIENDS.***

***THANK YOU FOR GIVE ME CHANCES TO PROVE AND IMPROVE MYSELF.***

Anwar M. Raouf

# Acknowledgments

First of all, I thank Allah, the gracious, Merciful, and compassionate, for providing me with the willingness, patience and strength to complete this project.

My special thanks and appreciation to the Ministry of Higher Education and Scientific Research of the Kurdistan Regional Government-Iraq, the Presidency of Sulaimani University, the Deanery College of Agricultural Engineering Sciences and the Department of Horticulture for giving me the chance to complete this study.

My gratitude is also extended to the Assistant Prof. ***Dr. Karzan T. Mahmud*** Dean of the College of Agricultural Engineering Sciences and Head of the Horticulture Department Assist. prof***. Dr. Rupak Towfiq Abdul-Razaq*** for facilitating all administrative issues.

I would like to express my sincere thanks with respect to my supervisor, Assistant Prof. Dr. Fakhraddin Mustafa Hamasalih for his continuous support, guidance, constructive remarks, and encouragement throughout this study.

My deep appreciation examination committee chairman ***Dr. Nawroz Abdulrazak Tahir*** and members, Prof. ***Dr. Shler Mahmood Taha,*** Assistant Prop. ***Dr. Mohammed Qader Khursheed,*** Assistant Prop. ***Dr. Jassim Mohammad Khalaf*** and Assistant Prop. ***Dr. Ibrahim Maaroof Noori*** for their helpful suggestions and constructive comments.

Special and unlimited thanks to dear brothers and friends (***Mr. Faraidun K. Ahmeed and Dr. Jamal M. Faraj***) they being beside me whenever and wherever and encouragement. throughout this study.

My thanks also for my friends and colleagues ***(Dr. Hassan Shekh Faraj, Dr. Sidiq A. Sidiq Kasnazany, Dr. Aram A. Muhammed, Dr. Zainab S. Lazm, Dr. Ali O. Mohammad, Dr. Chinur H. Mahmood, Dr. Hoshman O. Majed, Dr. Haidar A. Haidar, Dr. Salam M. Sulaiman, Dr. Luqman Gh. Karim, Dr. Emad O. Hamaali, Dr. Dana A. Abdulkhaliq, Dr. Rozhgar M. Ahmed, Dr. Mohammed A. Fattah, Dr. Peshaw M. Najmaddin, Dr. Zaid Kh. Khidhir, Dr. Kwestan A. Amin, Dr. Rukhosh J. Rashed, Dr. Faraidun M. Faraj, Dr. Bekhal M. Hamahassan, Dr. Himn Gh. Zahir, Dr. Miqdad K. Ali, Dr. Karwan M. Hamaxan, Dr. Muhamaeed T. Marrof, Dr. Didar A. Rahim, Mr. Rasul R. Azez, Mr. Azad A. Muhammed, Mrs. Chia Q. Rashid, Mr. Kamaran S. Rasul, Mr. Brwa Kh. Shakr, Mr. Zardasht K. Ali, Mr. Hemn A. Mustafa, Mr. Ayub K. Mahmood, Mr. Hawar S. Hama, Miss. Rezan M. Salh, Miss. Sonia H. Othman, Miss. Lanja H. Khal, Mrs. Hiran S. Karim, Mr. Tareq A. Ahmed, Miss. Kochar O. Salih, Mrs. Tishk H. Faraj, Mr. Zana M. Dabagh, Mr. Shvan S. Ramzi, Mr. Arivan and Mr. Deyar A. Hassan***) for their helps during the laboratory and field experimental works.

I would thank a library and general laboratory of our college and the Kurdistan Institution for Strategic Studies and Scientific Research (KISSR).

Anwar M. Raouf

# SUMMARY

The study was carried out during 2017-2019 growing seasons at four locations in Sulaimani governorate and one location in Halabja governorate, in the Iraqi Kurdistan region including; Sharbazher, Mergapan, Qaradagh, Barznja and Hawraman. A huge number (approximately 500) almond trees were observed for all locations, among them 38 trees were selected with the best morphological characteristics which were chosen 9, 3, 5, 7, and 14 trees depending on the locations, respectively. A simple experiment was conducted using RCBD (Randomized Complete Block Design) for this experiment and means was separated by Duncan's test.

In order to evaluate their tolerances to drought in glass house, an experiment was conducted at Department of Horticulture, College of Agricultural Engineering Sciences, University of Sulaimani, that seeds were taken from those genotype trees and stratified then sown in pots. A factorial RCBD experiment was used with two factors: genotypes and irrigation intervals. Therefore, thirty-eight seedling genotypes grown in pots under glasshouse condition were exposed to three irrigation intervals: 10, 20 and 40 days after 10 days from seedling emergence. Therefore, the number of treatment combinations was: 38 * 3 = 114 seedlings for each replicate and with a total 342 seedlings for the whole experiment. Analysis of variance was carried out and the means were compared according to L.S.D. (Least Significant Difference) test ($P \leq 0.05$) As a result of the study, the seedlings showed different levels of adaptation to drought that can be used to future breeding programs as rootstocks. The objectives of this study were to identify morphological, phytochemical and genetic diversity with relatedness among the most important almond genotypes in Sulaimani region which related to drought tolerance to and relationship between morphological, biochemical and molecular data.

The most important results can be summarized as follows:

**Regarding to the tree genotypes**

- There were significant effects of the genotypes on all parameters.
- Two genotypes at Barznja location recorded maximum values in annual shoot growth and diameter while genotype number 5 at Hawraman recorded the lowest.
- Genotype number 3 at Mergapan recorded the higher value in leaf area, while the genotype number 14 at Hawraman recorded the lowest.

- Hawraman genotypes recorded the higher and lower numbers of stomata in a square millimeter.
- Maximum and minimum values of chlorophyll concentration recorded by Sharbazher genotype 5 and Hawraman genotype 2, successively.
- Qaradagh genotype number 2 recorded the maximum number in nut width and length while genotype number 4 at the same location recorded the heighest in nut thickness and weight.
- Maximum proline value registered for genotype number 6 at Barznja while genotype number 3 registered the minimum at Hawraman.
- Phenolic and flavonoid contents in genotype number 5 at Qaradagh gave maximum values, while Hawraman genotype number 4 recorded the minimum values for phenolic, but the same genotype number at Sharbazher recorded the minimum value in flavonoid contents.
- Genotype number 5 at Hawraman and number 1 in Qaradagh recorded the highest and lowest values respectively in total saponin contents.
- At Qaradagh, the maximum and minimum values of condensed tannin contents were recorded for genotype 5 and 1, respectively.
- Antioxidant activity by (DPPH and ABTS) assay for genotype 1 at Qaradagh recorded the lowest value, while both genotypes 6 and 7 respectively gave the highest at Sharbazher.

**Regarding to the tree genotypes molecular analyses**

- Mean values for Random Amplified Polymorphic DNA (RAPD) for polymorphic bands was higher than that of Inter-Simple Sequence Repeat (ISSR) values.
- The minimum PIC values recorded for RAPD primes were higher than ISSR primer values and the maximum values were equal.
- Based on Jaccard similarity coefficients, the genetic distances were grouped into 3 major RAPD markers clusters, and the 15 ISSR markers, with four clades.
- According to genetic STRUCTURE analysis, all genotypes were divided into two groups in RAPD and ISSR markers.

**Regarding to seedling in glasshouse**

- Maximum values of seedling height and leaf number recorded by genotype number 9 at Hawraman.
- The higher values of seedling diameter, leaves area, vegetative weight, vegetative dry weight and root dry weight percentages were recorded by genotype number 3 at Qaradagh. Genotype number 12 at Hawraman recorded the lowest in seedling diameter, vegetative and root weight.
- Genotype number 4 at Qaradagh recorded the maximum value in root weight.
- Genotypes number 9 and 8 at Sharbazher recorded the minimum values at stomatal length and width, respectively.
- At Barznja, genotype number 2 recorded the maximum value in chlorophyll content.
- The highest value of proline recorded by genotype number 2 at Qaradagh.
- Genotype 2 at Sharbazher location recorded the maximum values for total phenolic and total flavonoid contents. While the, minimum values were recorded at the same location by genotypes 4 and 7, respectively.
- Genotypes number 13 and 14 at Hawraman recorded the maximum values of antioxidant activity by (DPPH and ABTS) assay, respectively.
- Seedling height, leaves number and area, vegetative and root weight, stomatal length and width recorded the highest values, in 10 days irrigation intervals while in 40 days irrigation intervals gave the lowest.
- Proline, total phenolic content, total flavonoid content and antioxidant activity by (DPPH and ABTS) assay recorded the maximum values in 40 days irrigation intervals inversely the minimum recorded in 10 days irrigation interval.
- Diameter of seedlings in 10 and 20 days irrigation intervals was equal and higher than those in 40 days intervals.
- Chlorophyll content in 20 days irrigation intervals gave the maximum while in 10 days irrigation intervals recorded the lowest.
- Percentages of vegetative and root dry weight recorded maximum value in 40 days irrigation intervals while 10 days gave the lowest.

# List of Contents

## List of Tables

# List of Figures

# List of Appendices

## List of Abbreviations

| | |
|---|---|
| **AA** | Ascorbic Acid |
| **ABTS** | 2,2’-azino-bis (3-ethylbenzothiazoline)-6-sulfonic acid |
| **AFLP** | Amplified Fragment Length Polymorphism |
| **B** | Barznja |
| **CE** | Catechin Equivalents |
| **CTAB** | Cetyl TriMethyl ammonium Bromide |
| **CTC** | Condensed Tannin Content |
| **DE** | Dry Plant Equivalents |
| **DNA** | Deoxyribonucleic Acid |
| **dNTP** | Deoxynucleotide Triphosphates |
| **DPPH** | 2,2-diphenyl-1-picrylhydrazyl |
| **GAE** | Gallic Acid Equivalents |
| **H** | Hawraman |
| **$IC_{50}$** | Half maximal inhibitory concentration |
| **ISSR** | Inter-Simple Sequence Repeat |
| **L. S. D.** | Least Significant Difference |
| **M** | Mergapan |
| **Mb** | Mega bases |
| **MSI** | Membrane Stability Index |
| **NADPH** | Nicotinamide Adenine Dinucleotide Phosphate Hydrogen |
| **PCA** | Principal Component Analysis |
| **PCR** | Polymerase Chain Reaction |
| **PEG** | Polyethylene Glycol |
| **PIC** | Polymorphic Information Content |
| **Q** | Qaradagh |
| **QE** | Quercetin Equivalent |
| **RAPD** | Random Amplified Polymorphic DNA |
| **RCBD** | Randomized Complete Block Design |
| **RFLP** | Restriction Fragment Length Polymorphism |
| **RNS** | Reactive Nitrogen Species |
| **ROS** | Reaction Oxygen Species |
| **RWC** | Relative Water Content |
| **S** | Sharbazher |
| **SC** | Saponin Content |
| **SE** | Saponin Equivalents |
| **SNP** | Single Nucleotide Polymorphisms |
| **SPAW** | Soil-Plant-Air-Water |
| **SRAP** | Sequence-Related Amplified Polymorphism |
| **SSR** | Simple Sequence Repeats |
| **TAC** | Total Antioxidant Capacity |
| **TBE** | Tris-Borate-EDTA |
| **TFC** | Total Flavonoid Content |
| **TPC** | Total Phenolic Content |
| **UPGMA** | Unweighted Pair-Group Method with Arithmetic |
| **USDA** | United States Department of Agriculture |

# CHAPTER ONE

# INTRODUCTION

Almond (*Prunus dulcis* (Mill.) D.A. Webb syn. *P. amygdalus* (L.) Batsch) considered as a commercial stone fruit plant, belongs to Rosaceae family ( Zhu, 2014; Sakar *et al.*, 2018). The genus of *Prunus* can be subdivided into three subgenera: Amygdalus, Prunophora and Cerasus. Wild almond is originated from arid mountainous region of Central Asia and deserts of western China, Iran, Turkistan, Afghanistan, Kurdistan, and North Africa (Browicz and Zohary, 1996; Xu *et al.*, 2004). Annual almond production for 2018/19 is estimated as 1.4 million metric tons (USDA, 2018). Botanists have observed over 30 species of almond (Grasselly, 1976; Ladizinsky, 1999). In addition, almond is commercially interested nut fruits among other stone fruits, because it is the only *Prunus* species grown for its seeds, and their juicy flesh or mesocarp, whereas the corky mesocarp is only used in animal feeding (Aguilar *et al.*, 1984) or as a manure (Alonso *et al.*, 2012), also, the fresh green almond can be as fruits in first few weeks in spring. Almond is often classified more as a nut than as a stone fruit, despite its very close genetic similarities with the other stone fruits, and mainly with peach (Socias i Company and Gradziel, 2017).

Recently, one of the main serious problems limiting productivity in all crops is drought during the growth and development of plants which cause economic, social and environmental problems throughout the world, however, almond is one of the vital plants that can grow under rain-fed condition in Iraq, particularly in the Kurdistan region. Almond is usually adapted to various abiotic stresses. Traditional approaches for germplasm characterization are based on morphological observations, it is affected by environmental conditions ( Casas *et al.*, 1999; Sorkheh *et al.*, 2007; Zeinalabedini *et al.*, 2008; Bouhadida *et al.*, 2009; Sorkheh *et al.*, 2009a). Using Morphological traits to the identification of almond plants are restricted, because of their environmental variations. Morphological traits, concern mostly the following: length, width, thickness of the nuts, shell to kernel percentage, annual shoot growth, annual shoot diameter, leaves area, vegetative dry weight, stomatal number, stomatal length and width in addition to chemical content, chlorophyll content, proline, phenols, flavonoids, saponins, tannins and antioxidants.

Natural chemical compounds produced by plants are also of special importance in regard to drought tolerance. These compounds include some heterogeneous group with numerous biological effects. The most components investigated in vegetables and fruit consist of compounds such as phenolics, flavonoids, tannins and saponins. Reactive oxygen species (ROS) can help molecular signaling in the plant cell, ROS is part of antioxidant reaction including both enzymatic (peroxidase, catalase) and non-enzymatic molecules (phenol, flavonoid, saponin and tannin) (Asada, 2006; Król *et al.*, 2014)

Genetically, almond is diploids, the chromosome number is (2n = 2x = 16) with genome size approximately 246 Mb (Sánchez-Pérez *et al.*, 2019). The modern molecular genetic tool can be applied to identify and characterize the relationships among them. Several similar studies have been performed regarding genetically recognized cultivars and wild species of almond, using several types of markers including RAPD, ISSR, SSR, SRAP and RFLP. Three decades ago, polymerase chain reaction (PCR) techniques with a random primer which used molecular markers showed the vital role in crop breeding, particularly in genetic diversity research and gene bank (Gupta and Varshney, 2000).

Genetic diversity is an important tool for the breeder, by which all genotypes can be differentiated, also it is very useful for the improvement of the chances of selection better segregates for various characters (Dwevedi and Lal, 2009).

In Iraqi Kurdistan region, not much information is available about almond genotypes regarding morphological and physiological traits (Kester *et al.*, 1991). Genetic constructions of dissimilar cultivars are mostly due to performances of cross-pollination among them (Arulsekar *et al.*, 1986; Socias i Company, 1990; Arús *et al.*, 2009; Socias i Company and Gradziel, 2017). In Kurdistan, no molecular evidence is obtainable concerning almond genotypes, therefore the application of genetic diversity for almond is of special interest particularly for determining the genotypes tolerant to water stress.

**Aims of the study**

The aims of the study could be summarized as follows:

- Selection of some almond genotypes which are adapted to the Sulaimani region.
- Evaluation of the behavior of some native wild almond genotypes as rootstocks.
- Estimation of genetic variability parameters with relationship among all almond genotypes.
- Determination of some phytochemical drought resistance related analyses.
- Comparison the studied almond genotypes for various morphological, biochemical and molecular traits.

# CHAPTER TWO
# LITERATURE REVIEW

## 2.1 Experimental Location

The study areas, Sulaimani and Halabja governorates are located in North-East of Iraq, graded as arid and semi-arid, hot and dry in summer and cold in winter (Najmaddin *et al.*, 2017). Sulaimani governorate total area is 17,023 $km^2$ located between 35° 33' 5386"N latitude and 45° 25' 5844"E longitude. Halabja governorate area about 1260 $km^2$ located between 35°10'59.22"N latitude and 45°58'59.05"E longitude (Alwaely *et al.*, 2015). Kurdistan region total area is 40.634 located between 36° 41' 03" N, latitude and 44° 38' 72" E longitude. Sulaimani governorate total area is 17,023 $km^2$ located between 35° 33' 5386"N latitude and 45° 25' 5844"E longitude. The total area of Iraq is 438 320 $km^2$ located between 33° 18' 46.0980" N latitude and 44° 21' 41.3568" E longitude. The climate in Iraq is mainly of the mainland, subtropical semi-arid type, with the north and north-eastern mountainous regions consuming Mediterranean weather. Rainfall is seasonal and be falls in the winter from December to February, excluding in the north of the country, where the rainy season is from November to April. Average annual rainfall is appraisal 216 mm but ranges from 1200 mm in the northeast to less than 100 mm in the south. Winters are cool to cold, with a day temperature of about 16°C dipping at night to 2°C with an opportunity of frost. Summers are dry and hot to highly hot, with a shade temperature of over 43°C during July and August, yet dipping at night to 26°C (Frenken, 2009).

## 2.2 Stone Fruits and Almond Cultivation

Stone fruits are species of the great generally distributed genus "*Prunus*" and belonges to Rosaceae family which includes; almond, peach, apricot, plum and cherries. All these have a main fruit plant species in *Prunus* originated in temperate regions in areas from eastern Europe to eastern Asia, but nowadays they are cultivated around the world for many purposes. The large numbers of *Prunus* species with a wide diversity and economic importance as well as the wild species with prevalent distribution caused the topics of wide basic and practical researches, and also the availability of germplasm and genomics resources. Nutritionally, stone fruits are sources of minerals and vitamins, and there is a growing interest in their possible value as nutraceuticals due to the incidence of antioxidant compounds (Kole and Abbott, 2012). Many of them have also been significantly important for breeding programs and the genetic improvement purposes, particularly almonds, which are commercially interested, other nut fruits as they are only species used for their seeds and flesh among the corky mesocarp are also used in animal feeding or as a manure (Aguilar *et al.*, 1984). Therefore, almond is often classified more as a nut than as stone fruit, despite its very close genetic similarities with the other stone fruits, and mainly with peach. Previous researcher (Socias i Company and Gradziel, 2017) reported the nutrient value of almond kernel per 100 g fresh weight (Table 2.1).

**Table 2.1 Nutrient value of almond nut (kernel) per 100 g fresh weight.**

| Nutrient | Value |
|---|---|
| Energy | 578 kcal |
| Protein | 21.26 g |
| Carbohydrate | 19.74 g |
| Fibre, total dietary | 11.8 g |
| Glucose | 4.54 g |
| Starch | 0.73 g |
| Calcium | 248 mg |
| Magnesium | 275 mg |
| Phosphorus | 474 mg |
| Potassium | 728 mg |
| Sodium | 1 mg |
| Folate, total | 29 mcg |
| Vitamin E | 25.87 mg |
| Saturated fatty acids | 3.88 g |
| Monounsaturated fatty acids | 32.16 g |
| Polyunsaturated fatty acids | 12.21 g |

Socias i Company (1990) demonstrated that the high heterosis of the cultivars is mostly due to the performances in cross-pollination among them and also it is recorded into a high diversity of forms with high genetic variability. In addition, growing almonds in different regions, have also been isolated during the period of progressing of characteristic ecotypes (Socias i Company and Gradziel, 2017).

## 2.3 Drought Tolerance and its Effects on Agriculture Production

Historically drought has caused direct and indirect economic, social and environmental problems throughout the world (Nagarajan, 2009). Generally, drought is an element of environmental change and clearly distinguished worldwide as shortness of precipitation that severely effects on plant growth and development with significant decreases in plant biomass accumulation (Eslamian and Eslamian, 2017). One of the most vital restrictions for agriculture production is water. A water stress response is a compound mixture of different factors such as molecular, biochemical, physiological and morphological leading to plant survive under drought conditions. Main consequences of water shortage in plants are a compact rate of cell division and expansion, stem elongation weakness of leaf size, root production, and stomatal distribution, the relation of plant water and nutrient with reduced crop productivity, and water use efficiency (Tribulato *et al.*, 2019).

Nowadays, global warming or climate change has caused a rise in temperature, less precipitation, increasing variability in rainfall and reducing recharge of underground aquifers in many areas (Pray *et al.*, 2011), also, may be worsening this situation in most agricultural regions. It is predicted that global warming will cause a massive drought and take over half the land surface on our planet in the next hundred years. Therefore, understanding the mechanisms for water deficit is quite important to the sustainability of agricultural production. Certainly, plants display a wide range of adaptations, to drought stress at different levels. Drought tolerant cultivars are also the most sustainable approach to reduce the pressure of the periodic drought. In our study, approaches of morphological, physiological, biochemical and molecular levels can be shown to improve drought tolerance for almond. These strategies will be necessary for crop production under generalized water limitation in the near future (Beatriz *et al.*, 2010). Drought can be considered as a set of climate pressures. Several phenomena can produce drought: heat shocks, water deficit, low air hygrometry, insolation and/or salinity. The combination of these phenomena leads to different types of drought. This diversity of drought had led to the selection of numerous types of tolerance mechanisms at a different level of life organization (molecule, cell, organ, plant ...). The study of these mechanisms can bring important information in the long-term purpose of crop breeding (Belhassen, 1997).

Water-deficit or drought stress conditions are especially unpredictable, but in some regions dry seasons are predictable. In the 21st century, those plants that may resist drought stress or resist water-deficit for a long time and preserve their validity and productions constitute one of the major research zones in agriculture. So, more studies are desired to understand the plant physiology under drought conditions. These studies will help us to enhance plant growth and production under water-deficit situation (Hasanuzzaman *et al.*, 2019).

One of the most plants in stone fruit grouses that generally known as a drought-tolerant plant is almond. thus, screening and recognizing drought-tolerant almond genotypes is required in order to improve and stabilize almond production under semi-arid and arid conditions (Karimi *et al.*, 2012). However, drought conditions limit the quantity and quality of almond production (Camposeo *et al.*, 2011).

## 2.4 Phenotypic Traits Used for Detecting Drought Tolerance

The quick and simple approach to evaluate drought stress is morphological observations that can be also used alongside with genotypes to crop improvement. Morphological characterization in conjunction with multivariate statistical methods such as principal component analysis (PCA) and cluster analysis are useful for genotype screening (Čolić *et al.*, 2012; Khadivi-Khub and Etemadi-Khah, 2015).

The majority of plants have some mechanisms for drought tolerance and rise of the water use efficiency. Drought tolerance mechanisms are generally recognized in agricultural crops but further studies have been shown in fruit trees. In addition to morphological mechanisms, other mechanisms such as osmotic regulation and changing root volume, fresh and dry weight to airborne organs will be affected on all parts of the plant. In spite of the research in this subject, the relative position of each one of these mechanisms has not been completely specified. Understanding the mechanisms of drought tolerance in the plant will ease decision making in the matter of irrigation management and improving the performance of efficient genotypes under drought stress conditions (Akbarpour *et al.*, 2017).

The foundations of drought tolerance may originate from other almond species which are extremely xerophytic counting *Prunus webbii*. This species may be used as a rootstock for peach, apricot, and almond (Baninasab and Rahemi, 2007). Gomes-Laranjo *et al.* (2005) reported that reduced water potential under drought conditions resulted in growth limitation, massive leaf abscission, and reduction in kernel weight of almonds. Several researchers have studied the adaptation of almond to water deficit from a different perspective, many morphological and physiological drought tolerance mechanisms have been recognized. These mechanisms include the ability for osmotic adjustment, reduced leaf area, changes in the adaptable properties of cells and tissues, reduced stem length, controlled stomatal regulation, leaf abscission and deeply penetrating of the root system. Almond trees in drought condition have a significantly compacted ratio of $CO_2$ assimilation. Almond has a lower water-use efficiency than other fruiting tree species. Most of these specialties make almond trees to survive prolonged droughts and have led to defining almonds as a drought-resistant species (Pirasteh-Anosheh *et al.*, 2016; Vats, 2018).

## 2.5 Drought Stress and Implication on Almond Genotypes

Improvement of the drought stress tolerance of plants is necessary due to the widespread incidence of drought damage to crop species. Applied regulated deficit irrigation on almond trees by (Romero *et al.*, 2004) illustrated that leaves area at the time of highest stress was significantly minor for water-stressed trees than for well-watered trees, the mean value of leaves area was (11.5 $m^2\,m^{-3}$) in watered tree while it was (9.6 $m^2\,m^{-3}$) in drought condition. Palasciano *et al.* (2005) reported leaf area, stomatal frequency, stomatal length and width to 5 Apulian wild almonds (*A. webbii*) and 15 cultivated almonds (*A. communis*), there were significant effect of genotypes on the characteristics, the lowest and highest values were (2.3-26.5) $cm^2$, (143.4-326) per $mm^2$, (19.3-30) µm and (9.4-14.9) µm for leaves area, stomatal frequency, stomatal length and width respectively. Damyar and Hassani (2006) evaluated 25 almond cultivars in Karaj region in Iran. Maximum and minimum annual shoot growth was (61.11 cm) and (30 cm).

The main significances of drought in crop plants are a compact rate of leaf size, cell division and expansion, root proliferation, stem elongation and disturbed stomatal oscillations, plant water and nutrient relation with reduced crop productivity, and water use efficiency (Farooq *et al.*, 2012).

When six almond genotypes and three soil water potentials (control, moderate and severe stress) were verified, the number of leaves per plant and consequently total leaf area and total leaf dry weight tended to be greater in control seedlings than in drought-treated seedlings, for all genotypes. In all genotypes (except one) leaf size (leaf area dividing by leaf number) was generally abridged by both water stress treatments. Total leaf dry weight, shoot dry weight and shoot growth in nearly all of the genotypes were significantly reduced by both water stress treatments. In both stress treatments, root dry weight for all genotypes reduced significantly. Different responses on greenness were shown by genotypes, leaf color for five genotypes increased immediately in the first week under severe stress, in the fifth week some of the genotypes did not show significant change when three genotypes showed minimum greenness especially in higher water stress. The genotypes had similar stomatal density, the lowest value was (406.7 $mm^{-2}$) recorded by (SH12) genotype. Stomata length ranged (15.1µm to 11.04 µm) among genotypes. Smallest stomatal width was (3.98µm) and there were no differences among the other genotypes. No significant difference observed in stomatal size and density among treatments and genotypes (Yadollahi *et al.*, 2011). Sorkheh *et al.* (2011) checked eight native

Iranian almond species *in vitro*, the result showed a significant effect of genotypes and drought for plantlet height, root length and dry weight of root.

Nikoumanesh *et al.* (2011a) used 62 Iranian almond cultivars and wild genotypes seedlings for two years and the data showed differences between plants in the first-year, maximum and a minimum of main trunks length was (34-293) cm, with diameter (3.37-36.41) mm and leaves area (2.09-11.72) $mm^2$. Camposeo *et al.* (2011) studied the effect of growing in climatic water shortage of some leaves and stomatal parameters of 15 adult trees and 5 seedling almonds under Mediterranean environments, the result showed significant differences among plants for all the inspected parameters in full summer, the uppermost and lowermost leaf area, stomatal frequency, stomatal length and width were (24.4-2.1 $cm^2$, 190.7-126.5 in $mm^2$, 27.5-23.1 µm and 13.8-9.5 µm), respectively. Čolić *et al.* (2012) evaluated (19) genotypes in northern Serbia, illustrated the highest and lowest values of leaf area (2.55-1.06 $cm^2$). Gikloo and Elhami (2012) studied two almond cultivars (Princess and Tuono) and showed that drought affected, leaf area with a significant difference between genotypes and the area in drought treatment ranged between (40.1-48.3) $mm^2$. No significant difference in leaf fresh weight was observed between cultivars but drought stress decreased leaves fresh weight from (8.33) g in three days irrigation intervals to (7.1) g in nine days irrigation intervals. However, drought stress declined leaf dry weight. Karimi *et al.* (2013a) tested the response of six almond genotypes to drought *in vitro* by using three different levels of PEG (polyethylene glycol) (0, 3.5 and 7% W/V), the results showed significant reduction in fresh weight of the explants, the lowest fresh weight recorded in 'Ferragnès' genotype explants on media containing 7% PEG. By increasing the level of PEG, the leaf number, mean leaf area, and total leaf area of the explants were reduced, the genotypes also significantly affected leaves numbers and area. Mehdigholi *et al.* (2013) studied leaf length and width for six Iranian almond populations, the leaves length values were (1.8 to 2.44) cm and the leaves width were (1.08 to 2.26) cm.

In a research about the effect of osmotic stress on eight wild almond species parameters grown *in vitro,* Rajabpoor *et al.* (2014) showed that leaf size, shoot growth and plant height varied significantly among all species by osmotic stress, however the researchers have noticed noticeable less effect of stress regard to shoot length and leaf number, but leaf size increased in the control compared to the highest osmotic stress level. Separately leaves fresh and dry weights also varied significantly by drought stress and species. For example, one of the species leaves fresh weight decreased from 638 mg for the controls to 272 mg at -1.2 Mpa. The weight decline in dry leaf was highest at higher osmotic stress, representative higher leaf water content for plants grown at maximum osmotic stress level. Shoot fresh and dry weights declined with growing osmotic stress level. Usually, there was an 80% lessening in shoot dry weight at drought stress. In a minimum level of osmosis, the stomatal density of the two species reduced with growing osmotic stress level but it was not significant. The stomatal population changed significantly between the control and drought-stressed plant.

Zokaee-Khosroshahi *et al.* (2014) used PEG to induced drought for 5 species of Iranian almond seedlings for about 6 months, morphological changes were observed, drought and genotypes treatments had significant effects on the fresh weight of plant tissues. Stressed seedlings had significantly lower stem, leaves number, root, and plant fresh weight values compared to the unstressed, while absolute values mixed by species. The main reduction in whole plant fresh weight (50.9%) caused mainly by the strong reduction in root fresh weight (75.8%), then the fresh weight of stems and roots were condensed minimally (8.4% and 1.4%) lessening, respectively. The dry weight of the plant part and subsequently the total plant declined as drought stress levels amplified. The results presented that genotypes had a significant effect on shoot length, on all measurement times. No significant variations in shoot lengths were observed in response to drought stress treatments. Total leaf area and leaf number affected by drought stress and produced significant reductions in all seedling species also both total leaf area and leaf number reduced in all species as the drought stress increased. Among genotypes, the value of total leaves area was between (11.4-268.1 $cm^2$) and the leaf number was between (7.3-31.7). Stomatal size (length and width) was significantly affected by genotypes. The highest (42.59 μm) and the lowermost (22.61 μm) length of stomatal aperture were recorded. Moreover, the greatest (19.51 μm) and the least (13.87 μm) stomatal width, among the examined species were measured. Stomatal number per square millimetre was significantly partial by genotypes and recorded the highest (251.51) and the lowest (197.85) stomata/$mm^2$. In addition, El Hamzaoui *et al.* (2014) carried out the morphological test for Moroccan cultivated almond alongside some foreign varieties, the result showed that the value of leaf length between ten origins (Morocco, Tunisia, Spain, France, Italy, Ukraine, Greece, Bulgaria, Syria and USA) were (73.3, 91.6, 88.1, 89.1, 90.5, 96.8, 79.5, 50.7, 88.0 and 86.0 mm), respectively.

Sepahvand *et al.* (2015) also evaluated morphological variables for almond trees among 155 genotypes, they found that the mean of one-year-old shoot thickness was (5.45 mm) and the highest value was (7 mm) and the lowest (3 mm). Furthermore, Fathi *et al.* (2017) observed the response of one-year-old almond seedling of five cultivars and genotypes grafted on GN15

'Garnem' rootstock plants exposed to three deficit-irrigations, counted moderate, severe stress and control were applied for 6 weeks. Severe drought-stressed plants had a significantly minor fresh weight of root, leaf, stem, and whole plant. Drought and genotypes significantly influenced the plant parts dry weight for all genotypes compared to the control group. Genotypes and drought stress have significantly affected total shoot length. So, it looks that this characteristic may be used as a drought stress indication in young seedlings of almond genotypes. Both examined factors drought stress and genotypes caused significant reductions in leaf number and total leaves area of plant. Stomatal density, length and width were significantly influenced by genotypes and drought, values of the previous parameter between control and the highest level of drought were (185.4-213.4 per $mm^2$, 20.92-24.8 µm and 11.32-12.118 µm) respectively. Morphology responses of drought stress *in vitro* for five commercial almond cultivars were observed by Akbarpour *et al.* (2017), the researchers were displayed that there were significant variances in all considered faces. The smallest number of established leaves was noticed in 6% PEG treatment and there were differences between genotypes, the number of developed leaves also gave significant differences between unstressed and stressed treatments. The number of developed leaves was significantly decreased when stress levels increased. The lowest number of developed leaves was noted in maximum PEG percentage treatment. The data showed that an increase in osmotic stress caused a significant decrease in plantlet height. Smallest plantlet heights among cultivars were recorded in high PEG treatment.

## 2.6 Influence of Almond Genotypes on Nuts Characteristics

The three important parts of almond fruits are kernels, shells and hulls, the kernel is the edible part with a high nutritional value, while shells and hulls are used as livestock feed and burned as fuel. In addition, physical characteristics of almond fruit are conditioned by both consumer and industrial requirements. It is clear that in almond seeds, mesocarp, endocarp and the kernel will have a similar shape with some variability, however, it can be seen that shell will determine in a high measurement the final shape of the kernel which will be developed inside. It is evident that the final shape of all the nuts of the almond tree was determined by the mother rather than the pollinator (Martínez-García *et al.*, 2019).

Many researchers investigated the physical properties of almond genotypes. Researchers demonstrated physical fruit traits for 45 genotypes in Moroccan almond as follows minimum and maximum of nut length (19.25 to 41.24 mm), nut width (15.90-27.19 mm), nut thickness (11.48-19.61 mm), nut weight (1.15-7.34 g) and shelling percentage (19.91-63.79 %) (Kodad *et al.*, 2015), In addition, Khadivi-Khub *et al.* (2016) studied 198 accessions of *Prunus scoparia* and data were recorded as maximum and minimum nut length, width, thickness and weight which were (17.81-8.48 mm, 12.97-5.94 mm, 11.01-6.97 mm and 2.6-1.21 g), respectively. Imani and Shamili (2017) pointed out that nut length, width, thickness, weight and kernel percentage of 60 almond cultivars and genotypes and the average values were (26.59-47.17 mm, 13.28-28.82 mm, 9.08-20 mm, 0.97-4.76 g and 23.52-68.88 %), respectively. Diversity in some nut physiognomies of seven populations of almond from the central and southern Zagros areas of Iran by contrast with three other almond species by (Rahimi *et al.*, 2017) was studied. The nut of *Prunus scoparia, Prunus elaeagnifolia, Prunus eburnean and Prunus dulcis* species was studied, nuts parameters were nut length (10.20-37.70) mm, nut width (8.20-21.80) mm, nut thickness (6.29-17.20) mm and nut weight (0.28-4.79) g, the data was collected from 72 species.

In addition, Rapposelli *et al.* (2018) documented 45 almond cultivars and marked some nut parameters including nut length, width and weight. The lowest and highest values were (2.43-4.05 cm, 1.80-3.34 cm and 1.33-7.47 g), respectively to the preceding nut characteristics. In Northern Morocco Sakar *et al.* (2018) recorded some characteristics for nine almond nuts and kernels traits (the highest and lowest for nut length, width and thickness), the data were (29.19-37.09, 24.89-20.84 and 17.29-14.16 mm), respectively. Melhaoui *et al.* (2018) measured some nut traits of four almond cultivars (Marcona, Fournat, Ferragnes and Ferraduel), cultivated in five different regions of North Eastern Morocco, the nut weight ranged between (2.65-4.41 g), nut width between (19.63-25.42 mm), nut length (26.15-41.15 mm), nut thickness was (13.65-16.86 mm). Martínez-García *et al.* (2019) tested shell and kernel shape in four almond cultivars for 3 years, the averages for shell length, width and thickness were (29.85, 22.07 and 19.95 mm), respectively. Fifty-four almond cultivars shelling percentage recorded by Fornés Comas *et al.* (2019), the highest percentage was (66.5 %) and the lowest was (17.32 %), the nut weight ranged from (2.2-12.7g).

## 2.7 Biochemical Study and Role in Drought Tolerance

Phytochemicals are chemical complexes produced naturally in plant parts. Many phytochemicals contribute to several biological procedures of the plant counting the construction of the flavor and color of plant foods. Phytochemicals have been divided into five major groups: phenolics, alkaloids, carotenoids, sulfur-containing phytochemicals and nitrogen-containing phytochemicals. They are a heterogeneous family of chemical compounds with many biological effects. The dietary elements of fruits and vegetables that are the most thoroughly investigated antioxidants are flavonoids, phenolic acids, lycopene, anthocyanins, vitamins A, B, C, sulfides, and tocopherols. Phytochemicals in fruits and vegetables, such as carotenoids, phenolic compounds and glucosinolates, may also have nutritive value. Phytochemicals form the backbone of oldest medicine, which uses plant provisions (leaves, seeds, stems, fruits, and roots) as a foundation of drugs. In recent years, a study on phytochemicals has improved all over the world and new reports such as useful food and nutraceuticals have been presented (Yahia, 2017).

Water availability, light/dark, nutrient, temperature and toxic compounds such as heavy metals are factors and biotic relations (pathogenic microbes, fungi, and insects), an environmental disorder which significantly distracts metabolism, development and yield, is deemed as stress conditions and cause stress answer in biological organization. Such forced stress is usually attended by a growth of the construction of reactive oxygen species (ROS) and reactive nitrogen species (RNS). In spite of their toxic nature and reactive, ROS and RNS are also opener ingredients of signal transduction passageways that elicit stress responses. Moreover, ROS and RNS are implicated in plant developmental operation and plant-microbe infection. However, ROS and RNS production could be improved by the antioxidants system to prohibit injury and cell death. During development, plants evaluate mechanisms to adapt to drought or even to resist dry periods. Inclusive researches have unraveled the molecule's mechanism of drought tolerance. Plants are equipped with greater levels of non-protein antioxidants and osmolytes that reprogram metabolisms and promote their antioxidant capacity. Generally, a sensitive plant also triggers its antioxidant system. Notwithstanding, in spite of this visible conflict, drought tolerance appears to be an operation of the antioxidant capacity recognized in response to drought. Moreover, not only the antioxidant activity is important during severe drought stress, but also intervene together with the improvement of water supply and resurrection from drought (Laxa *et al.*, 2019).

Almonds are a rich source of polyphenols. The most considerable type of polyphenols in almonds is firstly proanthocyanidins and secondly flavonoids and phenolic acids. Almond flavonoids include flavanols, flavanones and flavonols, in their oligomeric, monomeric and polymeric forms; this latter also called condensed tannins. Almond tannins or proanthocyanidins are a collection of flavan-3-ols with various degrees of polymerization. Although proanthocyanidin has been described in almond, data about their apportionment among cultivar and varieties are lacking (Xie *et al.*, 2012). Proline, on one hand, has a function as an osmolyte, and the other hand, it also works as a strong antioxidant with great ability to protect plant organs from oxidative injury as reflected in the form of reduced lipid peroxidation. So, proline is now considered as a non-enzymatic antioxidant, and plants need this biomolecule to scavenge ROS (Hasanuzzaman *et al.*, 2019). Rajabpoor *et al.* (2014) showed the effect of drought stress on eight wild almonds *in vitro*, in the maximum level of osmosis stress, proline ranged between (30.22-44.32 µg $g^{-1}$), and in minimum level between (19.34-29.25) µg $g^{-1}$). *In vitro*, five commercial almond cultivars evaluated by Akbarpour *et al.* (2017) and the morphology responses of drought stress were observed. The uppermost amount of proline content was noted in maximum polyethylene glycol (PEG) which was different significantly from the other treatments, also the difference between all cultivars was significant. Haider *et al.* (2018) found that proline contents accumulation significantly affected by water deficit conditions in 2-year-old peach (*Prunus persica* L) leaves, the value in control condition (watered plant) was (1.22 ng $g^{-1}$ FW) and in drought-treated plants in mild stress (1.71 ng $g^{-1}$ FW) and in severe stress (2.22 ng $g^{-1}$ FW). Proline under drought stress increased as compared to control. In addition, in leaves, water deficit can result in excess electron flow to the production of reactive oxygen species (ROS) which in turn damages leaf membranes and proteins (Demmig-Adams and Adams, 2002). While high-throughput luminol chemiluminescence methods are available to assay the composite ROS levels in tissues subjected to stress, the functions of ROS are multifaceted and include signaling (Mittler *et al.*, 2011), thereby complicating their interpretation. It is possible that metabolomics may provide valuable phenotypic information on the spectrum of antioxidants and photoprotectants in a genotype. For example, studies indicate that the xanthophyll carotenoids perform a critical photoprotectant role (Demmig-Adams and Adams, 2002), and methods for metabolite profiling of carotenoids are now available (Fraser *et al.*, 2007).

Chlorophyll is another factor in a plant that has a main role during the photosynthesis process. Isaakidis *et al.* (2004) showed that chlorophyll content of eight grafted almond trees (6-8) years old were measured for three consecutive years and chlorophyll measured by SPAD meter, the means of SPAD were in GN-22 rootstock had maximum levels of chlorophyll which were 42 SPAD and Drepanoto rootstocks record the lowest value which was 37 SPAD. These differences were significant among the rootstocks. Effect of drought and two almond cultivars shown by Gikloo and Elhami (2012), the result indicated that drought affected significantly on chlorophyll in 9 days irrigation intervals and the chlorophyll content was (46.7) mg/g, while in 3 days irrigation interval was (56.5) mg/g. Different levels of PEG were used on six almond genotypes, a higher level of PEG significantly reduced the total chlorophyll and on the other hand the genotypes significantly affected on the chlorophyll content. In evaluated almonds genotype *in vitro* the total chlorophyll significantly decreased by increasing the PEG percentage, also genotypes have a significant effect on the chlorophyll (Karimi *et al.*, 2013b). Some phytochemicals in almonds leaves were screened by (El Hawary *et al.*, 2014) who showed that saponin, flavonoid and tannin were presented in the almond leaves. Nahand *et al.* (2012) mentioned that almonds, cherry and apricot leaves contain chemical compounds such as polyphenolic compounds which were (3.49, 3.44 and 1.55%), respectively and tannin (1.2, 2.8 and 0.617%), respectively for prior tree leaves. The results of this study implemented by (Pirbalouti *et al.*, 2013) indicated a significant difference in total phenolic contents and antioxidants in almond leaves. 4 species and 5 cultivars were used to test antioxidant activity and total phenols. The total phenolic content of the extracts reached (19-30) mg GA/g dry weight extract. Results indicated that the degree of antioxidant activity in these plants varied substantially, Antioxidant value in *Amygdalus orientalis* species was nearly 25 times more than that of *Amygdalus lycioides*. Ibibia (2013) used different extraction methods to determine some phytochemicals in *Prunus amygdulus* L. leaves, using methanol extract, the result showed that tannin absented in almond leaves, and total phenolic content and total flavonoid content were (40.35 mg GA $g^{-1}$ and 26.54 mg QE $g^{-1}$), respectively. Edrah *et al.* (2013) screened peach *Prunus persica* L. and *Pistacia atlantica* leaves in Libyan origin using ethanol and aqueous extract. The result showed that in peach leaves flavonoids, tannins and saponin were absented in both extraction methods. Sivaci and Duman (2014) tested three almond varieties including (Texas, Ferragnes and Nonpareil), the total phenolic compounds values were detected in the leaves (2.03, 2.82 and 8.15 µg $mg^{-1}$), fresh weight respectively to prior genotypes. During October high value was achieved for total phenolic content compared to April and July. It was found that total antioxidant activity wide-ranging according to the season, plant part and genotypes. Total antioxidant activity DPPH in the leaves of almond varieties was significantly affected by varieties in the majority of months. Tiwari *et al.* (2015) tested total phenolic content and antioxidant capacity by phospho molybdenum method of plant leaf extracts by using different solvents, in methanol solvent total phenolic content in fresh leaves was (502.7 mg/l) and total antioxidant activity in fresh leaves was (162 mg/ml). Taghizadeh *et al.* (2015) evaluated some phytochemicals in ten selected genotypes of Mahaleb on barks, leaves and fruits, and displayed that phenolic compounds of the total phenolic content of leaves was between (7.25mg GA $g^{-1}$ DE to 23.13 mg GA $g^{-1}$ DE), the maximum and minimum total flavonoids content in Mahaleb leaves were (22.81 - 6.90 mg QE $g^{-1}$). The value of total antioxidant activity (DPPH assay) in mahaleb leaves was nearly (18 - 78% inhibition). In addition, Hussain *et al.* (2015) used methanol to extract some phytochemicals in leaves of peach *Prunus persica* L., the result displayed that flavonoids and saponins were strongly presented and tannins slightly presented. Joseph *et al.* (2016) also used some solvents to leaves extracts and analyzed some qualitative phytochemicals, in cherry leaves methanol solvent used for evaluating phenols, flavonoids, tannins and saponins. The total phenolic content was (6.26 mg GAE/g of extract), total flavonoids content was (3.86 mg QE/g of extract) and DPPH radical scavenging potency (IC50) was 56.00 µg/ml (Oyetayo and Bada, 2017). The screened qualitative composition of phytochemicals in wild cherry (*Prunus avium*) leaves by using ethanol extraction, explained that phenol and flavonoids were moderated in reaction, tannin was mild reaction and saponin was not detected.

Some phytochemical of cheery plum (*Prunus cerasifera*) leaves isolated by (Song *et al.*, 2017) who found that total phenol measured was 117.8 mg/g dry weight, the concentrations of antioxidant capacity (DPPH and ABTS) assay were nearly (420 and 350 µg/ml) respectively and soluble tannin contents were approximately 7.4% of the dry weight. Dziadek *et al.* (2019) stated that minimum and maximum values of total polyphenol are (6011.69-15318.42) mg/100 g DW in 10 cultivars of sweet cherry (*Prunus avium* L.) leaves with lowest and highest antioxidant activity by (DPPH and ABTS) assay (473.77-1754.72 and 680.95-1702.93 µmol Trolox/g dry weight), respectively.

## 2.8 Application of Biotechnology for Genetic Diversity

In recent years, molecular markers have been developed based on the more detailed knowledge of genome structure. Considerable emphasis has been laid on the use of molecular markers in practical breeding and genotype identification (Ovesná *et al.*, 2012). Molecular markers are molecules that could be used to trace desired genes in the examined genotypes. In fact, a piece of DNA or a protein can be used as a marker. Earlier approaches that made the selection of specific traits easier were based on the evaluation of morphological traits (Staub *et al.*, 1996). Characterization of genotype identity and traditional genetic relationships, of *Prunus* cultivars and species have been based on morphological and physiological traits. However, such traits are not always available for analysis and are affected by changing environmental conditions. Molecular markers developed for *Prunus* also offer a powerful tool to study the evolution of the genome its structure and determinants of genetic diversity (Wünsch and Hormaza, 2002).

Almond is an important crop in the most world region where summer droughts greatly decrease productivity. To drought tolerance, breeding programs focused on many cultivars and genotypes with different levels of tolerance have been found. Almond is a perfect model for studying drought stress responses of a woody plant because of its simplicity in a genomic organization (Campalans *et al.*, 2001). The valuation of genetic diversity and relationships among almond varieties are of a major position in the determination of gene pools, development of protection strategies and documentation of genetic resources (El Hamzaoui *et al.*, 2014). Developed methods for studying plant genetic structure are fundamentally morphological and molecular. Integral methods of these two kinds are a major priority in characterizing germplasm, which pools to a perfectible understanding of the genetic structure, diversity and version of essence collections (Basak *et al.*, 2019).

### 2.8.1 PCR-base techniques

The polymerase chain reaction (PCR) is a relatively simple technique that amplifies a DNA template to produce specific DNA fragments *in vitro*. PCR can achieve more sensitive detection and higher levels of amplification of specific sequences in less time than previously used methods. These features make the technique extremely useful, not only in basic research but also in commercial uses, including genetic identity testing, forensics, industrial quality control and *in vitro* diagnostics.

Kleppe *et al.* (1971) described a method for the first time using an enzymatic assay to replicate a short DNA template with primers *in vitro*. In addition, the PCR process was originally developed to amplify short segments of a longer DNA molecule in 1983 by Kary Mullis (Saiki *et al.*, 1985).

The development of a new technique to perform analysis with molecular markers has been the focus of many recent studies, and most of these are based on PCR amplification of genomic DNA (Babalola, 2003), in this technique, the target DNA is suspended in a reaction mixture consisting of distilled water, buffer the thermostable Taq polymerase and each of the four deoxynucleotide triphosphates (dNTPs), and also there are pairs of primers whose sequences are complementary to that of the DNA flanking the target region.

Polymerase chain reaction (PCR) cycle includes three important steps such as denaturation, primer annealing and primer extension, each step has different temperature and times depending on primers. PCR is used based on 94$^{\circ}$C temperature for 10 minutes is used to heat the lid of instruments and also the reaction mixture is heated to denature the double-stranded

DNA into single strands, and then cooled to an optimum temperature to facilitate primer annealing.

In the next step of a cycle, the temperature is reduced to approximately (40–60) $^{\circ}$C that the oligonucleotide primers are attached in a target DNA as primers which just need 1-2 minutes. The primer pair consists of a forward primer that binds to its complementary sequence upstream of the region to be amplified and a reverse primer that binds downstream, both with their 3 ends facing inward. During the primer extension, the DNA polymerase progressively adds dNTPs. Complementary to the target, to the 3 ends of each primer so that the target sequence is copied. The 5 ends of the primers define the length of the PCR product.

Eventually, new DNA begins to synthesis and the temperature is raised for about 70$^{\circ}$C for 1–2 minutes which is optimum for the DNA polymerase. Each step of the cycle should be optimized for each template and primer pair combination. These three steps constitute a PCR cycle usually 20–40 cycles, the amplified product may be analysed for size, quantity, sequence, etc., or used in further experimental procedures (Lorenz, 2012).

### 2.8.2 Determination of genetic diversity by molecular methods

DNA marker technologies were developed which offered unlimited loci over the entire genome and their actual sites on the chromosome. DNA markers have been used for identifying genotypes, studying genetic diversity, monitoring genetic events, elucidating evolutionary pathways and facilitating the manipulation of genes in breeding programs. However molecular markers of DNA sequence are readily detected and whose inheritance can be easily be monitored. The uses of molecular markers are based on the naturally occurring DNA polymorphism, which forms a basis for designing strategies to exploit for applied purposes. Molecular characterization can play a role in uncovering the history and estimating the diversity, distinctiveness and population structure. It can also serve as an aid in the genetic management of small populations, to avoid excessive inbreeding. A number of investigations have been described within and between-population diversity. Thus, the information about genetic relationships among almond genotypes and their pomological characteristics will be very useful in almond cross-breeding programs. Molecular markers have been used in laboratories since the late 1970s and they are applied across all the food and agricultural sectors. They are very versatile and can be used for a variety of purposes. Thus, they are used in genetic improvement and genetic diversity (Govindaraj *et al.*, 2015). Therefore, in this study, several techniques have been used to examine genetic diversity and relationships among genotypes including;

### 2.8.2.1 Random amplified polymorphic DNA (RAPD)

A single randomly chosen oligo nucleotide is required in the RAPD technique only. Single RAPD primers are able to hybridize to several hundred sites within the target DNA; however, not all of these hybridizations lead to the production of PCR fragments. The ability of RAPDs to produce multiple bands using a single primer means that a relatively small number of primers can be used to generate a very large number of fragments. These fragments are usually generated from different regions of the genome and hence multiple loci may be examined very quickly (Ovesná *et al.*, 2012). RAPD techniques have been positively used in *Prunus* for classifying cultivars ( Lu *et al.*, 1996; Casas *et al.*, 1999).

Many researchers used this marker such as; Randomly Amplified Polymorphic DNA technique used by (Gouta *et al.*, 2008) to study the genetic diversity among 58 almonds and one peach genotypes by using (12) primers. Polymorphic bands are ranged between (6-13) and highest genetic similarity coefficient among genotypes was (0.94), lowest value among almonds genotype was (0.45) and between almonds and peach was (0.33) and the genotypes were grouped to (6) clusters. El Hawary *et al.* (2014) used RAPD technique for documentation and investigation of genetic relatedness of three *Prunus amygdalus* cultivars. Twelve selected random primers were used to amplify the DNA. The lowest and highest values of a number of polymorphic bands were (1-6). Applied RAPD technique by using 16 random primers for sorting of 62 Iranian wild *Prunus* genotypes and some cultivated almonds were documented by (Nikoumanesh *et al.*, 2011b), their result showed genetic similarity coefficient was between (0.28-0.79), and the dendrogram showed 4 groups among the 62 genotypes. The number of polymorphic bands ranged between (11-25).

Sharma *et al.* (2012) used sixty RAPD primers to describe (32) almond genotypes in the University of Horticulture and Forestry, Solan, India. The lowest and highest number of polymorphic bands were (3-9), the PIC (Polymorphic Information Content) values were between (0.26-0.87). The genotypes divided into three clusters. Randomly Amplified Polymorphic DNA (RAPD) technique used to study the genetic relation among sixteen almond cultivars grown up at field gene bank in Jordan. Genetic similarity coefficient was (0.00-0.50). 14.2 % of the polymorphic band was observed among genotypes (Al-Ghzawi *et al.*, 2009). Martins *et al.* (2004) observed twenty-two regenerates from one clone used for analysis by ISSE and RAPD techniques, 64 RAPD and 10 ISSR primer were used and mean percentage polymorphic band for two techniques (4.16 and 6%), respectively. Thirteen randomly amplified polymorphic DNA (RAPD) markers were used to explain their genetic variability and relationships among (39) Iranian and foreign almond cultivars, maximum and minimum similarity coefficients were (0.29-0.89). and the dendrogram grouped the genotypes to five clusters, the polymorphic band was between (8-34) bands (Kiani *et al.*, 2006). Bartolozzi *et al.* (1998) used (37) primers to RAPD marker to recognize relatedness among 17 almond genotypes and one peach genotype gene and they found that similarity coefficient was between (0.27-0.67) and polymorphic bands arranged between (1-5). Establishment of genetic relationships among twenty-nine cultivars and three related wild species of almonds that widely grown in some countries (Iran, USA, Spain, Italy, France and Russia) was studied by using 80 RAPD primers (Sorkheh *et al.*, 2009b) who found that the value of PIC was (0.77), variance among population and within-population were (13.51-86.49) respectively. Genetic similarity coefficient was between (0.70-0.96). In Aula Dei experimental station Zaragoza, Spain, Casas *et al.* (1999) used seven primers to RAPD marker for (41) *Prunus* rootstocks, the dendrogram divided into three groups, the polymorphic band were between (7-15), genetic similarity coefficient arranged between (0.27-1.00). Amplification of genomic DNA for thirteen almond genotypes from different locations achieved using fifteen random decamer primers to RAPD marker, the dendrogram divided into two clusters, the polymorphic bands were between (2-8), genetic similarity coefficient arranged between (0.66-0.99) (Sharma and Sharma, 2010). Shiran *et al.* (2007) used RAPD marker to a genetic connection among 39 almonds cultivars which were assessed by using 42 random primers, the polymorphic bands arranged between (3-34) and PIC values were (0.47-0.94), genetic similarity coefficient arranged between (0.32-0.92).

#### 2.8.2.2 Inter simple sequence repeat (ISSR)

Inter simple sequence repeat (ISSR) PCR is a technique, which involves the use of microsatellite sequences as primers in a polymerase chain reaction to generate multilocus markers. ISSR markers are highly polymorphic and are useful in studies on genetic diversity, phylogeny, gene tagging, genome mapping and evolutionary biology. This Section provides an overview of the details of the technique and its application in genetics and plant breeding in a wide range of crop plants. It is a simple and quick method that combines most of the advantages of microsatellites (SSRs) and amplified fragment length polymorphism (AFLP) to the universality of random amplified polymorphic DNA (RAPD) (Yip *et al.*, 2007).

Several reviews prepared were documents around this area including; genetic diversity among nineteen almond cultivars evaluated by (MirAli and Nabulsi, 2003) using ISSR marker. Thirty-nine random primers were used; their polymorphic bands were between one to eight. Maximum and minimum similarity values among genotypes were (0.70 and 0.96). The dendrogram divided the cultivars into two main groups. Pinar *et al.* (2015) by using two markers (ISSR and RAPD) evaluated the levels of 95 almonds accessions genetic variability grown in Turkey. The numbers of primers were (4 and 13) respectively, the polymorphic bands ranged (2-10 and 1-8) respectively. Similarity coefficient of a gene was (0.90) in twice markers also the two techniques by the UPGMA method, the almonds accessions divided into nine clusters. Gregory (2004) reported that (15) ISSR primers were used for a parent of the first-generation hybrid population of two almonds cultivars (Nonpareil and Lauranne). Results document that the number of polymorphic bands of ISSR markers were (1-4) and also ISSR means of band per primer was (6.6). In addition, Otaghvari and Ghaffarian (2011) used ISSR techniques for the evaluation of genetic diversity for 19 almond genotypes by using 10 primers, the results showed that mean of the polymorphic band and gene similarity coefficient (10.1 and 0.82), respectively. Evaluation of the of genetic diversity of 48 samples of wild peach *Prunus mira* was executed from three regions in China using two markers, ISSR (9 primers) and RAPD (10 primers) by Tian *et al.* (2015), results showed that gene diversities were between (0.33-0.28) respectively, highest and lowest number of polymorphic bands for tow marker were (9-5) and (13-6) respectively, total variance percentage within and among population between (26.34-73.66) to ISSR and to RAPD (17.69-82.31) respectively, the two dendrograms grouped to 4 clusters, the minimum and maximum gene similarity coefficient were (0.71-0.88 and 0.76-0.90) to prior markers respectively, the genetic structure showed that *K* value to (ISSR = 5) and (RAPD = 3). ISSR marker used to show relationships among 29 species of almond, peach and associated species by (Sarhan *et al.*, 2015) using 21 primers. The result of number polymorphic genes, PIC, clusters and genetic similarity coefficient were (10-30), (0.59-0.56), (3) and (0.27-0.96) respectively.

#### 2.8.2.3 Simple sequence repeat (SSR)

PCR-based, simple sequence repeat (SSR) markers (microsatellites) become the marker of choice for fingerprinting and genetic diversity studies for a wide range of plants (Crawford, 2018). Microsatellites or SSR (Simple Sequence Repeats) or STR (Simple Tandem Repeats) consists of a stretch of DNA, a few nucleotides long – 2 to 6 base pairs (bp) – repeated several times in tandem. They are spread over a eukaryote genome. Microsatellites are of relatively small size, and can, therefore, be easily amplified using PCR from DNA extracted from a variety of sources including plant parts.

Because of their high polymorphism, abundance, and co-dominant inheritance, they are well suited for the assessment of genetic variability within crop species, and of the genetic relationships among species.

These SSR markers were used for the molecular characterization and identification of cultivars in different species including peach, almond, apricot, cherry and *Prunus* rootstocks using different methods for the analysis of the DNA, in addition, SSR markers were used also for genetic mapping in peach (Sosinski *et al.*, 2000; Dettori *et al.*, 2001; Etienne *et al.*, 2002; Bouhadida *et al.*, 2009; Dettori *et al.*, 2015), almond ( Joobeur *et al.*, 2000; Tavassolian *et al.*, 2010; Forcada *et al.*, 2015) and apricot (Dondini *et al.*, 2007; Pedryc *et al.*, 2009).

#### 2.8.2.4 Amplified fragment length polymorphism (AFLP)

The principle of AFLP is based on a selectively amplifying subset of restriction fragments from a complex mixture of DNA fragments obtained after digestion of genomic DNA with restriction end nucleases. Polymorphisms are detected from differences in the length of the amplified fragments by polyacrylamide gel electrophoresis (Karp *et al.*, 1998).

The AFLP technology is a powerful tool for the detection and evaluation of genetic variation in germplasm collections and in the screening of biodiversity as well as for fingerprinting studies (Costa *et al.*, 2016). AFLP markers have successfully been used for analyzing genetic diversity in some other plant species such as peanut. However, the technical background makes it possible to employ still more effective approaches to genome characterization (Kahvejian *et al.*, 2008).

#### 2.8.2.5 Single nucleotide polymorphisms (SNPs)

A new method which can be applied to genetic diversity and in population genetic analyses is SNPs, are easy which used to evaluate either functional or neutral variations (Clark *et al.*, 2005; Downing *et al.*, 2012; Cardoso *et al.*, 2015). SNPs are variations at single nucleotides which do not change the overall length of the DNA sequence in the region. Most SNPs are located in non-coding regions and have no direct impact on the phenotype of an individual. However, some introduce mutations in expressed sequences or regions which influence gene expression (promoters, enhancers) and may induce changes in protein structure or regulation. These SNPs have the potential to detect the functional genetic variation. Therefore, SNPs are used as an alternative to microsatellites in genetic diversity studies. Goonetilleke *et al.* (2018) reported that molecular marker (SNP) can be significantly important for analysis of genetic diversity of *Prunus*, particularly almond.

# CHAPTER THREE

# MATERIALS AND METHODS

## 3.1 Locations and Plant Material

Nearly five hundred trees were observed from two governorates (Sulaimani and Halabja) in Kurdistan region (North eastern – Iraq) (Figure 3.1). In the first visit in 2017, 99 trees were labelled from Sulaimani [Sharbazher (S), Mergapan (M), Qaradagh (Q) and Barznja (B)) and Halabja (Hawraman (H)] locations based on vigor, colors of the foliage and leaf area which are considered to reflect trees adaptations to drought (Zokaee-Khosroshahi *et al.*, 2014). In the second visit in 2018, the samples for first visit were reduced to (38) in all locations [S (9), M (3), Q (5), B (7) and H (14)] due to several difficulties, such as new reconstruction of farms and pests (Table 3.1).

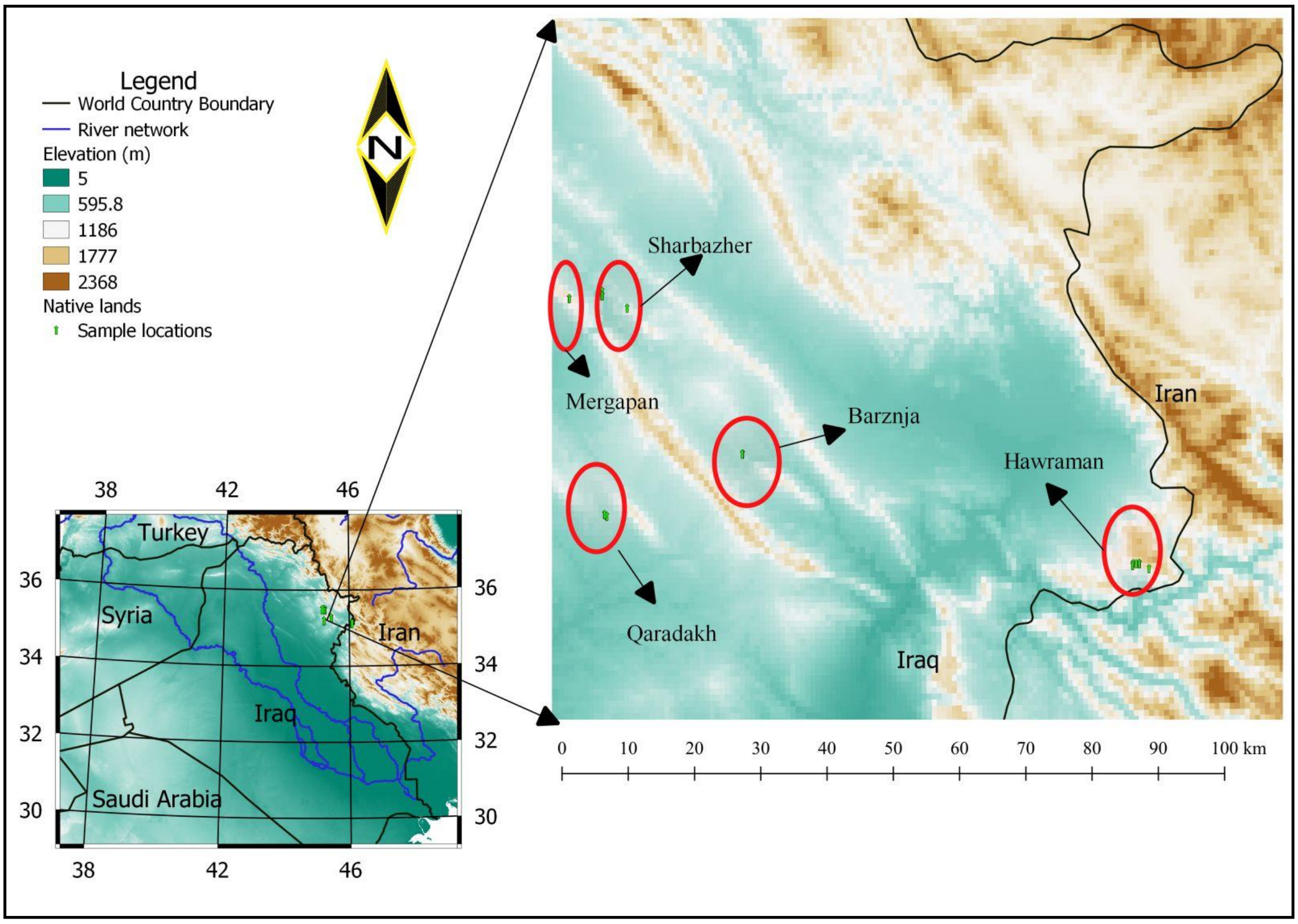

**Figure 3.1 Distribution of collection sites of the studied plant materials in Sulaimani and Halabja governorates.**

**Table 3.1 Genotypes names, locations, latitudes and altitudes.**

| Genotypes name | Locations | Latitudes | | Altitudes |
|---|---|---|---|---|
| | | N° | W° | |
| S-G1 to S-G9 | Sharbazher | 35°49′30″ | 45°18′93″ | 997.6 |
| M-G1 to M-G3 | Mergapan | 35°48′93″ | 45°13′47″ | 1148 |
| Q-G1 to Q-G5 | Qaradagh | 35°19′29″ | 45°19′69″ | 925.8 |
| B-G1 to B-G7 | Barznja | 35°27′69″ | 45°42′21″ | 1154 |
| H-G1 to H-G14 | Hawraman | 35°12′38″ | 46°07′80″ | 1402 |

## 3.2 Morphological Data Characteristics

The following characteristics have been investigated:

### 3.2.1 Annual shoot growth (cm)

Six current shoots growth were measured with a ruler, in November 2017.

### 3.2.2 Annual shoot diameter (mm)

In November 2017 six current shoot diameters were taken at the base using Vernier (electronic caliper) (Model: DMV-SL05, WORKZONE, Germany).

### 3.2.3 Leaves area ($cm^2$)

Six full expanded leaves were collected in (2017). Leaf area was taken according to (Sauceda-Acosta *et al.*, 2017). Digital camera (Nikon d7100) held steadily on a stand above the white paper that ruler affixed on it, the leaves were put behind the ruler and digital pictures were captured for all pictures, the same distance between the paper and cameras lens was fixed. Digimizer program v.4.5.2® (Medcalc Software 2015) was used to measurement of leaves area (Carvalho *et al.*, 2017).

### 3.2.4 Stomata numbers (Sto $mm^{-2}$)

Uncolored nail varnish used to make an impression of the abaxial of leaf (Palasciano *et al.*, 2005) and by digital camera (AmScope M100 China) leaf epidermis image was captured under a compound microscope, and then by same camera software, numbers of stomata in a known area were measured (Alkhatib *et al.*, 2019). Twelve leaves were taken and in 4 different parts of the leaves photos were captured.

### 3.2.5 leaves dry weight (%)

Six replicates of fresh leaf samples were selected, weighed and then dried in an oven at 70°C until the weight was stable, (Zokaee-Khosroshahi *et al.*, 2014) according to the following equation:

$$\text{Percent dry weight} = \frac{\text{Dry weight}}{\text{Fresh weight}} \times 100 \qquad (3.1)$$

### 3.2.6 Nut width, length and thickness (mm)

An electronical caliper (Model: DMV-SL05, WORKZONE, Germany) was used for measuring three dimensional growths of nuts as in (Figure 3.2) (Rharrabti and Sakar, 2016).

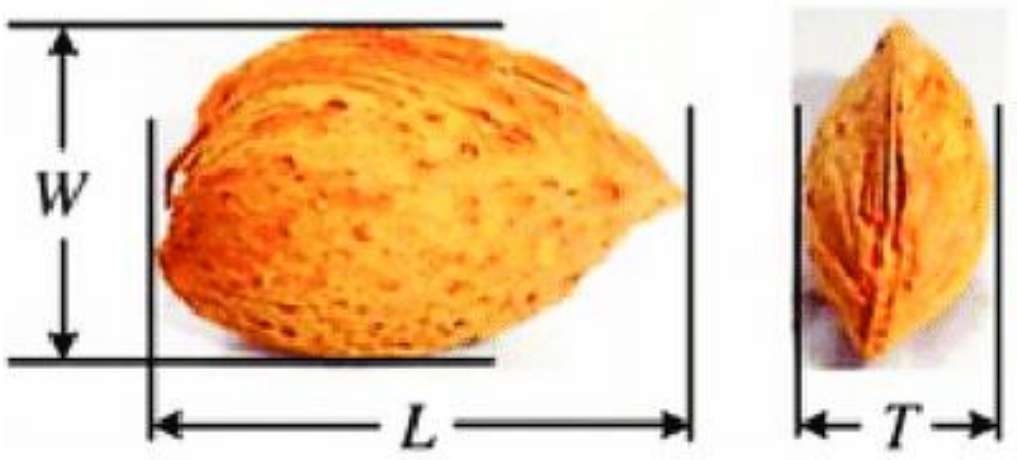

**Figure 3.2 Nut Thickness (T), Width (W) and Length (L).**

### 3.2.7 Nuts weight (g)

Average of six nut weights was taken using electronic sensitive balance.

### 3.2.8 Kernel (%)

The weight of the whole nut was taken and nut cracker was used to obtain the kernel weight, then the kernel ratio was taken according to the following formula:

$$\text{kernel percentage} = \frac{\text{Kernel weight}}{\text{Nut weight}} \times 100 \qquad (3.2)$$

## 3.3 Leaves Phytochemical Analyses

### 3.3.1 Chlorophyll concentration determination (SPAD)

Average of six leaves greenness was taken from different location on shoot and site of tree on a completely expanded leaf in situ using the Chlorophyll meter (SPAD-502; Minolta, Osaka, Japan) (Isaakidis *et al.*, (2004).

### 3.3.2 Proline content determination (μmol $g^{-1}$ FW)

According to Bates *et al.* (1973) with some modifications, proline was determined. Briefly, 0.25 g of each sample tissue was mixed with 5 ml sulfosalicylic acid (Appendix 1), and then centrifuged at 4000 rpm for 10 min, two ml of supernatant were taken and mixed with 2 ml acidic ninhydrin (Appendix 1) and 2 ml glacial acetic acid. Samples were heated at 100 ºC for 1 hour and cooled on ice bath. 4 ml of toluene was added to the mixture and mixed gently well. the samples left for 5 minutes to separate, aqueous phase was transferred into a new labelled tube and the mixture was measured spectrophotometrically at an absorbance of 520 nm (Model: UV-160, Shimadzu, Japan) against the blank containing toluene. The following formula used to calculate proline content.

$$\text{Proline} = \frac{\text{ABS of extract} - \text{ABS of blanck}}{\text{slope}} \times \frac{\text{Vol. of total extract}}{\text{Vol. of used extract}} \times \frac{1}{\text{Fresh weight}} \qquad (3.3)$$

### 3.3.3 Leaf samples preparation for analysis of some non-enzymatic antioxidants

For determining phenol, flavonoid, tannin and saponin in almond laves, ( Tabart *et al.*, 2007; Michiels *et al.*, 2012) methods were used for preparing the samples and as follows:
Leaf samples were taken from each genotype from deferent sides of the trees and shoot location, and snap-frozen using liquid nitrogen, frozen leaves were ground to fine powders using pestle and mortar. 1g of ground powder was taken put in 15 ml tube, then 10 ml 80% methanol was added to the tubes and mingled gently, after that, the mixture stored at room temperature for 24 hours. The samples were then shaken for 3 hours and then centrifuged at 4000 rpm for 15 minutes at 5˚C. The liquid supernatant was then transferred into a new clean labelled 15 ml tubes without contamination by the lower layer, Samples were kept at refrigerator at 4˚C. Extracted solution was used for non-enzymatic antioxidants assay as follows.

#### 3.3.3.1 Total phenolic content determination (TPC) (mg GAE $g^{-1}$ E)

Folin-Ciocalteu method that described by (Singleton and Joseph, 1965) with some modifications was used to determine total phenolic content (TPC) from the leaf extracts described above (Djeridane *et al.*, 2006; Guerreiro *et al.*, 2017; Rodrigues *et al.*, 2019), 50 µl of each extract was mixed with 4 ml of Folin-Ciocalteu reagent (Appendix 2) and allowed to react for 7 minutes. After that, 3.6 ml of 10% saturated $Na_2CO_3$ solution was added and left in the dark at 30˚C for 50 minutes. Regarding blank samples, the same previous step was repeated but just 50 µl of water was used instead of the samples. The absorbance of the reaction mixture was measured at 750 nm by spectrophotometer (Model: UV-160, Shimadzu, Japan). Gallic acid standard curve was utilized for calculation of total phenolic content (TPC) which expressed as milligrams of gallic acid equivalent (GAE) per gram of the plant extract (mg GAE $g^{-1}$ E) on a dry weight basis.

### 3.3.3.2 Total flavonoid content determination (TFC) (mg QE $g^{-1}$ E)

Rigane *et al.* (2017) method was applied for determining TFC as described below:
A mixture of 0.9 ml methanol (80%), 0.3 ml of 2% aluminum chloride ($AlCl_3$), 0.07 ml of (1M) potassium acetate ($CH_3COOK$) (preparation of the solutions showed in (Appendix 3), 1.7 ml of deionized water and 60 µl of each leaf sample extracts was used. Then the mixture was incubated at room temperature for 30 minutes. 60 µl of water was mixed with the same amount of chemicals in the previous step to adjust the blank. The absorbance of the reaction mixture was measured at 415 nm using a spectrophotometer (UV-160, Shimadzu, Japan). Total flavonoid content was determined from a standard curve using quercetin as a standard. Results were expressed as milligrams of quercetin equivalents per gram of the extract (mg QE $g^{-1}$ E).

### 3.3.3.3 Saponin content determination (SC) (mg SE $g^{-1}$ E)

Hiai *et al.* (1976) and Anh *et al.* (2018) method was used to evaluate saponin content of the leaves and as follows:
Half milliliter vanillin solution (4% w/v) was added to 80 µl of the leaf extract sample in cold water bath and then 3 ml of sulfuric acid (72% v/v) (Appendix 4) was gently added and kept again at the cold-water bath. The mixture was then warmed in a water bath at 60˚C for 10 minutes and then cooled in ice-cold water for 5 minutes. for to a blank sample, 80 µl of water was mixed with those chemical materials. Spectrophotometer (Model: UV-160, Shimadzu, Japan) was adjusted at 540 nm absorbance and samples were measured after blank samples were reading. The saponin content was determined from a standard curve using saponin as a standard. Results were expressed as milligrams of saponin equivalents (SE) per gram of the extract (mg SE $g^{-1}$ E).

### 3.3.3.4 Condensed tannin content determination (CTC) (mg CE $g^{-1}$ E)

The vanillin reaction, with some modifications was conducted to estimate the condensed tannin content ( Broadhurst and Jones, 1978; Agostini-Costa *et al.,* 2015). Briefly, 150 µl of each extract was mixed with 4 ml of 4% vanillin (Appendix 4) and 2 ml concentrated HCl with stirring 30 seconds. After that, the mixture was incubated at room temperature for about 20 minutes. Blanks were set up which contained 150 µl of water mixed with 2 ml of 4% vanillin and one milliliter concentrated HCl with stirring 30 seconds. Spectrophotometer (Model: UV-160, Shimadzu, Japan) was adjusted at 500 nm absorbance. Catechin was used as a reference. The condensed tannin content was expressed as milligrams of catechin equivalents (CE) per gram of the extract (mg CE $g^{-1}$ E).

#### 3.3.3.5 Antioxidants activity determination (DPPH assay) (Inhibition%)

The antioxidants capacity of each leaf extract was determined according to the DPPH (2, 2-diphenyl-1-picrylhydrazyl) method with some modifications which described by (AL-Ghazghazi *et al.*, 2014; Tahir *et al.*, 2019a). 1.7 ml of methanolic solution of DPPH ($6x10^{-5}$ M) solution (Appendix 5) was mixed with 30 µl of the sample. Blank samples containing 3 ml methanol was used, one milliliter of methanol was added to 1.7 ml of methanolic solution of DPPH ($6x10^{-5}$ M). All solutions incubated for 30 minutes at room temperature and then absorbances were adjusted at 517 nm. Different concentrations of ascorbic acid (2.0-12.0 µg $ml^{-1}$) were prepared and used as a reference. Antioxidant activity was measured as the inhibition percentage of the DPPH radical using the following equation:

$$\text{Inhibition} = \frac{\text{ABS 517 of control} - \text{ABS 517 of sample}}{\text{ABS 517 of control}} \times 100 \qquad (3.4)$$

#### 3.3.3.6 Antioxidants activity determination (ABTS assay) (Inhibition%)

According to ( Re *et al.*, 1999; Lee *et al.*, 2015), antioxidant activity was measured by radical ABTS (2,2'-azino-bis (3-ethylbenzothiazoline)-6-sulfonic acid) with some modifications. 20 µl of each sample extracted were mixed with 3 ml ABTS solution (Appendix 6), then the mixture was stirred vigorously and incubated under dark condition for 7 minutes at 24°C. 3 ml ethanol 95% was used as blank. The control prepared by 10 µl of distilled water mixed with 3 ml of ABTS solution, the absorbance adjusted at 734 nm, and the capability to inhibit the ABTS radical was calculated using the following formula:

$$\text{Inhibition} = \frac{\text{ABS 734 of control} - \text{ABS 734 of sample}}{\text{ABS 734 of control}} \times 100 \qquad (3.5)$$

### 3.3.3.7 Total antioxidant capacity determination (TAC) (mg AA $g^{-1}$ E)

Total antioxidant capacity (TAC) was evaluated with some modifications according (Prieto *et al.*, 1999Phatak and Hendre, 2013). 1.3 ml each of 28 mM sodium phosphate buffer (pH 7.4), 4 mM ammonium molybdate and 0.6 M sulfuric acid (Appendix 7) was used and mixed with 20 µl of each leaf extract sample. The mixture was incubated at 95˚C for 90 minutes in the oven. After samples were cooled at room temperature. Blank was contained 20 µl of water mixed with prior solution and absorbances were adjusted at 695 nm. Different concentrations of ascorbic acid (2.0-12.0 µg. $ml^{-1}$) were prepared and used as a reference and the results then were expressed as milligrams of ascorbic acid (AA) per gram of the extract (mg AA $g^{-1}$ E).

## 3.4 Molecular Biology Methods

### 3.4.1 Genomic DNA extraction from almond leaves

The CTAB (Cetyl TriMethyl ammonium Bromide) reagent used to isolate DNA from plant tissues facilitates the separation of polysaccharide during purification, extraction protocol modified by (Doyle and Doyle, 1987) was used to extract genomic DNA in higher education laboratory at Horticulture department. CTAB buffer was prepared as described in (Appendix8). Genomic DNA was isolated from young leaves of each almond leaves. (approximately 2-3 leaves) and snap-frozen using liquid nitrogen with pestle and mortar to grind the frozen leaf to a fine powder; 4 ml of CTAB buffer was added into the (15 ml) centrifuge tubes and mixed by vortex, then incubated in water bath at 60°C for at least 1 hour and mixed once after 15 minutes. After that, 4-5 ml chloroform (99%) was added and shaken for 2 minutes to form an emulsion. Samples were then centrifuged at 4000 rpm for 30 minutes at 15°C. The liquid supernatant was carefully transferred into a new clean labeled, 15 ml centrifuge tubes without contamination in the lower layer. 30 µl of RNAs (1.5 mg/ml) were added for each sample then incubation at 45°C for 60 minutes, and then directly 10 µl of proteinase K (10 mg/ml) was added with incubated at 45°C for 40 minutes. Next, same volume of sample chloroform (99%) was added and mixed well by inverting and samples were then centrifuged tubes at 4000 rpm for 30 minutes at 15°C. After that, the aqueous layer (supernatant) was carefully transferred into a clean labelled tube to avoid contamination by the lower layer. Estimated volume of the DNA samples and 0.08 volume of 7.5 M ammonium acetate was added and mixed with 0.54 volume of ice iso-propanol (99%) was also added for the precipitation of DNA and mixed by inverting 20 to 30 times by hand without shake. Tubes were stored in a freezer at -20°C overnight for a better yield. Samples were centrifuged at 4000 rpm for 30 minutes at 5°C, and then isopropanol supernatant was carefully poured off (discarded). One ml of ice-cold 70% ethanol was added to each tube and mixed gently. Samples were centrifuged again at 4000 rpm for 30 minutes at 5°C, the supernatant was then very carefully discarded without dislodging the pellet. To dry samples, tubes with pellet were left at room temperature for 1 hour. Suspend the dried pellet in 50-100 µl of deionized water. Samples were kept at refrigerator at 4°C for 1 day and then stored at -20°C for a long time. Quality of DNA was checked by running 2-3 µl on a 0.8% agarose gel (Ali *et al.*, 2019).

### 3.4.2 Agarose gel electrophoresis

PCR products were generally analyzed by running the product in a 1.6% (w/v) agarose gel in (1x) TBE (Tris-Borate-EDTA) buffer, 20 µl of ethidium bromide (EtBr) 500 µg/ml were added to the cooling gel. 1.0 µl of loading dye (1x diluted from 6x); was added to 20 µl of PCR Phusion amplified product and mixed. Samples were loaded into the wells with; 5 µl of Hyper ladder I were loaded alongside to enable accurate sizing of products. Gels were run at a constant voltage of 100V for 60 minutes (run VIEW Real Time Gel Visualization System), then imaged using a UV transilluminator. The images were captured by a digital imaging system (ENDURO™ GDS Touch Gel Documentation System) (Sharma *et al.*, 2012).

### 3.4.3 PCR genotyping analysis

All DNA lines were screened by PCR model (Applied Biosystems™ Veriti™ 96-Well Fast Thermal Cycler) that PCR reactions were prepared as described in (Table 3.2) and the temperature cycling conditions as in (Table 3.3). A list of primers is given in (Table 3.4)

### 3.4.4 Design primers

Primers were designed regarding many papers including ( Martins *et al.*, 2003; Martins *et al.*, 2004; Sharma *et al.*, 2012; Pinar *et al.*, 2015; Berindean *et al.*, 2016; Abodoma *et al.*, 2017; Saleh *et al.*, 2018) primers were Germanys made.

**Table 3.2 PCR reaction mixture for genotyping analysis reaction.**

| Materials | Volume µl |
|---|---|
| Master mix | 10 |
| Forward and reverse primer (20 pmol/ µl) | 0.7 |
| Genomic DNA (20-30ng/ µl) | 4 |
| Deionized Water $H_2O$ | 5.3 |
| Total volume | 20 |

**Table 3.3 PCR cycling for genotyping analysis steps.**

| Steps | Temperature | Time | Cycles |
|---|---|---|---|
| Initial denaturation | 94°C | 10 minutes | 1 |
| Denaturation | 94°C | 1 minute | 36 |
| Annealing temperature | depended on primer | 1 minute | |
| Extension | 72°C | 2 minutes | |
| Final Extension | 72°C | 10 minutes | 1 |

**Table 3.4 Primer names, sequences and annealing temperatures of RAPD and ISSR markers.**

| Primer Number | Primer Name | Primer Sequences 5 → 3 | Annealing Temperature (°C) |
|---|---|---|---|
| **RAPD** | | | |
| 1 | OPA-08 | GTGACGTAGG | 36 |
| 2 | OPA-10 | GTGATCGCAG | 36 |
| 3 | OPA-11 | CAATCGCCGT | 36 |
| 4 | OPA-16 | AGCCAGCGAA | 36 |
| 5 | OPB-11 | GTAGACCCGT | 36 |
| 6 | S075 | ACGGATCCTG | 36 |
| 7 | S084 | CAGACAAGCC | 36 |
| 8 | S085 | CTCTGTTCGG | 36 |
| 9 | S081 | TCGCCAGCCA | 36 |
| 10 | S093 | CCACCGCCAG | 36 |
| 11 | S078 | GGCTGCAGAA | 36 |
| 12 | S094 | AGAGATGCCC | 36 |
| 13 | S087 | GGTGCAGTCG | 36 |
| 14 | S088 | GGTCCTCAGG | 36 |
| 15 | S089 | CAGTTCGAGG | 36 |
| 16 | S090 | TACCGACACC | 36 |
| 17 | S091 | TCGGAGTGGC | 36 |
| 18 | S092 | ACTCAGGAGC | 36 |
| 19 | S095 | CAGTTCTGGC | 36 |
| 20 | S073 | CCAGATGCAC | 36 |
| **ISSR** | | | |
| 1 | 807 | AGAGAGAGAGAGAGAGT | 50 |
| 2 | 17898A | CACACACACACAAC | 55 |
| 3 | HB04 | GACAGACAGACAGACA | 60 |
| 4 | HB 8 | GAGAGAGAGAGAGG | 50 |
| 5 | HB10 | GAGAGAGAGAGACC | 50 |
| 6 | HB11 | GTGTGTGTGTGTCC | 50 |
| 7 | HB12 | CACCACCACGC | 50 |
| 8 | HB15 | GTGGTGGTGGC | 50 |
| 9 | AG7YC | AGAGAGAGAGAGAGYC | 55 |
| 10 | AGC6G | AGCAGCAGCAGCAGCAGCG | 55 |
| 11 | IS06 | GTGCGTGCGTGCGTGC | 60 |
| 12 | IS16 | DHBCGACGACGACGACGA | 60 |
| 13 | IS17 | BDBACAACAACAACAACA | 57 |
| 14 | IS19 | YHYGTGTGTGTGTG | 57 |
| 15 | ISSR.08 | ACACACACACACACACYA | 52 |

## 3.5 Pre-Drought Tolerance Test to Almond Genotypes in Glasshouse

### 3.5.1 Seed stratification and irrigation intervals

This experiment was conducted at Horticulture Department, College of Agricultural Engineering Sciences, University of Sulaimani during October 2018. Almond genotypes seeds were obtained from different locations and socked in sterilized distilled water for 24 hours at room temperature. For stratification treatments, seeds were mixed with moistened sand and stratified in pots that contained 3 kg sand in cold room under controlled conditions (cold room) (4±1°C) for 5 weeks (García-Gusano *et al.*, 2010; Yücedağ and Gultekin, 2011) during October 2018. Polyethylene plastic bags were used for sowing the stratified seeds that contained 3 kg loam and regularly the pots were irrigated by using sprinkler irrigation system for uniform plant establishment. In this stage, irrigation was carried out, irrigation treatments which were applied ten days after the emergence stage in three different irrigation intervals (II10, II20 and II40), since II10, II20 and II40 means that the pots irrigated every 10, 20, 40 days respectively, therefore the number of irrigations were 4, 2, and 1 time successively. The volume of water added to each experimental pot was estimated based on soil water depletion replenishment by gravimetric method. Available water for the soil was calculated after estimating each of the soil water content at -33 and -1500 kPa from special program (Soil-Plant-Air-Water) (SPAW) developed by (Saxton and Rawls, 2006).

### 3.5.2 Morphological data characteristics

#### 3.5.2.1 Seedling height (cm)

Seedling height was measured with a ruler, at the ends of experimental time from soil surface to the end of seedling.

#### 3.5.2.2 Seedling diameter (mm)

By using electronic caliper and 3 cm above the soil surface seedling diameter was measured.

#### 3.5.2.3 Leaves number

Total leaves that remain on the shoot were calculated.

#### 3.5.2.4 Leaves area ($cm^2$)

Look at the section 3.2.3

### 3.5.2.5 Vegetative growth weight (g)

At the end of experiment, the seedling was taken off and the total vegetative parts weighted by sensitive balance.

### 3.5.2.6 Vegetative dry weight (%)

All seedling vegetative parts were dried in the oven at 70°C until the weight was stable, by sensitive balance, the dried part weighted according to (Equation 3.1).

### 3.5.2.7 Root weight (g)

At the end of experiment, the seedling was taken off and the total root weighted by sensitive balance.

### 3.5.2.8 Root dry weight (%)

Seedling root parts were dried in the oven at 70°C until the weight was stable, by sensitive balance, the dried part weighted according to (Equation 3.1).

### 3.5.2.9 Stomatal length and width (µm)

Look at (3.2.4), the same method was used to capture a picture and by same software, the stomatal length and width were measured.

### 3.5.3 Chemical characteristic and analysis of some non-enzymatic antioxidants

### 3.5.3.1 Determination of chlorophyll concentration (SPAD)

Chlorophyll concentration was measured as described (3.3.1).

### 3.5.3.2 Determination of proline content (µmol $g^{-1}$ FW)

Proline was measured as described (3.3.2)

### 3.5.3.3 Determination of total phenolic content (TPC) (mg GAE $g^{-1}$ E)

Total phenolic content evaluated as showed in (3.3.3.1)

### 3.5.3.4 Determination of total flavonoid content (TFC) (mg QE $g^{-1}$ E)

Total flavonoid content evaluated as showed in (3.3.3.2)

### 3.5.3.5 Determination of antioxidants activity (DPPH assay) (Inhibition %)

Determination of antioxidant activity (DPPH) assay was measured as showed in (3.3.3.5)

### 3.5.3.6 Determination of antioxidants activity (ABTS assay) (Inhibition %)

Determination of antioxidant activity (ABTS) assay described as showed in (3.3.3.6)

## 3.6 Statistical Analysis

A simple RCBD was designed for genotypes experiment. ANOVA and comparison test among genotypes were performed by XLSTAT software (Version 2016.02.28451). Morphological data were converted to matrix data to create the PCA plot and dendrogram using Euclidean distance and Jaccard methods. And the scorable bands were coded manually as either present (1) or absent (0) (Tahir *et al.*, 2019b). Means were separated by Duncan's Multiple Range Test (DMRT).

In a pre-drought tolerance in glasshouse, a factorial (RCBD) was conducted. Two factors (Almond Genotypes and Irrigation Intervals) with three replicates each replicate contained 114 experimental unit (38 genotypes * 3 replicates), all possible comparisons among the means were carried out by using the least significant difference (L.S.D.) test at ($P \leq 0.05$) after they had shown significant differences in the general test.

# CHAPTER FOUR
# RESULTS AND DISCUSSION

## 4.1 Morphological Data of Genotype Trees

### 4.1.1 Morphological studies of vegetative characteristics

One of the most important approach to evaluate drought stress is morphological observations which are the quick and simple way for assessing plant genotypes under drought stress. In addition, measurements of morphological parameters including; vegetative growth (annual shoot growth, annual shoot diameter, leaves area and leaves dry weight), properties of stomata can be used alongside with genotypes to detect drought tolerance and crop improvement (Zokaee-Khosroshahi *et al.*, 2014). Thirty-eight almond genotypes were selected and the effects of genotypes for five morphological parameters showed significant differences among the genotypes. The highest values (B-G4 = 43.9 cm, B-G5 = 3.938 mm, M-G3 = 4.049 $cm^2$, H-G2 = 272.222 Sto per $mm^2$ and S-G1= 42.457%) were documented for recent shoot growth and diameter, leaves area, stomatal number per square millimeter and leaves dry weight percentage, respectively. On the other hand, the lowest values (H-G5 = 1.417 cm, H-G5 = 1.038 mm, H-G14 = 2.223 $cm^2$, H-G8 = 40.891 Sto per $mm^2$, and B-G5 = 24.676%) were recorded for the mentioned parameters as well, (Table 4.1). Our results clearly showed that genotypes have significant effects on stated parameters.

Minimum values in all parameters except leaves dry weight recorded in Hawraman location may be because genotypes of this location were differences genetically in our results in RAPD and ISSR test.

Bertolino *et al.* (2019) demonstrated that dissimilarity in density of stomata may get up due to genetic factors and/or different environmental factors, but highest and lowest value were recorded in same location (Hawraman) for that we suggest that the stomatal density was affected by genetic.

Generally, plants grown in arid and semi-arid environments are exposed to long periods of water deficit and developed adaptations in order to tolerate drought. In addition, main consequences of water shortage in plants are compact rates of cell division and expansion, stem elongation, leaf size, root production, stomatal distribution, the relation of plant water and nutrient with reduced crop productivity, and water use efficiency (Farooq *et al.*, 2012). Several researchers have studied the adaptation of almond to water deficit from different perspectives, many morphological and physiological drought tolerance mechanisms have been recognized. These mechanisms include the ability for osmotic adjustment, reduced leaf area, changes in the adaptable properties of cells and tissues, reduced stem length, control of stomatal regulation, leaf abscission and deeply penetrating of the root system (Pirasteh-Anosheh *et al.*, 2016; Vats, 2018).

A crucial physiological change of plant in drying environments is a rapid closure of stomata which described as the first line of defining water deficit and it is much quicker than other physiological changes. Water consumption and transpirational water loss reduced during stomatal closure. Because of the researcher believed that during drought stress, chemical signal will be sent to shoots from the root which is encouraged by stomatal closure. Thus, behavior and regulation of stomata play a vital role in plant tolerance during drought stress which has been found during signaling between root-to-shoot (Cherry, 1989; Wilkinson and Davies, 2002). Furthermore, the reduction in photosynthetic rate associated with stomatal closure due to changes in leaf water status is commonly observed in plants grown under water deficit conditions (Silva *et al.*, 2009). Researchers demonstrated that some morphological parameters such us; leaf area, frequency, stomatal length and width for 20 wild and cultivated almonds genotype were studied, significant effect of genotypes on some characteristics showed that the lowest and highest values were (2.3-26.5 $cm^2$), (143.4-326 per $mm^2$), (19.3-30 µm) and (9.4-14.9 µm) for leaves area, stomatal frequency, stomatal length and stomatal width, respectively (Palasciano *et al.*, 2005). Damyar and Hassani (2006) evaluated 25 almond cultivars at Karaj, Iran and found that maximum and minimum averages for shoot growth were 61.11 cm and 30, cm respectively.

Eventually, our result at different locations with regard to morphological data indicate that they responded to various genotypes according to their locations. Sardabi *et al.* (2006) showed that wild almond and some cultivar affected on the shoot height and growth, leaves area and number and stem diameter, also the genotypes leaves have difference stomatal numbers on lower surface area and the genotypes have less stomatal numbers which can tolerate water stress condition by defoliation of the leaves that thereafter avoids transpiration and evaporation.

Finally our result were closely resembled to (Zokaee-Khosroshahi *et al.,* 2014) result that showed that genotypes effected on shoot length, leaf area and vegetative dry weight.

**Table 4.1 Effect of almond tree genotypes on some vegetative growth characteristics.**

| Genotype | Annual shoot growth cm | | Annual shoot diameter mm | | Leaves area $cm^2$ | | Stomatal Sto per $mm^2$ | | Leaves dry weight % | |
|---|---|---|---|---|---|---|---|---|---|---|
| S-G1 | 15.917 | g-o | 1.815 | j-n | 3.319 | b | 196.931 | b-i | 42.457 | a |
| S-G2 | 23.000 | e-i | 2.697 | c-i | 3.247 | bc | 223.333 | b-f | 40.097 | b |
| S-G3 | 30.083 | cde | 2.802 | c-h | 3.181 | cd | 182.949 | d-k | 37.823 | b-e |
| S-G4 | 11.167 | j-q | 2.263 | f-k | 3.153 | cde | 176.046 | e-k | 39.441 | bc |
| S-G5 | 35.000 | a-d | 3.652 | ab | 3.108 | def | 139.444 | k | 37.044 | c-g |
| S-G6 | 19.417 | f-l | 2.612 | d-j | 3.083 | def | 192.593 | b-j | 38.172 | bcd |
| S-G7 | 21.417 | e-k | 2.710 | c-i | 3.056 | ef | 161.448 | g-k | 35.410 | e-j |
| S-G8 | 34.167 | a-d | 3.253 | a-d | 3.023 | f | 145.833 | jk | 37.264 | c-f |
| S-G9 | 18.917 | f-m | 2.517 | d-j | 3.068 | ef | 172.500 | f-k | 38.158 | bcd |
| M-G1 | 8.000 | n-q | 1.987 | h-m | 3.050 | ef | 232.837 | a-d | 40.221 | b |
| M-G2 | 18.583 | f-n | 2.958 | b-f | 3.067 | ef | 232.222 | a-d | 39.497 | bc |
| M-G3 | 13.083 | i-p | 2.310 | e-k | 4.049 | a | 175.139 | e-k | 36.643 | d-h |
| Q-G1 | 24.417 | d-h | 3.830 | a | 2.888 | g | 235.833 | abc | 38.804 | bcd |
| Q-G2 | 25.083 | d-g | 2.737 | c-i | 2.760 | hi | 233.333 | a-d | 36.887 | c-h |
| Q-G3 | 17.917 | f-n | 2.480 | d-j | 2.780 | h | 73.255 | l | 37.327 | c-f |
| Q-G4 | 25.333 | d-g | 2.168 | f-l | 2.784 | h | 189.444 | c-k | 34.830 | f-j |
| Q-G5 | 31.333 | b-e | 3.513 | abc | 2.756 | hi | 68.537 | l | 34.734 | f-j |
| B-G1 | 21.583 | e-j | 3.223 | a-d | 2.740 | hij | 198.056 | b-i | 30.070 | k |
| B-G2 | 35.000 | a-d | 3.125 | a-e | 2.667 | i-l | 197.923 | b-i | 30.219 | k |
| B-G3 | 40.500 | abc | 3.505 | abc | 2.706 | h-k | 209.286 | b-g | 33.694 | ij |
| B-G4 | 43.900 | a | 3.655 | ab | 2.651 | j-m | 207.500 | b-h | 30.370 | k |
| B-G5 | 37.250 | abc | 3.938 | a | 2.603 | klm | 151.270 | ijk | 24.676 | l |
| B-G6 | 38.250 | abc | 3.155 | a-d | 2.615 | klm | 173.611 | f-k | 33.585 | j |
| B-G7 | 39.833 | abc | 3.307 | a-d | 2.592 | lm | 180.794 | e-k | 29.394 | k |
| H-G1 | 9.167 | l-q | 1.923 | i-m | 2.554 | mn | 141.667 | k | 37.838 | b-e |
| H-G2 | 8.250 | m-q | 1.528 | k-o | 2.493 | no | 272.222 | a | 38.295 | bcd |
| H-G3 | 5.583 | opq | 1.283 | mno | 2.452 | op | 65.278 | l | 37.193 | c-f |
| H-G4 | 4.250 | pq | 1.422 | l-o | 2.449 | op | 222.222 | b-f | 34.462 | g-j |
| H-G5 | 1.417 | q | 1.038 | o | 2.409 | opq | 150.556 | ijk | 37.962 | b-e |
| H-G6 | 25.533 | d-g | 3.155 | a-d | 2.394 | o-r | 225.000 | a-e | 37.221 | c-f |
| H-G7 | 12.833 | i-p | 2.058 | g-m | 2.393 | o-r | 43.223 | l | 34.788 | fij |
| H-G8 | 10.667 | k-q | 2.172 | f-l | 2.347 | pqr | 40.891 | l | 33.658 | j |
| H-G9 | 4.250 | pq | 1.128 | no | 2.326 | qrs | 222.222 | b-f | 37.066 | c-g |
| H-G10 | 26.833 | def | 2.850 | b-g | 2.327 | qrs | 241.667 | ab | 34.410 | hij |
| H-G11 | 41.833 | ab | 3.292 | a-d | 2.291 | rs | 80.884 | l | 36.628 | d-h |
| H-G12 | 12.833 | i-p | 2.743 | c-i | 2.304 | qrs | 158.333 | h-k | 34.698 | f-j |
| H-G13 | 18.167 | f-n | 2.613 | d-j | 2.233 | s | 144.444 | jk | 36.259 | d-i |
| H-G14 | 13.917 | h-p | 2.183 | f-l | 2.223 | s | 213.889 | b-f | 38.197 | bcd |

### 4.1.2 Morphological study of nuts

Agro-morphological important traits in almond genotypes are nut phenotypic parameters including (width, length, thickness, weight, kernel percentage) for economic and health purposes. Therefore, identification of morphological traits can be discussed alongside with genetic diversity. To improve the gene pool, the physical traits are inappropriate because environmental factors have a direct influence on the developmental stages of the plant with regard to all traits. The diversity among genotypes are just limited (Terzopoulos and Bebeli, 2008). Table (4.2) shows the mean values of width, length, thickness, weight and kernel percentage of nuts for thirty-eight almond genotypes which were different significantly. Nut width of the almond genotypes ranged between 16.180 to 27.207 mm recorded by (B-G7 and Q-G5) respectively, and genotype H-G14 recorded minimum value to nut length it was (24.180 mm) and maximum value was (41.070 mm) recorded by (Q-G2), the nuts thickness between 11.487 mm recorded by H-G3 to 16.813 mm recorded by Q-G4, nut weight values start from (2.129 g) which recorded by (B-G7) to 7.517 g which recorded by (Q-G4) and kernel percentage ranged between 16.387 to 30.835% for genotypes (M-G1 and B-G7), respectively. In one hand genotype (B-G7) recorded lowest value in nut width and weight on another hand recorded maximum nut percentage that mean, it has a soft shell with full kernel in-shell.

The results nearly agree with (Esfahlan *et al.*, 2012) who showed that values of some almond nut parameters in 40 almond genotypes were different significantly. Nut weight ranged between 3.23 to 8.34 g, nut length from 30.5 to 43.6 mm, nut width from 18.3 to 29.4 mm, and nut thickness 15.00 to 22.33 mm. Kodad *et al.* (2015) recorded physical nut traits in 45 almond Moroccan genotypes, the minimum and maximum nut widths were (15.90-27.19 mm), nut lengths (19.25 to 41.24 mm), nut thicknesses (11.48-19.61 mm), nut weights (1.15-7.34 g) and shelling percentages (19.91-63.79%). Rapposelli *et al.*, (2018) also found significant effect of almond cultivars on nut weight, length and width which ranged (1.33-7.47 g, 2.43-4.05 cm and 1.80-3.34 cm) respectively for the previous parameters. In addition, differences in agronomical nut data might be due to the insentient characteristics of genotypes (Kumar and Ahmed, 2015). Furthermore, geographical locations with cross-pollination by insect could be another evidence of almond diversity (Kester and Gradziel, 1996; Woolley *et al.*, 2000).

The principal component analyses (PCA) plot (Fig. 4.1) showed the distribution of all genotypes and nut morphological data on the plot. The plot demonstrates a negative relationship between shell to the kernel and nut weight. It also displayed a positive linkage between nut width and thickness.

**Table 4.2 Effect of almond tree genotypes on some nut physical characteristics.**

| Genotype | Nut width mm | | Nut length mm | | Nut thickness mm | | Nut weight g | | Kernel % | |
|---|---|---|---|---|---|---|---|---|---|---|
| S-G1 | 23.413 | cde | 38.623 | a-d | 14.877 | c-g | 4.942 | cde | 19.085 | j-m |
| S-G2 | 20.177 | h-l | 33.350 | f-m | 14.807 | c-g | 4.029 | e-l | 17.933 | lmn |
| S-G3 | 19.197 | j-o | 32.133 | j-n | 13.867 | e-l | 4.611 | c-h | 21.328 | e-j |
| S-G4 | 20.610 | h-k | 34.423 | e-l | 15.033 | c-g | 4.981 | cde | 19.801 | h-l |
| S-G5 | 20.787 | g-j | 36.823 | b-f | 15.217 | b-e | 4.598 | c-h | 21.209 | f-j |
| S-G6 | 20.780 | g-j | 36.580 | c-g | 15.613 | a-d | 4.080 | e-l | 22.076 | e-h |
| S-G7 | 20.390 | h-k | 39.647 | abc | 14.373 | c-j | 5.005 | cde | 22.555 | def |
| S-G8 | 22.520 | d-g | 38.543 | a-d | 15.697 | abc | 4.247 | e-k | 21.904 | e-i |
| S-G9 | 17.643 | n-q | 36.400 | c-h | 12.980 | k-n | 3.754 | h-m | 24.568 | cd |
| M-G1 | 22.763 | def | 40.390 | ab | 14.620 | c-i | 4.911 | c-f | 16.387 | n |
| M-G2 | 21.920 | e-h | 32.863 | g-n | 15.060 | c-g | 4.353 | c-j | 16.488 | n |
| M-G3 | 23.630 | cde | 40.547 | ab | 14.823 | c-g | 6.311 | b | 16.971 | mn |
| Q-G1 | 21.223 | f-i | 30.743 | lmn | 14.947 | c-g | 4.463 | c-i | 19.676 | h-l |
| Q-G2 | 26.653 | a | 41.070 | a | 14.600 | c-i | 4.790 | c-g | 29.136 | ab |
| Q-G3 | 20.907 | g-j | 25.747 | o | 16.370 | ab | 3.720 | h-m | 27.740 | b |
| Q-G4 | 25.967 | ab | 35.520 | d-k | 16.813 | a | 7.517 | a | 16.409 | n |
| Q-G5 | 27.207 | a | 35.683 | d-j | 14.567 | c-i | 6.406 | b | 17.083 | mn |
| B-G1 | 19.340 | j-n | 29.727 | mn | 12.660 | l-o | 2.400 | op | 20.706 | f-j |
| B-G2 | 19.150 | j-o | 31.777 | k-n | 12.277 | mno | 3.443 | j-n | 25.289 | c |
| B-G3 | 24.563 | bc | 38.050 | a-e | 14.617 | c-i | 5.244 | cd | 18.220 | k-n |
| B-G4 | 24.043 | cd | 32.617 | h-n | 14.673 | c-h | 4.482 | c-i | 20.358 | f-k |
| B-G5 | 17.490 | opq | 28.973 | n | 12.907 | k-n | 3.504 | i-n | 20.357 | f-k |
| B-G6 | 16.543 | pq | 32.437 | i-n | 11.873 | no | 2.129 | p | 19.795 | h-l |
| B-G7 | 16.180 | q | 30.227 | mn | 13.380 | h-m | 2.173 | p | 30.835 | a |
| H-G1 | 19.113 | j-o | 31.537 | lmn | 12.990 | k-n | 3.818 | g-l | 23.575 | cde |
| H-G2 | 22.780 | def | 33.020 | g-m | 15.300 | bcd | 2.803 | m-p | 20.568 | f-k |
| H-G3 | 18.180 | m-p | 31.550 | lmn | 11.487 | o | 3.440 | j-n | 19.636 | i-l |
| H-G4 | 22.843 | c-f | 29.803 | mn | 15.133 | b-f | 2.731 | nop | 21.290 | e-j |
| H-G5 | 22.547 | d-g | 37.190 | b-e | 13.163 | j-n | 5.313 | c | 19.157 | j-m |
| H-G6 | 19.780 | i-m | 36.210 | c-i | 13.270 | i-m | 3.914 | f-l | 22.561 | def |
| H-G7 | 24.280 | cd | 36.197 | c-i | 14.223 | d-k | 4.300 | d-k | 21.997 | e-i |
| H-G8 | 22.963 | c-f | 30.423 | mn | 12.920 | k-n | 4.073 | e-l | 19.440 | jkl |
| H-G9 | 21.497 | fgi | 39.103 | a-d | 13.770 | f-l | 4.055 | e-l | 20.772 | f-j |
| H-G10 | 20.720 | h-k | 32.673 | h-n | 14.817 | c-g | 4.014 | e-l | 20.409 | f-k |
| H-G11 | 18.563 | l-o | 31.593 | l-n | 13.443 | h-m | 3.420 | j-n | 25.666 | c |
| H-G12 | 17.740 | n-q | 31.797 | k-n | 13.073 | j-n | 3.323 | k-o | 20.156 | g-l |
| H-G13 | 18.963 | k-o | 24.883 | o | 13.737 | g-l | 3.133 | l-o | 24.754 | c |
| H-G14 | 16.810 | pq | 24.180 | o | 12.240 | mno | 2.476 | op | 22.447 | d-g |

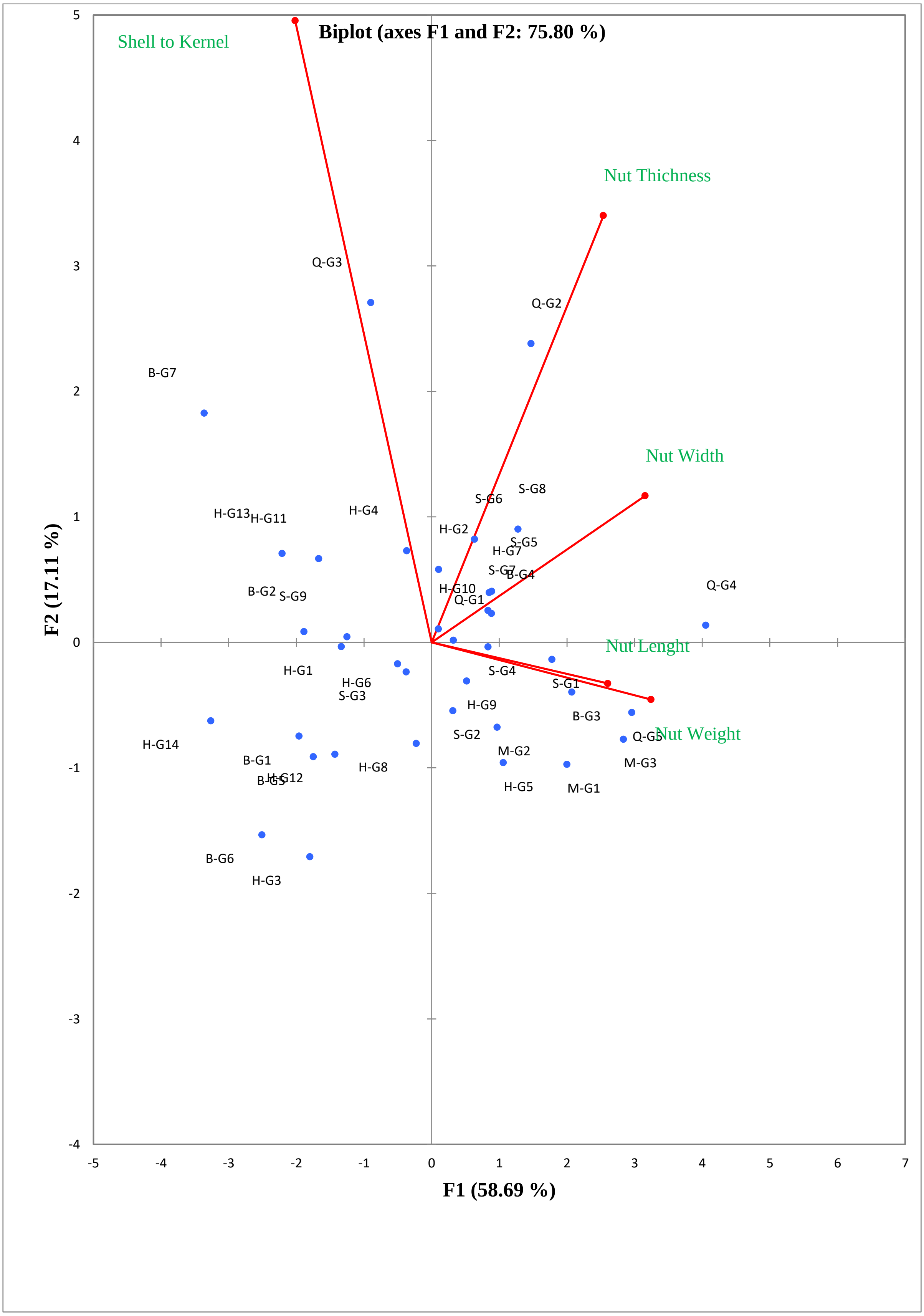

**Figure 4.1 PCA Plot among 38 genotypes accessions based on 5 nut characteristics at different locations.**

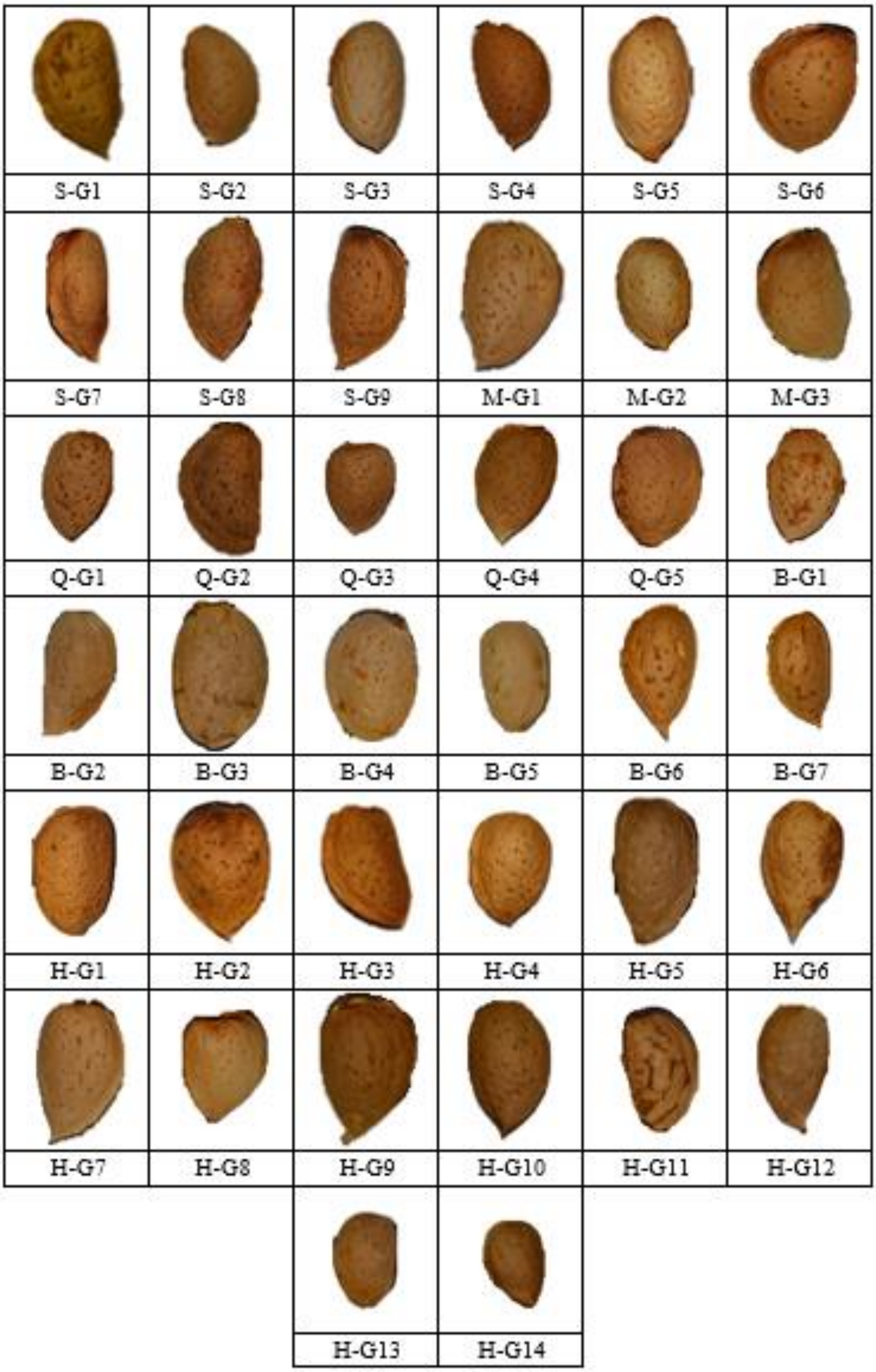

**Figure 4.2 Some morphological shapes of almond genotypes nut.**

## 4.2 Leaves Phytochemical Analyses

Investigation and determination of phytochemical analyses of leaves during drought stress are important factors for almond improvements in the world. Traditionally, Natural products derived from plants are used for health supplements (Zafar *et al.*, 2013), but recently, beneficial biological effects of specific phytochemicals and molecular mechanisms still remain as a subject of intense research. Therefore, phytochemical parameters have been proven to be good indicators of drought in different studies ( Šircelj *et al.*, 2007; Isah, 2019). In this study, some important phytochemical parameters were determined from leaves of different almond genotypes such as; chlorophyll, proline, phenolic content, flavonoids content, saponin, tannin, antioxidants activity (DPPH assay ABTS assay and total antioxidant capacity). Drought stress also prevents the photosynthesis of plants, by affecting chlorophyll mechanisms which results in chlorophyll content changes and also by damaging the photosynthetic (Mafakheri *et al.*, 2010).

### 4.2.1 Chlorophyll concentration

Chlorophyll is the most important green pigment, which is present in all green plants and responsible for the absorption of light to provide energy for photosynthesis. S-G5 genotype recorded significant maximum value for chlorophyll concentration (44.267 SPAD) while H-G2 possessed the lowest (20.233 SPAD) as shown in (Fig. 4.3). In Hawraman location, 13/14 genotypes were recorded lowest than (30 SPAD) which may be due to that genotypes of this location were different genetically from other location genotype. Sepehri and Golparvar (2011) detailed that chlorophyll content affected by plant genotypes and environmental conditions. We suggest that genotype have strong effect on chlorophyll. On another hand, in Sharbazher location the chlorophyll concentration recorded maximum value compared to other locations, nearly ranged between (35-44 SPAD), this may be due to higher value of proline accumulation. Accumulation of different sugars, active ions and amino acids similar to proline in plant is responsible for cells osmotic adjustment and osmotic adjustments keep cell expansion, turgor pressure and growth, and water flow during water shortage periods that protect leaves greenness, dipping the quantity of chlorophyll affected by the water shortage stress is related to rise of oxygen radicals in cells. Free radicals cause peroxidation and so chlorophyll pigments degradation. It looks that the decrease of chlorophyll concentration under drought is mostly because of the activity of chlorophylls enzyme, phenolic compounds and peroxidase, ensuing in degradation of chlorophyll (Salehi *et al.*, 2016). This result nearly agrees with (Isaakidis *et al.*, 2004) who recorded (37-42 SPAD) for deferent almond trees.

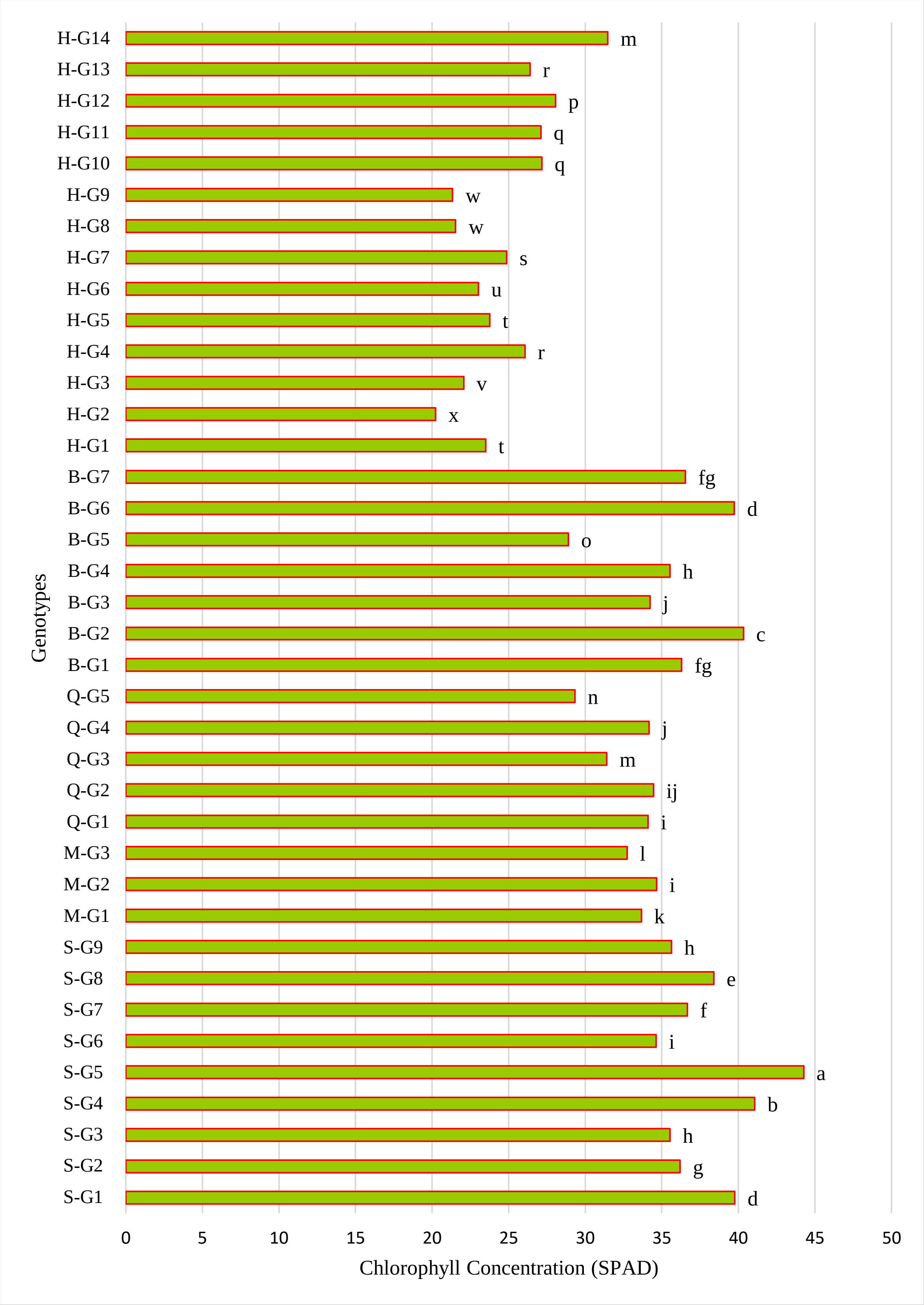

**Figure 4.3 Effect of almond genotypes on chlorophyll concentration.**

### 4.2.2 Proline content

Fig. (4.4) demonstrates that significant variability was observed among the genotypes in their ability for proline accumulation. Proline content in (B-G6) genotype was the highest (238.624 µmol $g^{-1}$ FW), while it is the lowest value (65.961 µmol $g^{-1}$ FW) in (H-G3) genotypes. In Barznja location, the proline content nearly ranged between (77-238 µmol $g^{-1}$ FW) and in Hawraman ranged between (65-114 µmol $g^{-1}$ FW), this result may be to the different in the summer temperature when in first the temperature is more than second location. Also, the genotypes at Hawraman were differed from other genotypes from all other studded locations. Proline content may be due to the presence of different genotypes at various geographical locations (Mafakheri *et al.*, 2010).

Amino acid synthesizes from the primary elements, the carbon and oxygen blained from air, hydrogen from water in the soil, forming carbon hydrate by means of photosynthesis and combining it with nitrogen which the plants obtain from the soil, leading to synthesis of amino acids are part of these proteins and have metabolic activity. Proline is an amino acid which plays a highly beneficial role in defending the plants from various stresses and supporting them to recover from stress more quickly (Hayat *et al.*, 2012). Proline production is one of the common physiological responses found in higher plants when they are exposed to adverse environmental conditions. In higher plants, accumulation of proline is an indication of disturbed physiological condition, which is activated by biotic or abiotic stress condition. Some factors such as drought, salinity, cold, heavy metals, or certain pathogens increased free proline content in plants. Therefore, the suitable analysis to screen physiological status and to assess stress tolerance of higher plants is the determination of free proline levels (Sunkar, 2010). The roles of proline under varying environments have been critically examined, many reports argued that proline content significantly increased when water is limited in the soil. Thus, accumulation of proline in the plant, particularly under various biotic and abiotic stresses can help in: first enhancing growth and other physiological characteristics of plants, second scavenges the ROS generated, thirdly, affects plant-water relations by maintaining turgidity of cells under stress and also increases the rate of photosynthesis. Fourthly, protects the plant from harmful radiation such as UV-B (Hayat *et al.*, 2012).

In drought conditions, proline is assumed to contribute to scavenge ROS, osmotic adjustment, membrane stability. Proline accumulation plays adaptive roles when drought stress accrues in plant, also it has been suggested to work as a compatible osmolyte and to be a carbon and nitrogen storage. On another hand it is been proposed to act as molecular chaperone stabilizing the structure of proteins, and proline accumulation can provide a way to buffer cytosolic pH and to balance cell redox status (Pessarakli, 2016). The requirement of amino acids is essential to increase yield and overall quality of crops.

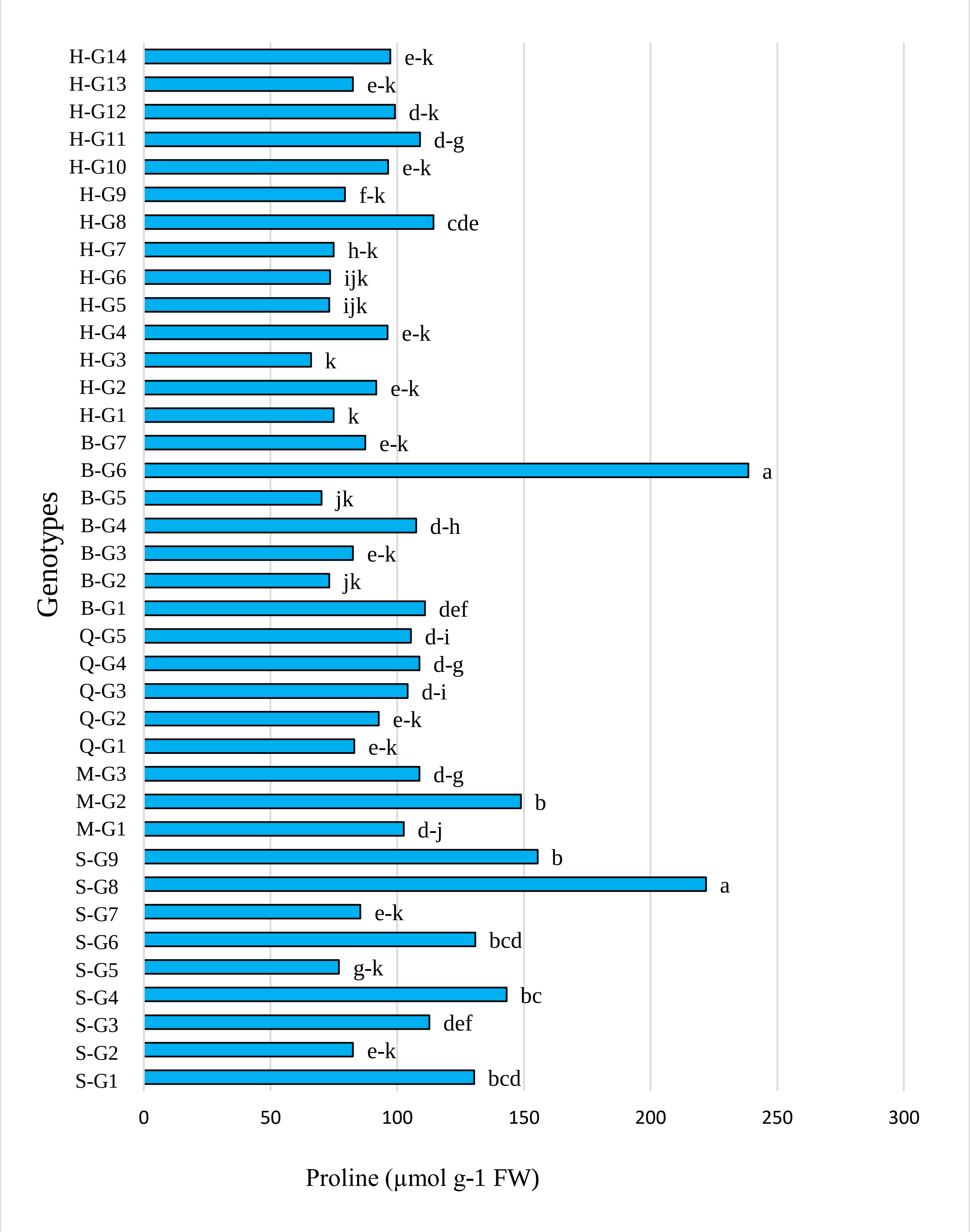

**Figure 4.4 Effect of almond genotypes on proline content.**

### 4.2.3 Total phenolic contents

Fig. (4.5) indicates that almond genotypes has significant effects on total phenolic content. (Q-G5) recorded the top value (11.070 mg GAE $g^{-1}$ E) compared to (2.144 mg GAE $g^{-1}$ E) as least for (S-G4) genotype. In our study, total phenolic content varied significantly according to genotypes. In Qaradagh locations significant differences was observed between genotypes and the total phenol content and it was ranged between (2.439-11.70 mg GAE $g^{-1}$ E) that mean the genotypes have strong effect on total phenol content compared to locations, this wide range also observed in (Sharbazher, Mergapan, Barznja, and Hawraman and there ranges were (2.144-6.754, 2.928-3.407, 2.509-5.022 and 2.293-7.431 mg GAE $g^{-1}$ E), respectively. Researchers reported that chemical variations between the populations and plant sections could be due to different genetic, environmental, geographical and morphological factors (Çirak *et al.*, 2011; Hamid *et al.*, 2011; Čolić *et al.*, 2017). Cosmulescu and Trandafir (2011) reported that there could be a correlation between phenolic content, season, genetic and ecological factors in walnut leaves.

Phenolic compounds are the major group of phytochemicals in almonds which are considered the most important compounds exhibiting antioxidant properties (Sivaci and Duman, 2014). They are secondary metabolites present in different parts of all plant species. The metabolism of phenolics compounds are related to the biochemical and morphological regulatory patterns of plants. Many types of stresses activated by ecological conditions, pathogens, and damages occurred to plants are known to induce and effect the generation of phenolic substances. Therefore, phenolics play critical roles in the defense mechanisms of plants. Particularly, they are good protective substances against different stresses (Jahanban-Esfahlan *et al.*, 2019).

Many researchers demonstrated that total phenolic components should be determined on the monthly basis for insistence. Sivaci and Duman (2014) demonstrated that total phenolic compound values detected in the leaves of three almond varieties (Texas, Ferragnes and Nonpareil) were (2.03, 2.82 and 8.15) µg $mg^{-1}$ fresh weight respectively, they emphasized that antioxidant activity and phenolics in almond leaves was varied by season.

In plants, phenolic accumulation is regularly a steady feature of plants under stress, which characterizes as a defense mechanism to plant with abiotic stresses. Phenolics act an important role in physiological processes to improve the tolerance of plants under stress conditions (Sharma *et al.*, 2019).

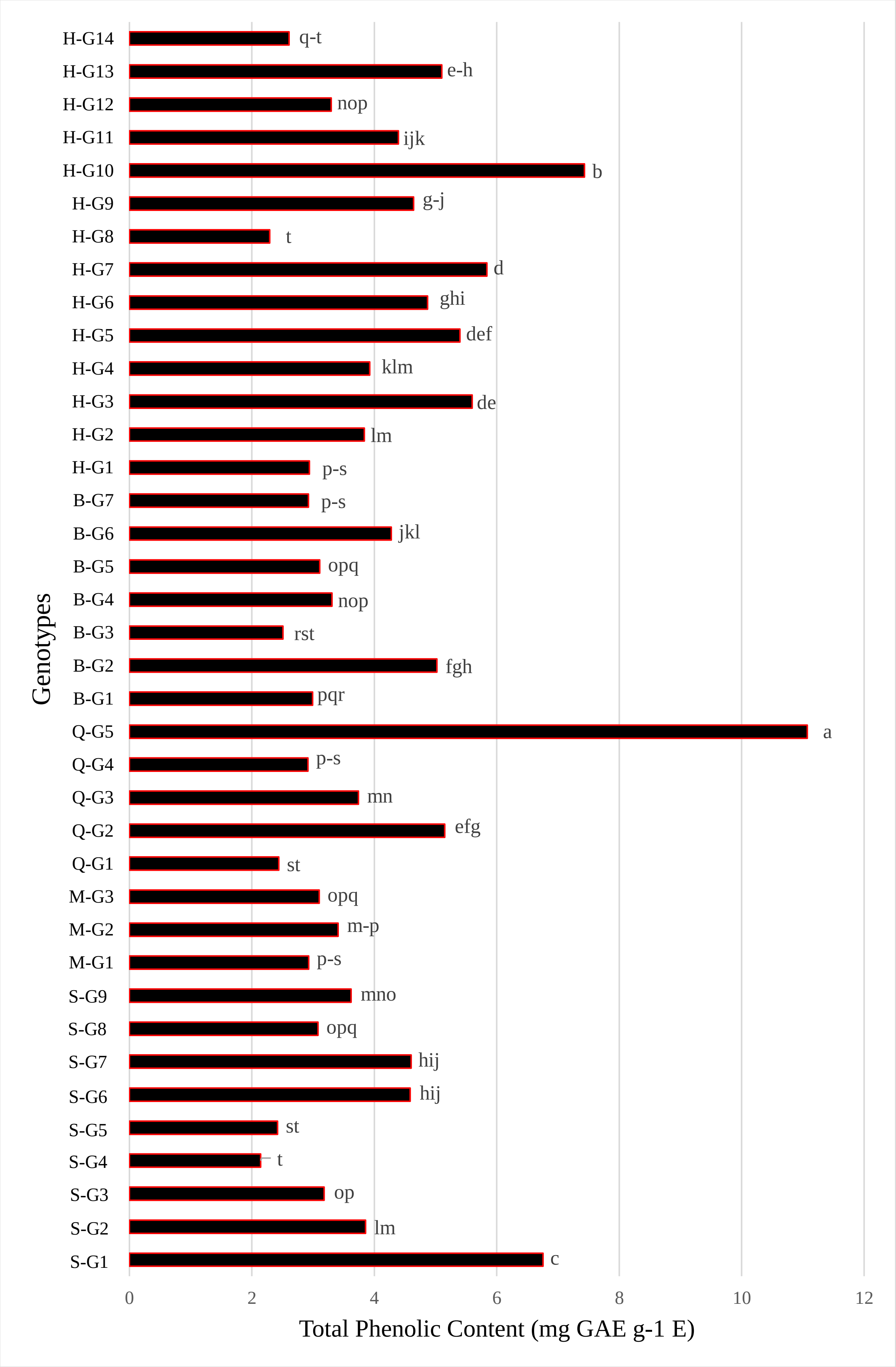

**Figure 4.5 Effect of almond genotypes on total phenolic content.**

### 4.2.4 Total flavonoids content

Fig (4.6) exhibits that Q-G5 was significantly superior in total flavonoids content (4.931 mg QE $g^{-1}$ E) which was maximum value compared to minimum (1.103 mg QE $g^{-1}$ E) for Q-G4 genotype, in same location the highest and lowest value were recorded that mean genotypes have super effect on total flavonoids content. Hughey *et al.* (2008) showed that flavonoid is under genetic control, they described that almonds Carmel varieties had 47% more flavonoids than Nonpareil. The composition of flavonoids in plants is influenced by several factors such as: variety, geographical region, ripeness stage, processing, storage, environmental conditions, exposure to pests and diseases, and UV radiation (Milbury *et al.*, 2006).

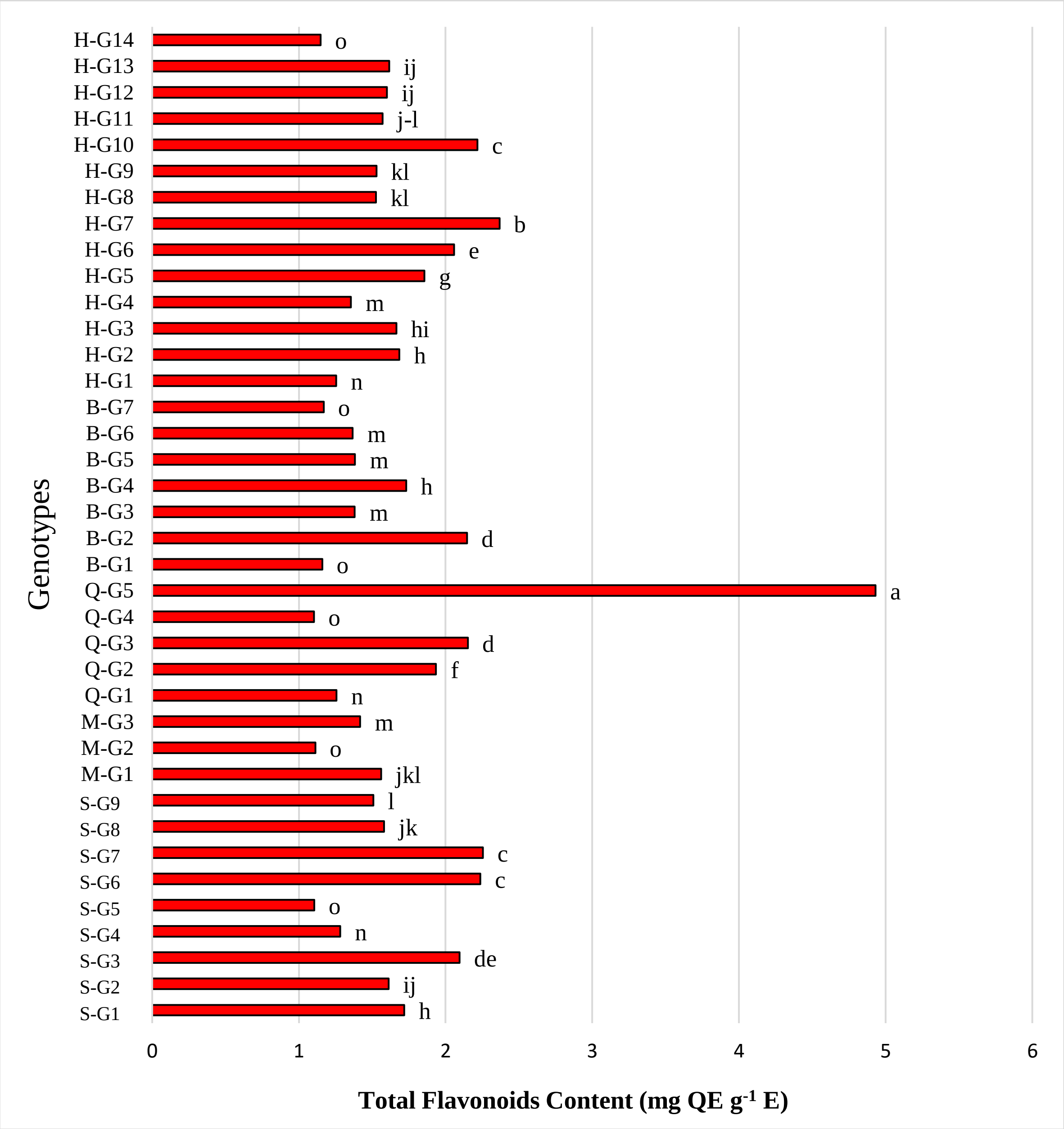

**Figure 4.6 Effect of almond genotypes on total flavonoids content.**

### 4.2.5 Saponin content

Fig. (4.7) illustrates statistical differences among genotypes with respect to saponin content, the highest value (38.005 mg SE $g^{-1}$ E) was recorded for H-G5 genotype while the lowest (10.314 mg SE $g^{-1}$ E) was for Q-G1 genotype.

The result clearly showed that genotypes have significant effect on saponin accumulation on plant when in all locations significant effect between genotypes was observed, for example in Hawraman location the saponin content ranged between (11.168-38.005 mg SE $g^{-1}$ E) and in Qaradagh location the values were lowest value (10.325 mg SE g-1 E) reached to (31.206 mg SE g-1 E).

Saponins are natural phytochemical products, spread generally in plants of different structures and functions. Saponins are derived from aglycone structure which consists of chemically complexes of varied groups of compounds including, triterpenoid and steroidal aglycones. It has an important role in plant ecology, ecosystem and also for a wide range of commercial prospects such as: cosmetic and pharmaceutical sectors, food applications. They have antioxidant properties. Cultivar and genotype reliant on dissimilarity in the saponin content of plants (Moses *et al.*, 2014). Generally, in many plants, saponins are made and stored during normal growth and development conditions. Many researchers demonstrated variations in distribution, composition and amounts of this natural substance among plant species, individual plants, organs and tissues during development and maturation. Seasonal variations may be a reflection of varying needs for plant protection.

In addition, in several plant species, the production of saponins is induced in response to abiotic and biotic stress including humidity, nutrient starvation, light, temperature and pathogen attack, consequently they influence both the quality and quantity of saponin content (Szakiel *et al.*, 2011; Costa *et al.*, 2014).

Saponins are one of the largest classes of plant natural products, the majority of the producing plant species are dicotyledonous and accumulate triterpenoid type saponins, while monocots mostly synthesize steroidal saponins (Bordbar et al., 2011).

Kumar *et al.* (2015) illustrated that drought caused growing saponin content in plants, it was linked to its defensive role against oxidative stress. In drought stress saponins work on the membrane permeability and it is the physiological responses to drought stress.

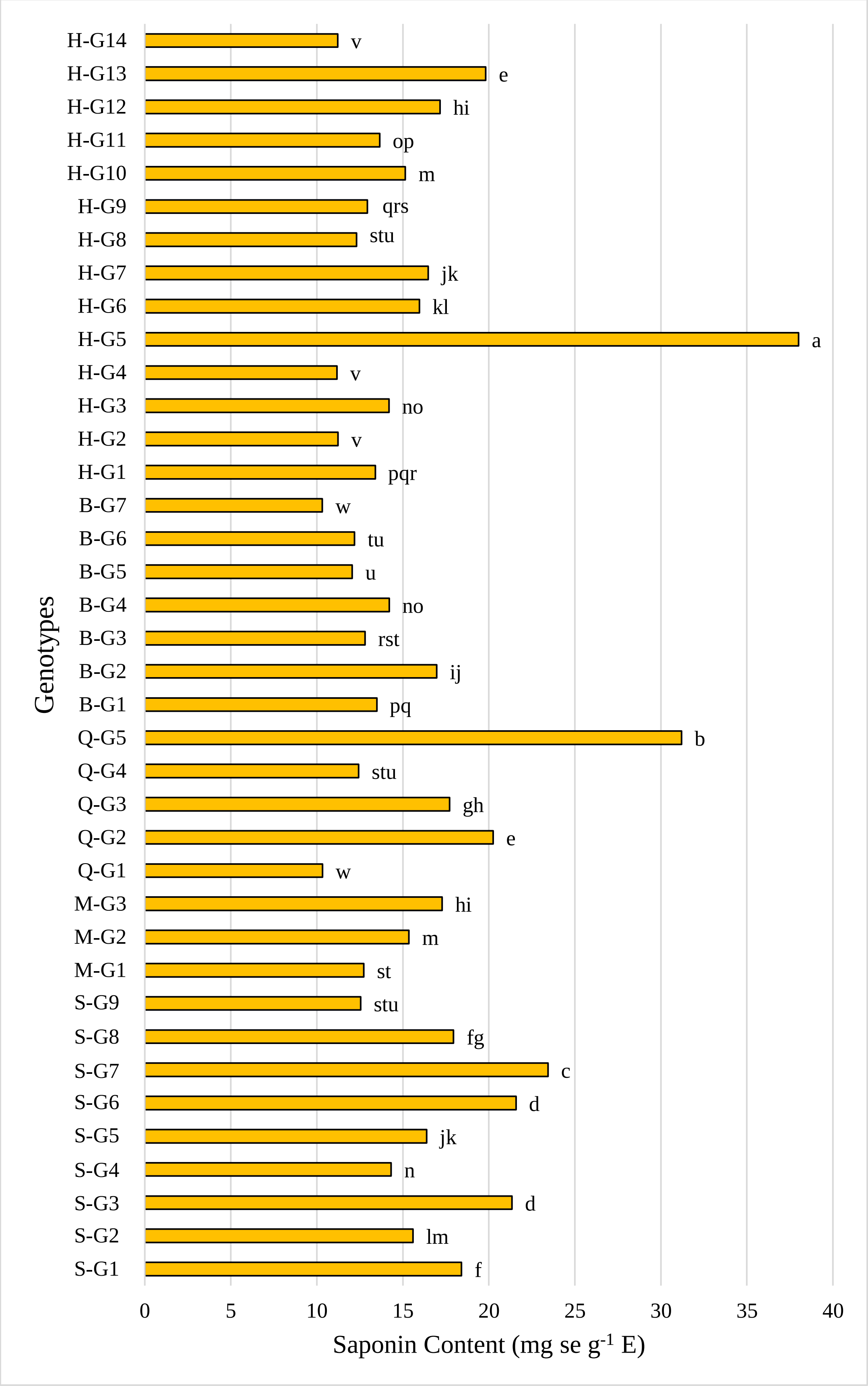

**Figure 4.7 Effect of almond genotypes on saponin content.**

### 4.2.6 Condensed tannin content

Fig (4.8) shows the significant variation in condensed tannin content among the genotypes under the genotypes effect, highest value of condensed tannin content was achieved from genotype number five at Qaradagh location (7.412 mg CE $g^{-1}$ E) but the lowest value was (2.062 mg CE $g^{-1}$ E) recoded in same location by genotype number one.

On one hand the genotype (Q-G5) recorded maximum value in total phenolic contents and total flavonoids content. Espinosa (2018) mentioned that condensed tannin polymerized products of flavan-3,4-diol and flavan-3-ol or a mix of both. On the other hand, genotype (Q-G1) in total phenolic contents recorded low value and it was not significant with the minimum value, this result agrees with that explained by Wu *et al.* (2016) when they described that tannin is polymerized from other polyphenols. Also, they mentioned that tannin content was mainly depended on genotypes, and the results were clearly showed this fact because in the same location (Qaradagh) the maximum and minimum values were recorded and genotypes recorded significant differences, this result may be to the effect of genotypes, and also in other locations the same result were observed. Madritch and Lindroth (2015) mentioned that tannin production can differ a hundred-fold together within and among species.

Tannin is a natural product which consists of condensed (CTs) and hydrolysable tannins (HTs). It is the second most abundant polyphenol after lignin that are common in most plant species (Kraus *et al.*, 2003). The main function of tannin is defending and protecting plants against biotic and abiotic stresses during the active growth of tissues and after tissue senescence and it is work as antioxidant. However, in plants, tannin production is genetically as well as environmentally controlled. Production of tannins has been shown to respond to environmental changes particularly (drought and temperature increase), also climate influences on chemical composition of tannins between green and senescent tissues (Top *et al.*, 2017). In addition, Tannin increased during limiting moisture in the soil.

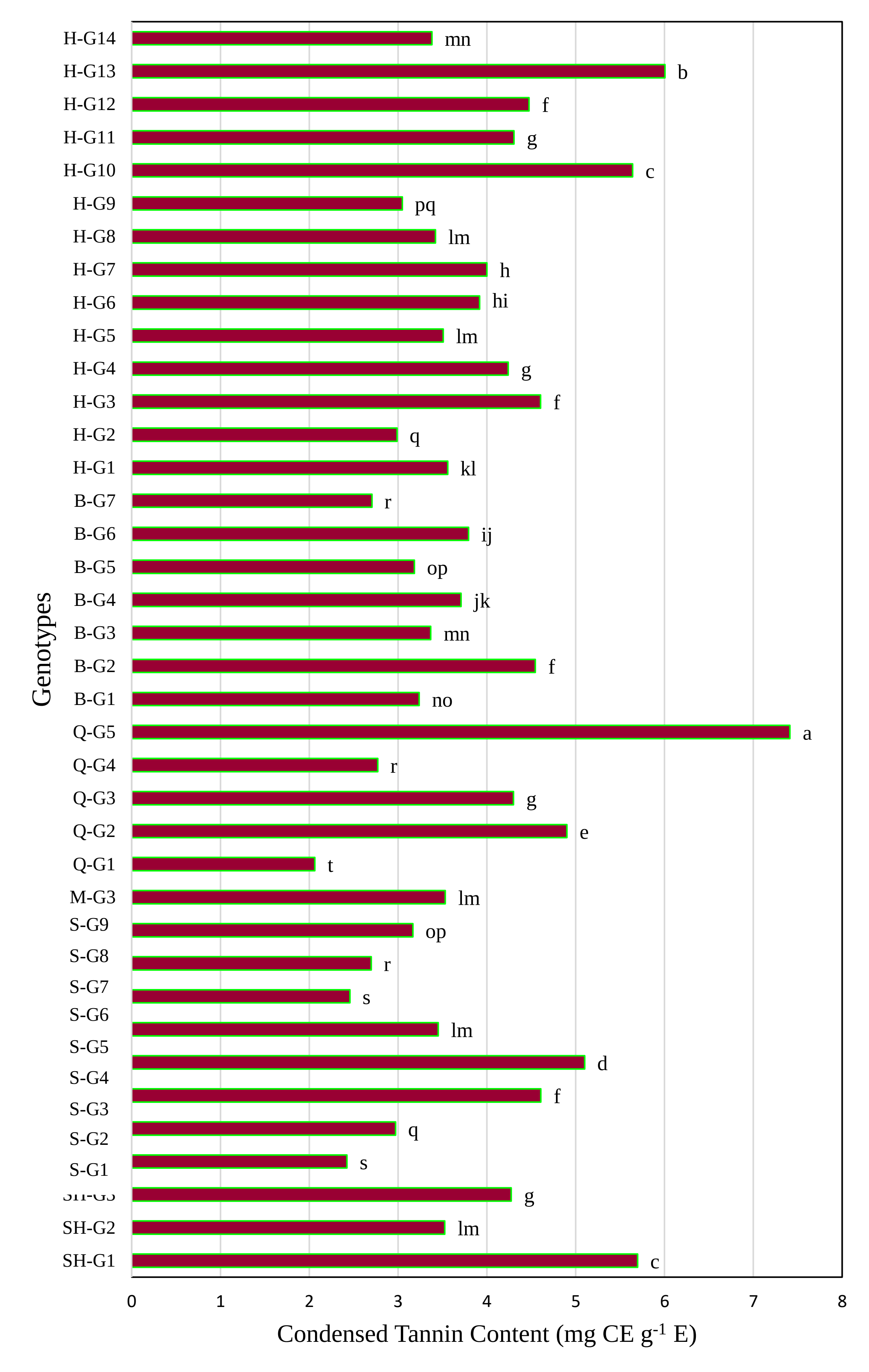

**Figure 4.8 Effect of almond genotypes on condensed tannin content.**

### 4.2.7 Determination of antioxidants activity

#### 4.2.7.1 Antioxidant activity by DPPH assay

Antioxidant activity by DPPH assay in almond leaves differed significantly among almond genotypes, the value (89.081 %inhibition) recorded by genotype (S-G6) was the highest value, (Q-G1) genotype recorded lowest (40.255 %inhibition) (Fig. 4.9). It is clear that genotype have significant effect on antioxidant activity, the genotypes were studded at Qaradagh location showed significant variation effect on antioxidant as well, in other locations the differences between antioxidant activity on genotypes were clear. The degree of activities of antioxidant systems under drought stress is exceptionally variable owing to variation in plant species, in the cultivars of the same species, development and the metabolism of the plant the duration and intensity of the stress (Vats, 2018).

#### 4.2.7.2 Antioxidant activity by ABTS assay

Fig (4.10) shows a significant effect of almond genotypes on antioxidant activity by ABTS assay in almond leaves. The values among genotypes ranged between (S-G7 = 98.077 %inhibition - Q-G1 = 46.698 %inhibition). We detected significant effect between genotypes in all locations may be because the ability of genotype to product antioxidant, Sivaci and Duman (2014) determined that antioxidant activity depended on the cultivar.

One of the most important phytochemicals which prevents the oxidation of molecules inside plant cell is antioxidants, that inhibit the initiation of oxidative chain reactions and delay the oxidation of lipids, consequently, they have positive effects on well-being of plant (Kasote *et al.*, 2015). It is clear that phonological traits are affected by drought and heat stress, which have a significant role in the adaptation of plants during different environmental factors (Basu and Maier, 2016). Generally, in almost all types of environmental stress conditions free reactive oxygen species (ROS) system could occur. Therefore, to prevent stress-induced oxidative damage, plants improve their antioxidant defense system to scavenge reactive oxygen species (ROS) (Haider *et al.*, 2018). In addition, to escape the toxic effects of free radicals, a complex of enzymatic and non-enzymatic antioxidant defense systems are effective in the plant (Kasote *et al.*, 2015).

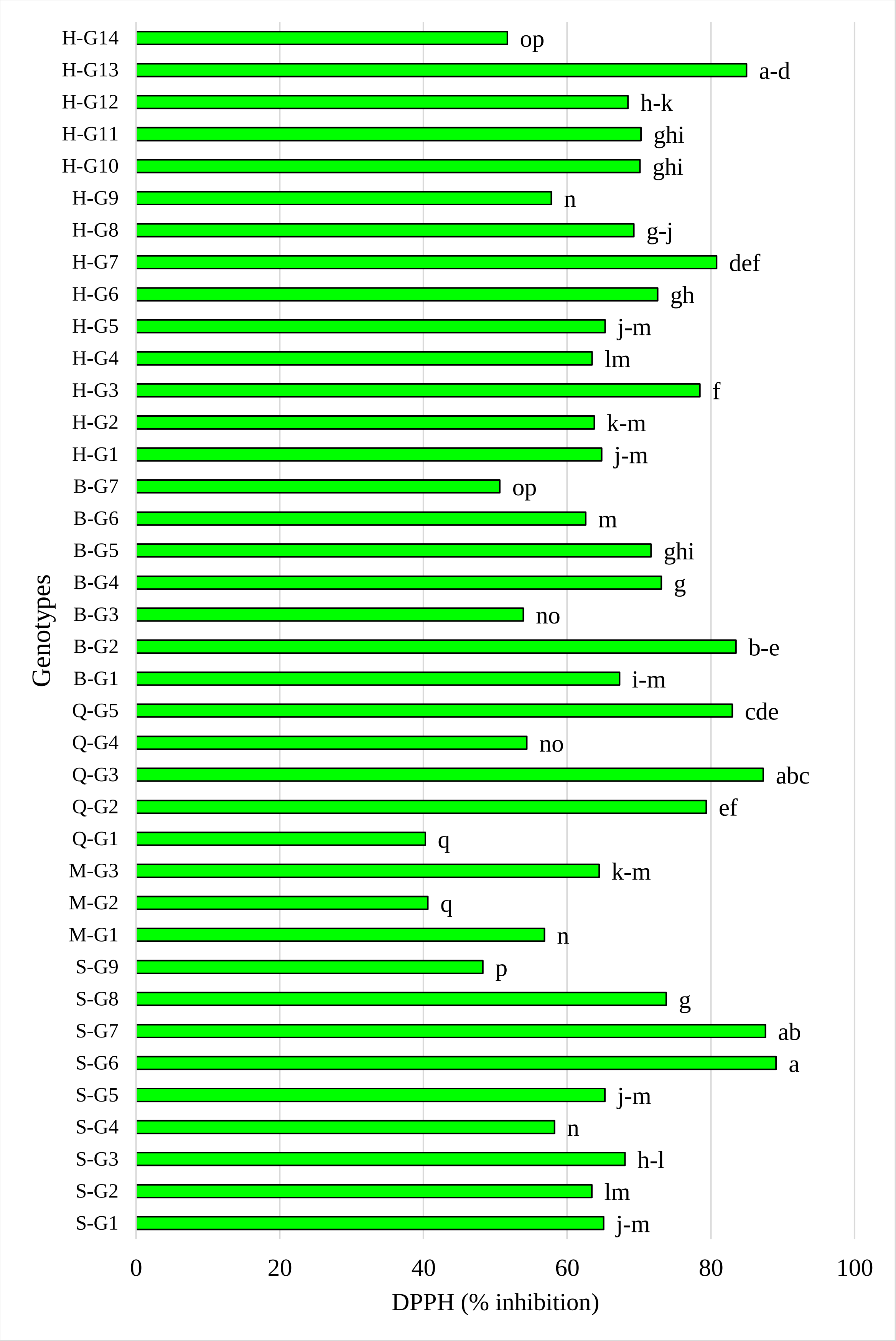

**Figure 4.9 Effect of almond genotypes on antioxidants activity by (DPPH) assay.**

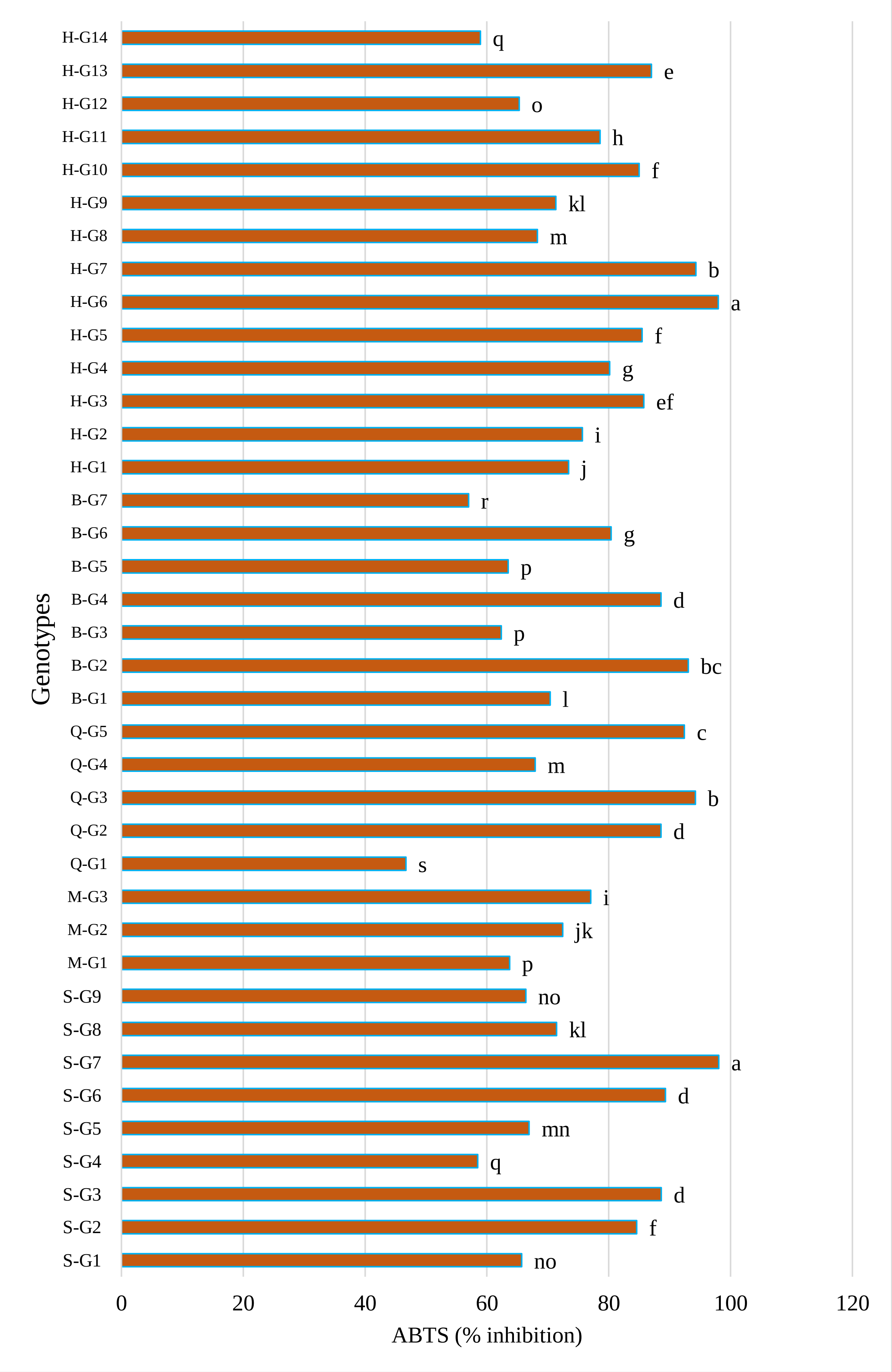

**Figure 4.10 Effect of almond genotypes on antioxidants activity (ABTS) assay.**

#### 4.2.7.3 Total antioxidant capacity

The values of total antioxidant capacity ranged from (0.338 - 0.004 mg AA $g^{-1}$ E) in which the highest TAC value was obtained from Q-G2 genotype while the lowest capacity recorded by (H-G9) genotype (Fig. 4.11), there were clear that genotypes have strong effect on total antioxidant capacity. Vats (2018) illustrated that genotypes have significant effect on antioxidant activity.

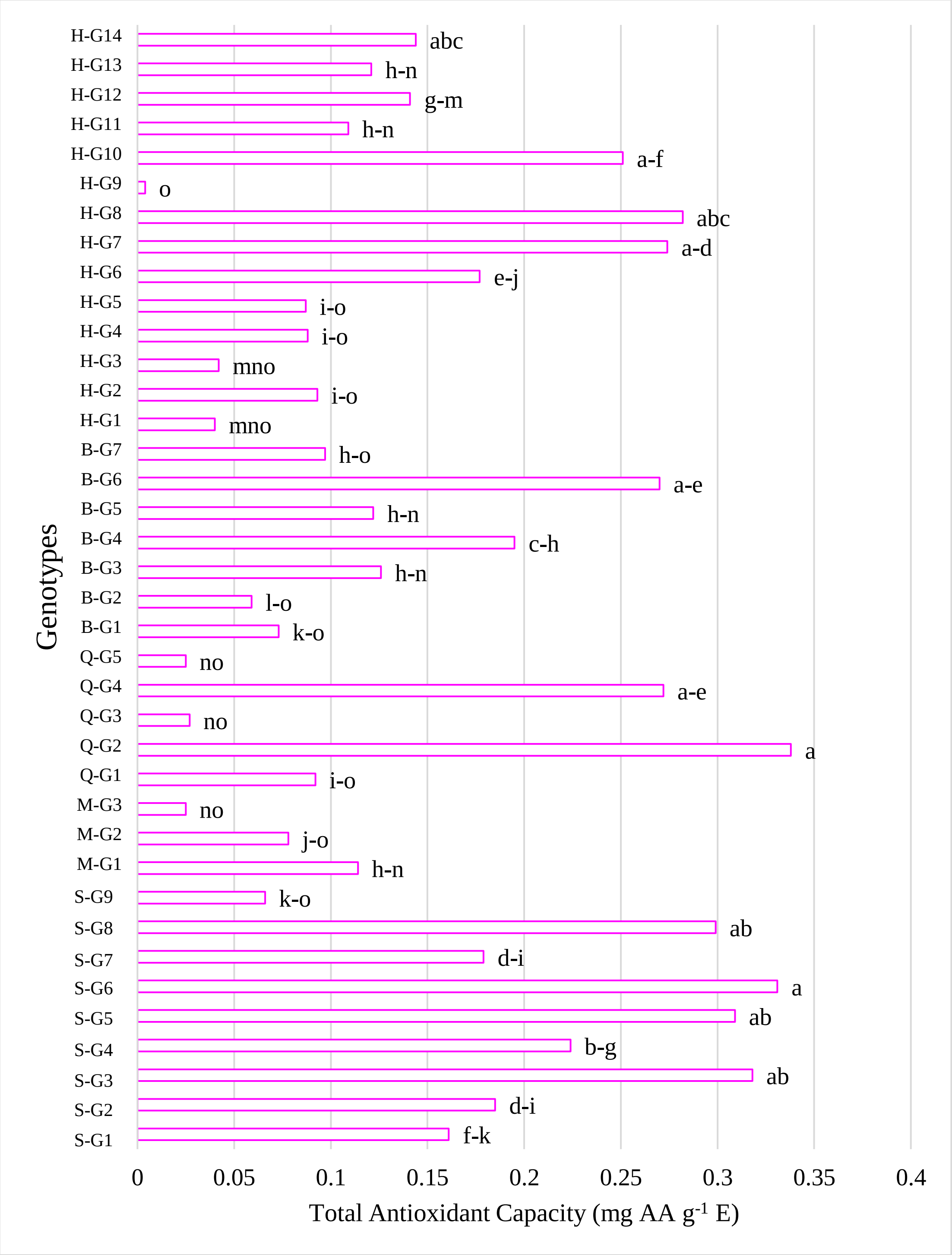

**Figure 4.11 Effect of almond genotypes on total antioxidant capacity.**

## 4.3 Characterization of Genetic Diversity and Relationship in Almond Genotypes by RAPD and ISSR Markers

### 4.3.1 Allelic variation in almond genotypes using RAPD and ISSR markers

Of the primers tested, 20 (out of 21) RAPD and 15 (out of 17) ISSR primers were confirmed to amplify fragments for their reproducibility and high polymorphism (Table 4.3). The maximum, minimum and mean values of polymorphic bands were (5, 15 and 9.5) for RAPD and (4, 12 and 8) for ISSR, respectively. The mean of major allele frequency for RAPD and ISSR were (0.18 and 0.29), respectively. Mean values of gene diversity were (0.92 and 0.82) for RAPD and ISSR, respectively. The PIC values were recorded for RAPD primes that ranged between (0.77 to 0.97), the mean was (0.92) and for ISSR primers between (0.35 to 0.96), the mean was (0.81). The PIC values discovered are nearly similar to those reported such as the PIC values for 16 RAPD primers exhibited by (Sharma *et al.*, 2012) ranged from (0.26 to 0.87). Mean value of PIC was reported (0.77) using 80 primers of RAPD to 29 almond cultivars (Sorkheh *et al.*, 2009b), while it was ranged 0.47 to 0.97 using 42 RAPD randomly primers to 39 almond varieties (Shiran *et al.*, 2007). In addition, for ISSR markers, range of PIC was from (0.59 to 0.69) using 21 primers applied to 29 *Prunus* species (Sarhan *et al.*, 2015). Another researcher used 9 ISSR primers in the peach plant and found that the PIC ranged between (0.71 to 0.88) (Tian *et al.*, 2015). Furthermore, ranges (0.13 to 0.47) and (0.12 to 0.47) of PIC values were verified after using (37 RAPD and 38 ISSR) random primers respectively for 45 peach cultivars (Sharma and Sharma, 2018). Regarding the polymorphic bands, (El Hawary *et al.*, 2014) demonstrated 2.8 mean value of the polymeric band for 10 primers. Also the polymorphic band mean was recorded as 8.36 (Gouta *et al.*, 2008). Abodoma *et al.* (2017) also reported that some Libyan almond polymorphic bands for using nine ISSR primer was 13.2 and mean value of polymeric band was 5.53. Used 13 primers by Cabrita *et al.* (2014) recorded polymorphic band of 4.23 for 13 RAPD primers (Pinar *et al.*, 2015) but 5 was recorded for 4 ISSR primers. Moreover, the average allele polymorphic was 18.6 per primer using ISSR primers applied to 29 *Prunus* species (Sarhan *et al.*, 2015).

**Table 4.3 Markers names, numbers of polymorphic bands, major allele frequencies, gene diversities and PIC values of 20 RAPD and 15 ISSR markers.**

| Marker | Number of polymorphic bands | Major allele frequency | Gene diversity | PIC |
|---|---|---|---|---|
| **RAPD** | | | | |
| OPA-08 | 5 | 0.18 | 0.89 | 0.88 |
| OPA-10 | 11 | 0.13 | 0.96 | 0.96 |
| OPA-11 | 11 | 0.05 | 0.97 | 0.97 |
| OPA-16 | 7 | 0.13 | 0.94 | 0.94 |
| OPB-11 | 12 | 0.21 | 0.92 | 0.92 |
| S075 | 15 | 0.08 | 0.96 | 0.96 |
| S084 | 10 | 0.13 | 0.94 | 0.94 |
| S085 | 12 | 0.24 | 0.92 | 0.92 |
| S081 | 8 | 0.32 | 0.87 | 0.86 |
| S093 | 11 | 0.45 | 0.78 | 0.77 |
| S078 | 12 | 0.21 | 0.91 | 0.91 |
| S094 | 9 | 0.08 | 0.96 | 0.96 |
| S087 | 8 | 0.24 | 0.87 | 0.86 |
| S088 | 10 | 0.16 | 0.95 | 0.95 |
| S089 | 8 | 0.08 | 0.96 | 0.95 |
| S090 | 7 | 0.13 | 0.93 | 0.93 |
| S091 | 7 | 0.24 | 0.89 | 0.88 |
| S092 | 9 | 0.16 | 0.93 | 0.93 |
| S095 | 10 | 0.16 | 0.93 | 0.93 |
| S073 | 8 | 0.16 | 0.93 | 0.92 |
| **Mean** | **9.5** | **0.18** | **0.92** | **0.92** |
| **ISSR** | | | | |
| 807 | 7 | 0.63 | 0.59 | 0.58 |
| 17898A | 11 | 0.08 | 0.96 | 0.96 |
| HB04 | 10 | 0.13 | 0.94 | 0.93 |
| HB8 | 7 | 0.08 | 0.96 | 0.96 |
| HB10 | 4 | 0.53 | 0.66 | 0.62 |
| HB11 | 7 | 0.13 | 0.92 | 0.91 |
| HB12 | 12 | 0.18 | 0.92 | 0.91 |
| HB15 | 4 | 0.79 | 0.37 | 0.35 |
| AG7YC | 11 | 0.08 | 0.97 | 0.96 |
| AGC6G | 8 | 0.11 | 0.95 | 0.94 |
| IS06 | 10 | 0.08 | 0.96 | 0.96 |
| IS16 | 6 | 0.50 | 0.69 | 0.66 |
| IS17 | 5 | 0.50 | 0.71 | 0.69 |
| IS19 | 9 | 0.16 | 0.92 | 0.91 |
| ISSR.08 | 9 | 0.34 | 0.86 | 0.85 |
| **Mean** | **8** | **0.29** | **0.82** | **0.81** |

### 4.3.2 Clustering and AMOVA analysis

Clustering analysis was performed for assessing the connection between almond genotypes, based on Jaccard similarity coefficients using the unweighted pair-group method (UPGMA). The dissimilarity coefficients ranged between 0.32 (B-G3 vs. B-G4), (M-G1 vs. M-G2) to 0.75 (H-G5 vs. Q-G1) (**Appendix 50**), all 38 Almond genotypes were clustered into 3 groups (A, B and C) with a mean dissimilarity (0.54) for 20 RADP markers (Fig 4.12) cluster A include (H-G3, S-G3, S-G6, H-G4, H-G8, S-G7, H-G11, H-G5, H-G9, H-G12, H-G7, H-G6, H-G2 and H-G10) only Q-G1 was observed in cluster B, the rest genotypes were found in cluster C. In addition, dissimilarity values were also observed between 0.19 (H-G13 vs. H-G12) to 0.78 (H-G5 vs. B-G6) by using 15 ISSR markers (Appendix 51) which clustered all genotypes into (A, B, C and D) with a mean dissimilarity (0.49) (Fig. 4.13), cluster A includes only H-G7, and cluster B (consists of all genotypes without cluster A, C, D) group C (H-G1, S-G6 and S-G7) cluster D, (H-G3, H-G5, H-G4, H-G6 and H-G14). In addition, both RADP and ISSR markers exhibited a dissimilarity between 0.32 (B-G3 vs. B-G4) to 0.72 (H-G1 vs. H-G9) (**Appendix 52**) and clustered all genotypes into 4 groups (A, B, C and D) with a mean dissimilarity (0.52) (Fig. 4.14) cluster A, includes H-G7, cluster B, (H-G4, H-G6, H-G8, H-G5, H-G11, H-G10, H-G12, H-G13, H-G2 and H-G9), cluster C (includes all genotypes without cluster A, B, and D) and cluster D, (H-G14, H-G3, S-G3, S-G6 and S-G7).

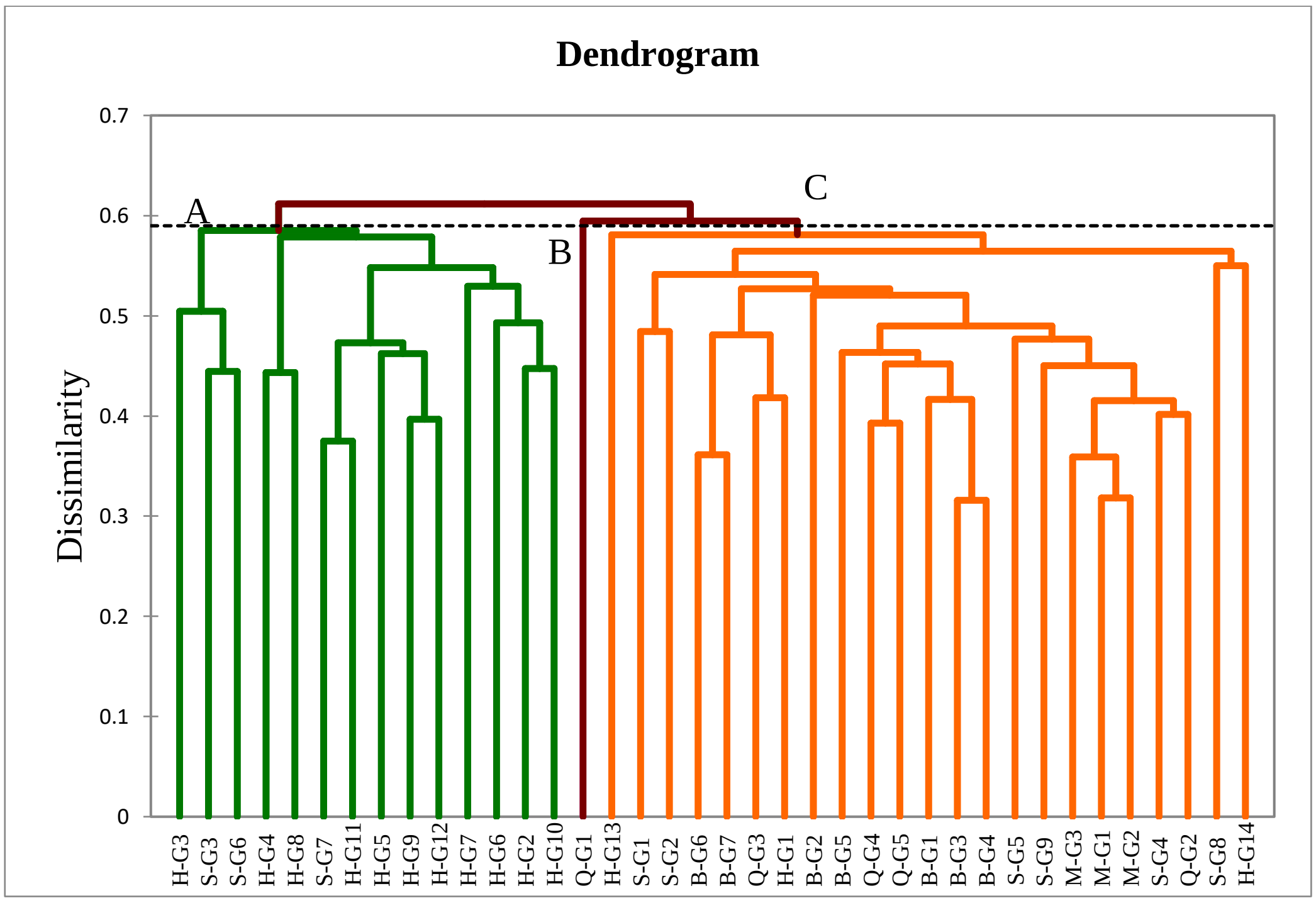

**Figure 4.12 Cluster tree created by UPGMA method based on 20 RAPD markers among 38 almond genotypes.**

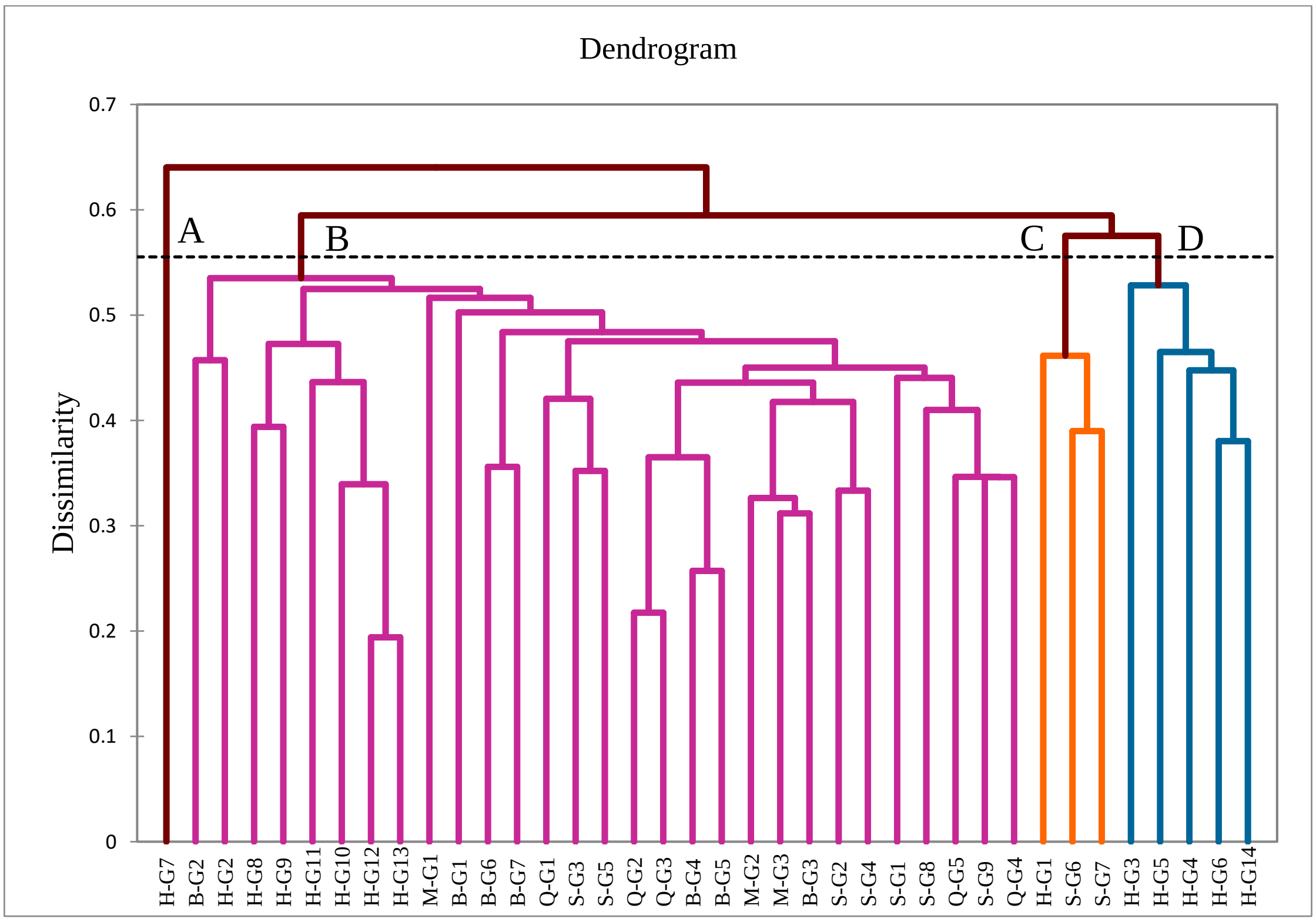

**Figure 4.13 Cluster tree created by UPGMA method based on 15 ISSR markers among 38 almond genotypes.**

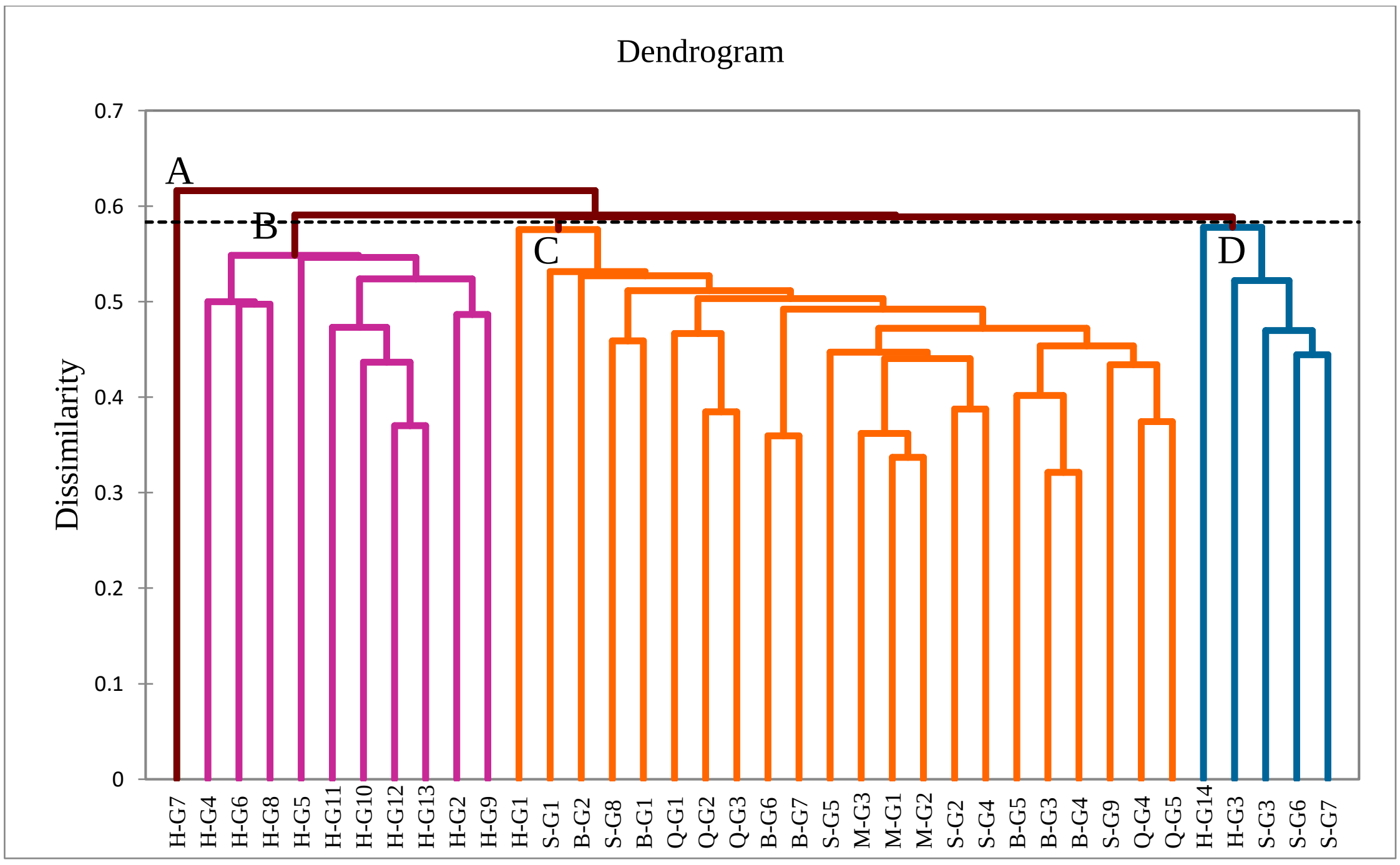

**Figure 4.14 Cluster tree created by UPGMA method based on 20 RAPD with 15 ISSR markers among 38 almond genotypes.**

Analysis of molecular variance (AMOVA) of the 38 almond genotypes in RAPD analysis demonstrated 88% of the total variation referred within the populations, while 12% of the variation referred credited to differences between populations (Table 4.4). In addition, the ISSR marker revealed high variance happened in the intra-populations 87% of the total variation, and merely 13% could be qualified to differences between sub-populations.

**Table 4.4 Analysis of molecular variance (AMOVA) of the five populations for 38 almond genotypes.**

| Source | df | SS | MS | Est. Var. | % | P-Value |
|---|---|---|---|---|---|---|
| RAPD | | | | | | |
| Among Pops | 4 | 307.18 | 76.80 | 5.29 | 12% | 0.001 |
| Within Pops | 33 | 1289.06 | 39.06 | 39.06 | 88% | 0.001 |
| Total | 37 | 1596.24 | | 44.35 | 100% | |
| ISSR | | | | | | |
| Among Pops | 4 | 188.16 | 47.04 | 3.391 | 13% | 0.001 |
| Within Pops | 33 | 754.31 | 22.86 | 22.86 | 87% | 0.001 |
| Total | 37 | 942.47 | | 26.25 | 100% | |

### 4.3.3 Genetic Structure for all genotypes using RAPD and ISSR markers

STRUCTURE method was used to collect evidence about population structure for almond genotypes depending on allele frequencies (Evanno *et al.*, 2005), therefore, in this work, according to delta K, genotypes were divided into two groups or sub-populations, group 1 (green line) and group 2 (red line) for RAPD and ISSR (Fig. 4.15 A and B), clusters were represented by colors, red line in RAPD and green line in ISSR consisted of Hawraman location, but the green line in RAPD and red line in ISSR represented other locations including (Sharbazher, Mergapan, Qaradagh and Barznja). In addition, a combination class of genotypes may refer to more than one background. For example, samples S-G3 and Q-G1 in RAPD markers (Fig. 4.15 A), and only sample S-G1 in ISSR marker (Fig. 4.15 B) can possibly have a complicated history linking intercrossing or practicably resulting from the gene flow between taxa. In addition, the high variability between genotypes may be due to consequences of changing climates within the locations. The true number of clusters (K) in a sample of individuals was observed for 20 RAPD and 15 ISSR markers, the peak started at 2 and the real K value with the highest value of K= 2 was also observed (Fig. 4.15 C and D), respectively.

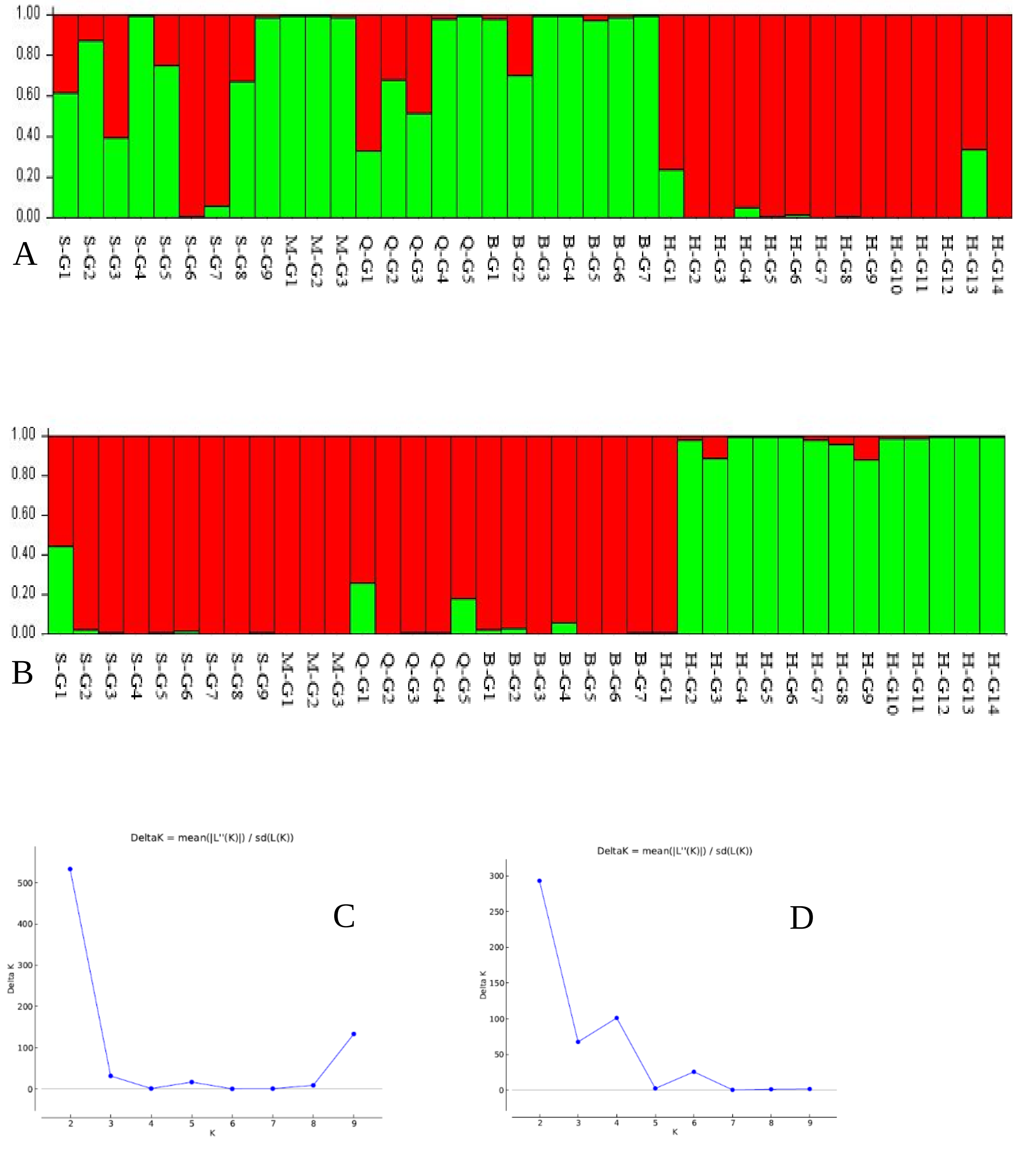

**Figure 4.15 Thirty-eight almond genotypes clustered into different sub-populations by STRUCTURE software. (A) for RAPD and (B) ISSR. Accessions are coordinated as per estimated membership coefficients (q) in K= 2 clusters. (C) for RAPD and (D) for ISSR, Determining the optimal value of K by the (ΔK) procedure described by** (Evanno *et al.*, 2005).

## 4.4 Determination of Drought Tolerance in Glasshouse for 38 Almond Genotypes

### 4.4.1 Morphological data analysis

Table (4.5) show significant differences for eight vegetative growth characteristics of 38 almond genotypes. The results show that the maximum values were 43.011 cm for height of H-G9, 3.296 mm for diameter of Q-G3, 31.556 for leaves number of H-G9, 2.140 $cm^2$ for leaves area of Q-G3, 6.291 g for weight of vegetative growth of Q-G3, 41.623% for percentage of vegetative dry weight of Q-G3, 8.528 g for root weight of Q-G4 and 29.068% for the percentage of root dry weight of Q-G3. On the other hand, the minimum values were as follows: 23.633 cm for height of H-G11, 2.500 mm for diameter of H-G12, 23.556 leaves number of M-G1, 1.553 $cm^2$ for leaves area of S-G1, 3.373 g for the vegetative weight of H-G12, 34.044% for the percentage of vegetative dry weight of B-G3, 3.662 g for the root weight of H-G12 and 22.126 % for the percentage of root dry weight of B-G4.

H-G9 recorded maximum value in leaves number, and may be caused by increased photosynthesis and effect on seedling growth to give the same genotype giving the maximum seedling height. Q-G3 recorded maximum value in seedling diameter and leaves area which may be because the same previous reason, and also higher value of leaves area may be caused by maximum vegetative growth weight and vegetative dry weight by genotype number 3 in Qaradagh location. Also, this growth effected on root dry weight because the photosynthesis product was transferred to root and lead Q-G3 recorded maximum values on root dry weight percentage.

Genotype number 12 at Hawraman recorded minimum value in vegetative and root dry weight, this may be caused by the ability of this genotype to collect dry mater which was lower than other genotypes and this result clearly showed that genotype has its effect on vegetative growth. Shoot length differs among almond genotypes revealed the overall vegetative growth of the tree in response to environmental conditions and due to the differences in their genetic makeup. Plant tried to cope with the water stress by reducing its leaf area in order to allow the conservation of energy, minimize the deleterious effects of loss of water and complete life cycle under stress conditions (Bohnert *et al.,* 1995), also the negative effect of water stress on the negative growth may be attributed to the inadequate water stress (Kramer and Boyer, 1995).

**Table 4.5 Effect of almond seedling genotypes on some vegetative and root growth characteristics.**

| Genotypes | Height cm | Diameter mm | Leaves number | Leaves area $cm^2$ | Vegetative growth weight g | Vegetative dry weight % | Root weight g | Root dry weight % |
|---|---|---|---|---|---|---|---|---|
| S-G1 | 29.839 | 2.941 | 27.333 | 1.553 | 4.093 | 38.257 | 5.728 | 27.489 |
| S-G2 | 35.533 | 2.986 | 26.333 | 1.617 | 4.036 | 37.076 | 5.701 | 25.768 |
| S-G3 | 32.989 | 2.673 | 27.667 | 1.673 | 4.547 | 35.685 | 5.728 | 23.650 |
| S-G4 | 40.789 | 2.884 | 29.000 | 1.754 | 4.406 | 36.160 | 5.378 | 26.148 |
| S-G5 | 32.933 | 3.107 | 25.333 | 1.841 | 3.967 | 36.485 | 5.309 | 26.381 |
| S-G6 | 34.222 | 2.909 | 29.667 | 1.965 | 4.797 | 39.732 | 6.730 | 23.658 |
| S-G7 | 35.644 | 2.960 | 29.000 | 1.997 | 3.969 | 37.465 | 5.963 | 26.405 |
| S-G8 | 34.500 | 3.162 | 25.444 | 2.027 | 5.033 | 37.882 | 6.447 | 25.373 |
| S-G9 | 33.511 | 3.054 | 27.667 | 2.028 | 4.876 | 36.063 | 6.457 | 23.376 |
| M-G1 | 28.656 | 2.959 | 23.556 | 1.988 | 3.952 | 37.277 | 6.431 | 24.084 |
| M-G2 | 35.722 | 2.714 | 27.667 | 2.028 | 4.730 | 36.364 | 6.047 | 22.453 |
| M-G3 | 35.222 | 3.027 | 27.000 | 2.027 | 4.810 | 37.404 | 7.523 | 25.669 |
| Q-G1 | 30.467 | 3.039 | 26.444 | 2.062 | 4.121 | 37.158 | 4.281 | 26.280 |
| Q-G2 | 30.422 | 3.042 | 27.111 | 2.096 | 4.436 | 36.763 | 5.108 | 23.533 |
| Q-G3 | 41.889 | 3.296 | 29.889 | 2.140 | 6.291 | 41.623 | 6.703 | 29.068 |
| Q-G4 | 36.567 | 3.274 | 28.222 | 2.129 | 5.272 | 38.127 | 8.528 | 22.839 |
| Q-G5 | 35.156 | 3.163 | 27.889 | 2.114 | 4.650 | 36.464 | 5.894 | 26.459 |
| B-G1 | 38.122 | 3.258 | 30.556 | 2.097 | 4.980 | 37.297 | 5.446 | 28.243 |
| B-G2 | 36.389 | 3.121 | 28.222 | 2.092 | 4.133 | 37.556 | 5.626 | 24.838 |
| B-G3 | 40.733 | 3.224 | 28.333 | 2.133 | 5.409 | 34.044 | 6.436 | 24.899 |
| B-G4 | 36.144 | 3.178 | 28.333 | 2.108 | 5.111 | 38.443 | 5.534 | 22.126 |
| B-G5 | 34.511 | 2.844 | 26.000 | 2.071 | 4.097 | 37.621 | 5.319 | 25.698 |
| B-G6 | 32.433 | 2.931 | 26.889 | 2.107 | 3.817 | 38.255 | 5.566 | 25.137 |
| B-G7 | 35.378 | 3.211 | 27.333 | 2.072 | 4.370 | 37.989 | 4.542 | 22.496 |
| H-G1 | 23.689 | 2.503 | 25.000 | 1.979 | 3.464 | 37.191 | 4.951 | 22.207 |
| H-G2 | 24.213 | 2.546 | 26.889 | 1.966 | 3.610 | 36.481 | 4.398 | 23.346 |
| H-G3 | 28.956 | 2.972 | 26.000 | 2.004 | 3.873 | 40.017 | 4.437 | 28.140 |
| H-G4 | 29.289 | 2.910 | 25.778 | 1.982 | 3.956 | 37.048 | 5.257 | 24.812 |
| H-G5 | 30.600 | 3.022 | 27.333 | 2.018 | 4.640 | 39.194 | 7.707 | 27.418 |
| H-G6 | 30.422 | 2.948 | 23.778 | 2.009 | 4.376 | 38.343 | 5.309 | 24.854 |
| H-G7 | 32.333 | 2.961 | 29.333 | 1.963 | 4.407 | 38.815 | 4.971 | 24.831 |
| H-G8 | 35.511 | 2.626 | 26.333 | 1.960 | 4.331 | 38.091 | 4.988 | 23.025 |
| H-G9 | 43.011 | 2.984 | 31.556 | 1.930 | 4.423 | 38.730 | 6.381 | 24.721 |
| H-G10 | 33.833 | 3.026 | 30.111 | 1.942 | 4.307 | 38.736 | 5.234 | 26.186 |
| H-G11 | 23.633 | 2.909 | 24.000 | 1.887 | 4.032 | 37.855 | 4.888 | 27.008 |
| H-G12 | 25.444 | 2.500 | 24.889 | 1.861 | 3.373 | 38.019 | 3.662 | 25.547 |
| H-G13 | 32.033 | 2.950 | 27.889 | 1.883 | 3.881 | 37.544 | 4.349 | 25.868 |
| H-G14 | 32.622 | 2.952 | 28.778 | 1.927 | 4.283 | 37.044 | 5.418 | 25.808 |
| **LSD** | **3.966** | **0.321** | **2.973** | **0.083** | **1.160** | **2.448** | **1.188** | **4.553** |

Table (4.6) indicates that 10 days irrigation intervals gave 29.675, 2.147 $cm^2$, 38.26% and 6.759 g for leaves number, leaves area, vegetative growth percentage and root weight, successively which were superior significantly to 20-day irrigation interval, whereas no significant differences between the two irrigation intervals were observed with regard to the rest of vegetative growth characteristics. On the other hand, 10-day irrigation intervals dominated significantly on 40-days with regard to all the studied characteristics except vegetative dry weight percentage and root dry weight percentage, while 40 days exceeded significantly to both 10 and 20 days with regard to root dry weight percentage, also 40 days dominated significantly on 20 days with respect to vegetative dry weight percentage. No significant differences were noticed for all the studied factors for both 20 and 40 days except vegetative and root dry weight percentages.

**Table 4.6 Effect of irrigation intervals on some vegetative and root growth characteristics of almond genotype seedlings.**

| Irrigation intervals | Height cm | Diameter mm | Leaves number | Leaves area $cm^2$ | Vegetative growth weight g | Vegetative dry weight % | Root weight g | Root dry weight % |
|---|---|---|---|---|---|---|---|---|
| 10 D | 34.401 | 3.056 | 29.675 | 2.147 | 4.787 | 38.258 | 6.759 | 20.453 |
| 20 D | 33.847 | 3.065 | 27.974 | 1.975 | 4.659 | 36.327 | 6.117 | 21.579 |
| 40 D | 31.491 | 2.782 | 24.342 | 1.802 | 3.727 | 38.334 | 4.048 | 33.429 |
| **LSD** | **1.114** | **0.090** | **0.785** | **0.023** | **0.326** | **0.688** | **0.334** | **1.279** |

Table (4.7) illustrates the effects of the interaction between almond seedling genotypes and irrigation period on eight vegetative and root growth characteristics. Results demonstrate that significant effects were observed between the treatments. The maximum values were (S-G4*20D = 48.933 cm, Q-G3*20D = 3.670 mm, H-G9*10D = 36.333, B-G5*10D = 2.317 $cm^2$, S-G8*10D = 7.257 g, Q-G3*40D = 44.588%, Q-G4*10D = 12.173 g and Q-G1*40D = 38.564%) for high, diameter, leave number, leaves area, vegetative growth weight, vegetative dry weight, root weight and root dry weight percentage of seedling, respectively. While the minimum values were (H-G2*40D = 22.673 g, H-G12*40D = 2.117 mm, S-G5*40D = 20, S-G3*40D = 1.347 $cm^2$, S-G3*40 D = 2.297 g, B-G3*20D = 31.056%, H-G12*40D = 2.333 g and H-G8*10D = 15.960%) for the former characteristics, respectively.

A main indicator to response of plants to drought stress is the reduction of growth (Akbarpour *et al.*, 2017). Actually, almond plant leaf is a good indicator to determine the effect of water stress, however, changes in shoot and root growth in response to water stress were different for all genotypes. Also, root dry weight and leaf area could potentially be good indicators for drought tolerance in addition to smaller stomatal size and lower specific leaf area. All genotypes managed to recover from moderate stress so almond seedlings could be tolerated well, however, sever water potential for longer periods will limit young plant growth (Yadollahi *et al.*, 2011). Generally, drought stress of some *Prunus* plants particularly almonds have effect on leaf characteristics, individual leaf under drought stress is smaller than ordinary leaf growth in normal condition. In addition, Ranney *et al.* (1991) found higher root to leaf area ratios in the more drought-tolerant *Prunus cerasus* cv. Meteor than the less drought-tolerant hybrid (*Prunus avium* × *Pseudocerasus*) cv. Colt., similar relationship between leaf size and drought adaptation has been found in ecotypes of *Cersis canadensis* (Abrams, 1988).

There were significant differences among three wild species of almond for most of the traits measured such as stem height, stem diameter and leaf area (Baninasab and Rahemi, 2007). Akbarpour et al. (2017) clarified that almond cultivars have genetic differences in their abilities to drought resistance. However, some cultivars were genetically more resistant to drought. This result may be due to genetic heterozygosity (Chalak *et al.*, 2007). Čolić *et al.* (2012) showed high morphological diversity in almond genotypes.

Gikloo and Elhami (2012) indicated variation among cultivars in response to drought stress. Drought tolerant cultivars have very small leaf area, the reduction in leaf area under drought stress could be considered as an avoidance mechanism which minimizes water losses in almond genotypes. Drought stress caused the reduction of leaf fresh weight. In general, water content of leaf decline under drought stress and finally reduce the plant fresh weight. Plants with high dry mass under drought stress may be measured as drought tolerant genotype (Zokaee-Khosroshahi *et al.*, 2014).

Our results agree with (Gikloo and Elhami, 2012) who stated that genotype can impact on leaves area. In addition, Karimi *et al.* (2013a) also assured that genotypes may significantly influences leaves numbers and area. Rajabpoor *et al.* (2014) demonstrated that the morphological data was nearly agree with our result.

Our morphological data may agree with (Yadollahi *et al.*, 2011) who argued that total leaf dry weight, shoot dry weight and shoot growth significantly reduced in drought for nearly all of almond genotypes. In addition, plantlet height, root length and dry weight of root were also significantly affected by genotypes during drought stress (Sorkheh *et al.*, 2011). Karimi *et al.* (2013a) also observed that drought stress declined leaf dry weight. After drought, the stomata close gradually with a decline of water-use efficiency and net photosynthesis (Hossain *et al.*, 2016).

Understanding of the morphophysiological traits may be used to generate new varieties with better yield under drought conditions. A crucial part of making crop stress-tolerant is the understanding of plant responses to different environments of drought stress (Yadav and Sharma, 2016).

**Table 4.7 Interaction between almond seedling genotypes and irrigation intervals on some vegetative and root growth characteristics.**

| Genotypes* Irrigation | Height cm | Diameter mm | Leaves number | Leaves area $cm^2$ | Vegetative growth weight g | Vegetative dry weight % | Root weight g | Root dry weight % |
|---|---|---|---|---|---|---|---|---|
| S-G1*10 D | 30.267 | 2.980 | 29.000 | 1.635 | 4.080 | 39.734 | 7.233 | 23.504 |
| S-G1*20 D | 29.833 | 3.067 | 26.667 | 1.627 | 4.180 | 34.901 | 6.393 | 24.916 |
| S-G1*40 D | 29.417 | 2.777 | 26.333 | 1.397 | 4.020 | 40.135 | 3.557 | 34.048 |
| S-G2*10 D | 37.000 | 3.290 | 28.000 | 1.738 | 4.407 | 38.705 | 6.780 | 19.893 |
| S-G2*20 D | 34.733 | 3.263 | 27.667 | 1.713 | 4.057 | 35.853 | 5.973 | 21.126 |
| S-G2*40 D | 34.867 | 2.403 | 23.333 | 1.400 | 3.643 | 36.671 | 4.350 | 36.284 |
| S-G3*10 D | 35.233 | 2.890 | 30.000 | 1.817 | 5.593 | 33.933 | 8.383 | 17.613 |
| S-G3*20 D | 34.067 | 2.710 | 30.333 | 1.854 | 5.750 | 37.470 | 5.587 | 25.151 |
| S-G3*40 D | 29.667 | 2.420 | 22.667 | 1.347 | 2.297 | 35.652 | 3.213 | 28.185 |
| S-G4*10 D | 35.400 | 3.060 | 30.667 | 1.813 | 5.543 | 38.999 | 6.750 | 22.636 |
| S-G4*20 D | 48.933 | 3.203 | 30.000 | 1.773 | 4.473 | 35.017 | 5.687 | 22.383 |
| S-G4*40 D | 38.033 | 2.390 | 26.333 | 1.677 | 3.200 | 34.465 | 3.697 | 33.425 |
| S-G5*10 D | 34.500 | 3.387 | 30.000 | 1.855 | 4.913 | 37.172 | 6.867 | 20.733 |
| S-G5*20 D | 32.567 | 3.113 | 26.000 | 1.865 | 4.040 | 35.728 | 6.363 | 21.449 |
| S-G5*40 D | 31.733 | 2.820 | 20.000 | 1.804 | 2.947 | 36.557 | 2.697 | 36.962 |
| S-G6*10 D | 35.100 | 2.867 | 36.000 | 1.985 | 5.393 | 40.304 | 8.483 | 18.288 |
| S-G6*20 D | 33.933 | 2.907 | 28.000 | 2.007 | 4.697 | 39.514 | 6.197 | 23.829 |
| S-G6*40 D | 33.633 | 2.953 | 25.000 | 1.903 | 4.300 | 39.376 | 5.510 | 28.857 |
| S-G7*10 D | 36.033 | 2.953 | 30.667 | 2.058 | 3.630 | 39.060 | 6.890 | 20.571 |
| S-G7*20 D | 36.500 | 2.960 | 30.333 | 1.980 | 4.407 | 34.568 | 7.230 | 22.084 |
| S-G7*40 D | 34.400 | 2.967 | 26.000 | 1.953 | 3.870 | 38.767 | 3.770 | 36.560 |
| S-G8*10 D | 40.833 | 3.657 | 31.000 | 2.198 | 7.257 | 37.782 | 8.947 | 20.123 |
| S-G8*20 D | 31.100 | 2.957 | 23.667 | 1.985 | 3.930 | 37.645 | 5.843 | 22.089 |
| S-G8*40 D | 31.567 | 2.873 | 21.667 | 1.897 | 3.913 | 38.219 | 4.550 | 33.907 |
| S-G9*10 D | 34.533 | 2.967 | 29.333 | 2.217 | 5.300 | 34.776 | 7.710 | 18.730 |
| S-G9*20 D | 34.700 | 3.153 | 28.667 | 1.957 | 5.193 | 34.158 | 7.633 | 20.742 |
| S-G9*40 D | 31.300 | 3.043 | 25.000 | 1.911 | 4.133 | 39.256 | 4.027 | 30.656 |
| M-G1*10 D | 29.100 | 2.993 | 24.000 | 2.275 | 4.073 | 36.800 | 7.503 | 16.891 |
| M-G1*20 D | 28.533 | 2.983 | 23.333 | 1.898 | 4.270 | 33.493 | 7.793 | 21.025 |
| M-G1*40 D | 28.333 | 2.900 | 23.333 | 1.792 | 3.513 | 41.537 | 3.997 | 34.337 |
| M-G2*10 D | 36.000 | 2.630 | 28.667 | 2.210 | 3.503 | 35.978 | 5.363 | 17.469 |
| M-G2*20 D | 35.667 | 2.793 | 28.333 | 1.975 | 5.620 | 34.399 | 8.293 | 17.774 |
| M-G2*40 D | 35.500 | 2.720 | 26.000 | 1.898 | 5.067 | 38.716 | 4.483 | 32.115 |
| M-G3*10 D | 36.500 | 3.060 | 27.333 | 2.193 | 5.053 | 38.048 | 7.570 | 23.432 |
| M-G3*20 D | 35.000 | 3.017 | 27.000 | 1.998 | 5.137 | 36.699 | 7.507 | 20.040 |
| M-G3*40 D | 34.167 | 3.003 | 26.667 | 1.890 | 4.240 | 37.466 | 7.493 | 33.534 |

***Continued***

**Table 4.7 Continued**

| Genotypes* Irrigation | Height cm | Diameter mm | Leaves number | Leaves area $cm^2$ | Vegetative growth weight g | Vegetative dry weight % | Root weight g | Root dry weight % |
|---|---|---|---|---|---|---|---|---|
| Q-G1*10 D | 26.000 | 3.037 | 28.000 | 2.182 | 3.927 | 34.076 | 3.360 | 18.506 |
| Q-G1*20 D | 32.733 | 3.073 | 27.000 | 2.022 | 4.247 | 36.376 | 6.643 | 21.769 |
| Q-G1*40 D | 32.667 | 3.007 | 24.333 | 1.983 | 4.190 | 41.021 | 2.840 | 38.564 |
| Q-G2*10 D | 26.000 | 2.980 | 28.667 | 2.238 | 4.467 | 31.495 | 6.417 | 16.505 |
| Q-G2*20 D | 34.967 | 3.080 | 28.000 | 2.057 | 4.527 | 38.596 | 5.957 | 20.248 |
| Q-G2*40 D | 30.300 | 3.067 | 24.667 | 1.992 | 4.313 | 40.198 | 2.950 | 33.846 |
| Q-G3*10 D | 42.900 | 3.303 | 33.000 | 2.277 | 7.240 | 39.832 | 7.820 | 24.721 |
| Q-G3*20 D | 42.867 | 3.670 | 32.667 | 2.104 | 7.070 | 40.448 | 6.903 | 24.193 |
| Q-G3*40 D | 39.900 | 2.913 | 24.000 | 2.040 | 4.563 | 44.588 | 5.387 | 38.290 |
| Q-G4*10 D | 39.933 | 3.597 | 32.000 | 2.295 | 6.953 | 36.508 | 12.173 | 16.354 |
| Q-G4*20 D | 36.033 | 3.533 | 26.667 | 2.067 | 4.713 | 39.472 | 7.570 | 23.752 |
| Q-G4*40 D | 33.733 | 2.693 | 26.000 | 2.027 | 4.150 | 38.401 | 5.840 | 28.412 |
| Q-G5*10 D | 35.867 | 3.213 | 30.667 | 2.271 | 5.220 | 39.381 | 6.687 | 21.978 |
| Q-G5*20 D | 35.000 | 3.037 | 27.000 | 2.080 | 4.497 | 34.333 | 6.467 | 21.951 |
| Q-G5*40 D | 34.600 | 3.240 | 26.000 | 1.991 | 4.233 | 35.678 | 4.530 | 35.448 |
| B-G1*10 D | 38.967 | 3.330 | 32.667 | 2.245 | 6.127 | 36.537 | 7.337 | 23.677 |
| B-G1*20 D | 39.167 | 3.663 | 31.000 | 2.057 | 5.377 | 37.650 | 4.947 | 23.051 |
| B-G1*40 D | 36.233 | 2.780 | 28.000 | 1.988 | 3.437 | 37.704 | 4.053 | 38.001 |
| B-G2*10 D | 38.667 | 3.097 | 31.333 | 2.285 | 3.317 | 36.947 | 5.557 | 16.050 |
| B-G2*20 D | 37.333 | 3.203 | 30.000 | 2.028 | 5.020 | 36.609 | 6.880 | 20.347 |
| B-G2*40 D | 33.167 | 3.063 | 23.333 | 1.961 | 4.063 | 39.112 | 4.440 | 38.116 |
| B-G3*10 D | 42.833 | 3.170 | 32.000 | 2.293 | 6.417 | 36.009 | 8.487 | 22.475 |
| B-G3*20 D | 40.633 | 3.110 | 26.667 | 2.093 | 4.573 | 31.056 | 5.167 | 17.792 |
| B-G3*40 D | 38.733 | 3.393 | 26.333 | 2.013 | 5.237 | 35.068 | 5.653 | 34.430 |
| B-G4*10 D | 37.400 | 3.313 | 30.333 | 2.275 | 4.823 | 39.768 | 5.780 | 19.626 |
| B-G4*20 D | 36.533 | 3.290 | 28.667 | 2.080 | 5.410 | 35.980 | 6.820 | 16.428 |
| B-G4*40 D | 34.500 | 2.930 | 26.000 | 1.969 | 5.100 | 39.580 | 4.003 | 30.324 |
| B-G5*10 D | 37.400 | 2.890 | 27.333 | 2.317 | 4.853 | 37.936 | 5.350 | 20.422 |
| B-G5*20 D | 35.833 | 2.917 | 27.000 | 2.035 | 4.460 | 35.172 | 6.733 | 19.556 |
| B-G5*40 D | 30.300 | 2.727 | 23.667 | 1.862 | 2.977 | 39.754 | 3.873 | 37.117 |
| B-G6*10 D | 32.367 | 2.937 | 28.333 | 2.295 | 4.447 | 39.648 | 5.643 | 20.632 |
| B-G6*20 D | 32.600 | 2.933 | 28.667 | 2.073 | 3.550 | 37.171 | 6.063 | 20.357 |
| B-G6*40 D | 32.333 | 2.923 | 23.667 | 1.951 | 3.453 | 37.946 | 4.990 | 34.423 |
| B-G7*10 D | 37.333 | 3.133 | 31.333 | 2.264 | 4.880 | 39.741 | 5.173 | 16.500 |
| B-G7*20 D | 36.867 | 3.180 | 28.000 | 2.054 | 4.200 | 35.700 | 4.567 | 19.832 |
| B-G7*40 D | 31.933 | 3.320 | 22.667 | 1.897 | 4.030 | 38.527 | 3.887 | 31.156 |
| H-G1*10 D | 24.300 | 2.430 | 24.667 | 2.242 | 3.733 | 38.433 | 6.193 | 16.410 |
| H-G1*20 D | 24.067 | 2.653 | 27.000 | 2.045 | 3.983 | 36.648 | 5.557 | 20.807 |
| H-G1*40 D | 22.700 | 2.427 | 23.333 | 1.650 | 2.677 | 36.490 | 3.103 | 29.403 |
| H-G2*10 D | 25.033 | 2.467 | 27.000 | 2.215 | 3.910 | 39.440 | 5.027 | 22.307 |
| H-G2*20 D | 24.933 | 2.627 | 27.667 | 2.000 | 3.740 | 33.522 | 4.557 | 16.637 |
| H-G2*40 D | 22.673 | 2.543 | 26.000 | 1.682 | 3.180 | 36.480 | 3.610 | 31.095 |

***Continued***

**Table 4.7 Continued**

| Genotypes* Irrigation | Height cm | Diameter mm | Leaves number | Leaves area $cm^2$ | Vegetative growth weight g | Vegetative dry weight % | Root weight g | Root dry weight % |
|---|---|---|---|---|---|---|---|---|
| H-G3*10 D | 32.267 | 3.133 | 26.333 | 2.178 | 4.070 | 39.864 | 5.443 | 22.482 |
| H-G3*20 D | 27.867 | 3.167 | 26.667 | 1.987 | 3.803 | 38.884 | 4.347 | 25.898 |
| H-G3*40 D | 26.733 | 2.617 | 25.000 | 1.848 | 3.747 | 41.304 | 3.520 | 36.039 |
| H-G4*10 D | 31.567 | 2.973 | 27.667 | 2.213 | 4.347 | 38.881 | 6.567 | 18.794 |
| H-G4*20 D | 30.300 | 3.097 | 27.000 | 1.973 | 4.150 | 33.336 | 4.830 | 22.994 |
| H-G4*40 D | 26.000 | 2.660 | 22.667 | 1.760 | 3.370 | 38.927 | 4.373 | 32.648 |
| H-G5*10 D | 31.767 | 3.180 | 30.667 | 2.192 | 4.667 | 38.804 | 10.337 | 17.333 |
| H-G5*20 D | 31.400 | 3.207 | 30.333 | 1.982 | 5.797 | 38.505 | 8.350 | 27.863 |
| H-G5*40 D | 28.633 | 2.680 | 21.000 | 1.879 | 3.457 | 40.273 | 4.433 | 37.059 |
| H-G6*10 D | 32.833 | 3.237 | 26.000 | 2.192 | 4.320 | 39.891 | 5.857 | 24.472 |
| H-G6*20 D | 31.333 | 3.263 | 24.333 | 1.995 | 6.253 | 37.138 | 7.663 | 22.221 |
| H-G6*40 D | 27.100 | 2.343 | 21.000 | 1.839 | 2.553 | 38.000 | 2.407 | 27.869 |
| H-G7*10 D | 34.667 | 3.008 | 32.333 | 2.173 | 4.177 | 42.149 | 5.437 | 19.632 |
| H-G7*20 D | 32.067 | 3.283 | 32.000 | 1.983 | 5.650 | 36.250 | 5.643 | 19.547 |
| H-G7*40 D | 30.267 | 2.590 | 23.667 | 1.734 | 3.393 | 38.047 | 3.833 | 35.313 |
| H-G8*10 D | 38.300 | 2.710 | 28.667 | 2.157 | 4.167 | 37.584 | 5.993 | 15.960 |
| H-G8*20 D | 36.567 | 2.570 | 28.333 | 1.967 | 4.693 | 37.448 | 5.093 | 20.703 |
| H-G8*40 D | 31.667 | 2.597 | 22.000 | 1.757 | 4.133 | 39.240 | 3.877 | 32.412 |
| H-G9*10 D | 45.400 | 3.150 | 36.333 | 2.136 | 5.160 | 41.650 | 8.777 | 24.116 |
| H-G9*20 D | 45.633 | 3.177 | 33.000 | 1.967 | 5.483 | 37.002 | 7.087 | 20.499 |
| H-G9*40 D | 38.000 | 2.627 | 25.333 | 1.687 | 2.627 | 37.536 | 3.280 | 29.547 |
| H-G10*10 D | 34.667 | 3.053 | 31.667 | 2.118 | 3.843 | 40.569 | 7.013 | 22.762 |
| H-G10*20 D | 33.933 | 3.157 | 31.333 | 1.965 | 4.753 | 38.281 | 5.183 | 22.507 |
| H-G10*40 D | 32.900 | 2.867 | 27.333 | 1.743 | 4.323 | 37.358 | 3.507 | 33.289 |
| H-G11*10 D | 24.667 | 3.137 | 25.333 | 2.122 | 4.800 | 38.907 | 7.360 | 25.819 |
| H-G11*20 D | 22.733 | 2.930 | 23.333 | 1.963 | 4.170 | 34.347 | 3.377 | 24.203 |
| H-G11*40 D | 23.500 | 2.660 | 23.333 | 1.575 | 3.127 | 40.311 | 3.927 | 31.003 |
| H-G12*10 D | 26.000 | 2.777 | 26.667 | 2.153 | 4.917 | 40.327 | 5.177 | 25.755 |
| H-G12*20 D | 25.233 | 2.607 | 24.667 | 1.959 | 2.577 | 36.645 | 3.477 | 22.681 |
| H-G12*40 D | 25.100 | 2.117 | 23.333 | 1.470 | 2.627 | 37.086 | 2.333 | 28.204 |
| H-G13*10 D | 35.367 | 3.153 | 33.000 | 2.140 | 4.333 | 40.087 | 4.857 | 21.352 |
| H-G13*20 D | 30.467 | 2.970 | 25.333 | 1.943 | 4.053 | 38.564 | 4.547 | 22.678 |
| H-G13*40 D | 30.267 | 2.727 | 25.333 | 1.567 | 3.257 | 33.981 | 3.643 | 33.575 |
| H-G14*10 D | 34.233 | 2.980 | 31.000 | 2.115 | 4.037 | 38.042 | 4.537 | 22.697 |
| H-G14*20 D | 33.533 | 2.940 | 30.667 | 1.930 | 4.557 | 35.835 | 7.533 | 18.874 |
| H-G14*40 D | 30.100 | 2.937 | 24.667 | 1.737 | 4.257 | 37.255 | 4.183 | 35.854 |
| LSD (0.05) | **6.869** | **0.556** | **4.838** | **0.143** | **2.009** | **4.241** | **2.057** | **7.886** |

**4.4.2 Some stomatal characteristics and chemical content in almond seedling leaf in the study to almond seedling drought tolerance experiment**

Table (4.8) exhibits that there were significant effects of genotypes on some stomatal characteristics and chemical contents of the leaves. Highest values were: (H-G6 = 25.373 µm, S-G4 = 8.878 µm, B-G2 = 37.122 SPAD, Q-G2 = 89.594 µmol.$g^{-1}$ FW, S-G2 = 2.884 mg GAE $g^{-1}$ E, S-G2 = 1.179 mg QE $g^{-1}$ E, H-G13 = 34.312% inhibition and H-G14 = 39.067% inhibition) for stomatal length, stomatal width, chlorophyll concentration, proline, total phenolic content, total flavonoid content, antioxidant activity by DPPH and ABTS assay of seedlings, respectively. While the minimum values were (S-G9 = 18.700 µm, S-G8 = 6.170 µm, Q-G2=19.817 SPAD, S-G3 = 52.322 µmol $g^{-1}$ FW, S-G4 = 0.741 mg GAE $g^{-1}$ E, S-G7 and B-G1 = 0.497 mg QE $g^{-1}$ E, Q-G1 = 27.636% inhibition and H-G6 = 24.043% inhibition) for previous characteristics, respectively.

Genotypes have the significant effect on stomatal size, in Sharbazher location for example, maximum and minimum values of stomatal width were recorded, Palasciano *et al.* (2005) showed that stomatal characteristics (stomatal size and density) depended only on genotype.

Chlorophyll concentration significantly changed between genotypes, in Barznja location the values were ranged between (21.022-37.122 SPAD), these may be caused by the differences of genotypes. Sepehri and Golparvar (2011) detailed that chlorophyll content affected by plant genotypes.

Significant variability was detected among the genotypes in their capability for proline accumulation, these variations may be due to genotype ability or sensitivity to stress, Zamani *et al.* (2002) described that in different stresses, proline production rate in plants tolerance is not similar.

Total phenol content, total flavonoid content, antioxidant activity by DPPH and ABTS assay were changed significantly which may be due to the ability of genotype to synthesize these products, researchers (Hughey *et al.*, 2008; Sivaci and Duman, 2014; Čolić *et al.*, 2017; Vats, 2018) demonstrated that synthesis of prior phytochemical were under genetic control or different between the genotypes or cultivar.

**Table 4.8 Effect of almond seedling genotypes on some stomatal and leaf chemical content characteristics.**

| Genotypes | Stomatal length µm | Stomatal width µm | Chlorophyll concentration SPAD | Proline µmol $g^{-1}$ FW | TPC mg GAE $g^{-1}$ E | TFC mg QE $g^{-1}$ E | DPPH % inhibition | ABTS % inhibition |
|---|---|---|---|---|---|---|---|---|
| S-G1 | 20.335 | 6.494 | 31.028 | 62.493 | 1.916 | 0.779 | 30.807 | 27.777 |
| S-G2 | 19.957 | 6.604 | 28.344 | 59.788 | 2.884 | 1.179 | 30.452 | 30.105 |
| S-G3 | 19.350 | 7.257 | 26.650 | 52.322 | 1.216 | 0.815 | 30.667 | 31.812 |
| S-G4 | 23.209 | 8.878 | 24.311 | 55.850 | 0.741 | 0.663 | 30.194 | 32.117 |
| S-G5 | 22.908 | 7.807 | 25.311 | 53.204 | 2.184 | 0.768 | 31.347 | 32.723 |
| S-G6 | 20.521 | 7.031 | 35.544 | 70.194 | 1.972 | 0.680 | 34.278 | 33.193 |
| S-G7 | 22.651 | 7.501 | 33.167 | 70.841 | 1.142 | 0.497 | 32.196 | 35.668 |
| S-G8 | 21.168 | 6.170 | 26.494 | 81.011 | 1.646 | 0.573 | 31.792 | 32.503 |
| S-G9 | 18.700 | 6.702 | 30.861 | 72.487 | 0.901 | 0.688 | 29.765 | 33.008 |
| M-G1 | 20.145 | 6.973 | 26.594 | 72.428 | 1.234 | 0.722 | 32.896 | 30.374 |
| M-G2 | 22.438 | 6.757 | 30.889 | 71.546 | 1.338 | 0.822 | 28.399 | 32.282 |
| M-G3 | 20.737 | 7.248 | 34.303 | 63.433 | 2.275 | 0.727 | 31.310 | 30.961 |
| Q-G1 | 19.082 | 6.728 | 33.628 | 73.780 | 2.261 | 0.837 | 27.636 | 25.883 |
| Q-G2 | 22.213 | 7.607 | 19.817 | 89.594 | 2.193 | 0.754 | 30.214 | 32.811 |
| Q-G3 | 21.001 | 7.414 | 23.706 | 74.074 | 2.087 | 0.737 | 31.236 | 34.853 |
| Q-G4 | 18.762 | 6.368 | 27.272 | 73.016 | 1.981 | 0.756 | 30.220 | 31.126 |
| Q-G5 | 24.156 | 7.184 | 21.744 | 81.717 | 1.039 | 0.629 | 33.789 | 36.911 |
| B-G1 | 23.436 | 8.535 | 32.422 | 85.303 | 1.255 | 0.497 | 30.226 | 36.886 |
| B-G2 | 24.790 | 7.467 | 37.122 | 59.671 | 2.275 | 0.711 | 31.234 | 30.239 |
| B-G3 | 25.300 | 7.453 | 21.022 | 76.896 | 1.694 | 0.667 | 30.133 | 33.769 |
| B-G4 | 21.994 | 6.438 | 30.989 | 86.596 | 2.325 | 0.666 | 30.834 | 30.019 |
| B-G5 | 23.304 | 6.891 | 22.100 | 62.316 | 1.974 | 0.723 | 31.095 | 33.500 |
| B-G6 | 25.070 | 6.328 | 32.372 | 78.424 | 1.782 | 0.645 | 30.637 | 35.046 |
| B-G7 | 21.463 | 6.691 | 24.456 | 59.259 | 1.707 | 0.825 | 28.611 | 33.907 |
| H-G1 | 21.818 | 6.226 | 33.367 | 80.541 | 1.025 | 0.597 | 32.266 | 35.179 |
| H-G2 | 23.071 | 7.052 | 25.767 | 71.840 | 1.362 | 0.752 | 31.126 | 35.969 |
| H-G3 | 24.113 | 7.474 | 21.350 | 77.837 | 2.093 | 0.623 | 31.642 | 32.174 |
| H-G4 | 23.887 | 7.884 | 29.694 | 82.128 | 1.468 | 0.643 | 30.504 | 36.568 |
| H-G5 | 23.786 | 6.969 | 23.222 | 71.958 | 2.411 | 0.613 | 31.343 | 35.554 |
| H-G6 | 25.373 | 7.519 | 20.083 | 79.541 | 1.733 | 0.594 | 31.561 | 24.043 |
| H-G7 | 21.158 | 7.183 | 29.989 | 71.958 | 1.541 | 0.694 | 33.145 | 28.695 |
| H-G8 | 24.319 | 7.562 | 21.378 | 69.724 | 1.826 | 0.756 | 29.829 | 33.828 |
| H-G9 | 21.542 | 6.865 | 32.022 | 74.544 | 1.540 | 0.700 | 30.733 | 35.057 |
| H-G10 | 22.419 | 6.844 | 35.850 | 80.364 | 1.828 | 0.652 | 30.549 | 38.064 |
| H-G11 | 21.589 | 6.955 | 28.078 | 75.426 | 2.050 | 0.890 | 32.384 | 38.482 |
| H-G12 | 21.801 | 6.856 | 28.100 | 83.480 | 1.833 | 0.791 | 32.502 | 38.148 |
| H-G13 | 22.989 | 6.639 | 22.800 | 75.779 | 1.513 | 0.820 | 34.312 | 38.422 |
| H-G14 | 24.926 | 7.731 | 26.072 | 63.727 | 1.019 | 0.688 | 29.762 | 39.067 |
| **LSD** | **1.271** | **0.670** | **1.282** | **9.340** | **0.139** | **0.018** | **0.968** | **1.737** |

Table (4.9) shows that 10 days irrigation intervals dominated significantly on 40 days with respect to stomatal length but with no significant difference with 20 days giving the mean values (22.664 vs. 22.618 and 21.468 µm) for 10, 20 and 40 days, respectively, while 10 days was superior significantly to both 20 and 40 days with regard to stomatal width (9.680 vs. 7.00 and 4.658 µm) for 10, 20, and 40 days, successively. Chlorophyll concentration of 20 days was superior significantly to 10 days but with no significant difference with 40 days (28.879 vs. 27.165 and 27.476) SPAD for 20, 10 and 40 days, alternatively. Proline, total phenolic content, total flavonoid content and antioxidant activity (DPPH and ABTS) assays for 40 days irrigation interval (107.695 µmol $g^{-1}$ FW, 2.354 mg GAE $g^{-1}$ E, 0.806 mg QE $g^{-1}$ E, 32.948% inhibition and 35.919% inhibition), respectively was superior significantly to both 10 and 20 days. Also, 20 days irrigation interval exceeded significantly that of 10 days for the previous mentioned chemical contents.

In 10 days irrigation interval, stomatal width recorded maximum values with significant differences on 20 and 40-days irrigation intervals. These indicate that the pore of stomata was opened and it is normal when the water increases in soil the transpiration will be faster.

Yellowing leaves of almond genotypes with reduction in chlorophyll concentration may refer to as visual symptoms of extreme cellular damages under severe drought stress. Chlorophyll level in a genotype will determine its relative tolerance and it decreased under water stress, because the plant lost the abilities to balance between the production of reactive oxygen species and the antioxidant defense under drought condition (Reddy *et al.*, 2004), they are responsible for scavenging of single oxygen, causing accumulation of ROS via enhanced leakage of electrons to oxygen. They induce oxidative stress in proteins, membrane lipids and other cellular components (Farooq *et al.,* 2009).

Irrigation intervals have significantly affected on (proline content, total phenol content, total flavonoids content, antioxidant activity by DPPH and ABTS assay). Under drought stress, proline accumulation in the explants was found to be a general response of almond to drought stress, which is significantly increased (Karimi *et al.*, 2012). Habibi (2018) noticed that proline, total phenolic and flavonoids have inclination to increase in plant with drought stress.

**Table 4.9 Effect of irrigation intervals on some stomatal and leaf chemical content characteristics.**

| Irrigation intervals | Stomatal length µm | Stomatal width µm | Chlorophyll concentration SPAD | Proline µmol $g^{-1}$ FW | TPC mg GAE $g^{-1}$ E | TFC mg QE $g^{-1}$ E | DPPH % inhibition | ABTS % inhibition |
|---|---|---|---|---|---|---|---|---|
| 10 D | 22.664 | 9.680 | 27.165 | 42.226 | 1.154 | 0.629 | 29.724 | 29.918 |
| 20 D | 22.618 | 7.000 | 28.879 | 66.797 | 1.645 | 0.710 | 30.614 | 34.167 |
| 40 D | 21.468 | 4.658 | 27.476 | 107.695 | 2.354 | 0.806 | 32.948 | 35.919 |
| **LSD** | **0.357** | **0.188** | **0.360** | **2.624** | **0.039** | **0.005** | **0.272** | **0.488** |

Table (4.10) indicate that maximum values for genotypes and irrigation intervals combinations were (S-G1*20D = 28.272 µm, B-G1*10D = 12.720 µm, S-G6*20D = 48.783 SPAD, Q-G2*40D = 144.797 µmol. $g^{-1}$ FW, S-G2*40D = 3.913 mg GAE $g^{-1}$ E, S-G2*40D = 1.910 mg QE $g^{-1}$ E, H-G13*40D = 36.938% inhibition and B-G1*40D = 41.969% inhibition) for stomatal length, stomatal width, chlorophyll Concentration, proline, total phenol content, total flavonoid content, DPPH and ABTS of seedling, respectively. On the other hand, the minimum values were (S-G1*10D = 14.097 µm, B-G5*40D = 3.173 µm, Q-G5*40D = 12.883 SPAD, S-G3*10D = 27.866 µmol $g^{-1}$ FW, B-G1*10D = 0.145 mg GAE $g^{-1}$ E, S-G7*10D = 0.342 mg QE $g^{-1}$ E, Q-G1*10D = 25.848% inhibition and H-G6*10D = 20.353% inhibition) for the previous characteristics, respectively.

Yadav and Sarma (2016) reported that stomatal closure and gas exchange limitation may be related to drought stress. Drought stress is characterized by reduced leaf water potential, decrease in water content, closure of stomata, turgor loss, and reduction in cell enlargement and growth. Severe water stress can hamper photosynthesis, disrupt the overall metabolism and eventually causes plant necrosis.

Many researchers demonstrated that important mechanisms in leaves or roots to indicate tolerance of almond trees to water stress are osmotic adjustment, stomata conductance decrease, transpiration of water loss, acceleration of leaf shedding and root to shoot dry weight ratio (Yadollahi *et al.*, 2011). Observation of changes in some physiological processes such as stomatal changes in size and conductance, or photosynthetic rate in juvenile plants may be typical for progressive stages of short-term drought.

During water stress, leaf stomatal cells of almond plants have strong mechanisms to keep the leaves active and even turgid. Almond plant leaf can use some mechanisms of resistance to drought such as stomatal closure reduction in leaf size and total leaf area. However, the sensitivity or activity of mechanisms may differ from a genotype to another.

**Table 4.10 Interaction between almond seedling genotypes and irrigation intervals on some stomatal and leaf chemical content characteristics.**

| Genotypes* Irrigation | Stomatal length µm | Stomatal width µm | Chlorophyll concentration SPAD | Proline µmol $g^{-1}$ FW | TPC mg GAE $g^{-1}$ E | TFC mg QE $g^{-1}$ E | DPPH % inhibition | ABTS % inhibition |
|---|---|---|---|---|---|---|---|---|
| S-G1*10 D | 14.097 | 8.637 | 36.767 | 47.619 | 1.387 | 0.755 | 29.143 | 25.142 |
| S-G1*20 D | 28.272 | 5.910 | 27.900 | 61.552 | 1.871 | 0.776 | 30.266 | 26.676 |
| S-G1*40 D | 18.637 | 4.937 | 28.417 | 78.307 | 2.490 | 0.806 | 33.013 | 31.512 |
| S-G2*10 D | 19.073 | 7.428 | 22.833 | 38.977 | 2.170 | 0.741 | 29.079 | 25.554 |
| S-G2*20 D | 20.428 | 7.197 | 26.517 | 58.201 | 2.568 | 0.885 | 30.235 | 30.953 |
| S-G2*40 D | 20.368 | 5.187 | 35.683 | 82.187 | 3.913 | 1.910 | 32.041 | 33.808 |
| S-G3*10 D | 18.638 | 8.650 | 18.383 | 27.866 | 0.698 | 0.783 | 29.197 | 29.257 |
| S-G3*20 D | 21.913 | 7.655 | 30.917 | 50.617 | 1.361 | 0.818 | 30.643 | 32.091 |
| S-G3*40 D | 17.500 | 5.465 | 30.650 | 78.483 | 1.590 | 0.843 | 32.161 | 34.089 |
| S-G4*10 D | 24.497 | 10.498 | 29.650 | 38.272 | 0.199 | 0.588 | 28.192 | 28.600 |
| S-G4*20 D | 23.805 | 9.178 | 21.000 | 52.028 | 0.611 | 0.690 | 29.938 | 33.652 |
| S-G4*40 D | 21.325 | 6.958 | 22.283 | 77.249 | 1.411 | 0.712 | 32.452 | 34.100 |
| S-G5*10 D | 20.174 | 9.462 | 36.333 | 31.922 | 1.814 | 0.758 | 29.393 | 29.964 |
| S-G5*20 D | 25.885 | 8.120 | 17.700 | 50.088 | 1.968 | 0.765 | 30.474 | 33.459 |
| S-G5*40 D | 22.665 | 5.840 | 21.900 | 77.601 | 2.769 | 0.782 | 34.174 | 34.748 |
| S-G6*10 D | 19.116 | 8.016 | 28.383 | 44.621 | 1.762 | 0.611 | 33.779 | 29.912 |
| S-G6*20 D | 23.643 | 7.893 | 48.783 | 70.723 | 2.019 | 0.705 | 33.891 | 34.659 |
| S-G6*40 D | 18.803 | 5.183 | 29.467 | 95.238 | 2.136 | 0.725 | 35.164 | 35.007 |
| S-G7*10 D | 21.582 | 8.701 | 24.833 | 38.448 | 0.588 | 0.342 | 31.576 | 31.557 |
| S-G7*20 D | 28.112 | 8.690 | 35.133 | 62.963 | 1.259 | 0.557 | 31.733 | 35.375 |
| S-G7*40 D | 18.260 | 5.112 | 39.533 | 111.111 | 1.580 | 0.594 | 33.279 | 40.072 |
| S-G8*10 D | 26.544 | 8.426 | 32.550 | 49.206 | 1.225 | 0.406 | 30.195 | 30.206 |
| S-G8*20 D | 17.725 | 6.072 | 23.200 | 70.018 | 1.546 | 0.603 | 31.220 | 32.198 |
| S-G8*40 D | 19.233 | 4.012 | 23.733 | 123.810 | 2.166 | 0.709 | 33.962 | 35.103 |
| S-G9*10 D | 23.193 | 8.882 | 28.150 | 38.624 | 0.593 | 0.679 | 28.271 | 27.945 |
| S-G9*20 D | 15.253 | 6.087 | 41.717 | 75.661 | 0.989 | 0.680 | 29.852 | 34.462 |
| S-G9*40 D | 17.653 | 5.137 | 22.717 | 103.175 | 1.120 | 0.704 | 31.171 | 36.618 |
| M-G1*10 D | 22.585 | 8.742 | 23.100 | 41.799 | 0.497 | 0.676 | 31.263 | 27.325 |
| M-G1*20 D | 20.010 | 7.068 | 39.567 | 54.497 | 1.182 | 0.689 | 31.413 | 31.241 |
| M-G1*40 D | 17.840 | 5.110 | 17.117 | 120.988 | 2.024 | 0.801 | 36.012 | 32.555 |
| M-G2*10 D | 23.768 | 9.578 | 35.400 | 44.974 | 0.225 | 0.655 | 26.858 | 28.664 |
| M-G2*20 D | 20.132 | 6.007 | 32.300 | 67.372 | 0.975 | 0.796 | 27.262 | 32.282 |
| M-G2*40 D | 23.415 | 4.685 | 24.967 | 102.293 | 2.813 | 1.015 | 31.076 | 35.900 |
| M-G3*10 D | 18.765 | 8.340 | 38.167 | 43.915 | 1.473 | 0.656 | 29.870 | 29.116 |
| M-G3*20 D | 23.708 | 7.667 | 24.875 | 66.490 | 2.312 | 0.755 | 30.796 | 30.317 |
| M-G3*40 D | 19.738 | 5.737 | 39.867 | 79.894 | 3.039 | 0.771 | 33.264 | 33.450 |
| Q-G1*10 D | 18.698 | 8.115 | 24.383 | 44.621 | 1.923 | 0.781 | 25.848 | 23.896 |
| Q-G1*20 D | 20.493 | 6.632 | 46.417 | 70.723 | 2.226 | 0.788 | 26.822 | 25.919 |
| Q-G1*40 D | 18.053 | 5.437 | 30.083 | 105.996 | 2.634 | 0.943 | 30.239 | 27.833 |
| Q-G2*10 D | 25.140 | 10.732 | 24.850 | 47.972 | 1.185 | 0.629 | 29.434 | 30.067 |
| Q-G2*20 D | 25.085 | 7.590 | 16.083 | 76.014 | 1.877 | 0.781 | 29.730 | 33.639 |
| Q-G2*40 D | 16.415 | 4.498 | 18.517 | 144.797 | 3.517 | 0.852 | 31.477 | 34.729 |

***Continued***

**Table 4.10 Continued**

| Genotypes* Irrigation | Stomatal length µm | Stomatal width µm | Chlorophyll concentration SPAD | Proline µmol $g^{-1}$ FW | TPC mg GAE $g^{-1}$ E | TFC mg QE $g^{-1}$ E | DPPH % inhibition | ABTS % inhibition |
|---|---|---|---|---|---|---|---|---|
| Q-G3*10 D | 25.752 | 11.332 | 15.700 | 31.217 | 1.572 | 0.664 | 30.066 | 30.969 |
| Q-G3*20 D | 20.377 | 6.322 | 20.050 | 61.552 | 2.184 | 0.709 | 30.496 | 35.691 |
| Q-G3*40 D | 16.873 | 4.588 | 35.367 | 129.453 | 2.505 | 0.839 | 33.146 | 37.898 |
| Q-G4*10 D | 18.727 | 9.245 | 29.817 | 44.444 | 1.164 | 0.713 | 28.630 | 29.625 |
| Q-G4*20 D | 21.087 | 5.847 | 25.300 | 74.427 | 1.929 | 0.721 | 29.462 | 30.964 |
| Q-G4*40 D | 16.472 | 4.012 | 26.700 | 100.176 | 2.851 | 0.834 | 32.567 | 32.791 |
| Q-G5*10 D | 26.693 | 10.060 | 24.917 | 46.384 | 0.369 | 0.525 | 32.054 | 34.244 |
| Q-G5*20 D | 25.040 | 7.403 | 27.433 | 79.189 | 0.953 | 0.660 | 33.373 | 37.849 |
| Q-G5*40 D | 20.733 | 4.090 | 12.883 | 119.577 | 1.794 | 0.702 | 35.939 | 38.641 |
| B-G1*10 D | 26.823 | 12.720 | 44.583 | 46.208 | 0.145 | 0.481 | 29.776 | 27.869 |
| B-G1*20 D | 21.802 | 7.633 | 31.783 | 79.189 | 1.753 | 0.490 | 29.820 | 40.819 |
| B-G1*40 D | 21.683 | 5.252 | 20.900 | 130.511 | 1.867 | 0.519 | 31.083 | 41.969 |
| B-G2*10 D | 24.653 | 11.020 | 33.383 | 47.619 | 1.806 | 0.541 | 30.613 | 27.088 |
| B-G2*20 D | 26.375 | 7.322 | 36.500 | 58.730 | 2.149 | 0.702 | 30.842 | 31.557 |
| B-G2*40 D | 23.342 | 4.058 | 41.483 | 72.663 | 2.869 | 0.888 | 32.249 | 32.071 |
| B-G3*10 D | 22.133 | 10.518 | 20.233 | 46.208 | 1.061 | 0.640 | 28.592 | 25.576 |
| B-G3*20 D | 27.498 | 7.925 | 17.017 | 77.954 | 1.549 | 0.643 | 29.227 | 37.605 |
| B-G3*40 D | 26.268 | 3.915 | 25.817 | 106.526 | 2.473 | 0.718 | 32.579 | 38.128 |
| B-G4*10 D | 18.922 | 9.387 | 24.150 | 47.795 | 1.957 | 0.651 | 30.114 | 26.066 |
| B-G4*20 D | 23.332 | 6.488 | 33.533 | 82.363 | 2.350 | 0.664 | 30.425 | 31.976 |
| B-G4*40 D | 23.730 | 3.440 | 35.283 | 129.630 | 2.668 | 0.684 | 31.964 | 32.015 |
| B-G5*10 D | 22.887 | 10.395 | 21.717 | 46.561 | 1.535 | 0.638 | 29.618 | 25.441 |
| B-G5*20 D | 21.280 | 7.103 | 24.050 | 61.728 | 1.993 | 0.743 | 30.474 | 37.005 |
| B-G5*40 D | 25.747 | 3.173 | 20.533 | 78.660 | 2.393 | 0.787 | 33.192 | 38.054 |
| B-G6*10 D | 24.560 | 8.545 | 25.183 | 40.917 | 1.369 | 0.608 | 29.012 | 27.307 |
| B-G6*20 D | 25.075 | 6.667 | 41.400 | 57.672 | 1.603 | 0.607 | 30.291 | 38.004 |
| B-G6*40 D | 25.575 | 3.773 | 30.533 | 136.684 | 2.375 | 0.719 | 32.609 | 39.826 |
| B-G7*10 D | 22.680 | 9.807 | 21.833 | 43.739 | 1.082 | 0.745 | 27.559 | 27.669 |
| B-G7*20 D | 18.160 | 6.697 | 23.617 | 59.259 | 1.275 | 0.861 | 28.271 | 36.778 |
| B-G7*40 D | 23.548 | 3.568 | 27.917 | 74.780 | 2.763 | 0.869 | 30.002 | 37.273 |
| H-G1*10 D | 22.935 | 9.522 | 28.217 | 41.975 | 0.662 | 0.455 | 30.066 | 33.544 |
| H-G1*20 D | 20.475 | 5.767 | 30.050 | 77.954 | 0.963 | 0.735 | 32.027 | 35.350 |
| H-G1*40 D | 22.045 | 3.390 | 41.833 | 121.693 | 1.449 | 0.600 | 34.704 | 36.643 |
| H-G2*10 D | 27.713 | 11.837 | 27.250 | 45.503 | 1.039 | 0.740 | 30.459 | 33.780 |
| H-G2*20 D | 23.833 | 5.838 | 29.900 | 73.192 | 1.219 | 0.702 | 30.916 | 35.940 |
| H-G2*40 D | 17.667 | 3.480 | 20.150 | 96.825 | 1.829 | 0.814 | 32.004 | 38.187 |
| H-G3*10 D | 26.003 | 10.653 | 24.033 | 41.446 | 1.939 | 0.589 | 30.195 | 31.337 |
| H-G3*20 D | 23.475 | 7.405 | 20.150 | 71.958 | 2.078 | 0.608 | 31.275 | 32.312 |
| H-G3*40 D | 22.862 | 4.365 | 19.867 | 120.106 | 2.262 | 0.673 | 33.457 | 32.874 |
| H-G4*10 D | 24.187 | 11.158 | 23.967 | 45.150 | 1.026 | 0.603 | 29.193 | 32.910 |
| H-G4*20 D | 24.082 | 7.075 | 34.433 | 85.185 | 1.368 | 0.630 | 30.683 | 38.256 |
| H-G4*40 D | 23.393 | 5.420 | 30.683 | 116.049 | 2.008 | 0.697 | 31.637 | 38.538 |

***Continued***

**Table 4.10 Continued**

| Genotypes* Irrigation | Stomatal length µm | Stomatal width µm | Chlorophyll concentration SPAD | Proline µmol $g^{-1}$ FW | TPC mg GAE $g^{-1}$ E | TFC mg QE $g^{-1}$ E | DPPH % inhibition | ABTS % inhibition |
|---|---|---|---|---|---|---|---|---|
| H-G5*10 D | 21.497 | 9.168 | 17.767 | 37.213 | 1.623 | 0.540 | 29.862 | 33.613 |
| H-G5*20 D | 24.860 | 8.055 | 27.000 | 63.139 | 2.004 | 0.620 | 30.809 | 35.400 |
| H-G5*40 D | 25.000 | 3.685 | 24.900 | 115.520 | 3.606 | 0.678 | 33.357 | 37.648 |
| H-G6*10 D | 27.020 | 11.473 | 23.150 | 45.679 | 1.425 | 0.508 | 30.903 | 20.353 |
| H-G6*20 D | 23.965 | 6.795 | 20.300 | 66.490 | 1.773 | 0.633 | 30.910 | 23.790 |
| H-G6*40 D | 25.135 | 4.290 | 16.800 | 126.455 | 2.000 | 0.641 | 32.871 | 27.985 |
| H-G7*10 D | 22.763 | 11.287 | 29.467 | 39.683 | 0.808 | 0.619 | 31.233 | 25.950 |
| H-G7*20 D | 20.887 | 6.405 | 31.183 | 53.263 | 1.199 | 0.690 | 32.687 | 28.584 |
| H-G7*40 D | 19.825 | 3.858 | 29.317 | 122.928 | 2.617 | 0.775 | 35.516 | 31.552 |
| H-G8*10 D | 23.772 | 11.545 | 26.833 | 42.152 | 1.380 | 0.699 | 28.361 | 32.866 |
| H-G8*20 D | 23.967 | 6.775 | 15.517 | 58.730 | 2.059 | 0.776 | 29.954 | 33.262 |
| H-G8*40 D | 25.220 | 4.365 | 21.783 | 108.289 | 2.039 | 0.793 | 31.171 | 35.356 |
| H-G9*10 D | 22.790 | 9.275 | 38.833 | 39.683 | 1.308 | 0.510 | 28.912 | 34.191 |
| H-G9*20 D | 20.118 | 6.002 | 32.300 | 57.143 | 1.613 | 0.604 | 29.797 | 34.374 |
| H-G9*40 D | 21.718 | 5.318 | 24.933 | 126.808 | 1.699 | 0.986 | 33.488 | 36.607 |
| H-G10*10 D | 21.858 | 8.675 | 29.350 | 44.092 | 1.083 | 0.636 | 29.992 | 34.361 |
| H-G10*20 D | 21.467 | 6.443 | 34.450 | 76.720 | 1.921 | 0.648 | 30.210 | 39.891 |
| H-G10*40 D | 23.933 | 5.415 | 43.750 | 120.282 | 2.479 | 0.673 | 31.445 | 39.938 |
| H-G11*10 D | 21.682 | 9.608 | 27.533 | 35.273 | 1.790 | 0.781 | 31.209 | 35.772 |
| H-G11*20 D | 21.623 | 6.375 | 21.583 | 66.138 | 1.939 | 0.953 | 32.095 | 39.575 |
| H-G11*40 D | 21.463 | 4.882 | 35.117 | 124.868 | 2.423 | 0.937 | 33.849 | 40.100 |
| H-G12*10 D | 21.330 | 8.872 | 24.700 | 43.386 | 0.864 | 0.598 | 30.597 | 36.131 |
| H-G12*20 D | 19.135 | 6.887 | 37.117 | 75.838 | 1.312 | 0.763 | 32.011 | 37.607 |
| H-G12*40 D | 24.938 | 4.808 | 22.483 | 131.217 | 3.323 | 1.013 | 34.899 | 40.706 |
| H-G13*10 D | 24.048 | 7.490 | 16.217 | 42.504 | 0.536 | 0.752 | 32.058 | 34.275 |
| H-G13*20 D | 23.688 | 6.992 | 26.883 | 74.427 | 1.533 | 0.797 | 33.940 | 39.932 |
| H-G13*40 D | 21.232 | 5.437 | 25.300 | 110.406 | 2.469 | 0.910 | 36.938 | 41.059 |
| H-G14*10 D | 23.918 | 10.057 | 29.650 | 40.917 | 0.554 | 0.619 | 28.342 | 38.742 |
| H-G14*20 D | 23.405 | 7.998 | 23.750 | 59.083 | 1.031 | 0.718 | 29.057 | 38.915 |
| H-G14*40 D | 27.455 | 5.137 | 24.817 | 91.182 | 1.472 | 0.726 | 31.887 | 39.543 |
| LSD (0.05) | **2.201** | **1.161** | **2.220** | **16.177** | **0.241** | **0.031** | **1.677** | **3.008** |

Drought reduces plant growth and development, which causes to produce smaller organs, and decrease flower production (Hossain *et al.*, 2016). Water stress inhibits cell division less than cell enlargement. Compactness of plant growth is affected by many biochemical and physiological processes, such as translocation, photosynthesis, ion uptake, respiration, carbohydrates, growth promoters and nutrient metabolism (Jaleel *et al.*, 2008). Yadollahi *et al.*, (2011) and Zokaee-Khosroshahi *et al.* (2014) clearly showed that lower stomatal size could be linked to resistance to drought in almond genotype. The difference in stomatal aperture is crucial to the ability of adaptation of the genotypes and an important factor in stress response is aperture size (Zhu *et al.*, 2005). Taiz and Zeiger (2002) mentioned that during periods of water deficit, abscisic acid synthesis rises in the roots and transports to the shoot through, and finally cause stomatal closure.

Yellowing leaves of almond genotypes with reduction in chlorophyll concentration may refer to as visual symptoms of extreme cellular damages under severe drought stress. In addition, researchers demonstrated that there is no doubt that measurements of chlorophyll fluorescence can also provide useful information about leaf photosynthetic performance with respect to water deficit (Baker and Rosenqvist, 2004). In addition, this evaluation provides a rapid and non-invasive technique to monitor crop growth and performance. Decrease in relative water content (RWC) of leaves initially induces stomatal closure, imposes a decrease in the supply of $CO_2$ to the mesophyll cells and, consequently, results in decrease in the rate of leaf photosynthesis and chlorophyll fluorescence parameters (Lu and Zhang, 1998). One of the most abundant pigments in the plant is chlorophyll, which plays an active role in the photosynthetic process, so it has significant effect on growth of plants. During different stresses, accumulation and chlorophyll contents was significantly decreased in mild (-16.66%) and severe (-40%) stress as compared to control (Haider *et al.*, 2018). Sepehri and Golparvar (2011) stated that plant chlorophyll content affected by drought stress depends on environmental conditions and plant genotypes, some varieties possessed compact chlorophyll content and some amplified. Generally, there are several reports on chlorophyll decay under drought stress. Structural damages to chloroplasts by photo degradation and or elevated ROS formation of the pigments may led to damage of chlorophylls in the leaves under drought stress conditions (Karimi *et al.*, 2013b).

Under drought stress, proline accumulation in the explants was found to be a general response of almond to drought stress, which is significantly increased**.** Some researchers demonstrated that accumulation of proline in the almond explants associated with changes in levels of both RWC and MSI under drought stress (Karimi *et al.*, 2012). In addition, proline accumulation in plant leaves may be linked to reducing in RWC or structural damages (Taylor, 1996). Ozden *et al.* (2009) suggested that proline acts as an osmoregulatory, an osmo-protector or a regulator of the redox potential of cells under drought stress, is believed that proline accumulation helps plant tolerate drought stress conditions. Therefore, it can be seen that drought-tolerant almond genotypes can tolerate prolonged dehydration periods. In the sensitive genotypes (‘Mamaei’ and ‘Ferragnès’) more proline was accumulated. Therefore, it can be concluded that to evaluate the drought stress in almonds, proline is considered as a best physiological marker. However, it can be suggesting that other mechanisms may be used beside proline accumulation and osmotic regulation to screen drought-tolerant of almond genotypes. Taghizadeh *et al.* (2015) showed that variation of total phenolic in the Mahaleb genotypes may be due to genetic variations.

Plants tolerate different abiotic stresses by accumulating polyphenols compound due to response to these stresses and helping it to acclimatize to negative environments Hence, the phenols concentration in tissue is an indicator to predict the level of stress tolerance which varies significantly in different plants. Phenolic compounds impact the growth of plants and development, such as seed biomass accumulation, germination, and enhanced plant metabolism (Sharma *et al.,* 2019). Also Habibi (2018) observed that the proline, total phenolic and flavonoids have a tendency to rise in plants with mild drought stress.

The grade to which the activities of antioxidants rise in drought-stressed plants seems to be extremely variable between plant species, even between different genotypes of the same species (Pessarakli, 2016).

Endogenous flavonoids considerably increase in plants exposed to abiotic and biotic stresses such as drought, wounding, metal toxicity and nutrient deficiency (Hossain *et al.,* 2016).

Karimi *et al.* (2012) showed that drought-tolerant almond genotypes can be screened using *in-vitro* methods and these genotypes showed less decrease of growth characteristics and healthier stability under drought stress and presented the ability of dehydration without injury. Proline accumulation in the almond leaves is a common reply to drought stress and its concentration perhaps is not associated with drought tolerance of the plant, though almonds are susceptible to oxidative stresses and structural damages for the drought stress. Still, dehydration may be tolerated via osmoregulation mechanisms. It can, therefore, be inferred that the tolerance of almond genotypes to drought can result from a combination of certain physiological features.

Karimi *et al.* (2012) showed that under drought stress, membrane stability index (MSI) was deteriorated and also relative water content (RWC) significantly decreased with cell dehydration which brought about some malfunctions of cell metabolism of drought-tolerant almond genotypes. MSI decline lead to reactive oxygen species (ROS) that damages cell membrane and other cell structures.

Laxa *et al.* (2019) reported that in agriculture, water deficiency affects plant performance and production, plant-genotype-specific characteristics, stress intensity and duration, are three important performances to plant survival of the severe drought stress periods, in addition, it can also depend on the speed and efficiency of recovery to determine plant performance. During drought stress, both reactive oxygen species (ROS) and reactive nitrogen species (RNS) are activated which in turn affect the redox regulatory state of the cell. This redox regulatory with antioxidant system can have a significant role in drought tolerance. Furthermore, the significance of the antioxidant system in surviving severe phases of dehydration is further supported by the strong antioxidant system usually encountered in resurrection plants.

To understand the mechanisms of drought-adaptive plants, remarkable improvement has been made, despite the intricacy of drought resistance process. The morpho-physiological alterations are considered as a main factor of adaptation through drought resistance. The molecular mechanisms that regulate the expression of genes are responsible for controlling these changes in the process of adaptation. Plant species vary in their drought adaptation. These variations can be used as a main source for research study in drought adaption. The improvements of both drought resistance and yields in cultivated varieties can be achieved by significant exploiting of these natural variations through the understanding of their basic mechanisms and then encourage choosing these traits (Basu *et al.*, 2016).

# Conclusions

The most important conclusions of this study can be stated as follows:

- **Regarding to the tree genotypes**

1. There are very different genotypes in Sulaimani and Halabja governorates.
2. Genotypes and location are effective nearly on all parameters.
3. Almond genotypes responded to water deficit in the form of changes in various morphological, physiological and phytochemical traits.
4. Sharbazher and Qaradagh genotypes leaves contain phytochemical more than other genotypes.
5. Proline in Mergapan genotypes are more than other locations.
6. The higher values of antioxidant activity are observed in Sharbazher genotypes.

- **Regarding to the tree genotypes molecular**

1. Sharbazher, Mergapan, Qaradagh and Barznja genotypes are more related genetically.
2. Hwaraman genotypes genetically are different from other locations genotypes.
3. Number of polymorphic bands, gene diversity and polymorphic information content in RAPD marker are higher than ISSR marker.

- **Regarding to seedling pre-drought tolerance**

1. The tolerance of drought is different among genotypes.
2. Genotypes are different significantly in all parameters in glasshouse.
3. Irrigation interval are affected significantly on seedling growth.

# Recommendations

Based on the conclusions mentioned previously, the following important recommendations can be noted:

1. Collecting more different almond genotypes in other locations for future researches.
2. Conducting quantitative traits loci analysis and genome-wide associated to determine QTL that associated with drought tolerance in almond genotypes with using various types of markers including SNIPs, SRAPs, ALFPs, and SSRs.
3. Using different irrigation intervals in field and greenhouse and using PEG as a moisture reducer.
4. We observed different flowering times in almond, so we suggest this phenomenon is important and it is a best indicator to almond tolerant to late frosts.
5. Using the stress tolerance genotypes by breeder to make a new species.
6. Checking almond nut phytochemically.
7. Applying supplemental irrigation after rainfall-interruption.
8. Testing the ground level area where almond genotype widespread.
9. Fertilizing with some nutrients such as potassium and boron.
10. Applying some physiological indices (e.g. stress tolerance index STI, stress intensity SI, stress susceptibility index SSI, stress tolerance TOL and mean productivity MP) to identify tolerant genotypes that produce high yield under drought and normal condition.
11. Measuring the length of root seedlings.

# APPENDIXCES

## Appendix 1. Preparation of solution to proline determination.

Sulfosalicylic acid (3%): Dissolve 3 g 5-sulfosalicylic acid (2-hydroxy-5-sulfobenzoic acid) in 80 mL distilled water and make up to 100 ml. Solution can be stored at room temperature for weeks.
Acidic ninhydrin: 1.25 g ninhydrin (1,2,3-indantrione monohydrate), 30 mL glacial acetic acid, 20 mL of 6 M orthophosphoric acid, dissolve by vortexing and gentle warming. Solution can be stored at 4◦C for up to 1 week.

## Appendix 2. Preparation of solutions to total phenolic content determination.

Folin–Ciocalteu phenol reagent (10 x) for 100 ml:
10 mL of Folin–Ciocalteu phenol reagent are added to 90 ml of $dH_2O$.
$Na_2CO_3$ 10% for 100 ml:
10 of $Na_2CO_3$ are dissolved in 90 ml $dH_2O$. Store at 15-27 ºC.

## Appendix 3. Preparation of solutions to total flavonoids content determination.

Methanol 80% for 100 ml:
80 ml of methanol is added to 20 ml of $dH_2O$. Store at 5-27 ºC
Aluminum chloride hexahydrate ($AlCl_3$) 2% for 100 ml:
2 g of $AlCl_3$ are dissolved in $dH_2O$ and the volume is made up to 100 ml. Store at 15-27 ºC
Potassium acetate ($CH_3COOK$) 1 M for 100 ml:
9.815g of CH3COOK are dissolved in distilled water and the volume is made up to 100 ml to give a 1 M CH3COOK solution. Store at 15-27 ºC

## Appendix 4. Preparation of solutions to saponin content determination.

Sulfuric acid 72% for 100 ml
Add 72 ml of concentrated sulfuric acid to 28 ml $dH_2O$.
Vanillin 4 % for 100 ml:
4 g of vanillin are dissolved in methanol and the volume is made up to 100 ml.

## Appendix 5. Preparation of solution to antioxidants activity (DPPH assay) determination.

0.06(6 x$10^{-5}$ M) DPPH = X/ (394.32*0.1), X = 0.0024 g
0.24f DPPH is dissolved in 100 ml of methanol (95%).

**Appendix 6. Preparation of solution to antioxidants activity (ABTS assay) determination.**

ABTS solution (7 mM): 0.0576 g of ABTS is dissolved in 15 ml of distilled water. Store this solution in amber flask (-20°C).
Potassium persulfate (140 mM): Mix 378.4 mg of the salt with 10 ml of distilled water. Store in amber flask (-20°C).
ABTS solution: the day before the experiment mix in another amber flask 10 ml of the ABTS solution with 176 µl of the potassium persulfate and leave it at room temperature for 16 hours.

**Appendix 7. Preparation of solution to antioxidants activity Total antioxidant capacity determination.**

Sodium phosphate (28 mM):
Mix 19 ml of 0.2 M of $NaH_2PO_4$ with 81 ml of 0.2 M of $Na_2HPO_4$ to make a solution of Na phosphate of 0.2 M, then add 100 ml of distilled water to mix, mix 6 ml of sodium phosphate 0.2 M with 36 ml of $dH_2O$ to make a solution of 28 mM of sodium phosphate
Ammonium Molybdate (4 mM):
0.5 g of ammonium molybdate are dissolved with 100 ml $dH_2O$
Sulfuric acid (0.6 M):
3.33l of $H_2SO_4$ is added to 96.67 ml of $dH_20$.

**Appendix 8. CTAB buffer preparation.**

CTAB Buffer

- 10 ml 1 M Tris HCl pH 8.0
- 28 ml 5 M NaCl
- 4 ml of 0.5 M EDTA
- 2 g of CTAB (cetyltrimethyl ammonium bromide)
- Brought total volume to 100 ml with $ddH_2O$.

Tris HCl (1 M) pH 8.0

- 12.1 g Tris dissolved in about 70 ml of $dH_2O$.
- brought pH down to 8.0 by added concentrated HCl
- Brought total volume to 100 ml with $dH_2O$.

EDTA (0.5 M)

- 18.6 g EDTA Added to nearly 70 ml $dH_2O$
- Added 1.6-1.8 g of NaOH pellets
- Adjust pH to 8.0 by with a few more pellets.
- Brought total volume to 100 ml with $dH_2O$.

NaCl (5 M)

- 29.2 g of NaCl
- 70 ml $dH_2O$
- Dissolved and brought to 100 ml.

Ammonium acetate (7.5 M)

- 57.81 g ammonium acetate
- ~50 ml of dH2O
- Brought to 100 ml total volume

**Appendix 9. Mean squares of variance analysis for some vegetative characters for almond tree genotypes.**

| S. O. V. | df | Annual Shoot Growth (cm) | Annual Shoot Diameter (mm) | Leaves area ($cm^2$) | Stomatal Sto $mm^2$ | Leaves Dry Weight (%) |
|---|---|---|---|---|---|---|
| Block | 2 | 19.2235 | 0.0206 | 0.0192 | 811.1914 | 3.6270 |
| Genotype | 37 | 426.2271 | 1.7447 | 0.4489 | 10246.9928 | 36.3560 |
| Error | 74 | 30.5061 | 0.1760 | 0.0033 | 665.4954 | 1.8318 |

* significant at P≤*0.05*

**Appendix 10. Mean squares of variance analysis for some nut characters for almond tree genotypes.**

| S. O. V. | df | Nut Width (mm) | Nut Length (mm) | Nut Thickness (mm) | Nut Weight (g) | Shell to Kernel (%) |
|---|---|---|---|---|---|---|
| Block | 2 | 3.2818 | 4.4335 | 0.0891 | 0.3984 | 33.9786 |
| Genotype | 37 | 23.4046 | 55.1264 | 4.6209 | 3.9952 | 0.0956* |
| Error | 74 | 0.8790 | 3.8833 | 0.4883 | 0.2561 | 1.4718 |

* significant at P≤ *0.05*

**Appendix 11. Mean squares of variance analysis for some phytochemical characters for almond tree genotypes.**

| S. O. V. | df | Chlorophyll Concentration (SPAD) | proline (µmol $g^{-1}$ FW) | TPC (mg GAE $g^{-1}$ E) | TFC (mg QE $g^{-1}$ E) | SC (mg SE $g^{-1}$ E) |
|---|---|---|---|---|---|---|
| Block | 2 | 0.0117 | 76.5667 | 0.1274 | 0.0580 | 0.0618 |
| Genotype | 37 | 121.0265 | 4155.8349 | 8.7896 | 1.2737 | 92.4654 |
| Error | 74 | 0.0512 | 227.8586 | 0.0850 | 0.0014 | 0.1236 |

* significant at P≤ *0.05*

**Appendix 12. Mean squares of variance analysis for some phytochemical characters for almond tree genotypes.**

| S. O. V. | df | CTC (mg CE $g^{-1}$ E) | DPPH (% inhibition) | ABTS (% inhibition) | TAC (mg AA g-1 E) |
|---|---|---|---|---|---|
| Block | 2 | 0.1349 | 63.0797 | 1.4114 | 0.0155 |
| Genotype | 37 | 3.6509 | 466.1277 | 507.5353 | 0.0297 |
| Error | 74 | 0.0090 | 6.0540 | 0.7899 | 0.0026 |

* significant at P≤ *0.05*

**Appendix 13. Mean squares of variance analysis for vegetative characters for almond seedling in glasshouse.**

| S. O. V. | df | High (cm) | diameter (mm) | leaves number | leaves area ($cm^2$) | vegetative weight (gm) | vegetative Dry weight (%) | Root weight (gm) | Root Dry weight (%) |
|---|---|---|---|---|---|---|---|---|---|
| Block | 2 | 39.2518* | 0.1856* | 21.4064* | 0.0248* | 5.8922* | 4.0244* | 0.5393* | 144.5508* |
| Genotype | 37 | 195.1084* | 0.3694* | 31.7206* | 0.1739* | 2.9779* | 15.6832* | 8.8978* | 28.0809* |
| Irrigation | 2 | 272.2054* | 2.9457* | 846.0468* | 3.3913* | 38.1885* | 147.4918* | 228.8277* | 5891.3717* |
| Genotype × Irrigation | 74 | 14.6829* | 0.1240* | 10.5122* | 0.0220* | 1.8061* | 12.2447* | 3.7545* | 23.2954* |
| Error | 226 | 18.2257* | 0.1195* | 9.0407* | 0.0079* | 1.5591* | 6.9447* | 1.6347* | 24.0211* |

* significant at *P≤ 0.05*

**Appendix 14. Mean squares of variance analysis for some stomatal characters and chemical content in almond seedling in glasshouse.**

| S. O. V. | df | Stomatal length (µm) | Stomatal width (µm) | Chlorophyll Concentration (spad) | Proline (µmol g$^{-1}$ FW) | TPC (mg GAE g$^{-1}$ E) | TFC (mg QE g$^{-1}$ E) | DPPH (% inhibition) | ABTS (% inhibition) |
|---|---|---|---|---|---|---|---|---|---|
| Block | 2 | 1.1210* | 2.0587* | 1.9253* | 150.4942* | 0.0405* | 0.0014* | 1.9920* | 8.8904* |
| Genotype | 37 | 31.6592* | 3.1584* | 214.4135* | 790.2284* | 2.1317* | 0.1264* | 19.2036* | 109.1196* |
| Irrigation | 2 | 52.3571* | 719.8827* | 95.0872* | 124689.9592* | 41.4904* | 0.8956* | 316.1300 | 1085.6938* |
| Genotype × Irrigation | 74 | 24.2890* | 2.9954* | 145.9363* | 407.4386* | 0.3469* | 0.0376* | 1.1490* | 10.9011* |
| Error | 226 | 1.8709* | 0.5210* | 1.9041* | 101.0943* | 0.0224* | 0.0004* | 1.0863* | 3.4961* |

* significant at P≤ *0.05*

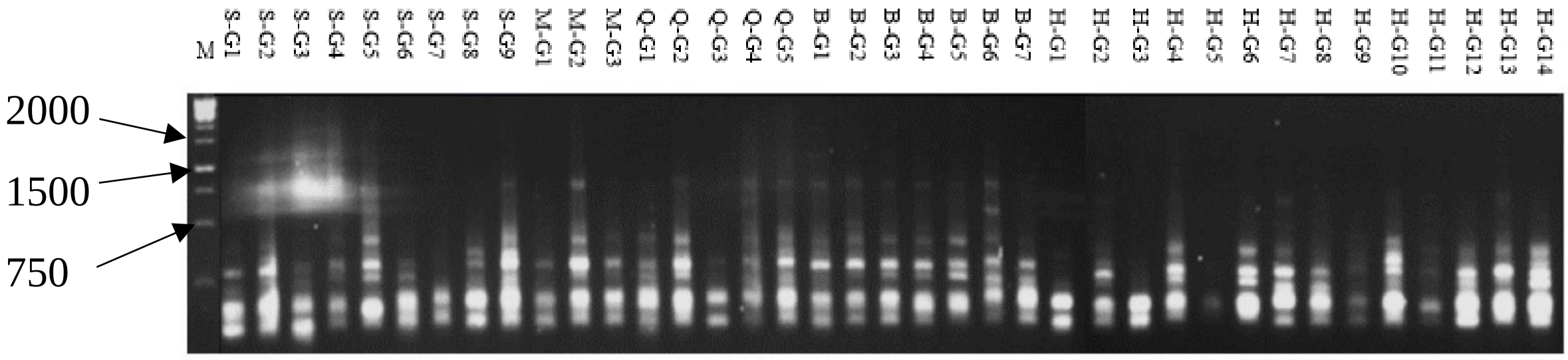

**Appendix 15. RAPD amplification profile for primer OPA-08 on almond genotypes. M: 1 kb DNA ladder.**

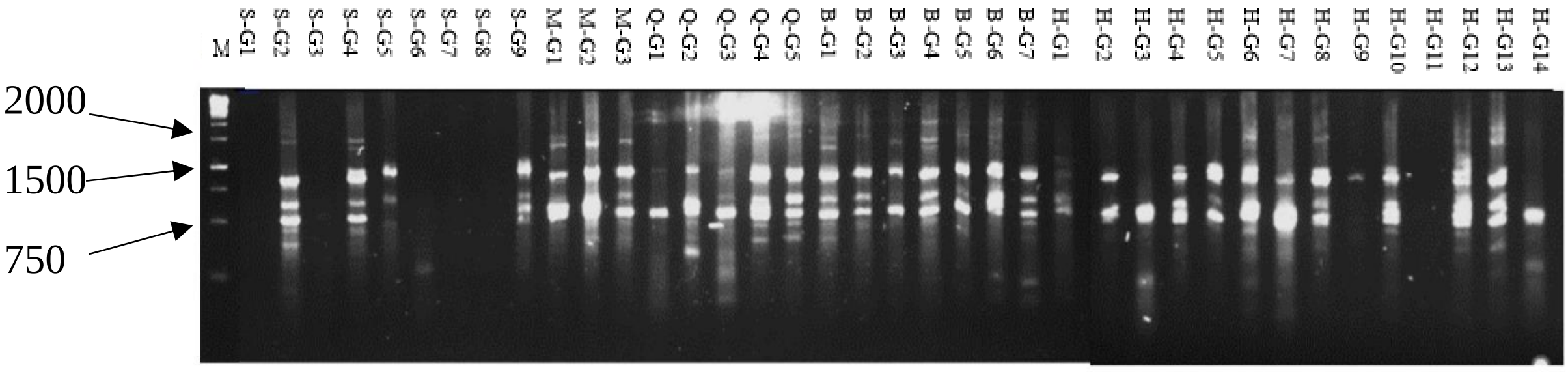

**Appendix 16. RAPD amplification profile for primer OPA-10 on almond genotypes. M: 1 kb DNA ladder.**

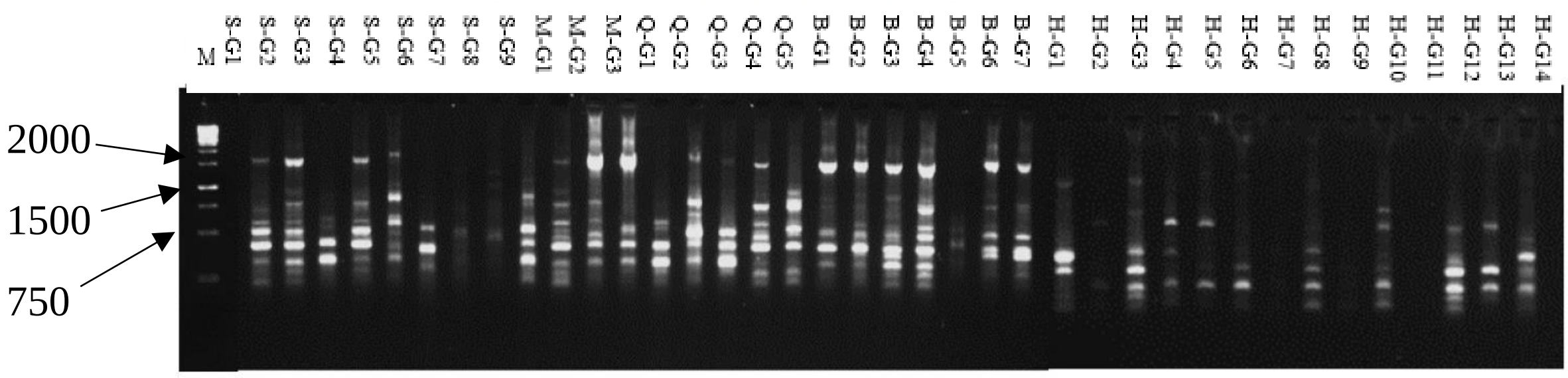

**Appendix 17. RAPD amplification profile for primer OPA-11 on almond genotypes. M: 1 kb DNA ladder.**

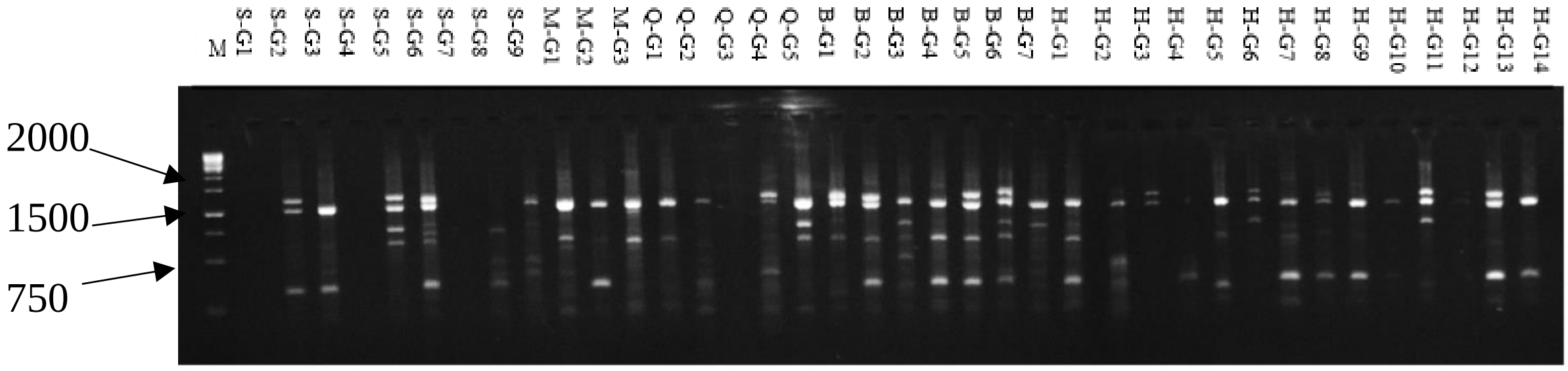

**Appendix 18. RAPD amplification profile for primer OPA-16 on almond genotypes. M: 1 kb DNA ladder.**

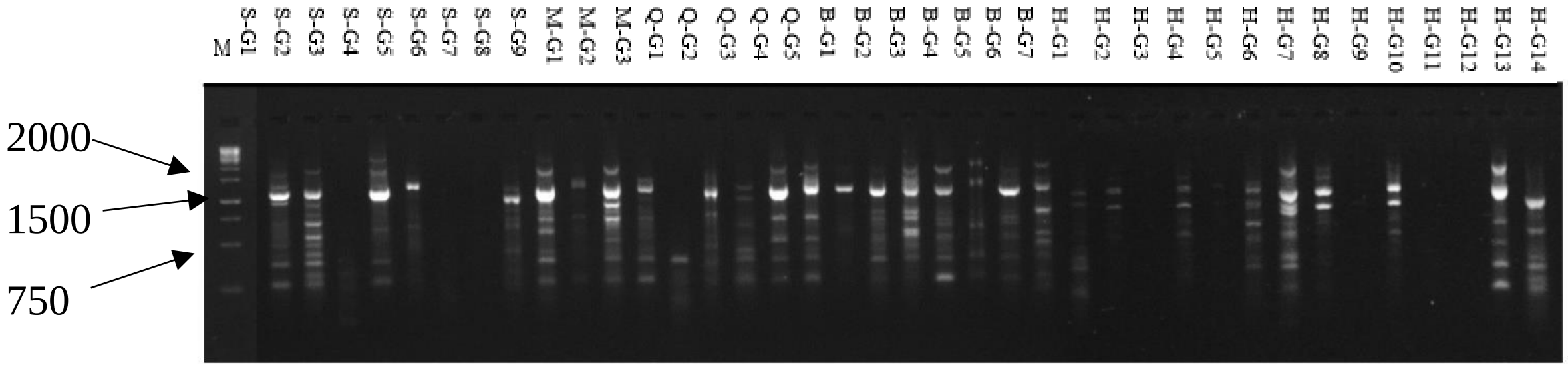

**Appendix 19. RAPD amplification profile for primer OPB-11 on almond genotypes. M: 1 kb DNA ladder.**

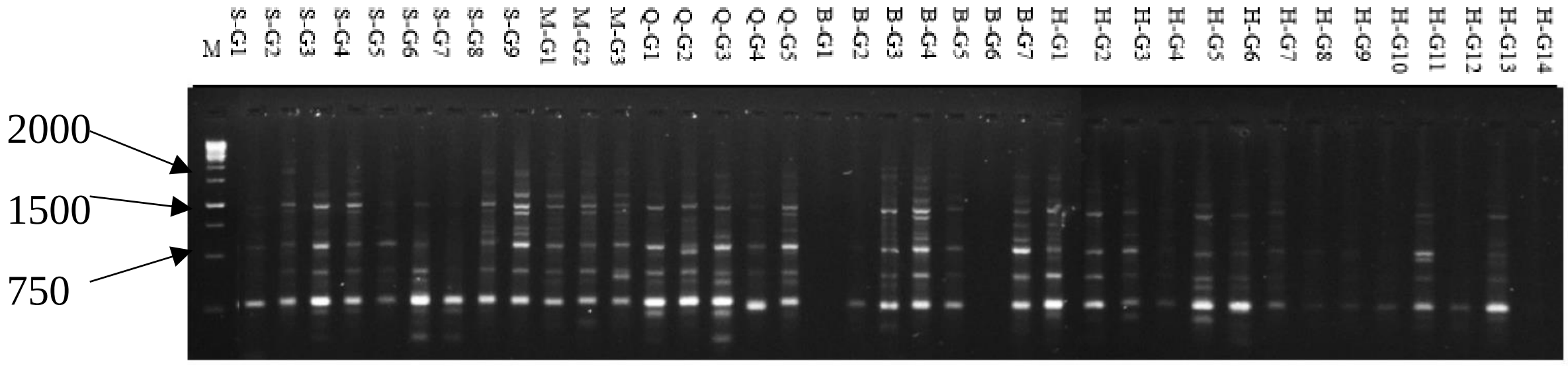

**Appendix 20. RAPD amplification profile for primer S075 on almond genotypes. M: 1 kb DNA ladder.**

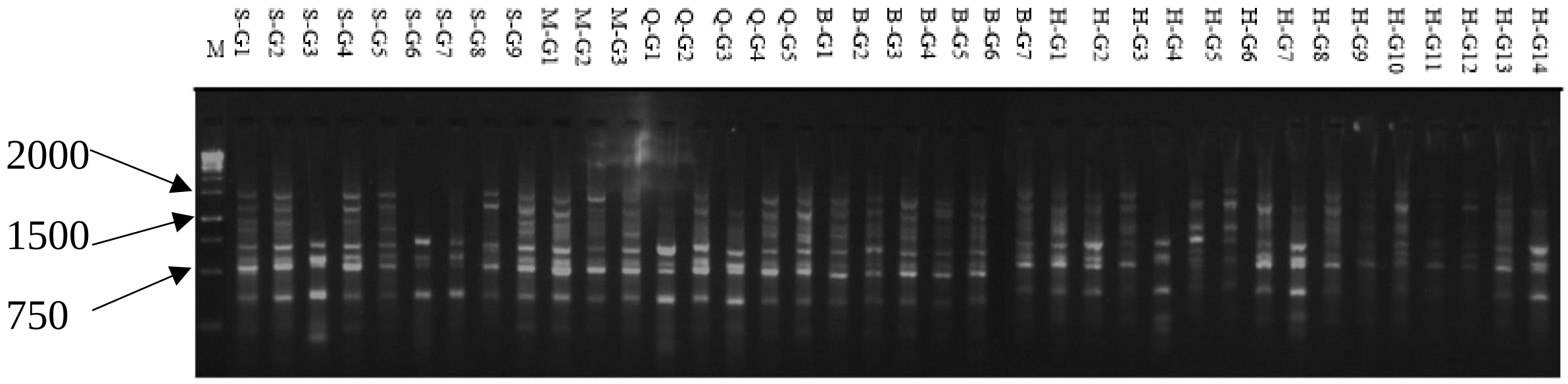

**Appendix 21. RAPD amplification profile for primer S084 on almond genotypes. M: 1 kb DNA ladder.**

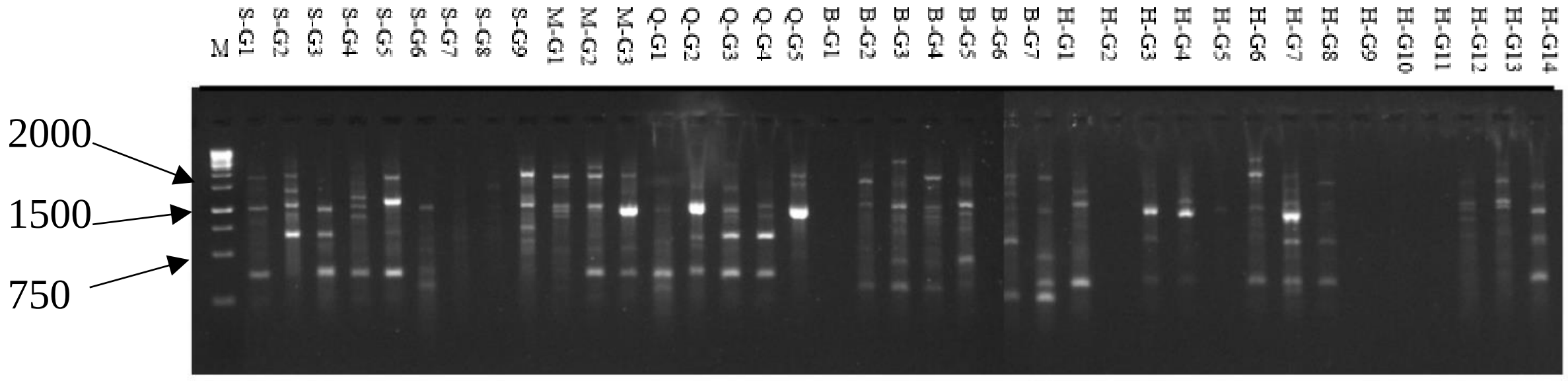

**Appendix 22. RAPD amplification profile for primer S085 on almond genotypes. M: 1 kb DNA ladder.**

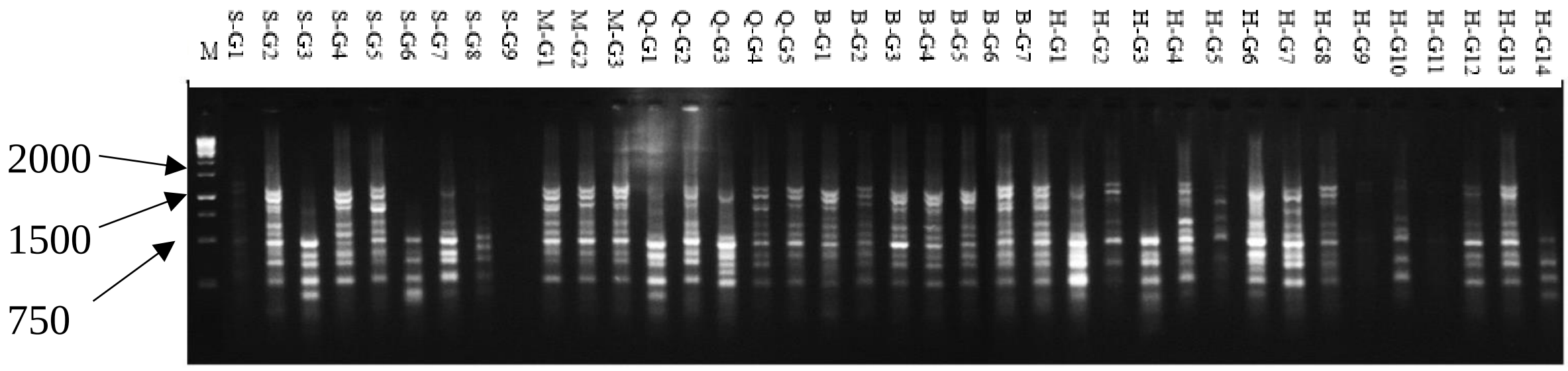

**Appendix 23. RAPD amplification profile for primer S081 on almond genotypes. M: 1 kb DNA ladder.**

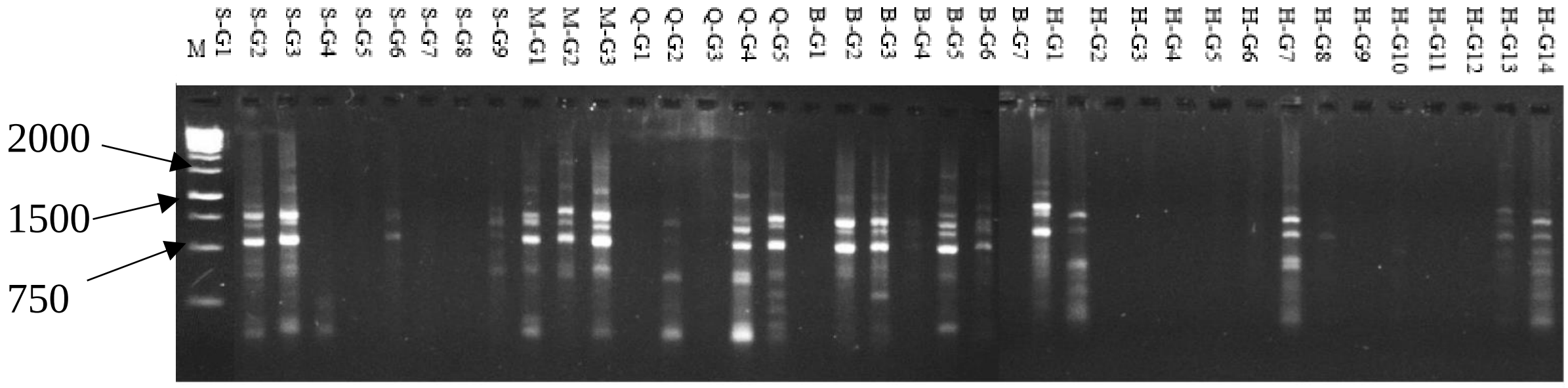

**Appendix 24. RAPD amplification profile for primer S093 on almond genotypes. M: 1 kb DNA ladder.**

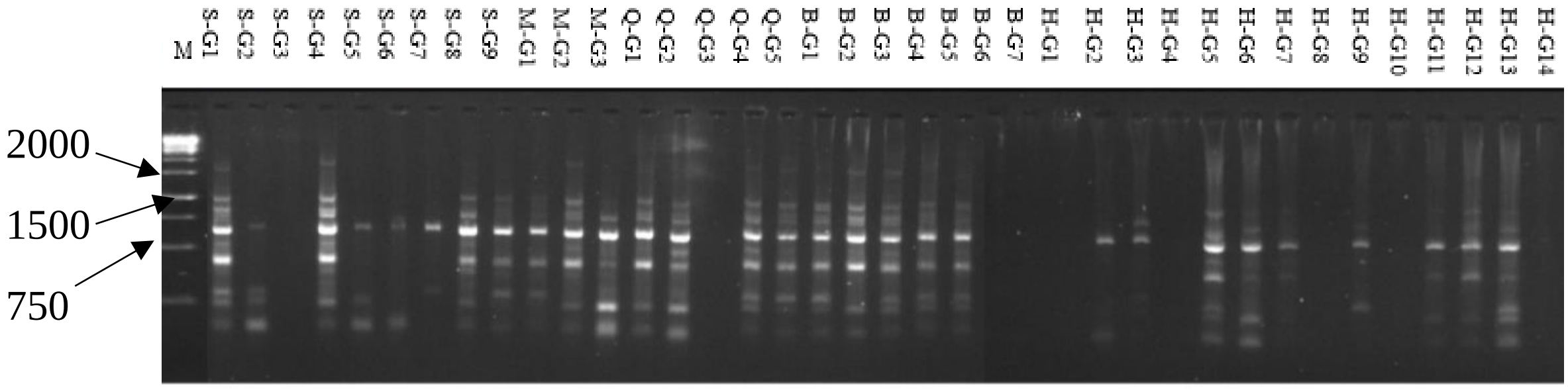

**Appendix 25. RAPD amplification profile for primer S078 on almond genotypes. M: 1 kb DNA ladder.**

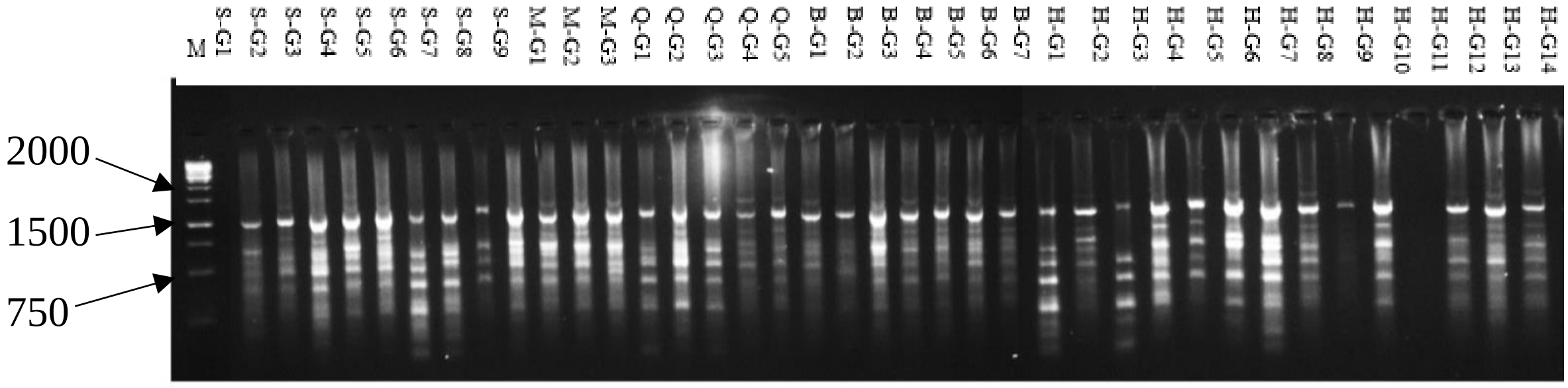

**Appendix 26. RAPD amplification profile for primer S094 on almond genotypes. M: 1 kb DNA ladder.**

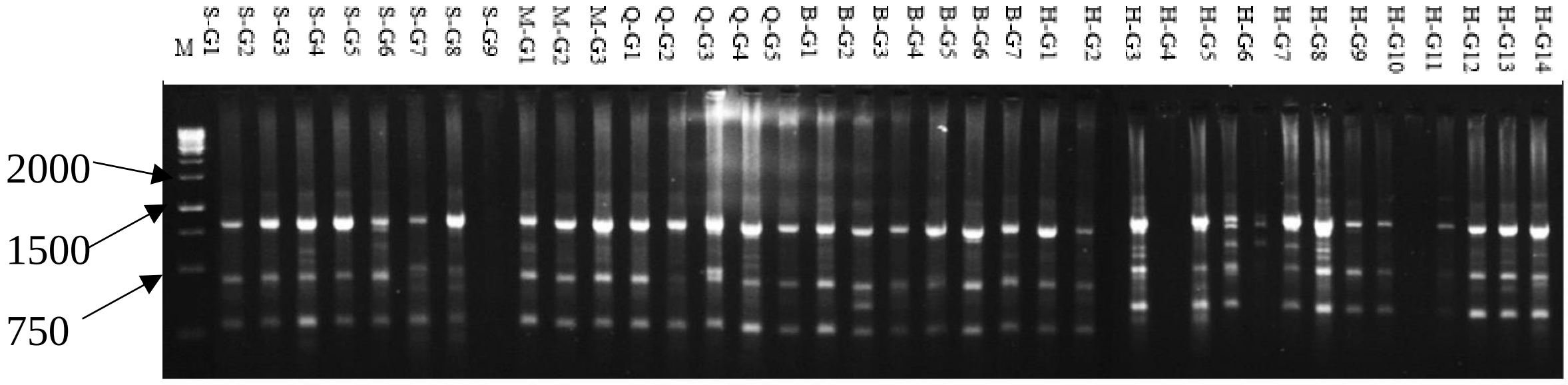

**Appendix 27. RAPD amplification profile for primer S087 on almond genotypes. M: 1 kb DNA ladder.**

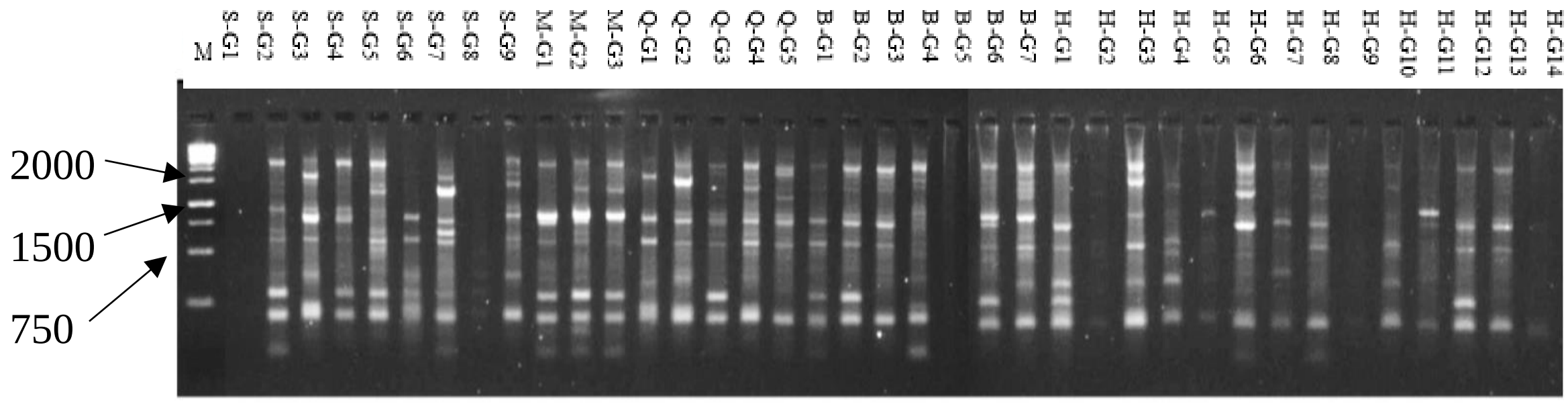

**Appendix 28. RAPD amplification profile for primer S088 on almond genotypes. M: 1 kb DNA ladder.**

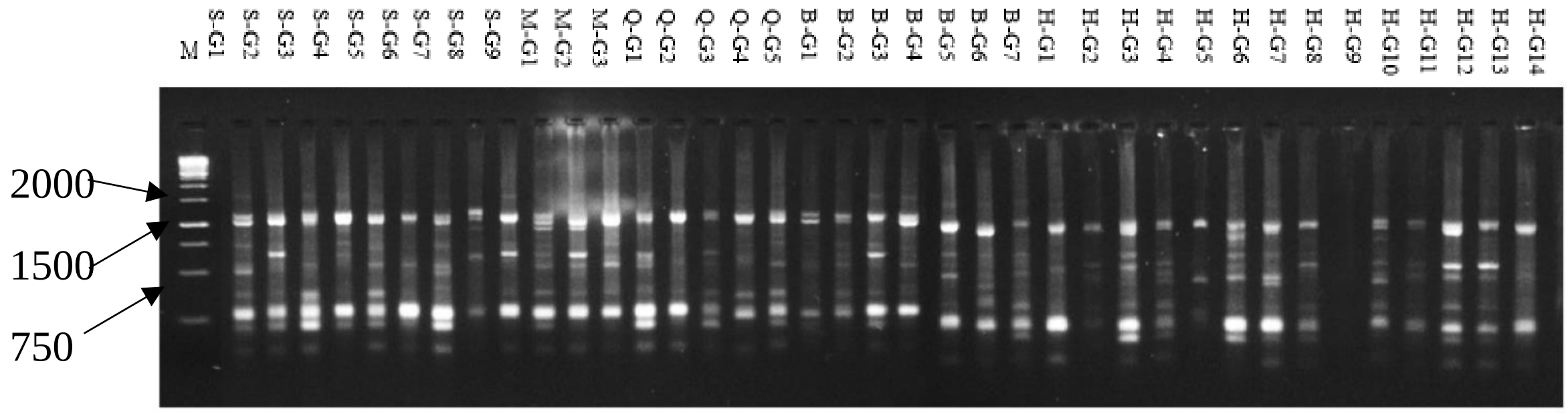

**Appendix 29. RAPD amplification profile for primer S089 on almond genotypes. M: 1 kb DNA ladder.**

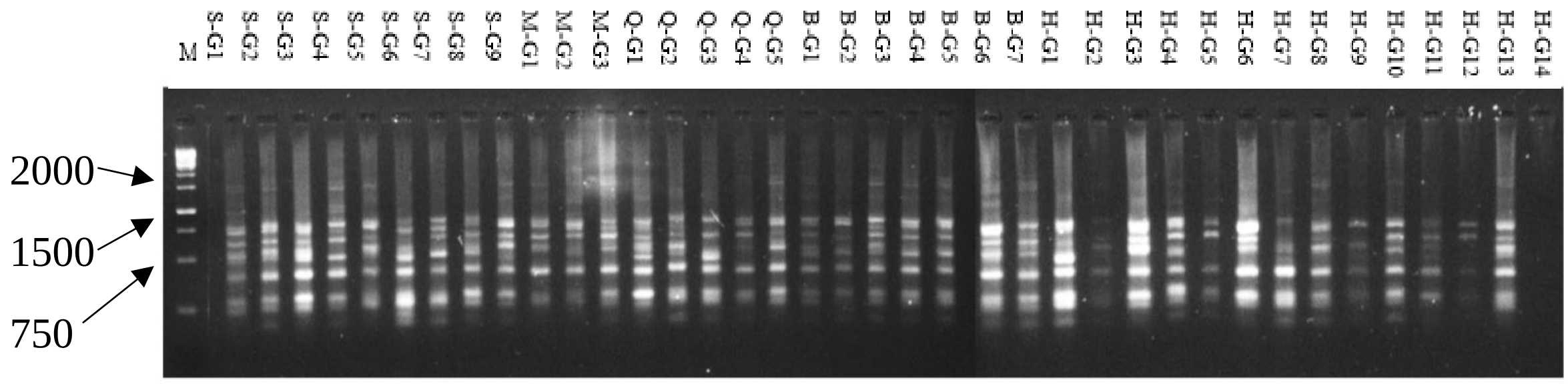

**Appendix 30. RAPD amplification profile for primer S090 on almond genotypes. M: 1 kb DNA ladder.**

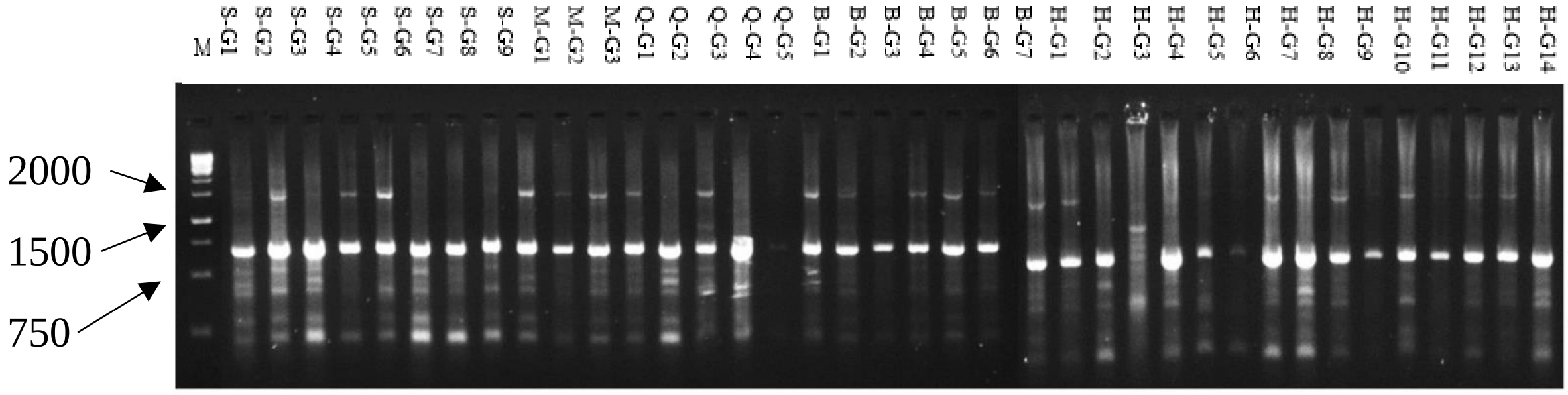

**Appendix 31. RAPD amplification profile for primer S091 on almond genotypes. M: 1 kb DNA ladder.**

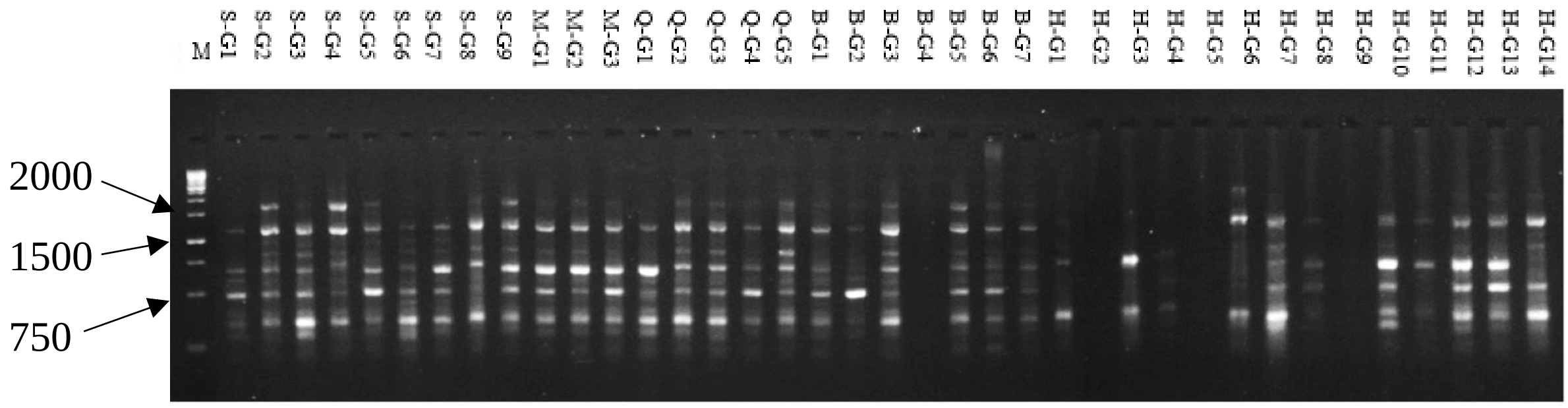

**Appendix 32. RAPD amplification profile for primer S092 on almond genotypes. M: 1 kb DNA ladder.**

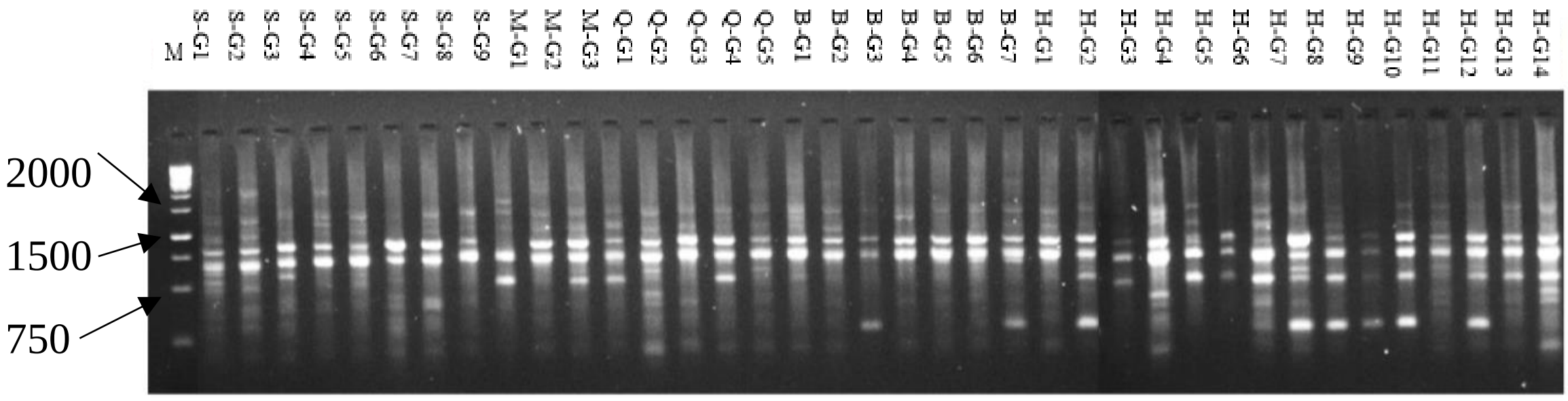

**Appendix 33. RAPD amplification profile for primer S095 on almond genotypes. M: 1 kb DNA ladder.**

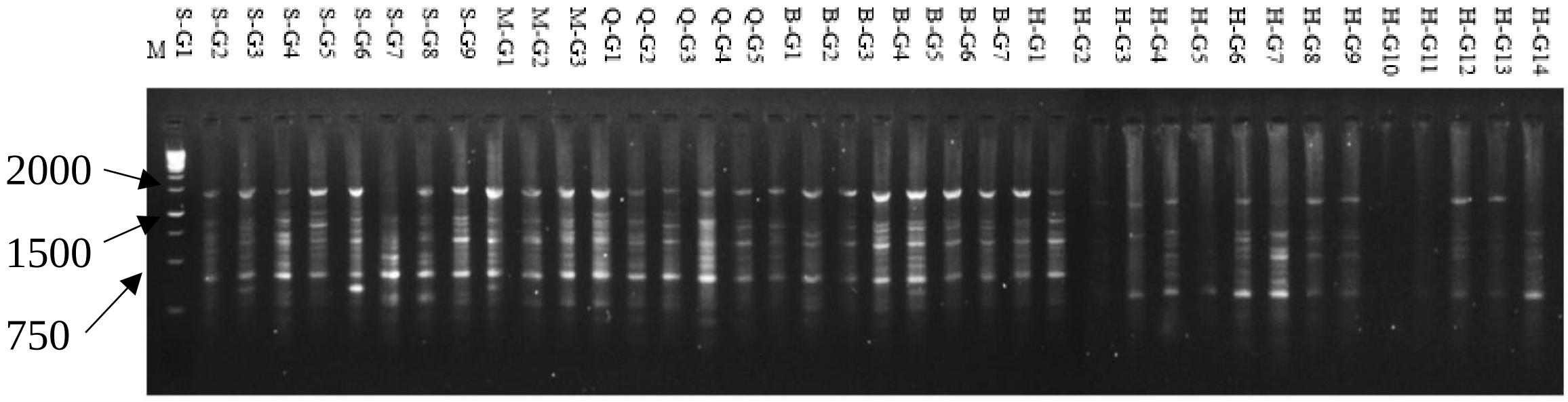

**Appendix 34. RAPD amplification profile for primer S073 on almond genotypes. M: 1 kb DNA ladder.**

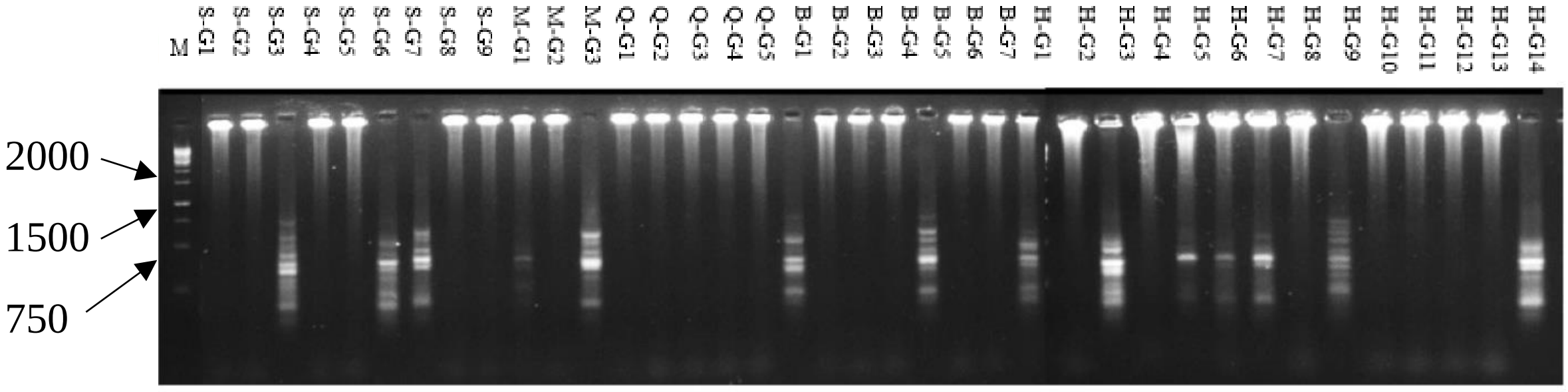

**Appendix 35. ISSR amplification profile for primer 807 on almond genotypes. M: 1 kb DNA ladder.**

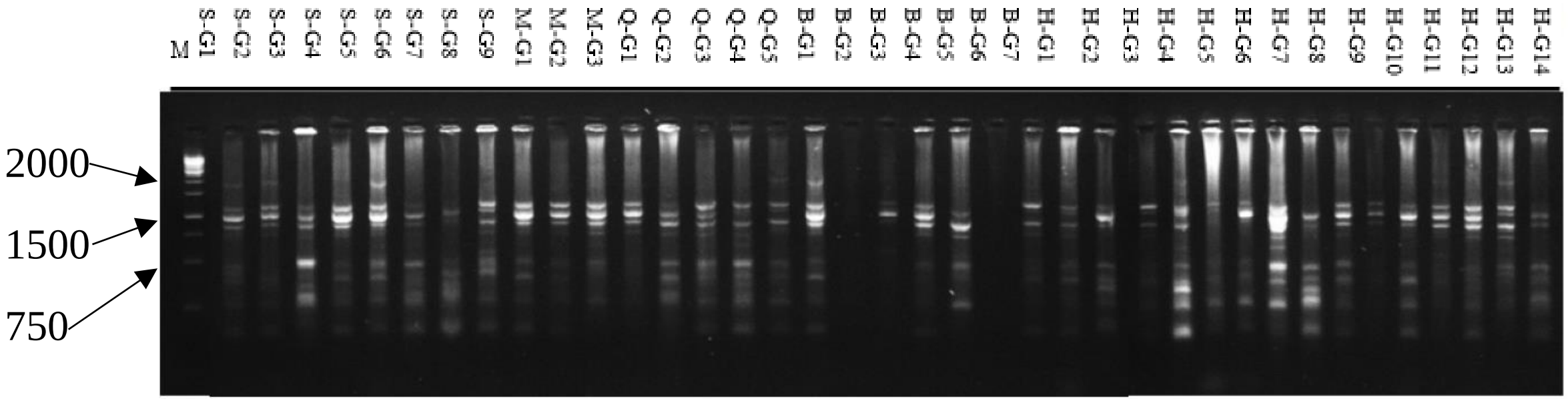

**Appendix 36. ISSR amplification profile for primer 17898A on almond genotypes. M: 1 kb DNA ladder.**

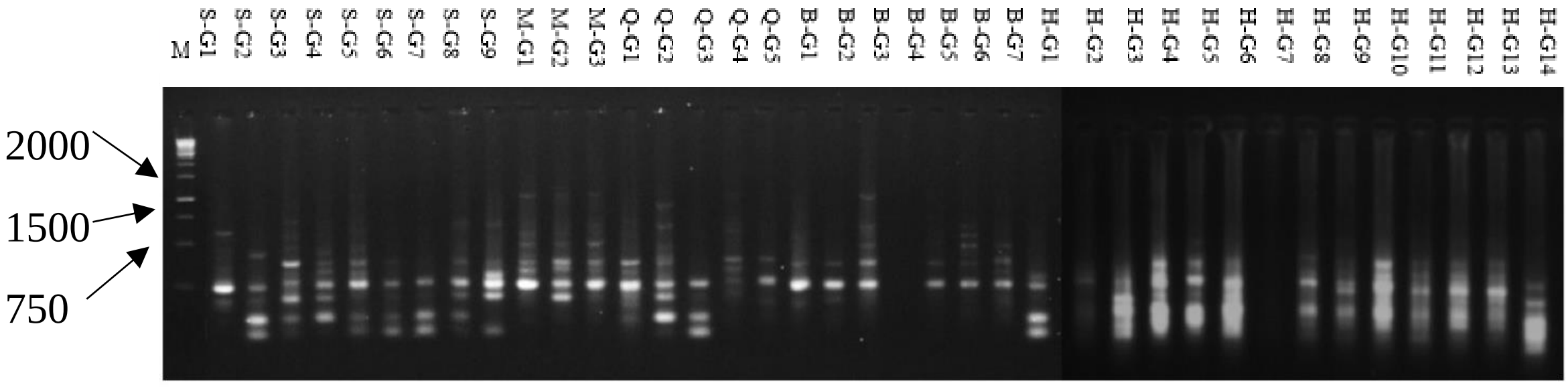

**Appendix 37. ISSR amplification profile for primer HB04 on almond genotypes. M: 1 kb DNA ladder.**

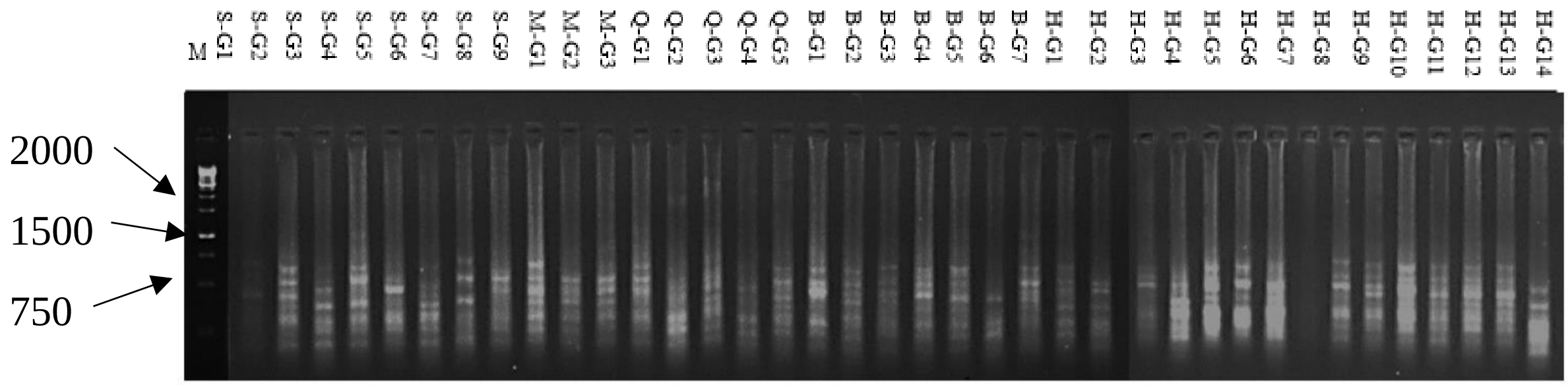

**Appendix 38. ISSR amplification profile for primer HB 8 on almond genotypes. M: 1 kb DNA ladder.**

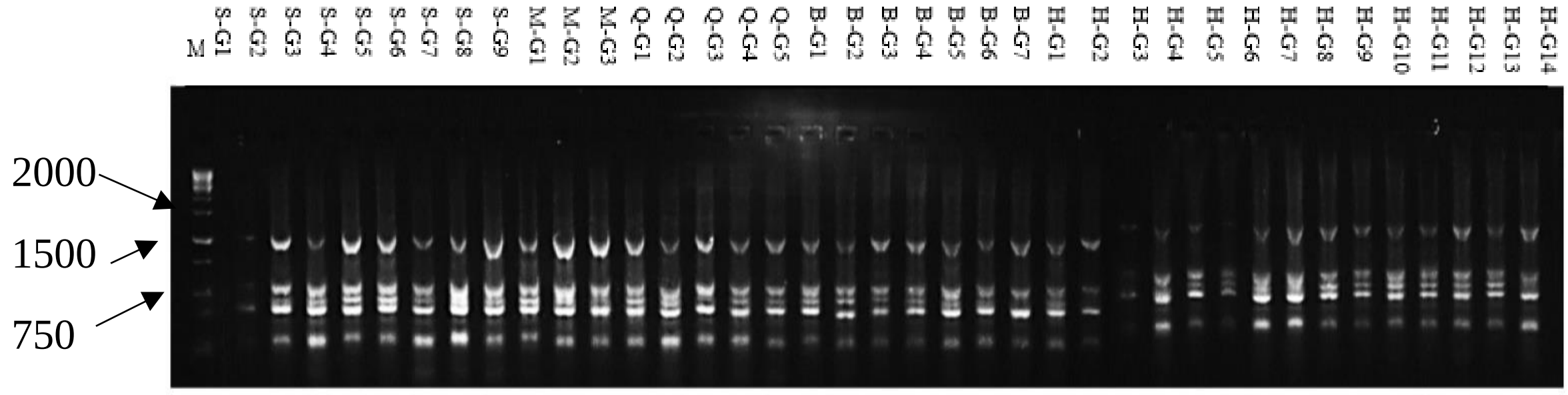

**Appendix 39. ISSR amplification profile for primer HB 10 on almond genotypes. M: 1 kb DNA ladder.**

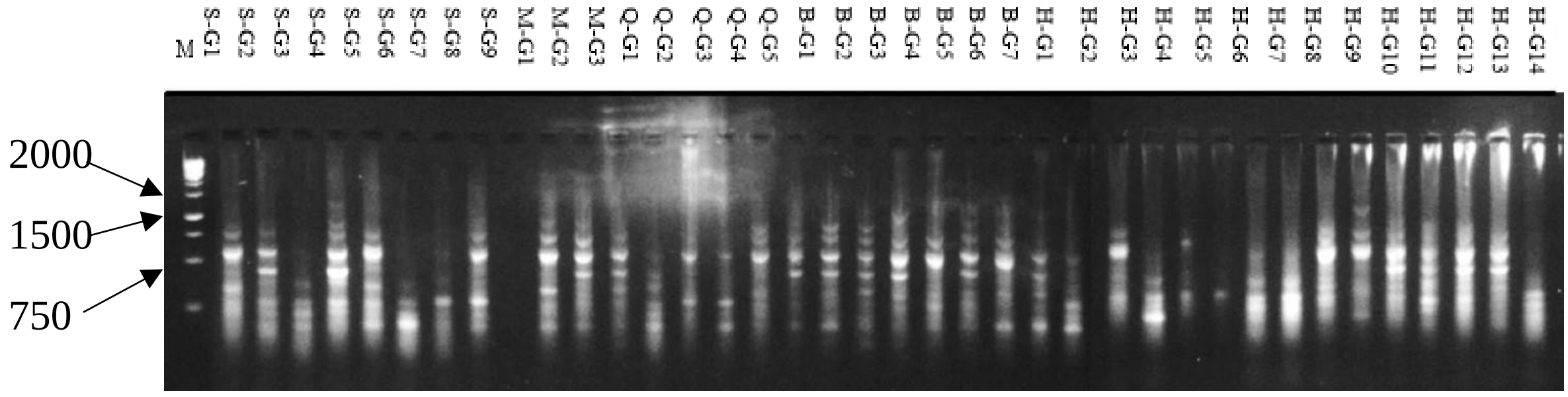

**Appendix 40. ISSR amplification profile for primer HB 11 on almond genotypes. M: 1 kb DNA ladder.**

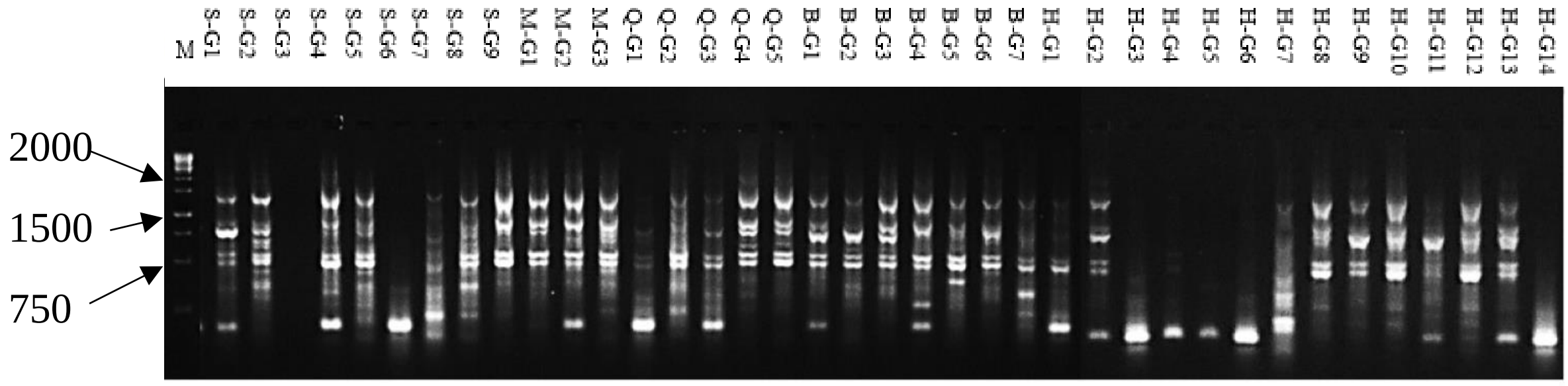

**Appendix 41. ISSR amplification profile for primer HB 12 on almond genotypes. M: 1 kb DNA ladder.**

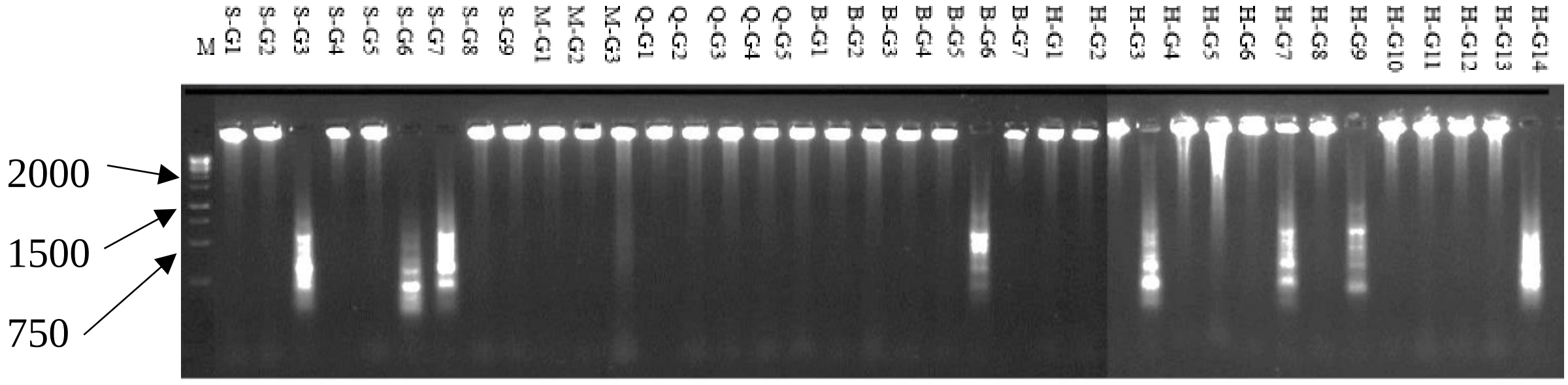

**Appendix 42. ISSR amplification profile for primer HB 15 on almond genotypes. M: 1 kb DNA ladder.**

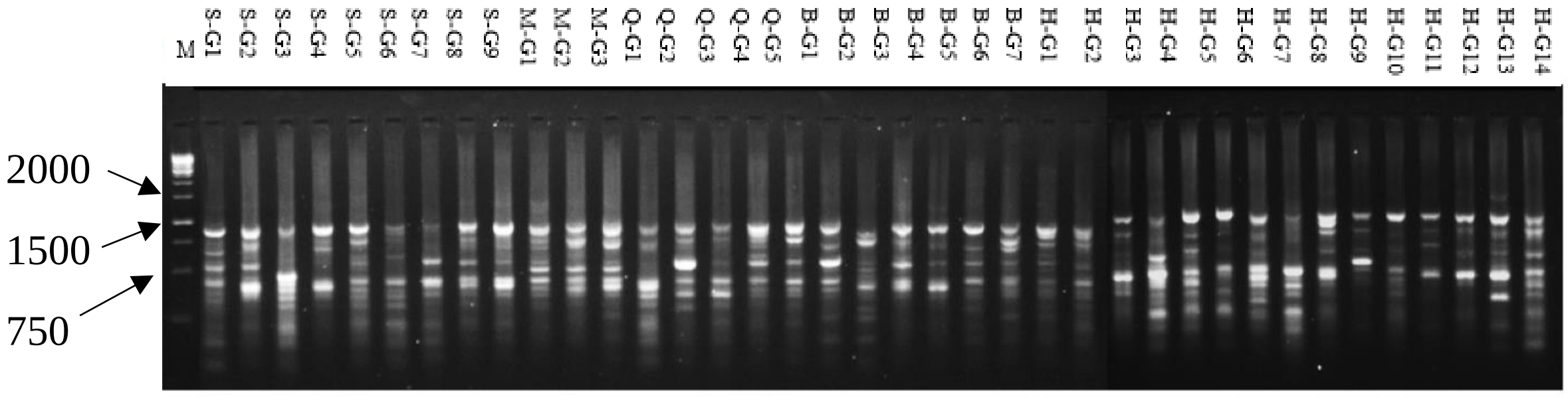

**Appendix 43. ISSR amplification profile for primer AG7YC on almond genotypes. M: 1 kb DNA ladder.**

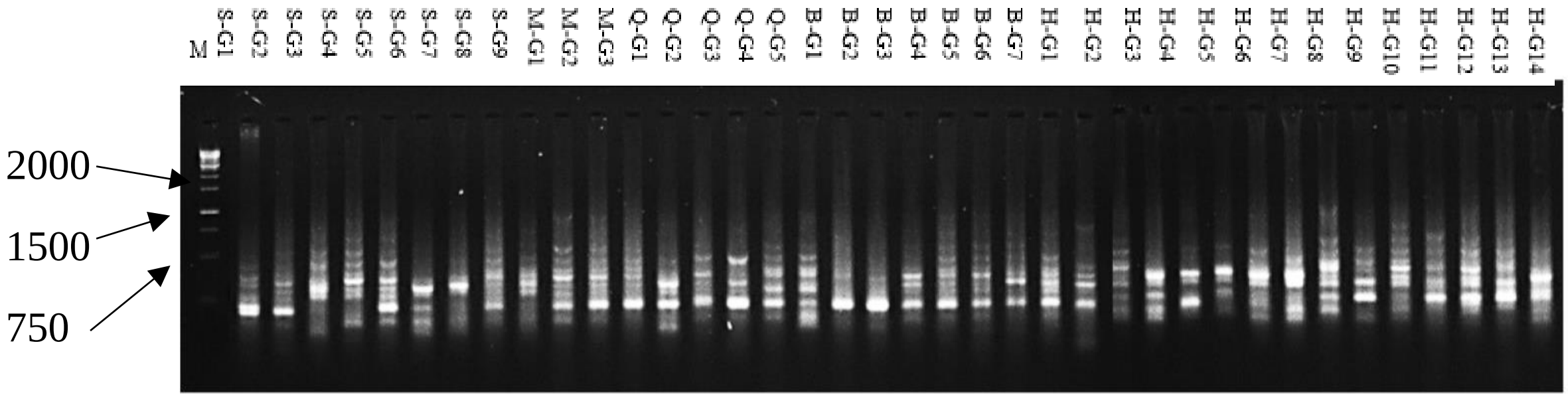

**Appendix 44. ISSR amplification profile for primer AGC6G on almond genotypes. M: 1 kb DNA ladder.**

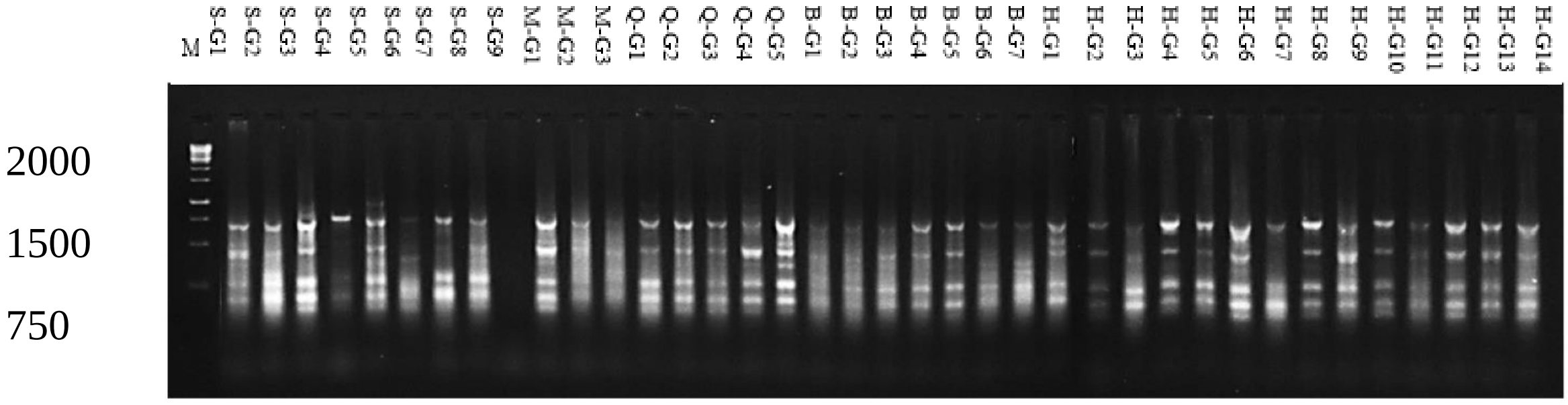

**Appendix 45. ISSR amplification profile for primer IS06 on almond genotypes. M: 1 kb DNA ladder.**

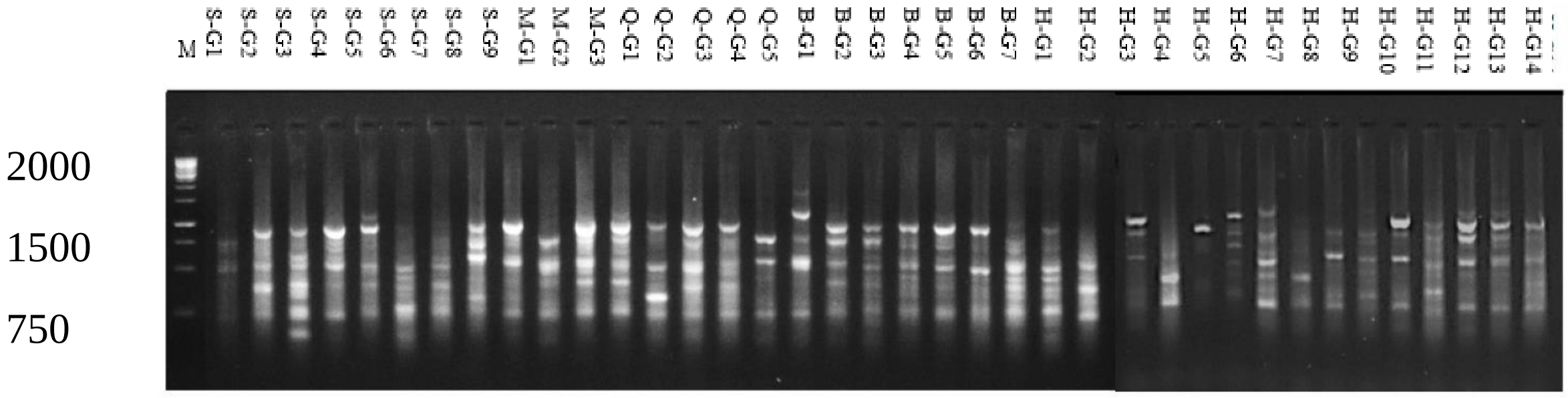

**Appendix 46. ISSR amplification profile for primer IS16 on almond genotypes. M: 1 kb DNA ladder.**

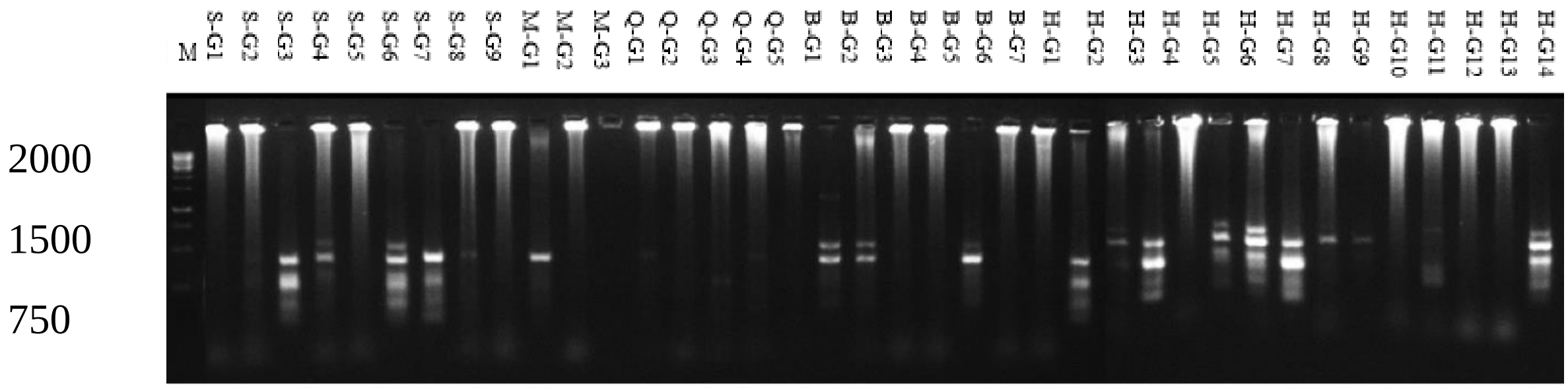

**Appendix 47. ISSR amplification profile for primer IS17 on almond genotypes. M: 1 kb DNA ladder.**

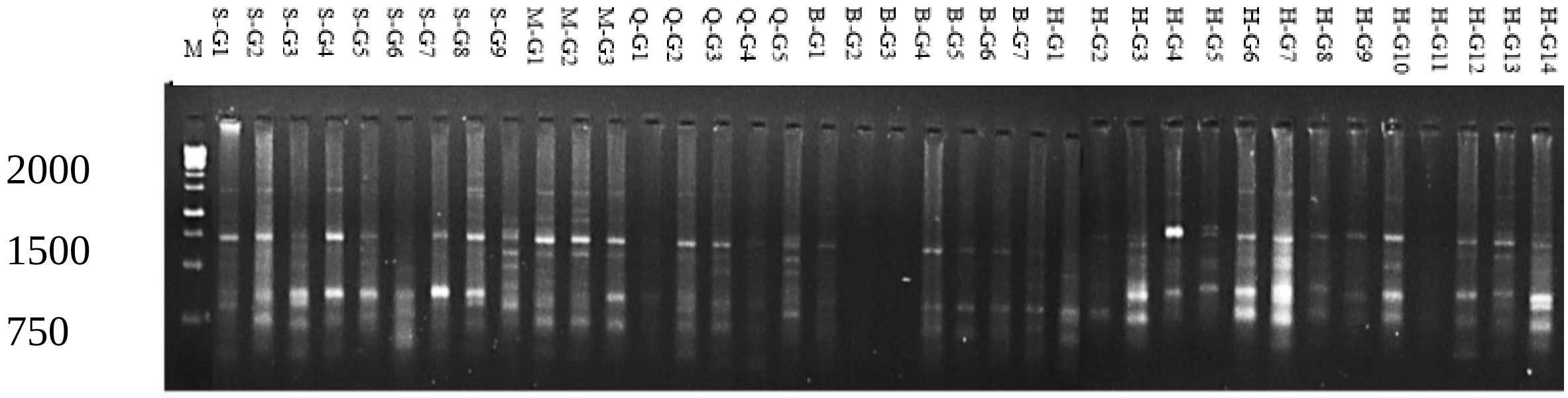

**Appendix48. ISSR amplification profile for primer IS19 on almond genotypes. M: 1 kb DNA ladder.**

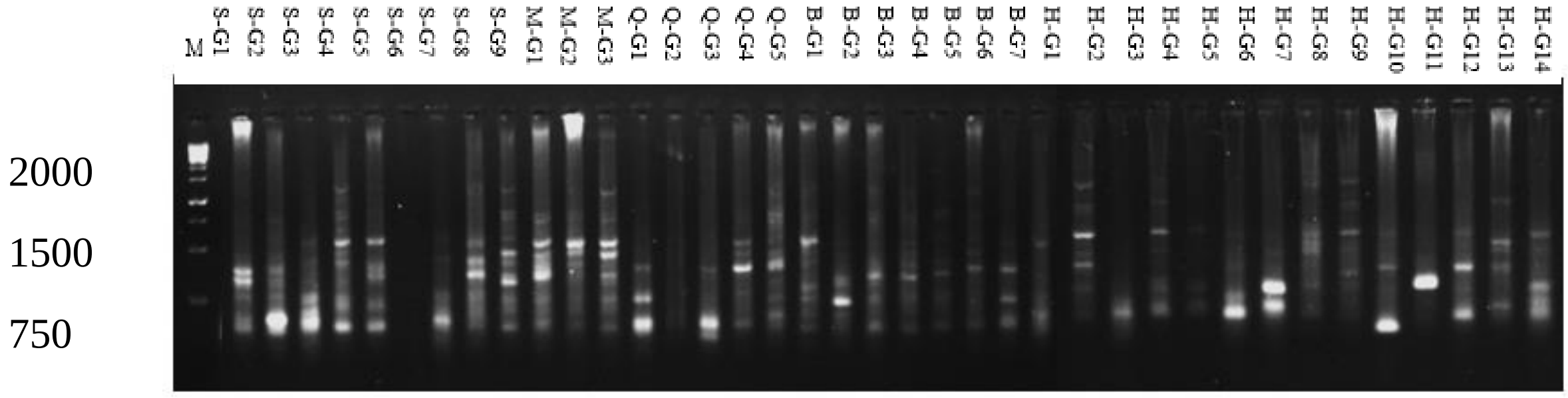

**Appendix 49. ISSR amplification profile for primer ISSR.08 on almond genotypes. M: 1 kb DNA ladder.**

**Appendix 50. Jaccard similarity coefficients applied among 38 almond genotypes by using 20 RAPD markers.**

| | S-G1 | S-G2 | S-G3 | S-G4 | S-G5 | S-G6 | S-G7 | S-G8 | S-G9 |
|---|---|---|---|---|---|---|---|---|---|
| S-G1 | | | | | | | | | |
| S-G2 | 0.48 | | | | | | | | |
| S-G3 | 0.51 | 0.51 | | | | | | | |
| S-G4 | 0.48 | 0.42 | 0.42 | | | | | | |
| S-G5 | 0.60 | 0.49 | 0.58 | 0.51 | | | | | |
| S-G6 | 0.64 | 0.58 | 0.44 | 0.59 | 0.63 | | | | |
| S-G7 | 0.59 | 0.55 | 0.45 | 0.56 | 0.62 | 0.47 | | | |
| S-G8 | 0.55 | 0.61 | 0.54 | 0.55 | 0.60 | 0.71 | 0.53 | | |
| S-G9 | 0.54 | 0.49 | 0.52 | 0.48 | 0.48 | 0.63 | 0.60 | 0.47 | |
| M-G1 | 0.60 | 0.45 | 0.59 | 0.45 | 0.43 | 0.65 | 0.61 | 0.55 | 0.41 |
| M-G2 | 0.53 | 0.45 | 0.54 | 0.41 | 0.44 | 0.63 | 0.58 | 0.51 | 0.38 |
| M-G3 | 0.55 | 0.53 | 0.54 | 0.37 | 0.50 | 0.59 | 0.55 | 0.58 | 0.48 |
| Q-G1 | 0.59 | 0.63 | 0.59 | 0.63 | 0.58 | 0.61 | 0.62 | 0.64 | 0.60 |
| Q-G2 | 0.58 | 0.58 | 0.49 | 0.40 | 0.51 | 0.53 | 0.55 | 0.59 | 0.50 |
| Q-G3 | 0.51 | 0.53 | 0.44 | 0.48 | 0.60 | 0.57 | 0.57 | 0.53 | 0.51 |
| Q-G4 | 0.52 | 0.50 | 0.53 | 0.43 | 0.54 | 0.60 | 0.56 | 0.58 | 0.50 |
| Q-G5 | 0.56 | 0.49 | 0.62 | 0.45 | 0.48 | 0.68 | 0.63 | 0.54 | 0.48 |
| B-G1 | 0.60 | 0.54 | 0.63 | 0.49 | 0.57 | 0.70 | 0.59 | 0.46 | 0.50 |
| B-G2 | 0.57 | 0.61 | 0.64 | 0.53 | 0.59 | 0.67 | 0.61 | 0.61 | 0.48 |
| B-G3 | 0.63 | 0.55 | 0.61 | 0.51 | 0.52 | 0.66 | 0.60 | 0.58 | 0.53 |
| B-G4 | 0.58 | 0.47 | 0.57 | 0.47 | 0.49 | 0.64 | 0.59 | 0.56 | 0.49 |
| B-G5 | 0.59 | 0.56 | 0.62 | 0.51 | 0.50 | 0.63 | 0.60 | 0.49 | 0.47 |
| B-G6 | 0.60 | 0.56 | 0.50 | 0.38 | 0.60 | 0.63 | 0.62 | 0.57 | 0.51 |
| B-G7 | 0.54 | 0.50 | 0.49 | 0.43 | 0.55 | 0.63 | 0.62 | 0.61 | 0.54 |
| H-G1 | 0.57 | 0.62 | 0.52 | 0.52 | 0.64 | 0.67 | 0.68 | 0.54 | 0.56 |
| H-G2 | 0.62 | 0.63 | 0.63 | 0.59 | 0.65 | 0.63 | 0.58 | 0.55 | 0.62 |
| H-G3 | 0.71 | 0.59 | 0.54 | 0.61 | 0.67 | 0.47 | 0.51 | 0.68 | 0.62 |
| H-G4 | 0.63 | 0.63 | 0.56 | 0.52 | 0.62 | 0.61 | 0.63 | 0.65 | 0.63 |
| H-G5 | 0.72 | 0.61 | 0.59 | 0.61 | 0.68 | 0.59 | 0.49 | 0.69 | 0.68 |
| H-G6 | 0.66 | 0.65 | 0.63 | 0.56 | 0.61 | 0.63 | 0.61 | 0.62 | 0.65 |
| H-G7 | 0.65 | 0.63 | 0.60 | 0.58 | 0.61 | 0.60 | 0.59 | 0.66 | 0.60 |
| H-G8 | 0.62 | 0.64 | 0.60 | 0.54 | 0.69 | 0.67 | 0.65 | 0.64 | 0.64 |
| H-G9 | 0.63 | 0.53 | 0.59 | 0.60 | 0.65 | 0.59 | 0.49 | 0.65 | 0.60 |
| H-G10 | 0.70 | 0.61 | 0.63 | 0.55 | 0.57 | 0.63 | 0.49 | 0.56 | 0.66 |
| H-G11 | 0.67 | 0.61 | 0.52 | 0.56 | 0.62 | 0.57 | 0.38 | 0.60 | 0.66 |
| H-G12 | 0.67 | 0.54 | 0.54 | 0.55 | 0.61 | 0.57 | 0.43 | 0.66 | 0.64 |
| H-G13 | 0.63 | 0.54 | 0.64 | 0.55 | 0.54 | 0.70 | 0.57 | 0.68 | 0.61 |
| H-G14 | 0.62 | 0.62 | 0.52 | 0.56 | 0.61 | 0.63 | 0.64 | 0.55 | 0.57 |

***Continued***

**Appendix 50.  Continued**

| | M-G1 | M-G2 | M-G3 | Q-G1 | Q-G2 | Q-G3 | Q-G4 | Q-G5 | B-G1 |
|---|---|---|---|---|---|---|---|---|---|
| S-G1 | | | | | | | | | |
| S-G2 | | | | | | | | | |
| S-G3 | | | | | | | | | |
| S-G4 | | | | | | | | | |
| S-G5 | | | | | | | | | |
| S-G6 | | | | | | | | | |
| S-G7 | | | | | | | | | |
| S-G8 | | | | | | | | | |
| S-G9 | | | | | | | | | |
| M-G1 | | | | | | | | | |
| M-G2 | 0.32 | | | | | | | | |
| M-G3 | 0.36 | 0.36 | | | | | | | |
| Q-G1 | 0.57 | 0.58 | 0.55 | | | | | | |
| Q-G2 | 0.45 | 0.42 | 0.39 | 0.51 | | | | | |
| Q-G3 | 0.52 | 0.45 | 0.50 | 0.51 | 0.48 | | | | |
| Q-G4 | 0.54 | 0.42 | 0.44 | 0.55 | 0.45 | 0.52 | | | |
| Q-G5 | 0.49 | 0.45 | 0.51 | 0.59 | 0.57 | 0.55 | 0.39 | | |
| B-G1 | 0.48 | 0.50 | 0.51 | 0.62 | 0.54 | 0.55 | 0.47 | 0.46 | |
| B-G2 | 0.53 | 0.53 | 0.47 | 0.58 | 0.52 | 0.58 | 0.49 | 0.55 | 0.47 |
| B-G3 | 0.51 | 0.47 | 0.52 | 0.55 | 0.55 | 0.56 | 0.46 | 0.41 | 0.42 |
| B-G4 | 0.46 | 0.39 | 0.48 | 0.62 | 0.53 | 0.55 | 0.48 | 0.43 | 0.41 |
| B-G5 | 0.43 | 0.45 | 0.51 | 0.59 | 0.48 | 0.53 | 0.47 | 0.45 | 0.51 |
| B-G6 | 0.52 | 0.48 | 0.46 | 0.67 | 0.45 | 0.49 | 0.50 | 0.50 | 0.47 |
| B-G7 | 0.53 | 0.48 | 0.50 | 0.64 | 0.54 | 0.47 | 0.49 | 0.45 | 0.52 |
| H-G1 | 0.59 | 0.56 | 0.57 | 0.63 | 0.57 | 0.42 | 0.59 | 0.61 | 0.58 |
| H-G2 | 0.63 | 0.60 | 0.61 | 0.71 | 0.56 | 0.62 | 0.64 | 0.68 | 0.61 |
| H-G3 | 0.68 | 0.60 | 0.61 | 0.65 | 0.54 | 0.57 | 0.58 | 0.69 | 0.71 |
| H-G4 | 0.57 | 0.55 | 0.53 | 0.65 | 0.53 | 0.52 | 0.56 | 0.65 | 0.62 |
| H-G5 | 0.65 | 0.57 | 0.60 | **0.75** | 0.59 | 0.64 | 0.65 | 0.69 | 0.61 |
| H-G6 | 0.62 | 0.55 | 0.49 | 0.64 | 0.50 | 0.60 | 0.59 | 0.63 | 0.64 |
| H-G7 | 0.65 | 0.61 | 0.62 | 0.61 | 0.57 | 0.57 | 0.54 | 0.61 | 0.63 |
| H-G8 | 0.58 | 0.57 | 0.50 | 0.70 | 0.53 | 0.52 | 0.53 | 0.63 | 0.62 |
| H-G9 | 0.61 | 0.55 | 0.56 | 0.71 | 0.61 | 0.63 | 0.60 | 0.65 | 0.62 |
| H-G10 | 0.60 | 0.57 | 0.58 | 0.67 | 0.58 | 0.60 | 0.59 | 0.66 | 0.60 |
| H-G11 | 0.64 | 0.60 | 0.57 | 0.74 | 0.58 | 0.61 | 0.62 | 0.64 | 0.59 |
| H-G12 | 0.60 | 0.56 | 0.56 | 0.69 | 0.61 | 0.57 | 0.60 | 0.66 | 0.62 |
| H-G13 | 0.54 | 0.56 | 0.56 | 0.65 | 0.60 | 0.63 | 0.54 | 0.54 | 0.57 |
| H-G14 | 0.63 | 0.57 | 0.61 | 0.58 | 0.53 | 0.47 | 0.59 | 0.60 | 0.56 |

***Continued***

**Appendix 50.  Continued**

| | B-G2 | B-G3 | B-G4 | B-G5 | B-G6 | B-G7 | H-G1 | H-G2 | H-G3 | H-G4 |
|---|---|---|---|---|---|---|---|---|---|---|
| S-G1 | | | | | | | | | | |
| S-G2 | | | | | | | | | | |
| S-G3 | | | | | | | | | | |
| S-G4 | | | | | | | | | | |
| S-G5 | | | | | | | | | | |
| S-G6 | | | | | | | | | | |
| S-G7 | | | | | | | | | | |
| S-G8 | | | | | | | | | | |
| S-G9 | | | | | | | | | | |
| M-G1 | | | | | | | | | | |
| M-G2 | | | | | | | | | | |
| M-G3 | | | | | | | | | | |
| Q-G1 | | | | | | | | | | |
| Q-G2 | | | | | | | | | | |
| Q-G3 | | | | | | | | | | |
| Q-G4 | | | | | | | | | | |
| Q-G5 | | | | | | | | | | |
| B-G1 | | | | | | | | | | |
| B-G2 | | | | | | | | | | |
| B-G3 | 0.55 | | | | | | | | | |
| B-G4 | 0.54 | **0.32** | | | | | | | | |
| B-G5 | 0.52 | 0.44 | 0.44 | | | | | | | |
| B-G6 | 0.56 | 0.48 | 0.46 | 0.50 | | | | | | |
| B-G7 | 0.58 | 0.46 | 0.48 | 0.51 | 0.36 | | | | | |
| H-G1 | 0.59 | 0.62 | 0.63 | 0.61 | 0.50 | 0.47 | | | | |
| H-G2 | 0.64 | 0.68 | 0.66 | 0.61 | 0.61 | 0.67 | 0.63 | | | |
| H-G3 | 0.65 | 0.66 | 0.63 | 0.67 | 0.61 | 0.66 | 0.63 | 0.62 | | |
| H-G4 | 0.61 | 0.60 | 0.60 | 0.58 | 0.54 | 0.53 | 0.54 | 0.62 | 0.60 | |
| H-G5 | 0.69 | 0.69 | 0.60 | 0.68 | 0.59 | 0.62 | 0.69 | 0.53 | 0.61 | 0.61 |
| H-G6 | 0.60 | 0.63 | 0.59 | 0.62 | 0.58 | 0.64 | 0.57 | 0.52 | 0.53 | 0.53 |
| H-G7 | 0.65 | 0.64 | 0.63 | 0.60 | 0.64 | 0.63 | 0.59 | 0.60 | 0.61 | 0.59 |
| H-G8 | 0.56 | 0.67 | 0.62 | 0.63 | 0.52 | 0.58 | 0.52 | 0.61 | 0.61 | 0.44 |
| H-G9 | 0.59 | 0.64 | 0.57 | 0.64 | 0.62 | 0.64 | 0.74 | 0.47 | 0.57 | 0.66 |
| H-G10 | 0.64 | 0.66 | 0.60 | 0.62 | 0.62 | 0.66 | 0.61 | 0.45 | 0.59 | 0.49 |
| H-G11 | 0.69 | 0.63 | 0.57 | 0.63 | 0.60 | 0.61 | 0.68 | 0.61 | 0.60 | 0.54 |
| H-G12 | 0.66 | 0.63 | 0.61 | 0.65 | 0.64 | 0.62 | 0.63 | 0.45 | 0.55 | 0.57 |
| H-G13 | 0.59 | 0.52 | 0.54 | 0.56 | 0.62 | 0.59 | 0.66 | 0.66 | 0.66 | 0.62 |
| H-G14 | 0.63 | 0.56 | 0.57 | 0.56 | 0.49 | 0.61 | 0.55 | 0.65 | 0.65 | 0.62 |

***Continued***

**Appendix 50.  Continued**

| | H-G5 | H-G6 | H-G7 | H-G8 | H-G9 | H-G10 | H-G11 | H-G12 | H-G13 | H-G14 |
|---|---|---|---|---|---|---|---|---|---|---|
| S-G1 | | | | | | | | | | |
| S-G2 | | | | | | | | | | |
| S-G3 | | | | | | | | | | |
| S-G4 | | | | | | | | | | |
| S-G5 | | | | | | | | | | |
| S-G6 | | | | | | | | | | |
| S-G7 | | | | | | | | | | |
| S-G8 | | | | | | | | | | |
| S-G9 | | | | | | | | | | |
| M-G1 | | | | | | | | | | |
| M-G2 | | | | | | | | | | |
| M-G3 | | | | | | | | | | |
| Q-G1 | | | | | | | | | | |
| Q-G2 | | | | | | | | | | |
| Q-G3 | | | | | | | | | | |
| Q-G4 | | | | | | | | | | |
| Q-G5 | | | | | | | | | | |
| B-G1 | | | | | | | | | | |
| B-G2 | | | | | | | | | | |
| B-G3 | | | | | | | | | | |
| B-G4 | | | | | | | | | | |
| B-G5 | | | | | | | | | | |
| B-G6 | | | | | | | | | | |
| B-G7 | | | | | | | | | | |
| H-G1 | | | | | | | | | | |
| H-G2 | | | | | | | | | | |
| H-G3 | | | | | | | | | | |
| H-G4 | | | | | | | | | | |
| H-G5 | | | | | | | | | | |
| H-G6 | 0.61 | | | | | | | | | |
| H-G7 | 0.60 | 0.50 | | | | | | | | |
| H-G8 | 0.60 | 0.51 | 0.55 | | | | | | | |
| H-G9 | 0.50 | 0.57 | 0.64 | 0.57 | | | | | | |
| H-G10 | 0.50 | 0.46 | 0.48 | 0.52 | 0.54 | | | | | |
| H-G11 | 0.48 | 0.57 | 0.59 | 0.59 | 0.54 | 0.49 | | | | |
| H-G12 | 0.43 | 0.52 | 0.59 | 0.57 | 0.40 | 0.41 | 0.41 | | | |
| H-G13 | 0.58 | 0.65 | 0.55 | 0.60 | 0.62 | 0.58 | 0.59 | 0.47 | | |
| H-G14 | 0.66 | 0.66 | 0.62 | 0.58 | 0.70 | 0.65 | 0.65 | 0.64 | 0.61 | |

**Appendix 51. Jaccard similarity coefficients applied among 38 almond genotypes by using 15 ISSR markers.**

| | S-G1 | S-G2 | S-G3 | S-G4 | S-G5 | S-G6 | S-G7 | S-G8 | S-G9 |
|---|---|---|---|---|---|---|---|---|---|
| S-G1 | | | | | | | | | |
| S-G2 | 0.34 | | | | | | | | |
| S-G3 | 0.54 | 0.48 | | | | | | | |
| S-G4 | 0.52 | 0.33 | 0.47 | | | | | | |
| S-G5 | 0.46 | 0.41 | 0.35 | 0.35 | | | | | |
| S-G6 | 0.62 | 0.46 | 0.48 | 0.55 | 0.54 | | | | |
| S-G7 | 0.57 | 0.41 | 0.56 | 0.52 | 0.57 | 0.39 | | | |
| S-G8 | 0.50 | 0.48 | 0.41 | 0.43 | 0.42 | 0.60 | 0.51 | | |
| S-G9 | 0.45 | 0.40 | 0.51 | 0.38 | 0.41 | 0.61 | 0.58 | 0.47 | |
| M-G1 | 0.58 | 0.51 | 0.53 | 0.44 | 0.45 | 0.64 | 0.60 | 0.49 | 0.49 |
| M-G2 | 0.48 | 0.41 | 0.52 | 0.45 | 0.41 | 0.63 | 0.61 | 0.46 | 0.43 |
| M-G3 | 0.49 | 0.37 | 0.51 | 0.42 | 0.43 | 0.57 | 0.54 | 0.47 | 0.44 |
| Q-G1 | 0.52 | 0.48 | 0.44 | 0.51 | 0.40 | 0.58 | 0.57 | 0.46 | 0.45 |
| Q-G2 | 0.52 | 0.37 | 0.46 | 0.47 | 0.39 | 0.46 | 0.41 | 0.38 | 0.44 |
| Q-G3 | 0.52 | 0.34 | 0.52 | 0.45 | 0.40 | 0.48 | 0.40 | 0.46 | 0.43 |
| Q-G4 | 0.39 | 0.42 | 0.51 | 0.46 | 0.44 | 0.61 | 0.58 | 0.35 | 0.35 |
| Q-G5 | 0.42 | 0.49 | 0.57 | 0.52 | 0.44 | 0.67 | 0.62 | 0.41 | 0.35 |
| B-G1 | 0.55 | 0.52 | 0.54 | 0.47 | 0.50 | 0.63 | 0.67 | 0.46 | 0.47 |
| B-G2 | 0.52 | 0.52 | 0.56 | 0.45 | 0.50 | 0.69 | 0.65 | 0.50 | 0.52 |
| B-G3 | 0.43 | 0.36 | 0.55 | 0.51 | 0.50 | 0.54 | 0.51 | 0.45 | 0.40 |
| B-G4 | 0.46 | 0.34 | 0.56 | 0.47 | 0.48 | 0.54 | 0.47 | 0.44 | 0.41 |
| B-G5 | 0.53 | 0.42 | 0.55 | 0.48 | 0.55 | 0.55 | 0.55 | 0.45 | 0.48 |
| B-G6 | 0.48 | 0.50 | 0.60 | 0.53 | 0.50 | 0.68 | 0.62 | 0.50 | 0.47 |
| B-G7 | 0.49 | 0.53 | 0.53 | 0.49 | 0.42 | 0.59 | 0.59 | 0.49 | 0.47 |
| H-G1 | 0.61 | 0.53 | 0.60 | 0.58 | 0.59 | 0.49 | 0.43 | 0.62 | 0.59 |
| H-G2 | 0.54 | 0.54 | 0.64 | 0.60 | 0.58 | 0.71 | 0.68 | 0.52 | 0.58 |
| H-G3 | 0.63 | 0.58 | 0.54 | 0.63 | 0.61 | 0.48 | 0.62 | 0.65 | 0.63 |
| H-G4 | 0.53 | 0.46 | 0.53 | 0.55 | 0.53 | 0.57 | 0.62 | 0.57 | 0.60 |
| H-G5 | 0.64 | 0.55 | 0.68 | 0.66 | 0.65 | 0.58 | 0.57 | 0.61 | 0.67 |
| H-G6 | 0.55 | 0.46 | 0.59 | 0.57 | 0.58 | 0.44 | 0.53 | 0.57 | 0.60 |
| H-G7 | 0.59 | 0.65 | 0.65 | 0.63 | 0.66 | 0.68 | 0.66 | 0.67 | 0.61 |
| H-G8 | 0.49 | 0.40 | 0.57 | 0.46 | 0.51 | 0.61 | 0.57 | 0.47 | 0.50 |
| H-G9 | 0.56 | 0.53 | 0.60 | 0.54 | 0.61 | 0.72 | 0.66 | 0.56 | 0.50 |
| H-G10 | 0.45 | 0.46 | 0.55 | 0.48 | 0.48 | 0.71 | 0.64 | 0.47 | 0.44 |
| H-G11 | 0.55 | 0.50 | 0.61 | 0.56 | 0.57 | 0.77 | 0.72 | 0.59 | 0.54 |
| H-G12 | 0.45 | 0.41 | 0.57 | 0.55 | 0.53 | 0.68 | 0.62 | 0.51 | 0.49 |
| H-G13 | 0.43 | 0.41 | 0.57 | 0.51 | 0.46 | 0.66 | 0.60 | 0.48 | 0.47 |
| H-G14 | 0.56 | 0.58 | 0.49 | 0.64 | 0.53 | 0.51 | 0.57 | 0.58 | 0.62 |

***Continued***

**Appendix 51.  Continued**

| | M-G1 | M-G2 | M-G3 | Q-G1 | Q-G2 | Q-G3 | Q-G4 | Q-G5 | B-G1 |
|---|---|---|---|---|---|---|---|---|---|
| S-G1 | | | | | | | | | |
| S-G2 | | | | | | | | | |
| S-G3 | | | | | | | | | |
| S-G4 | | | | | | | | | |
| S-G5 | | | | | | | | | |
| S-G6 | | | | | | | | | |
| S-G7 | | | | | | | | | |
| S-G8 | | | | | | | | | |
| S-G9 | | | | | | | | | |
| M-G1 | | | | | | | | | |
| M-G2 | 0.36 | | | | | | | | |
| M-G3 | 0.41 | 0.32 | | | | | | | |
| Q-G1 | 0.58 | 0.46 | 0.53 | | | | | | |
| Q-G2 | 0.52 | 0.45 | 0.43 | 0.41 | | | | | |
| Q-G3 | 0.56 | 0.50 | 0.49 | 0.40 | 0.22 | | | | |
| Q-G4 | 0.48 | 0.42 | 0.46 | 0.51 | 0.40 | 0.43 | | | |
| Q-G5 | 0.55 | 0.49 | 0.48 | 0.46 | 0.46 | 0.47 | 0.35 | | |
| B-G1 | 0.51 | 0.46 | 0.52 | 0.60 | 0.53 | 0.55 | 0.44 | 0.48 | |
| B-G2 | 0.53 | 0.42 | 0.45 | 0.48 | 0.57 | 0.58 | 0.55 | 0.47 | 0.48 |
| B-G3 | 0.52 | 0.33 | 0.31 | 0.43 | 0.38 | 0.43 | 0.42 | 0.40 | 0.48 |
| B-G4 | 0.54 | 0.46 | 0.45 | 0.52 | 0.30 | 0.33 | 0.37 | 0.46 | 0.48 |
| B-G5 | 0.59 | 0.51 | 0.54 | 0.60 | 0.38 | 0.45 | 0.45 | 0.51 | 0.42 |
| B-G6 | 0.59 | 0.41 | 0.46 | 0.56 | 0.50 | 0.50 | 0.38 | 0.47 | 0.54 |
| B-G7 | 0.63 | 0.49 | 0.49 | 0.51 | 0.45 | 0.42 | 0.46 | 0.46 | 0.53 |
| H-G1 | 0.60 | 0.61 | 0.58 | 0.63 | 0.49 | 0.47 | 0.62 | 0.61 | 0.66 |
| H-G2 | 0.62 | 0.56 | 0.57 | 0.54 | 0.59 | 0.54 | 0.60 | 0.49 | 0.61 |
| H-G3 | 0.62 | 0.64 | 0.61 | 0.60 | 0.53 | 0.62 | 0.62 | 0.67 | 0.63 |
| H-G4 | 0.64 | 0.58 | 0.59 | 0.53 | 0.53 | 0.51 | 0.54 | 0.54 | 0.58 |
| H-G5 | 0.65 | 0.73 | 0.62 | 0.64 | 0.55 | 0.55 | 0.65 | 0.60 | 0.70 |
| H-G6 | 0.59 | 0.60 | 0.54 | 0.57 | 0.48 | 0.48 | 0.60 | 0.61 | 0.57 |
| H-G7 | 0.65 | 0.68 | 0.64 | 0.66 | 0.67 | 0.63 | 0.67 | 0.64 | 0.59 |
| H-G8 | 0.54 | 0.51 | 0.50 | 0.58 | 0.49 | 0.47 | 0.45 | 0.52 | 0.49 |
| H-G9 | 0.66 | 0.61 | 0.58 | 0.67 | 0.61 | 0.58 | 0.53 | 0.57 | 0.57 |
| H-G10 | 0.61 | 0.46 | 0.48 | 0.45 | 0.55 | 0.53 | 0.47 | 0.43 | 0.49 |
| H-G11 | 0.58 | 0.50 | 0.56 | 0.54 | 0.63 | 0.65 | 0.56 | 0.55 | 0.60 |
| H-G12 | 0.64 | 0.49 | 0.55 | 0.51 | 0.52 | 0.49 | 0.46 | 0.48 | 0.53 |
| H-G13 | 0.61 | 0.48 | 0.52 | 0.46 | 0.48 | 0.46 | 0.40 | 0.46 | 0.52 |
| H-G14 | 0.64 | 0.66 | 0.64 | 0.55 | 0.56 | 0.57 | 0.63 | 0.59 | 0.63 |

***Continued***

**Appendix 51. Continued**

| | B-G2 | B-G3 | B-G4 | B-G5 | B-G6 | B-G7 | H-G1 | H-G2 | H-G3 | H-G4 |
|---|---|---|---|---|---|---|---|---|---|---|
| S-G1 | | | | | | | | | | |
| S-G2 | | | | | | | | | | |
| S-G3 | | | | | | | | | | |
| S-G4 | | | | | | | | | | |
| S-G5 | | | | | | | | | | |
| S-G6 | | | | | | | | | | |
| S-G7 | | | | | | | | | | |
| S-G8 | | | | | | | | | | |
| S-G9 | | | | | | | | | | |
| M-G1 | | | | | | | | | | |
| M-G2 | | | | | | | | | | |
| M-G3 | | | | | | | | | | |
| Q-G1 | | | | | | | | | | |
| Q-G2 | | | | | | | | | | |
| Q-G3 | | | | | | | | | | |
| Q-G4 | | | | | | | | | | |
| Q-G5 | | | | | | | | | | |
| B-G1 | | | | | | | | | | |
| B-G2 | | | | | | | | | | |
| B-G3 | 0.41 | | | | | | | | | |
| B-G4 | 0.54 | 0.33 | | | | | | | | |
| B-G5 | 0.57 | 0.42 | 0.26 | | | | | | | |
| B-G6 | 0.54 | 0.42 | 0.43 | 0.53 | | | | | | |
| B-G7 | 0.55 | 0.47 | 0.49 | 0.56 | 0.36 | | | | | |
| H-G1 | 0.62 | 0.57 | 0.54 | 0.53 | 0.63 | 0.55 | | | | |
| H-G2 | 0.46 | 0.44 | 0.56 | 0.59 | 0.61 | 0.59 | 0.66 | | | |
| H-G3 | 0.69 | 0.53 | 0.63 | 0.58 | 0.67 | 0.65 | 0.63 | 0.64 | | |
| H-G4 | 0.58 | 0.54 | 0.53 | 0.62 | 0.59 | 0.52 | 0.66 | 0.55 | 0.59 | |
| H-G5 | 0.70 | 0.58 | 0.59 | 0.63 | **0.78** | 0.71 | 0.64 | 0.58 | 0.57 | 0.49 |
| H-G6 | 0.68 | 0.50 | 0.50 | 0.53 | 0.71 | 0.62 | 0.59 | 0.58 | 0.46 | 0.46 |
| H-G7 | 0.67 | 0.65 | 0.66 | 0.65 | 0.68 | 0.61 | 0.70 | 0.68 | 0.62 | 0.69 |
| H-G8 | 0.57 | 0.48 | 0.42 | 0.45 | 0.59 | 0.58 | 0.61 | 0.49 | 0.58 | 0.57 |
| H-G9 | 0.62 | 0.57 | 0.54 | 0.57 | 0.61 | 0.55 | 0.69 | 0.52 | 0.69 | 0.61 |
| H-G10 | 0.47 | 0.48 | 0.49 | 0.56 | 0.51 | 0.48 | 0.68 | 0.47 | 0.67 | 0.54 |
| H-G11 | 0.48 | 0.48 | 0.65 | 0.66 | 0.56 | 0.63 | 0.74 | 0.50 | 0.63 | 0.60 |
| H-G12 | 0.58 | 0.51 | 0.51 | 0.59 | 0.53 | 0.54 | 0.70 | 0.51 | 0.66 | 0.44 |
| H-G13 | 0.59 | 0.47 | 0.50 | 0.55 | 0.54 | 0.56 | 0.67 | 0.52 | 0.60 | 0.49 |
| H-G14 | 0.70 | 0.64 | 0.62 | 0.64 | 0.72 | 0.64 | 0.63 | 0.63 | 0.49 | 0.44 |

***Continued***

**Appendix 51. Continued**

| | H-G5 | H-G6 | H-G7 | H-G8 | H-G9 | H-G10 | H-G11 | H-G12 | H-G13 | H-G14 |
|---|---|---|---|---|---|---|---|---|---|---|
| S-G1 | | | | | | | | | | |
| S-G2 | | | | | | | | | | |
| S-G3 | | | | | | | | | | |
| S-G4 | | | | | | | | | | |
| S-G5 | | | | | | | | | | |
| S-G6 | | | | | | | | | | |
| S-G7 | | | | | | | | | | |
| S-G8 | | | | | | | | | | |
| S-G9 | | | | | | | | | | |
| M-G1 | | | | | | | | | | |
| M-G2 | | | | | | | | | | |
| M-G3 | | | | | | | | | | |
| Q-G1 | | | | | | | | | | |
| Q-G2 | | | | | | | | | | |
| Q-G3 | | | | | | | | | | |
| Q-G4 | | | | | | | | | | |
| Q-G5 | | | | | | | | | | |
| B-G1 | | | | | | | | | | |
| B-G2 | | | | | | | | | | |
| B-G3 | | | | | | | | | | |
| B-G4 | | | | | | | | | | |
| B-G5 | | | | | | | | | | |
| B-G6 | | | | | | | | | | |
| B-G7 | | | | | | | | | | |
| H-G1 | | | | | | | | | | |
| H-G2 | | | | | | | | | | |
| H-G3 | | | | | | | | | | |
| H-G4 | | | | | | | | | | |
| H-G5 | | | | | | | | | | |
| H-G6 | 0.41 | | | | | | | | | |
| H-G7 | 0.71 | 0.55 | | | | | | | | |
| H-G8 | 0.55 | 0.48 | 0.62 | | | | | | | |
| H-G9 | 0.68 | 0.61 | 0.62 | 0.39 | | | | | | |
| H-G10 | 0.62 | 0.53 | 0.52 | 0.41 | 0.51 | | | | | |
| H-G11 | 0.67 | 0.63 | 0.68 | 0.53 | 0.59 | 0.46 | | | | |
| H-G12 | 0.60 | 0.51 | 0.62 | 0.42 | 0.45 | 0.32 | 0.43 | | | |
| H-G13 | 0.57 | 0.48 | 0.62 | 0.36 | 0.51 | 0.36 | 0.43 | **0.19** | | |
| H-G14 | 0.50 | 0.38 | 0.54 | 0.56 | 0.63 | 0.56 | 0.66 | 0.49 | 0.51 | |

**Appendix 52. Jaccard similarity coefficients applied among 38 almond genotypes by using 20 RAPD and 15 ISSR markers.**

| | S-G1 | S-G2 | S-G3 | S-G4 | S-G5 | S-G6 | S-G7 | S-G8 | S-G9 |
|---|---|---|---|---|---|---|---|---|---|
| S-G1 | | | | | | | | | |
| S-G2 | 0.43 | | | | | | | | |
| S-G3 | 0.52 | 0.50 | | | | | | | |
| S-G4 | 0.50 | 0.39 | 0.44 | | | | | | |
| S-G5 | 0.55 | 0.46 | 0.51 | 0.45 | | | | | |
| S-G6 | 0.63 | 0.54 | 0.46 | 0.58 | 0.60 | | | | |
| S-G7 | 0.59 | 0.50 | 0.48 | 0.55 | 0.60 | 0.44 | | | |
| S-G8 | 0.53 | 0.56 | 0.50 | 0.51 | 0.54 | 0.67 | 0.52 | | |
| S-G9 | 0.51 | 0.45 | 0.52 | 0.45 | 0.45 | 0.63 | 0.59 | 0.47 | |
| M-G1 | 0.59 | 0.48 | 0.57 | 0.45 | 0.44 | 0.65 | 0.61 | 0.53 | 0.44 |
| M-G2 | 0.51 | 0.43 | 0.54 | 0.43 | 0.43 | 0.63 | 0.59 | 0.49 | 0.40 |
| M-G3 | 0.53 | 0.47 | 0.53 | 0.39 | 0.47 | 0.58 | 0.54 | 0.54 | 0.46 |
| Q-G1 | 0.56 | 0.57 | 0.53 | 0.58 | 0.51 | 0.60 | 0.60 | 0.57 | 0.54 |
| Q-G2 | 0.56 | 0.50 | 0.48 | 0.43 | 0.46 | 0.50 | 0.51 | 0.51 | 0.48 |
| Q-G3 | 0.51 | 0.46 | 0.47 | 0.47 | 0.52 | 0.54 | 0.52 | 0.50 | 0.48 |
| Q-G4 | 0.47 | 0.47 | 0.53 | 0.44 | 0.50 | 0.61 | 0.57 | 0.50 | 0.44 |
| Q-G5 | 0.51 | 0.49 | 0.60 | 0.47 | 0.46 | 0.68 | 0.63 | 0.49 | 0.43 |
| B-G1 | 0.58 | 0.53 | 0.60 | 0.48 | 0.54 | 0.68 | 0.62 | 0.46 | 0.49 |
| B-G2 | 0.55 | 0.57 | 0.61 | 0.50 | 0.55 | 0.68 | 0.62 | 0.57 | 0.50 |
| B-G3 | 0.55 | 0.47 | 0.59 | 0.51 | 0.51 | 0.61 | 0.57 | 0.53 | 0.48 |
| B-G4 | 0.54 | 0.42 | 0.57 | 0.47 | 0.49 | 0.61 | 0.56 | 0.52 | 0.46 |
| B-G5 | 0.57 | 0.51 | 0.60 | 0.50 | 0.52 | 0.60 | 0.58 | 0.48 | 0.47 |
| B-G6 | 0.56 | 0.54 | 0.53 | 0.44 | 0.56 | 0.65 | 0.62 | 0.55 | 0.50 |
| B-G7 | 0.52 | 0.51 | 0.50 | 0.45 | 0.50 | 0.62 | 0.61 | 0.57 | 0.52 |
| H-G1 | 0.59 | 0.59 | 0.55 | 0.54 | 0.62 | 0.61 | 0.61 | 0.57 | 0.58 |
| H-G2 | 0.59 | 0.59 | 0.63 | 0.59 | 0.62 | 0.66 | 0.62 | 0.54 | 0.60 |
| H-G3 | 0.68 | 0.59 | 0.54 | 0.62 | 0.64 | 0.48 | 0.55 | 0.67 | 0.62 |
| H-G4 | 0.59 | 0.56 | 0.55 | 0.53 | 0.58 | 0.60 | 0.63 | 0.62 | 0.62 |
| H-G5 | 0.69 | 0.59 | 0.62 | 0.63 | 0.67 | 0.59 | 0.52 | 0.66 | 0.68 |
| H-G6 | 0.62 | 0.58 | 0.62 | 0.57 | 0.60 | 0.56 | 0.58 | 0.60 | 0.63 |
| H-G7 | 0.63 | 0.64 | 0.62 | 0.60 | 0.63 | 0.63 | 0.61 | 0.66 | 0.60 |
| H-G8 | 0.57 | 0.55 | 0.59 | 0.51 | 0.62 | 0.65 | 0.62 | 0.58 | 0.59 |
| H-G9 | 0.60 | 0.53 | 0.60 | 0.58 | 0.64 | 0.64 | 0.55 | 0.62 | 0.57 |
| H-G10 | 0.61 | 0.55 | 0.60 | 0.53 | 0.53 | 0.67 | 0.55 | 0.53 | 0.58 |
| H-G11 | 0.63 | 0.57 | 0.55 | 0.56 | 0.60 | 0.65 | 0.51 | 0.59 | 0.62 |
| H-G12 | 0.59 | 0.49 | 0.55 | 0.55 | 0.58 | 0.61 | 0.50 | 0.61 | 0.59 |
| H-G13 | 0.55 | 0.46 | 0.62 | 0.54 | 0.51 | 0.68 | 0.58 | 0.60 | 0.55 |
| H-G14 | 0.60 | 0.60 | 0.51 | 0.59 | 0.58 | 0.59 | 0.62 | 0.56 | 0.59 |

***Continued***

**Appendix 52. Continued**

| | M-G1 | M-G2 | M-G3 | Q-G1 | Q-G2 | Q-G3 | Q-G4 | Q-G5 | B-G1 |
|---|---|---|---|---|---|---|---|---|---|
| S-G1 | | | | | | | | | |
| S-G2 | | | | | | | | | |
| S-G3 | | | | | | | | | |
| S-G4 | | | | | | | | | |
| S-G5 | | | | | | | | | |
| S-G6 | | | | | | | | | |
| S-G7 | | | | | | | | | |
| S-G8 | | | | | | | | | |
| S-G9 | | | | | | | | | |
| M-G1 | | | | | | | | | |
| M-G2 | 0.34 | | | | | | | | |
| M-G3 | 0.38 | 0.34 | | | | | | | |
| Q-G1 | 0.57 | 0.53 | 0.54 | | | | | | |
| Q-G2 | 0.48 | 0.43 | 0.41 | 0.47 | | | | | |
| Q-G3 | 0.54 | 0.47 | 0.50 | 0.47 | 0.39 | | | | |
| Q-G4 | 0.52 | 0.42 | 0.45 | 0.54 | 0.43 | 0.49 | | | |
| Q-G5 | 0.52 | 0.47 | 0.50 | 0.54 | 0.52 | 0.52 | 0.37 | | |
| B-G1 | 0.49 | 0.48 | 0.52 | 0.61 | 0.54 | 0.55 | 0.46 | 0.47 | |
| B-G2 | 0.53 | 0.49 | 0.46 | 0.54 | 0.54 | 0.58 | 0.51 | 0.52 | 0.48 |
| B-G3 | 0.52 | 0.41 | 0.44 | 0.50 | 0.48 | 0.51 | 0.45 | 0.41 | 0.45 |
| B-G4 | 0.49 | 0.41 | 0.47 | 0.58 | 0.45 | 0.47 | 0.44 | 0.44 | 0.44 |
| B-G5 | 0.50 | 0.48 | 0.52 | 0.60 | 0.44 | 0.50 | 0.46 | 0.48 | 0.48 |
| B-G6 | 0.55 | 0.46 | 0.46 | 0.63 | 0.47 | 0.49 | 0.46 | 0.49 | 0.49 |
| B-G7 | 0.57 | 0.49 | 0.50 | 0.59 | 0.50 | 0.45 | 0.48 | 0.45 | 0.52 |
| H-G1 | 0.60 | 0.58 | 0.58 | 0.63 | 0.54 | 0.44 | 0.60 | 0.61 | 0.62 |
| H-G2 | 0.63 | 0.58 | 0.59 | 0.65 | 0.57 | 0.59 | 0.63 | 0.62 | 0.61 |
| H-G3 | 0.65 | 0.62 | 0.61 | 0.63 | 0.54 | 0.59 | 0.60 | 0.68 | 0.67 |
| H-G4 | 0.60 | 0.56 | 0.55 | 0.60 | 0.53 | 0.52 | 0.55 | 0.61 | 0.61 |
| H-G5 | 0.65 | 0.64 | 0.61 | 0.71 | 0.58 | 0.61 | 0.65 | 0.66 | 0.65 |
| H-G6 | 0.61 | 0.57 | 0.51 | 0.61 | 0.50 | 0.58 | 0.59 | 0.62 | 0.61 |
| H-G7 | 0.65 | 0.64 | 0.63 | 0.63 | 0.61 | 0.60 | 0.59 | 0.62 | 0.61 |
| H-G8 | 0.57 | 0.55 | 0.50 | 0.65 | 0.52 | 0.50 | 0.50 | 0.59 | 0.57 |
| H-G9 | 0.63 | 0.58 | 0.57 | 0.69 | 0.61 | 0.62 | 0.58 | 0.62 | 0.61 |
| H-G10 | 0.60 | 0.53 | 0.54 | 0.59 | 0.56 | 0.58 | 0.55 | 0.57 | 0.56 |
| H-G11 | 0.62 | 0.56 | 0.56 | 0.66 | 0.60 | 0.63 | 0.60 | 0.60 | 0.59 |
| H-G12 | 0.61 | 0.54 | 0.56 | 0.62 | 0.58 | 0.54 | 0.55 | 0.59 | 0.59 |
| H-G13 | 0.57 | 0.53 | 0.55 | 0.57 | 0.55 | 0.57 | 0.48 | 0.50 | 0.55 |
| H-G14 | 0.63 | 0.61 | 0.62 | 0.57 | 0.54 | 0.51 | 0.61 | 0.60 | 0.59 |

***Continued***

**Appendix 52.  Continued**

| | B-G2 | B-G3 | B-G4 | B-G5 | B-G6 | B-G7 | H-G1 | H-G2 | H-G3 | H-G4 |
|---|---|---|---|---|---|---|---|---|---|---|
| S-G1 | | | | | | | | | | |
| S-G2 | | | | | | | | | | |
| S-G3 | | | | | | | | | | |
| S-G4 | | | | | | | | | | |
| S-G5 | | | | | | | | | | |
| S-G6 | | | | | | | | | | |
| S-G7 | | | | | | | | | | |
| S-G8 | | | | | | | | | | |
| S-G9 | | | | | | | | | | |
| M-G1 | | | | | | | | | | |
| M-G2 | | | | | | | | | | |
| M-G3 | | | | | | | | | | |
| Q-G1 | | | | | | | | | | |
| Q-G2 | | | | | | | | | | |
| Q-G3 | | | | | | | | | | |
| Q-G4 | | | | | | | | | | |
| Q-G5 | | | | | | | | | | |
| B-G1 | | | | | | | | | | |
| B-G2 | | | | | | | | | | |
| B-G3 | 0.49 | | | | | | | | | |
| B-G4 | 0.54 | **0.32** | | | | | | | | |
| B-G5 | 0.54 | 0.43 | 0.37 | | | | | | | |
| B-G6 | 0.55 | 0.46 | 0.45 | 0.51 | | | | | | |
| B-G7 | 0.57 | 0.46 | 0.48 | 0.53 | 0.36 | | | | | |
| H-G1 | 0.60 | 0.60 | 0.60 | 0.58 | 0.54 | 0.50 | | | | |
| H-G2 | 0.57 | 0.59 | 0.63 | 0.60 | 0.61 | 0.65 | 0.64 | | | |
| H-G3 | 0.67 | 0.61 | 0.63 | 0.63 | 0.63 | 0.66 | 0.63 | 0.63 | | |
| H-G4 | 0.56 | 0.58 | 0.57 | 0.60 | 0.56 | 0.53 | 0.59 | 0.59 | 0.60 | |
| H-G5 | 0.69 | 0.65 | 0.60 | 0.66 | 0.66 | 0.65 | 0.68 | 0.55 | 0.59 | 0.57 |
| H-G6 | 0.64 | 0.57 | 0.56 | 0.58 | 0.63 | 0.63 | 0.58 | 0.55 | 0.50 | 0.50 |
| H-G7 | 0.66 | 0.64 | 0.65 | 0.62 | 0.65 | 0.62 | 0.63 | 0.63 | 0.62 | 0.63 |
| H-G8 | 0.56 | 0.59 | 0.55 | 0.56 | 0.54 | 0.58 | 0.55 | 0.56 | 0.59 | 0.50 |
| H-G9 | 0.60 | 0.62 | 0.56 | 0.61 | 0.62 | 0.61 | **0.72** | 0.49 | 0.62 | 0.64 |
| H-G10 | 0.58 | 0.59 | 0.56 | 0.59 | 0.59 | 0.60 | 0.64 | 0.46 | 0.63 | 0.51 |
| H-G11 | 0.61 | 0.57 | 0.60 | 0.64 | 0.59 | 0.62 | 0.71 | 0.57 | 0.61 | 0.56 |
| H-G12 | 0.63 | 0.58 | 0.57 | 0.63 | 0.61 | 0.59 | 0.66 | 0.47 | 0.59 | 0.52 |
| H-G13 | 0.59 | 0.50 | 0.52 | 0.56 | 0.59 | 0.58 | 0.67 | 0.60 | 0.63 | 0.57 |
| H-G14 | 0.66 | 0.59 | 0.59 | 0.59 | 0.57 | 0.62 | 0.58 | 0.64 | 0.59 | 0.56 |

***Continued***

**Appendix 52.  Continued**

| | H-G5 | H-G6 | H-G7 | H-G8 | H-G9 | H-G10 | H-G11 | H-G12 | H-G13 | H-G14 |
|---|---|---|---|---|---|---|---|---|---|---|
| S-G1 | | | | | | | | | | |
| S-G2 | | | | | | | | | | |
| S-G3 | | | | | | | | | | |
| S-G4 | | | | | | | | | | |
| S-G5 | | | | | | | | | | |
| S-G6 | | | | | | | | | | |
| S-G7 | | | | | | | | | | |
| S-G8 | | | | | | | | | | |
| S-G9 | | | | | | | | | | |
| M-G1 | | | | | | | | | | |
| M-G2 | | | | | | | | | | |
| M-G3 | | | | | | | | | | |
| Q-G1 | | | | | | | | | | |
| Q-G2 | | | | | | | | | | |
| Q-G3 | | | | | | | | | | |
| Q-G4 | | | | | | | | | | |
| Q-G5 | | | | | | | | | | |
| B-G1 | | | | | | | | | | |
| B-G2 | | | | | | | | | | |
| B-G3 | | | | | | | | | | |
| B-G4 | | | | | | | | | | |
| B-G5 | | | | | | | | | | |
| B-G6 | | | | | | | | | | |
| B-G7 | | | | | | | | | | |
| H-G1 | | | | | | | | | | |
| H-G2 | | | | | | | | | | |
| H-G3 | | | | | | | | | | |
| H-G4 | | | | | | | | | | |
| H-G5 | | | | | | | | | | |
| H-G6 | 0.53 | | | | | | | | | |
| H-G7 | 0.64 | 0.52 | | | | | | | | |
| H-G8 | 0.58 | 0.50 | 0.58 | | | | | | | |
| H-G9 | 0.56 | 0.59 | 0.63 | 0.51 | | | | | | |
| H-G10 | 0.55 | 0.49 | 0.50 | 0.47 | 0.53 | | | | | |
| H-G11 | 0.56 | 0.59 | 0.62 | 0.57 | 0.56 | 0.48 | | | | |
| H-G12 | 0.50 | 0.52 | 0.60 | 0.51 | 0.42 | 0.38 | 0.42 | | | |
| H-G13 | 0.57 | 0.58 | 0.58 | 0.50 | 0.58 | 0.50 | 0.53 | 0.37 | | |
| H-G14 | 0.61 | 0.56 | 0.59 | 0.57 | 0.68 | 0.62 | 0.66 | 0.59 | 0.57 | |

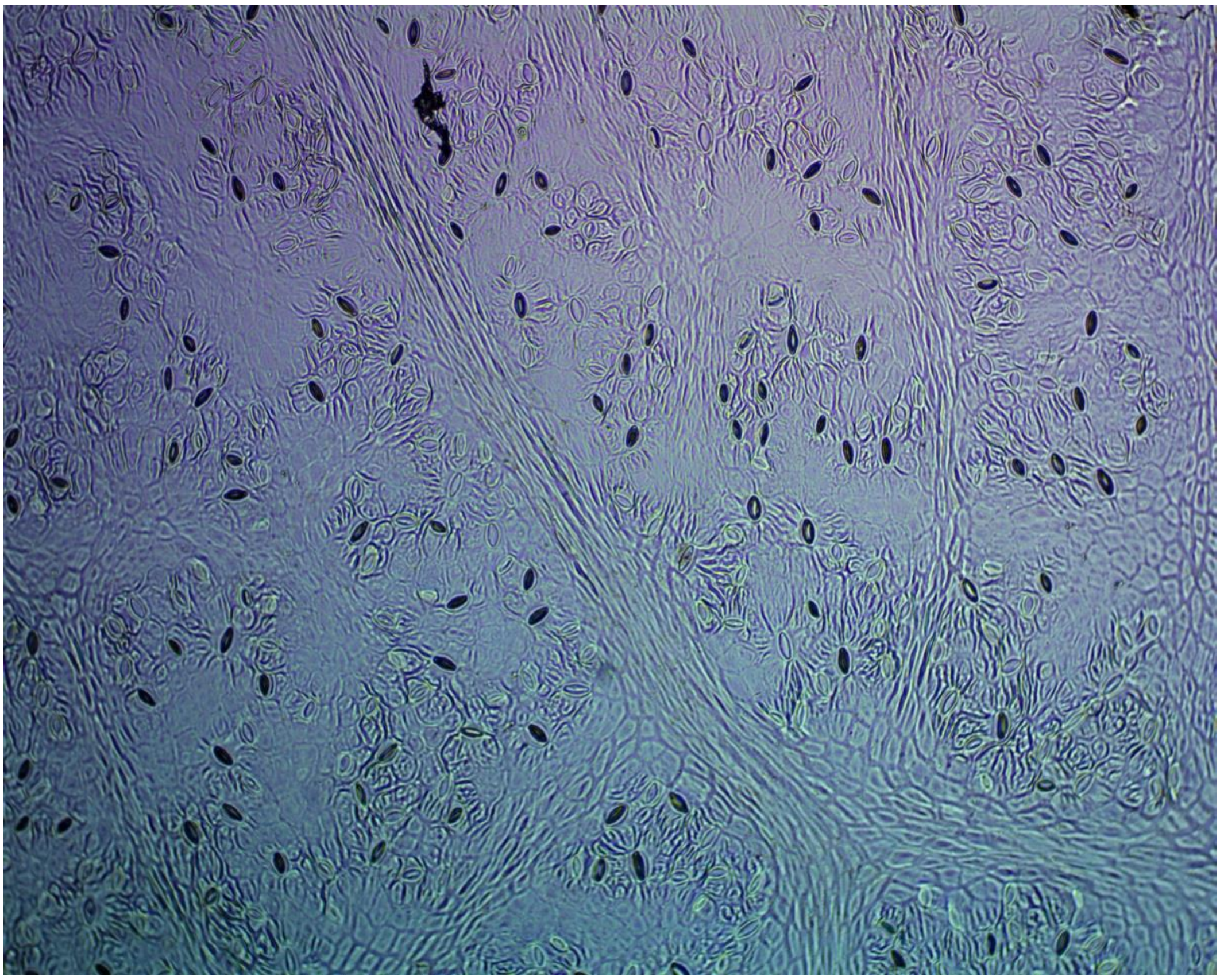

**Appendix 53. Almond stomata under microscope.**

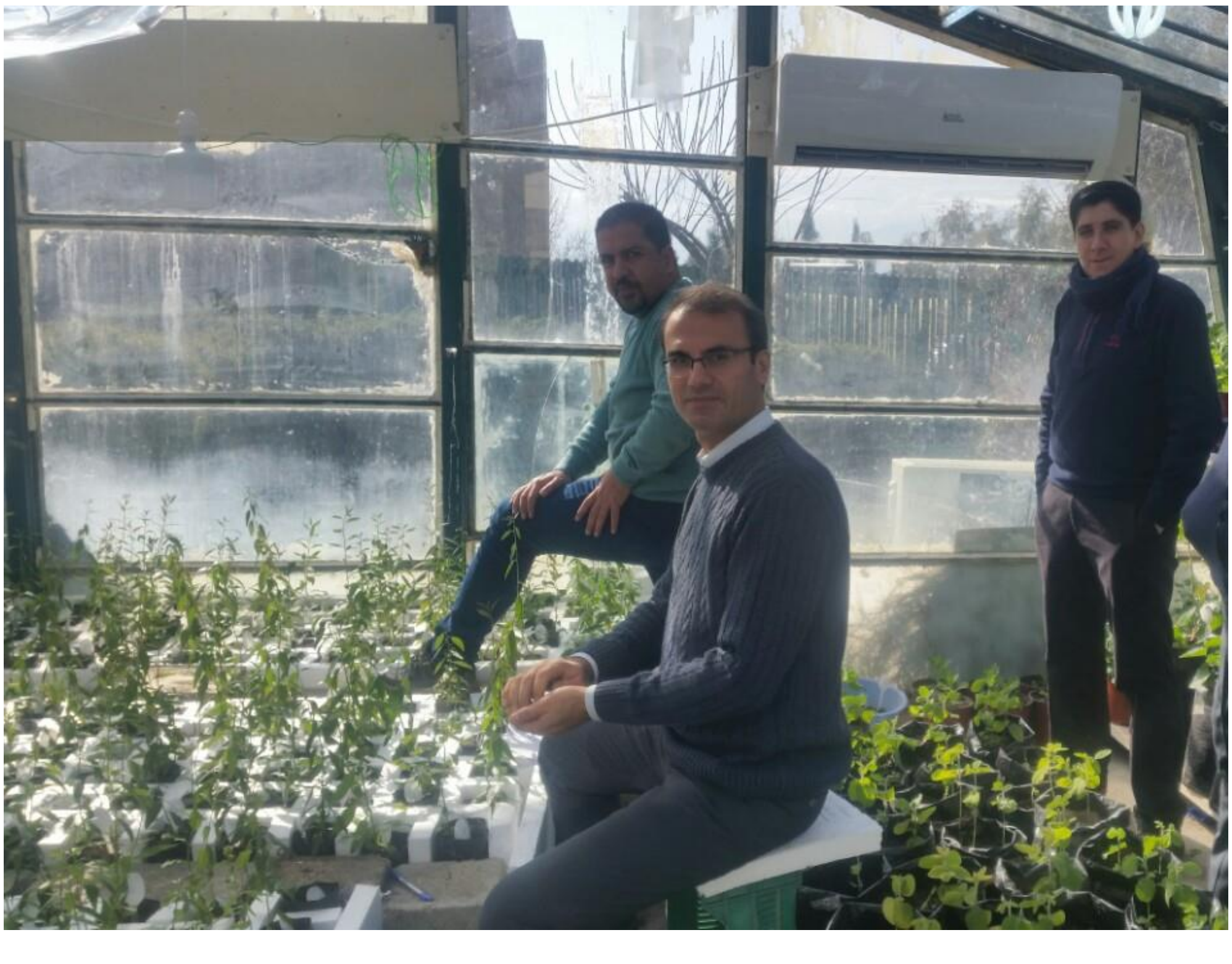

**Appendix 54. Pre-drought tolerance experiment for almond seedling in the glasshouse.**

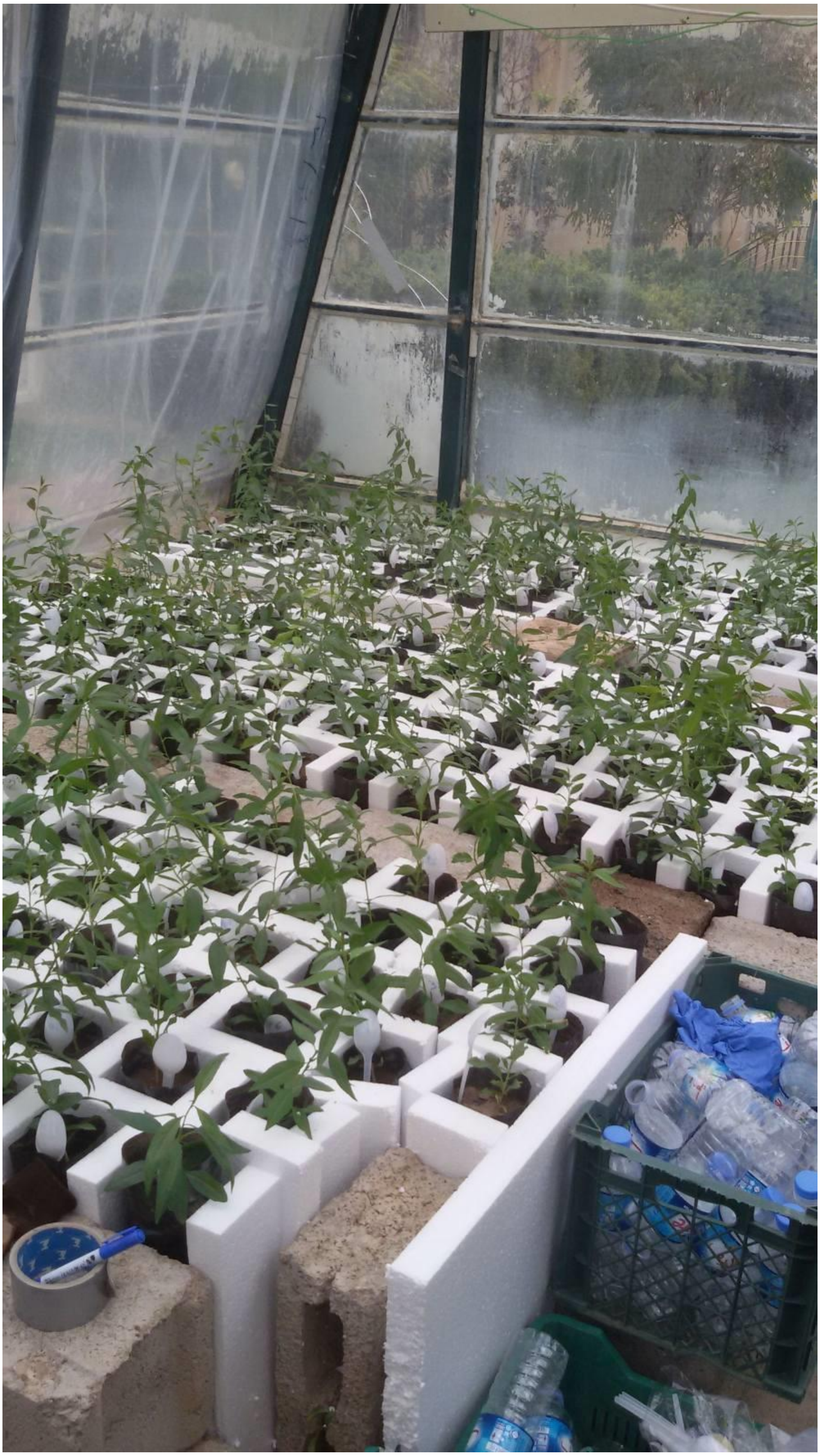

**Appendix 55 Almond seedling in the glasshouse.**

حكومة أقليم كوردستان

وزارة التعليم العالي و البحث العلمي

جامعة السليمانية

كلية علوم الهندسة الزراعية

# تقييم تحمل الجفاف لبعض الطرز الوراثية للوز مظهرياً و كيميائياً و جزيئياً في محافظة السليمانية

رسالة

مقدمة إلى مجلس كلية علوم الهندسة الزراعية في جامعة السليمانية

كجزء من متطلبات نيل شهادة دكتوراه فلسفة في العلوم الزراعية

(البستنة - انتاج الفاكهة الديمية)

من قبل

## انور محمد رؤوف محمود

بكالوريوس في البستنة/كلية الزراعة/جامعة السليمانية (1999)

ماجستير في البستنة/كلية الزراعة/جامعة السليمانية (2010)

باشراف

**د. فخرالدين مصطفى حمه صالح**

أستاذ المساعد

2020 م 1441 هـ

# الخلاصه

نفذت هذه الدراسة خلال موسمي النمو (2017-2018 و 2018-2019) في اربعة مناطق ضمن محافظة السليمانية ، و الاخرى في محافظة حلبجة في اقليم كوردستان العراق ، والتي شملت شهربازار و ميركةبان و قرداغ و برزنجة و هورامان. لوحظ اعداد هائلة من اشجار اللوز مايقارب 500 شجرة في جميع المناطق و من ضمن تلك الاشجار تم اختيار 38 شجرة في المناطق المذكورة (9 ، 3 ، 5 ، 7 ، 14) شجرة وعلى التوالي و على اساس الصفات المظهرية. استخدم تصميم RCBD بسيط في هذه التجربة و قورنت المتوسطات حسب اختبار دنكن.

لغرض تقييم تحمل الجفاف للطرز الوراثية المختارة اجريت تجربة اخرى في البيت الزجاجي التابع لقسم البستنة كلية علوم الهندسة الزراعية - جامعة السليمانية وذلك بزراعة البذور المأخوذة من الطرز الوراثية وبعد اجراء التنضيد عليها زرعت البذور في اكياس و استخدم تجربة عاملية ضمن تصميم القطاعات العشوائية الكاملة و كان العاملين عبارة عن الطرز الوراثية و فترات الري. لذا زرعت شتلات الـثمان و لثلاثون طرزاً في اكياس داخل البيت الزجاجي وتعرضت الشتلات الى ثلاث معاملات مدة الري : 10 ، 20 ، 40 بعد 10 ايام من بزوغ البادرات. و بذلك فان عدد المعاملات التوافقية كان 38 x 3 = 114 معاملة لكل مكرر ، و كان العدد الكلي للشتلات 342 شتلة. تم تحليل النتائج و قورنت المتوسطات حسب اختبار اقل فرق معنوي L.S.D. على مستوى 0.05.

اظهرت نتائج الدراسة ان هناك اختلاف في مستويات تكيف الشتلات للجفاف والتي يمكن استخدامها في المستقبل في برامج التربية لانتاج الاصول. تهدف الدراسة تعيين التغيرات المظهرية ، الكيمياء النباتية و التنوع الوراثي والمتعلقة بطرز الوراثية المهمة في منطقة السليمانية والتي تتعلق بمقاومة الجفاف والعلاقة بين الصفات المظهرية والكيموحيوية والوراثة الجزئية.

يمكن تلخيص اهم النتائج كما يلي

**فيما يتعلق بالطرز الوراثية للاشجار**

- لوحظ ان الطرز الوراثية اثرت بشكل معنوي على جميع الصفات المدروسة.
- الطرازان الوراثيان في برزنجة اعطيا اعلى قيمة لطول وقطر النمو السنوي للافرع والطراز الوراثي رقم 5 في هورامان اعطى ادنى قيمية لكلا الصفتين المذكورتين.
- الطراز الوراثي رقم 3 في ميركةبان سجل اعلى قيمة للمساحة الورقية ، بينما سجل الطراز الوراثي رقم 14 في هورامان اقل قيمة.
- سجل اعلى و ادنى قيم لعدد الثغور في المليمتر الواحد في طرز منطقة هورامان.
- ان اعلى و ادنى تركيز للكلوروفيل سجل في الطرازين الوراثيين 5 في شهربازار و 2 في هورامان على التوالي.
- الطراز الوراثي رقم 2 في قرداغ اعطى اعلى قيمة في صفتي عرض وطول النواه والطراز رقم 4 التابع لنفس المنطقة اعطى اعلى قيم في صفتي سمك و وزن النواه.
- اعلى تركيز للبرولين سجل في الطراز الوراثي رقم 6 في برزنجة وادنى تركيز كان في الطراز الوراثي رقم 3 في منطقة هورامان.
- الطراز الوراثي رقم 5 في قرداغ اعطى اعلى قيم لمحتوى الفينولات الكلية و الفلافونويدات بينما الطراز الوراثي رقم 4 في منطقة هورامان سجلت اقل قيمة للفينولات الكلية ، في حين الطراز الوراثي رقم 4 لمنطقة شهربازار سجل اقل محتوى من الفلافونويدات.
- ان اعلى و اقل قيم لمحتوى الكلي لسابونين كانا في الطرازين الوراثيين رقم 5 في هورامان و 1 في قرداغ على التوالي.
- ان اعلى و ادنى قيم لمحتوى التانين كانا في الطرازين الوراثيين 5 و 1 على التوالي و في منطقة قرداغ.

- ان نشاط مضادات الاكسدة والمقدرة بطريقة التقدير الحيوي (DPPH و ABTS) لطراز الوراثي رقم 1 في منطقة قرداغ كان اقل قيمة بينما اعطى الطرازين الوراثيين رقم 6 و 7 في منطقة شهربازار اعطيا اعلى قيمة و على التوالي.

**فيما يتعلق بالطرز الوراثيية الجزيئية**

- الحزم المختلفة الشكل في تعاظم المادة الوراثة متعددة المظاهر عشوائياً (RAPD) كان اعلى مما هو في التوابع الترادفية البسيطة الداخلية (ISSR).
- محتوى معلومات متعدة الاشكال كانت اقل في (RAPD) مقارنة بـ (ISSR) ولكن كانتا متساويين في اعلى قيمة.
- اعتماداً على معاملات التشابه جاكارد البعد الوراثي في (RAPD) وزع على 3 مجاميع و (ISSR) وزع على اربع مجاميع.
- اعتماداً على تحليل STRUCTURE جميع الطرز تنقسم الى قسمين.

**فيما يتعلق بالشتلات داخل البيت الزجاجي**

- اعلى قيمة لطول الشتلات و عددالاوراق سجل في الطراز الوراثي رقم 9 في هورامان.
- اعلى قيمة لقطر ساق الشجرة و المساحة الورقية و وزن المجموع الخضري و النسبة المؤية للمجموع الخضري و الجذري سجل في الطراز الوراثي رقم 3 في قرداغ. والطراز الوراثي رقم 12 في هورامان اعطت اعلى قيمة في قطر ساق الشتلات و وزن المجموع الخضري و الجذري.
- الطراز الوراثي رقم 4 في قرداغ اعطى اعلى وزن للمجحوع الجذري.
- في الطرازين الوراثيين 9 و 8 في منطقة شهربازار سجل اقصر طول وعرض للثغور على التوالي.
- الطراز الوراثي رقم 2 في برزنجة سجل اعلى تركيز للكلوروفيل.
- اعلى تركيز للبرولين كان عند الطراز الوراثي رقم 2 في قرداغ.
- الطراز الوراثي رقم 2 من شهربازار اعطى اعلى تركيز للفينولات الكلية والفلافونويدات ، وادنى تركيز كان في نفس المنطقة في الطرازين الوراثيين 4 و 7 على التوالي.
- الطرازين الوراثيين 13 و 14 في هورامان اعطتا اعلى قيمة لنشاط مضدات الاكسدة بطريقتي (DPPH وABTS) على التوالي.
- ارتفاع الشتلات و عددالاوراق والمساحة الورقية ووزن كل من المجموع الخضري و المجموع الجذري و طول و عرض الثغور كانوا اعلى القيم عند مدة الري 10 ايام و ادنى قيمة سجل في مدة الري 40 يوماً.
- اعلى القيم من تركيز البرولين و الفينولات الكلية و الفلافونويدات و نشاطات مضادات الاكسدة بكلا الطريقتين (DPPH وABTS) مرافق لمدة الري 40 يوماً و بالعكس ادنى القيم كان فى فترة الري 10 ايام.
- قطر السيقان للشتلات في فترة الري 10 و 20 ايام كانتا متساويتان واعلى من ما سجل في فترة الري 40 ايام.
- سجل اعلى تركيز الكلوروفيل في فترة الري 20 يوماً في حين في فترة الري 10 ايام سجل اقل قيمة.
- النسبة المؤية للمادة الجافة للمجموع الخضري والجذري في معاملات الري لمدة 40 يوماً اعطى اعلى نسبة و ادنى نسبة سجل في فترة الري 10 ايام.

حکومەتی هەرێمی کوردستان
وه زارەتی خویندنی بالاّ و توێژینەوه ی زانستی
زانکۆی سلێمانی
کۆلێجی زانستە ئەندازیارییە کشتوکاڵییەکان

# هەڵسەنگاندنی بەرگەگرتنی هەندێك بۆماوه شێوەی بادەم بۆ بێئاوی لە ڕێگەی خەسڵەتەکانی گەشە و پێکهاتەی کیمیایی و گەردیلەییەوه لە پارێزگای سلێمانی

**نامەیەکە**

**پێشکەش کراوه بە کۆلێجی زانستە ئەندازیارییە کشتوکاڵییەکان لە زانکۆی سلێمانی وەك بەشێك لە پێداویستیەکانی بەدەستهێنانی بڕوانامەی دکتۆرای فەلسەفە لە زانستە ئەندازیاییە کشتوکاڵییەکاندا**

(باخداری - بەرهەمهێنانی میوەی دێمەکار)

لەلایەن

**ئەنوەر محەمەد ڕەئوف مەحمود**

بەکالۆریۆس لە باخداری/ کۆلێجی کشتوکاڵ/ زانکۆی سلێمانی (1999)

ماستەر لە باخداری/ کۆلێجی کشتوکاڵ/ زانکۆی سلێمانی (2010)

بە سەرپەرشتی

**د. فەخرەدین موستەفا حەمە ساڵح**

پرۆفیسۆری یاریدەدەر

**2719 ك** **2020 م**

## پوختە

ئەم توێژینەوەیە جێبەجێ کرا لە وەرزی گەشەی ساڵی (2017-2018 و 2018-2019) لە چوار ناوچەی سەر بە پارێزگای سلێمانی و یەك ناوچە سەر بە پارێزگای هەڵەبجە لە هەرێمی کوردستانی عێراق. ناوچەکان بریتیبوون لە شارباژێر، مێرگەپان، قەرەداغ، بەرزنجە، هەورامان. ژمارەیەکی زۆر (نزیکەی پێنج سەد) داری بادەم بینرا لە هەموو ناوچەکان و لە کۆتایدا 38 دار هەڵبژێردرا کە باشترین خەسڵەتیان هەبوو بەم شێوەیە دابەش بوون 9، 3، 5، 7، 14 بە سەر ئەو ناوچانەی کەپێشتر باس کراوە یەك لەدوای یەك. بۆ ئەم توێژینەوەیە دیزاینی قاڵبی هەرەمەکی تەواوی سادە هەڵبژێردرا و ناوەندەکان جیاکرانەوە بە تاقیکردنەوەی دەنکن. بەمەبەستی هەڵسەنگاندنی بەرگەگرتنی بێ ئاوی لەناوایاندا ، توێژینەوەیەکی دیکە جێبەجێ کرا لە خانووی شووشەدا لەبەشی باخداری کۆلێجی زانستە ئەندازیاریە کشتوکاڵیەکانی زانکۆی سلێمانی ، ئەویش پاش وەرگرتنی تۆو لەم بۆماوەشێوانە و شێکەشی کردنیان و ڕواندنیان لە ناو کیسەی نایلۆندا. دیزایینی بلۆکی هەرەمەکی تەواو جێ بە جێ کرا کە دوو فاکتەری لەخۆگرتبوو ئەوانیش بۆماوە شێوەو ماوەی ئاودان بوون بە سێ دوبارەبونەوە. لەناو خانووی شووشە شەتڵی سی و هەشت بۆماوەشێوە کە لەناو کیسەدا ڕوێندرابوون و پاش 10 ڕۆژ لە چەکەرە کردنیان مامەڵەی ئاودان بۆ ماوەی نێوان ئاودانیان 10، 20، 40 ڕۆژ جێبەجێکرا ، بەمەش ژمارەی مامەڵە پەیوەندیدارەکان 38 x 3 = 114 و کۆی گشتی شەتڵەکان 342 شەتڵ بوو. شیتەڵکاری بۆتێکڕای مامەڵەکان کراو بەراورد کرا بە ڕێگەی کەمترین جیاوازی بەهادار (L.S.D) وەلەژێر ئاستی بەهای 0.05.

لەئەنجامی ئەم توێژینەوەیەوە دەردەکەوێت کە شەتڵەکان خەسڵەتی بەرگەگرتنی جیاوازیان پیشان دا بۆ بێ ئاوی کە دەتوانین لە ئایندەدا بەکاری بهێنین بۆ پرۆگرامی پەروەردەکردن و دۆزینەوەی بنچینەی باش. ئامانجمان لەم توێژینەوەیە دۆزینەوەی گۆڕانکاریە شێوەیەکان و کیمیاییەکان و بۆماوەییەکان بوو وە بەستنەوەی بە بۆماوەشێوەکانی ناوچەی سلێمانی کە پەیوەستە بە دێمەکاریەوە و پەیوەندی داتای شێوەی و پێکهاتەی کیمیایی و بۆماوەییەوە.

گرنگترین ئەنجامەکان ئەتوانرێت کورتبکرێتەوە بەمشێوەیە

**ئەوەی پەیوەندیدارە بە بۆماوە شێوەی دارەکانەوە**

- بۆماوەشێوە کاریگەری واتاداری هەیە لەسەر هەموو ئەو خەسڵەتانەی کە لەم توێژینەوەیەدا وەرگیرابوون.
- دوو بۆماوە شێوە لە بەرزنجە زۆرترین گەشەی لقی ساڵانەو تیرەی لقی ساڵانەیان بەدەست هێنابوو وە بۆماوە شێوەی ژمارە 5 لە هەورامان کەمترین نرخی تۆمار کردبوو.
- بۆماوەشێوەی ژمارە 3 لە مێرگەپان گەورەترین ڕووبەری گەڵای تۆمارکردبوو وە بۆماوەشێوەی ژمارە 14 لە هەورامان کەمترین بوو.
- لەناو بۆماوەشێوەکانی هەوراماندا زۆرترین و کەمترین ژمارەی دەمیلە لە یەك میلیمەتر دوجادا تۆمار کرابوو.
- زۆرترین و کەمترین نرخی ناوەرۆکی ڕەنگی کلۆرۆفیل لە هەریەك لە بۆماوەشێوەی ژمارە 5 شارباژێر و ژمارە 2 هەورامان یەکلەدوای یەك تۆمارکرا بوو.
- بۆماوەشێوەی ژمارە 2 لە قەرەداغ لەهەردوو خەسڵەتی پانی و درێژی بادەم زۆرتین نرخی تۆمارکردبوو وە هەر لەهەمان ناوچە بۆماوەشێوەی ژمارە 4 بەرزترین نرخی تۆمارکردبوو لە ئەستوری و کێشی بادەمەکان.
- بەرزترین بڕی پرۆلین لە بۆماوەشێوەی ژمارە 6ی بەرزنجە تۆمارکرا بوو و نزمترین لە بۆماوەشێوەی ژمارە 3ی هەورامان.
- فینۆڵ و فلاڤۆنۆید لە بۆماوەشێوەی 5ی قەرەداغ بەرزترین بوو لەکاتێکدا بۆماوەشێوەی ژمارەی 4 لە هەورامان کەمترین فینۆڵ و وەهەمان ژمارەی بۆماوەشێوە لە ناوچەی شارباژێر کەمترین فلاڤۆنۆیدی تیا دوزرا بووەوە.
- بۆماوەشێوەی ژمارە 5 لەهەورامان وە 1 لەقەرەداغ یەك لەدوای یەك بەرزترین و نزمترین ساپۆنینی کۆکراوەیان لەخۆ گرتبوو.

- له قەرەداغ کەمترین و زۆرترین ماددەی تانین لە بۆماوەشێوەی ژمارە 5 و 1 یەك لەدوای یەك تۆمار کرا بوو.
- چالاکی دژەئۆکسانەکان بە هەردوو ڕێگای (DPPH و ABTS) لە بۆماوەشێوەی ژمارە 1ی قەرەداغ کەمترین بوو ، لە بۆماوەشێوەی ژمارە 6 و 7ی یەك لە دوای یەکی شارباژێڕ بەرزترین تۆمارکرا بوو.

**ئەوەی پەیوەندیدارە بە گەردیلەیی بۆماوە شێوەی دارەکانەوە**

- ناوەندە نرخ لە باندە شێوە جیاوازەکاندا لە (RAPD) زیاتربوو وەك لە (ISSR).
- ناوەرۆکی زانیاری شێوە جیاواز لە (RAPD) کەمتر بوو وەك لە (ISSR) بەڵام لە بەرزترینداچوونیەك بوون.
- پاڵپشت بە مامەڵەی لەیەکچوونی جاکارد دوری بۆماوەی لە (RAPD) کرا بە 3 کۆمەڵەو لە (ISSR) کرا بە 4 کۆمەڵە.
- پشتبەست بە شیتەڵکاری STRUCTURE هەموو بۆماوەشێوەکان ئەبن بە دووبەشەوە.

**ئەوەی پەیوەستە بەشەتڵەکانی ناو خانووی شوشە**

- بەرزترین بەرزی شەتڵ و ژمارەی گەڵا لە بۆماوەشێوەی ژمارە 9ی هەورامان تۆمارکرابوو.
- بەرزترین تیرەی قەد و ڕووبەری گەڵا و کێشی سەوزە گەشە و ڕێژەی سەدی کێشی ووشککراوەی سەوزەبەش و ووشککراوەی کۆمەڵەی ڕەگ بەدەستهات لە بۆماوەشێوەی ژمارە 3 قەرەداغ. بۆماوە شێوەی ژمارە 12ی هەورامان کەمترین تیرەی قەدی شەتڵ ، کێشی سەوزە گەشە و کۆمەڵەی ڕەگی بەدەستهێنا.
- بۆماوەشێوەی ژمارە 4 لە قەرداغ زۆرترین کێشی ڕەگی بەدەست هێنابوو.
- بۆماوەشێوەی ژمارە 9 و 8 لە شارباژێڕ کەمترین درێژی و پانی دەمیلەکانیان یەك لەدوای یەك تۆمارکردبوو.
- لەبەرزنجە بۆماوەشێوەی ژمارە 2 بەرزترین بڕی کلۆرۆفیلی لەخۆگرتبوو.
- بەرزترین بڕی پرۆلین لە بۆماوەشێوەی ژمارە 2 قەرەداغەوە بەدەستهات.
- بۆماوەشێوەی ژمارە 2ی شارباژێڕ بەرزترین بڕ لە فینۆڵ و فلافۆنۆیدی تێدابوو ، هەروەها نزمترین لە هەمان ناوچە تۆمار کرابوو لە هەریەك لە بۆماوەشێوەی 4 و 7 یەك لەدوای یەك.
- بۆماوەشێوەی ژمارە 13 و 14ی هەورامان بەرزترین نرخیان بەدەستهێنابوو لە چالاکی دژەئۆکسان بە هەردوو ڕێگای (DPPH و ABTS) یەك لەدوای یەك.
- بەرزی شەتڵەکان و ژمارەی گەڵا و ڕووبەری گەڵا و کێشی هەریەك لە سەوزە گەشە و کۆمەڵەی ڕەگ و درێژی و پانی دەمیلەکان بەرزترین نرخیان تۆمارکردبوو لە 10 ڕۆژ ماوەی ئاودان‌دا و کەمترین نرخ لە 40 ڕۆژ ئاودان‌دا تۆمار کرابوو
- پرۆلین و کۆکراوەی فینۆڵ و فلافۆنۆید و چالاکی دژەئۆکسان بەهەردوو ڕێگەی (DPPH and ABTS) بەرزترین نرخیان تۆمارکردبوو لە 40 ڕۆژ ماوەی ئاودان و پێچەوانەکی کەکەمترین بوو لە 10 ڕۆژ ئاودان تۆمار کرابوو.
- تیرەی قەدەکان لە 10 و 20 ڕۆژ ماوەی ئاودان‌دا چونیەکبوون و گەورەتربوو لە 40 ڕۆژ ماوەی ئاوداندن.
- کلۆرۆفیل لە 20 ڕۆژ ماوەی ئاودان‌دا بەرزترین نرخی تۆمارکردبوو و لە 10 ڕۆژ کەمترین.
- ڕێژەی سەدی ووشکراوەی هەریەك لە سەوزە بەش و کۆمەڵەی ڕەگ لە 40 ڕۆژ ماوەی ئاودان بەرزترین بوو وە لە 10 ڕۆژ ماوەی ئاودان نزمترین بوو.